\documentclass[a4paper,12pt]{article}
\usepackage{amsmath,amssymb}
\usepackage{subfigure}
\usepackage{graphicx}
\usepackage{rotating}
\usepackage[utf8]{inputenc}
\usepackage[T1]{fontenc}
\usepackage{textcomp}
\usepackage{hyperref}
\usepackage[ backend=bibtex,%  backend=biber, % 
            hyperref=true,
            style=numeric-comp,
            maxcitenames=300,
            maxbibnames=500,
            block=none,
            sorting=none
            ]{biblatex}
\usepackage{amssymb,amsfonts,textcomp}
\usepackage{color}
\usepackage{array}
\usepackage{supertabular}
\usepackage{hhline}

\makeatletter
\newcommand\arraybslash{\let\\\@arraycr}
\makeatother
\renewcommand{\vec}[1]{\mathbf{#1}}
\newcommand{\sigmaHH}{\sigma_{hh}}
\newcommand{\sigmaHHMC}{\sigma_{hh, MC}}

\newcommand{\RHH}{R_{hh}}

\newcommand{\be}{\begin{equation}}
\newcommand{\ee}{\end{equation}}
\newcommand{\beq}{\begin{eqnarray}}
\newcommand{\eeq}{\end{eqnarray}}
\newcommand{\de}{{\sc Delphes}~3 }
\newcommand{\amc}{{\sc MG5\textunderscore}a{\sc MC@NLO}}

\newcommand{\py}{{\sc Pythia}~8}

\usepackage[toc,page]{appendix}
\usepackage[affil-it]{authblk}
\author{Alexandra Carvalho\footnote{alexandra.oliveira@cern.ch}}
\affil{National Institute for Chemical Physics and Biophysics, 10143 Tallinn, Estonia}
\author{Florian Goertz\footnote{fgoertz@mpi-hd.mpg.de}}
\affil{Max-Planck-Institut f{\"u}r Kernphysik, Saupfercheckweg 1, 69117 Heidelberg, Germany}
\author{Ken Mimasu\footnote{ken.mimasu@uclouvain.be}}
\affil{Centre for Cosmology, Particle Physics and Phenomenology (CP3), Universite catholique de Louvain, Chemin du Cyclotron, 2, B-1348 Louvain-la-Neuve, Belgium}
\author{Maxime Gouzevitch\footnote{mgouzevi@cern.ch}}
\affil{Institut de Physique Nucleaire de Lyon, Universite de Lyon, Universite Claude Bernard
Lyon 1, CNRS-IN2P3, Villeurbanne, France}
\author{Anamika Aggarwal}
\affil{National Institute for Subatomic Physics (NIKHEF),
Science Park 105, 1098 XG Amsterdam, Netherlands}
\affil{Radboud University, Heyendaalseweg 135, 6525 AJ Nijmegen, Netherlands}
\title{On the reinterpretation of non-resonant searches for Higgs boson pairs}

\begin{document}
\maketitle
%\tableofcontents
\begin{abstract}

The detection of production of a pair of Higgs bosons before the end of LHC operation would be a clear evidence of New Physics (NP). 
As searches for non-resonant production of Higgs pairs are being designed it is of particular importance to be able to conveniently present current experimental results in terms of limits on  the most 'model-independent' fashion possible.  
In this article we provide an analytic parametrization of the {\it differential} Higgs-pair production
%of the Higgs-pair production cross section 
at the LHC in the effective field theory (EFT) extension of the SM.  
%coefficients and to derive corresponding projected sensitivities - taking into account shape information.
%In this article, we provide such 
It results from a fit to the theory prediction for the $gg \to hh$ cross section at the 13\,TeV at the LHC.
%in proton-proton collisions differential in $m_{hh}$ and $|cos\,\theta^*|$. %in terms of the relevant
%effective couplings, parametrizing physics beyond the standard model in general terms. 
Subsequently the resulting formula is used for a reweighing technique that allows to recast to recast exclusion bounds from ATLAS and CMS HH$\to\gamma\gamma\,b\bar{b}$ searches to any point of the considered EFT parameter space. 
%derived assuming certain benchmark
%points in parameter space with a given kinematics, to points with modified kinematic distributions.
We demonstrate with a fast simulation of the LHC detectors that with this approach it is possible to cover the continuously of the EFT parameter space, taking correctly into account the efficiencies of signal selections, without the necessity of rerunning a large number of full detector simulations. % monte-carlo. % analysis with a good approximation. % for an arbitrary n. %umber of EFT parameter points. 
Finally, the resulting exclusion bounds are confronted to several explicit models such as setups with additional scalars, including 2HDM, vector-like fermions, and minimal composite
Higgs models, mapped to the EFT. 
% like the scalar extension of the Higgs and fermion sector. 
%efficiently with constraints from the LHC. 
%We will demonstrate this procedure, using recent ATLAS and CMS results,
%for various new physics models, such as setups with additional scalars, including 2HDM, vector-like fermions, and minimal composite
%Higgs models, providing also a map between their explicit parameters and effective couplings after electroweak symmetry breaking.
\end{abstract}

\section{Introduction}
Examining the production of pairs of Higgs bosons $h$ is
a crucial task in the long term strategy of the LHC.
Analysing this process offers a unique direct window
on the Higgs potential, whose behaviour lies at the heart 
of many unresolved questions on nature, such as the
hierarchy problem, the question of vacuum stability \cite{Krive:1976sg,Lindner:1988ww,Degrassi:2012ry},
the feasibility of electroweak baryogenesis \cite{Grojean:2004xa,Goertz:2015dba,Huang:2015tdv}, 
and the potential dynamical trigger for electroweak symmetry 
breaking (EWSB) \cite{Giudice:2007fh,Grober:2010yv,Carmona:2014iwa}
 - just to name a few.
Of particular importance regarding the search for physics beyond the Standard Model (SM)
are differential observables, such as the distribution
of the invariant mass of the Higgs pairs $m_{hh}$ %, or the Higgs $p_T$ or $|cos\,\theta^*|$ distributions in the case of $hh$ production
(see, {\it e.g.}, \cite{Carvalho:2015ttv,Azatov:2015oxa,Goertz:2014qta}), since they 
are especially sensitive to effects of new physics (NP). %, often growing with the energy.

As we eventually want to confront corresponding experimental results with theory,
we need to assume a theoretical framework, which in our case
will be the effective field theory (EFT) extension of the SM, 
supplementing it with higher dimensional
operators. This captures the effects of any 
yet undiscovered physics beyond the SM, given that it is separated from the latter by
a mass gap and in particular is significantly heavier than the scales 
probed experimentally. In that sense, the setup is `model independent', 
providing the most general parametrization of nature, based on what we observe
at low energies. On the other hand, any known UV model (residing at scales well beyond the
electroweak scale) can be mapped to such an
effective Lagrangian, rendering a formulation of experimental results in terms of bounds
on its parameters an extremely useful bridge between data and theory.
Having at hand analytic formulae which translate between observables in a realistic
collider environment and effective couplings will thus be of utmost importance for the experimental results to be properly interpreted.

In this paper we will present on the one hand such a fit of the theory prediction for the
Higgs-pair production cross section in gluon fusion at the LHC in terms of effective couplings, 
considering both the total cross section as well as the ones differential in $m_{hh}$ and 
$|cos\,\theta^*|$. At Leading Order (LO) the totality of the HH process is described by those two variables\footnote{The latter results can in particular be used to generalize the parameter-space 
scan of Ref. \cite{Carvalho:2015ttv} (where the simulations were restricted to sub-regions of the parameter space) 
to arbitrary combinations of EFT couplings.}. 
Beyond that, we will employ the formula to reinterpret exclusion bounds, derived assuming certain benchmark
points in parameter space with a given kinematics, for points with modified kinematic distributions. In fact,
different distributions in variables, such as $m_{hh}$, will lead to different efficiencies of selection cuts,
which can be accounted for, knowing the distributions in terms of effective couplings. Without such an
analytic knowledge, a full monte-carlo (MC) simulation would actually be required for each parameter-space
point, which is not feasible for a dense scan of the parameters.
Finally, we will use the interpreted results to confront explicit models, mapped to an effective Lagrangian, 
efficiently with constraints from the LHC. 
%In particular, we will translate recent ATLAS and CMS bounds 
%to various NP models, 
To perform this exercise we collect from literature setups with additional scalars, including two-Higgs-doublet models (2HDM), 
vector-like fermions, and minimal composite Higgs models and provide a collection of relations between their 
explicit parameters and effective couplings after EWSB.

\section{Setup}
The theory, on which our analysis is based, is the EFT extension of the SM
with $D=6$ operators ${\cal O}_i$
with coefficients $\sim c_i$,  
suppressed by the NP scale $\Lambda^2$ (assuming for the moment a 
linearly realized SM gauge symmetry).\footnote{The validity of 
this most general parametrization of heavy NP has been scrutinized in detail in \cite{Contino:2016jqw}.}
The terms relevant for the production of 
Higgs pairs in gluon fusion to leading order (LO) are contained in the effective 
Lagrangian \cite{Goertz:2014qta}
\begin{equation}
\label{eq:LEFT}
  \begin{split}
	{\cal L} = {\cal L}_{\rm SM} &+ \frac{c_H}{2\Lambda^2}(\partial^\mu |H|^2)^2 - \frac{c_6}{\Lambda^2} \lambda_{SM} |H|^6
	\\ 
	&- \left( \frac{c_t}{\Lambda^2}y_t |H|^2 \bar Q_L H^c t_R + \text{h.c.} \right)
	\\
	&+ \frac{\alpha_s c_g}{4 \pi \Lambda^2} |H|^2 G_{\mu\nu}^a G^{\mu\nu}_a\,.
  \end{split}
\end{equation}
Here, ${\cal L}_{\rm SM}$ is the SM Lagrangian, $\lambda_{SM}=m_h^2/2v$ is the SM value of the tri-linear Higgs 
self coupling, with $m_h\approx 125$\,GeV the mass of the Higgs boson, $y_t$ is the top-quark Yukawa coupling, 
and $\alpha_s$ is the strong coupling constant.\footnote{We neglect changes in the bottom Yukawa coupling, as it is already 
constrained in a way that it plays at most a subleading role in Higgs pair production (see, {\it e.g.}, 
\cite{ATLAS:2016pkl}).} 

After EWSB (in unitary gauge) $H \to 1/\sqrt2\, (0, v+h )^T$, where 
$v/\sqrt 2 \equiv \langle |H| \rangle \approx 174$\,GeV is the vacuum expectation value of the Higgs field, 
the terms relevant for the $gg \to hh$ process become \cite{Goertz:2014qta,Azatov:2015oxa,Carvalho:2015ttv}
\begin{equation}
\begin{split}
{\cal L}_h = 
\frac{1}{2} \partial_{\mu}\, h \partial^{\mu} h - \frac{1}{2} m_h^2 h^2 -
  {\kappa_{\lambda}}\,  \lambda_{SM} v\, h^3 
- \frac{ m_t}{v}(v+   {\kappa_t} \,   h  +  \frac{c_{2}}{v}   \, h\,  h ) \,( \bar{t}_L t_R + h.c.) \\ 
+ \frac{1}{4} \frac{\alpha_s}{3 \pi v} (   c_g \, h -  \frac{c_{2g}}{2 v} \, h\, h ) \,  G^{\mu \nu}G_{\mu\nu}\,.
\label{eq:L}
\end{split}
\end{equation}
In fact, the operators in (\ref{eq:LEFT}) modify the trilinear Higgs self coupling and the top yukawa coupling, parametrized
via multiplicative factors $\kappa_\lambda$ and $\kappa_t$, where in the SM  $\kappa_\lambda = \kappa_t = 1$.
Beyond that, they induce additional contact interactions between two Higgs bosons and two fermions and between 
a gluon pair and one or two Higgs bosons, parametrized by $c_2,c_g,$ and $c_{2g}$, which vanish in the SM, 
{\it i.e.}, $c_2=c_g=c_{2g}=0$. The relations between these effective couplings and the coefficients of the
operators in Eq.~(\ref{eq:LEFT}) can be obtained straightforwardly (see, {\it e.g.,} \cite{Goertz:2014qta})
and imply the constraint $c_g=-c_{2g}$. This correlation is however broken in the non-linear realization 
of EWSB and is not assumed a priori in this work, where we treat all couplings in Eq.~(\ref{eq:L}) as free.
In the next section, we will provide a parametrization of the Higgs-pair production cross sections
in terms of the effective couplings entering the effective Lagrangian (\ref{eq:L}).

\section{Fit to the (differential) \boldmath$ gg \to hh$ cross section}
\label{sec:diffit}

We will now describe our fit to the cross section for Higgs pair production in gluon fusion at the 13 TeV LHC 
in the EFT defined above.\footnote{\label{fn:sim} 
We do not consider other production mechanisms in this work, such as vector-boson fusion, which will be 
sub-dominant in the bulk of the EFT parameter space. %The $hh$ cross section is simulated via a \textsc{MadGraph5\_aMC@NLO} implementation provided by the authors of~\cite{Hespel:2014sla} (see \cite{Carvalho:2015ttv} and below) and we employ the NLO set of the PDF4LHC parton densities~\cite{Butterworth:2015oua, Dulat:2015mca, Harland-Lang:2014zoa, Ball:2014uwa}.
} We consider the total cross section but as well as the ones differential in $m_{hh}$ and $|cos\,\theta^*|$, with $\theta^*$
being the polar angle of either of the Higgs bosons with respect to the beam axis (featuring the same $|cos\,\theta^*|$ at
parton level).

Examining the different contributions to the full amplitude, it is easy to see that the 
$gg \to hh$ cross section $\sigma_{hh} \equiv \sigma(gg \to hh)$ (normalized to its SM value $\sigma_{hh}^{SM}$)
can be parametrized to LO in terms of the parameters in (\ref{eq:L}) as \cite{Carvalho:2015ttv}
\begin{equation}
\label{eq:fit}
\begin{split}
\RHH \equiv \frac{\sigma_{hh}}{\sigma_{hh}^{SM}} = {\rm Poly}(\vec{A}) = \
& A_1\, \kappa_t ^4 + A_2\, c_2^2 + (A_3\, \kappa_t^2 + A_4\, c_g^2)\, \kappa_{\lambda}^2 
+ A_5\, c_{2g}^2 + ( A_6\, c_2 + A_7\, \kappa_t \kappa_{\lambda} )\kappa_t^2 \\
& + (A_8\, \kappa_t \kappa_{\lambda} + A_9\, c_g \kappa_{\lambda} ) c_{2} + A_{10}\, c_2 c_{2g} 
+ (A_{11}\, c_g \kappa_{\lambda} + A_{12}\, c_{2g})\, \kappa_t^2 \\
& + (A_{13}\, \kappa_{\lambda} c_g + A_{14}\, c_{2g} )\, \kappa_t \kappa_{\lambda} + A_{15}\, c_{g} c_{2g} \kappa_{\lambda}
\end{split}\,.
\end{equation}
In Ref.~\cite{Carvalho:2015ttv}, the parameters $A_i$, that form a vector $\vec{A}\equiv (A_i)$ of coefficients of the 
polynomial ${\rm Poly}(\vec{A})$,
were extracted from simulations to describe the total cross section. In this paper, we push the logic further to describe the differential cross section. We slice the kinematic space into bins with sufficient granularity. In each bin, Eq. \ref{eq:fit} holds, but the coefficients become bin dependent,  $A_i \to A_i^j$ , following the matrix-element (ME) integration and the PDF evolution. We define the cross section in a bin $j$ by $\sigma_{hh, j} \equiv  \sigma_{hh}\times {\rm Frac}^j$, where ${\rm Frac}^j$ is the fraction of events contained in bin $j$. The differential $hh$ production ratio now becomes
\begin{eqnarray}
\RHH^j \equiv \frac{\sigmaHH}{\sigmaHH^{\rm SM}} \frac{{\rm Frac}^j}{{\rm Frac}_{\rm SM}^j} = {\rm Poly}(\vec{A}^j)\,.
\label{eq:cx}
\end{eqnarray}

We simulate the differential cross section for various points in the five-dimensional
parameter space, spanned by $\{\kappa_\lambda,\kappa_t,c_2,c_g,c_{2g}\}$, to determine $\vec{A}^j$. At LO accuracy the ME information differential in the invariant mass of the Higgs pair ($m_{hh}$) and in $|cos\,\theta^{*}|$  completely defines the 2$\to$2 process. At any subsequent simulation step - after the Higgs decays - the process evolution is independent of the NP at high energies, up to detector level. Therefore the results based on the ME-level information are applicable to reconstruct NP shapes at LO at detector level, provided that the former is known on an event-by-event basis.%\footnote{
%As the Higgs boson is a narrow particle
%While NP
%could change the Higgs branching ratios, an effect we will comment on later, the distributions in terms of the reconstructed 
%Higgs bosons are clearly independent of the Higgs decays. }

After an optimization procedure, we converged to an optimal list of 59 bins in $m_{hh}$ that allowed for a precise reweighing of an ensemble of generated events to any NP shape. Since the distribution in the $cos\,\theta^*$ variable is rather flat, only 4 bins are considered\footnote{Here, and in the following, we omit the absolute value signs around $cos\,\theta^*$ for brevity.}. % with boundaries at $cos\,\theta^*$ = 0,0.4,0.6,0.8, and 1. 
For the binning in $m_{hh}$ we choose 10 GeV intervals up to 700 GeV, where the bulk of the cross section is observed, and use a more coarse binning above. Explicitly, our
binning is given by

\begin{equation}
\begin{tabular}{lp{10cm}} % 
  $m_{hh}$ \mbox{[GeV]} = & [250, 260, 270, 280, 290, 300, 310, 320, 330, 340,\\ 
   & \ 350, 360, 370, 380, 390, 400, 410, 420, 430, 440,\\
   & \ 450, 460, 470, 480, 490, 500, 510, 520, 530, 540,\\
   & \ 550, 560, 570, 580, 590, 600, 610, 620, 630, 640,\\
   & \ 650, 660, 670, 680, 690, 700,\\ 
   & \ 750, 800, 850, 900, 950, 1000, 1100, 1200, 1300, 1400, \\ 
   & \ 1500, 1750, 2000, 13000] \\[1mm]
   \qquad $ cos\,\theta^*$ = & [0,0.4,0.6,0.8,1] \qquad .\\
\end{tabular}
\label{tab:bins} 
\end{equation} 

Following the procedure established in~\cite{Carvalho:2015ttv}, the components of $\vec{A}^j$ are extracted by maximizing the likelihood simultaneously for all the coefficients, employing an ensemble of MC simulated samples, 
scanning the model parameters, \emph{i.e.}, minimizing 
\begin{equation} 
\log L(\vec{A}^j) = - \sum_{i} \left(  \frac{T^{(i,j)}-\sigmaHH^i(\vec{A}^j)}{\delta T^{(i,j)}} \right)^2\,,
\label{eq:lik}
\end{equation} 
where the index $i$ runs over different points in the $\{\kappa_\lambda,\kappa_t,c_2,c_g,c_{2g}\}$  parameter space.
Here, $T^{(i,j)} \equiv \sigmaHHMC^i\, {\rm Frac}_i^j$, with $\sigmaHHMC^i$ the total cross section calculated via MC simulation and ${\rm Frac}_i^j$ the corresponding bin fraction (also taken from MC), while $\sigmaHH^i(\vec{A}^j)\equiv \RHH^j\, \sigmaHH^{\rm SM}\,{\rm Frac}_{\rm SM}^j$ is the differential cross section parametrization following Eq.~\ref{eq:cx}.
Only the statistical (MC) uncertainty on the cross section in each bin, $\delta T^{(i,j)}$, is considered. For the SM point we assume a non-zero value of $10^{-4}$ to regularize the likelihood. 

A careful choice of NP parameters is considered in the simulation to improve the convergence of the procedure. In particular it is important to avoid negative $\RHH^j$ values due to statistical fluctuations and, in general, a very poor population
of certain bins (due to the limited total statistics).
We thus consider simulated datasets such that the minimization of equation~\ref{eq:lik} can be done in parameter-subspaces (S$k$) with lower dimensionality that nevertheless contain the full information and guarantee 
appropriate statistics
for all bins\footnote{In principle, a system of 15 equations for each bin, relating the components of $\vec{A}^j$ to $\RHH^j$, evaluated at 15 particular sets of values of the effective couplings, could be solved recursively. This method however
leads again to significantly less stable results compared to a broader fit in the parameter space, see below.}.  The list of subspaces and the coefficients that they serve to determine are compiled in Table \ref{tab:datasets}. 

The simulation of events is performed within the \amc\ framework~\cite{Alwall:2014hca} where the entire event generation process is automated~\cite{Christensen:2009jx}. We make use of UFO model files~\cite{Degrande:2011ua} extracted from the Lagrangian~\ref{eq:LEFT}, as constructed by the authors of Ref. \cite{Hespel:2014sla}. 
For each NP point we simulate 50,000 events, while for the SM benchmark we generate a sample
of 13,000,000 events. 

\begin{table}[t]
\centering
{\scriptsize
\begin{tabular}{cccc}
\hline
Subspace & Parameters  & Coefficients to be determined & scan \\
\hline 
S1 & $\kappa_\lambda,\, \kappa_t$                  & $A_1^j,\,A_3^j,\, A_7^j$ & kl = E-4, +-2.5, +-5, +- 7.5, +- 10, +- 12.5, +- 15 \\
   &                                               &                          & kt from 0.5 to 2.5 with steps of 0.5 \\ \hline 
S2 & $\kappa_\lambda,\, c_2,\, c_g,\, c_{2g}$                   & $A_2^j,\,A_5^j,\, A_{10}^j,\, A_{15}^j$      & c2 = +-1, +-3 , c2g = +-0.5, +-1.5 \\
   & &  & kl = 1 +-5, +- 10 and  (cg,c2g) = +-1 (independently) \\ \hline
\hline 
S3 & $\kappa_\lambda,\, c_2,\, c_g$                & $A_4^j, A_9^j$           & kl = 1 +-5, +- 10 ,  cg = +-1 and c2 = +-1, +-3  \\ \hline  
S4 & $\kappa_\lambda,\, \kappa_t,\, c_2$           & $A_6^j,\, A_8^j$         & kl = 1 +-5, +- 10, kt from 0.5 to 2.5 with steps of 1.0 and c2 = +-1, +-3 \\ \hline 
S5 & $\kappa_\lambda,\, \kappa_t,\, c_{2g},\, c_g$ & $A_{12}^j,\, A_{14}^j$   & kl = 1 +-5, +- 10, kt from 0.5 to 2.5,  (cg,c2g) = +-1 (independently) \\ \hline
\hline 
\multicolumn{4}{c}{Additional points to better precise some coefficients (fixing $\kappa_\lambda = \kappa_t = e^{-4}$) }  \\ \hline \hline
$A_2^j$ &                                              & \multicolumn{2}{c}{c2 from -3 to 3 in steps of 0.5} \\  
$A_{10}^j$ &                                              & \multicolumn{2}{c}{c2 = +-1, +-3 and c2g = +-0.5, +-1.5}  \\
$A_{15}^j$ &                                              & \multicolumn{2}{c}{(cg , -c2g) = +-0.5, +-1.5 (independently) }  \\
%\hline
%S1 & $\kappa_\lambda,\, \kappa_t$  & $A_1^j,\,A_3^j,\, A_7^j$ \\
%\hline
%S2 & $\kappa_\lambda,\, \kappa_t,\, c_2$ & $A_2^j,\, A_6^j,\, A_8^j$\\
%\hline
%S3 & $\kappa_\lambda,\, c_2,\, c_g$  & $A_4^j, A_9^j$\\
%\hline
%S4 & $\kappa_\lambda,\, c_2,\, c_{2g}$  & $A_5^j A_{10}^j,\, A_{15}^j$ \\
%\hline
%S5 & $\kappa_\lambda,\, \kappa_t,\, c_{2g}$ & $A_{12}^j A_{14}^j$\\
%\hline
%S6 & $\kappa_\lambda,\, \kappa_t,\, c_g$ & $A_{11}^j,\,A_{13}^j$\\
\hline
\hline
\end{tabular}}
\caption{\footnotesize \label{tab:datasets} 
Definition of the lower dimensional subsets (S$k$) to find the coefficients of equation~\ref{eq:cx}. }
\end{table}

Our results for the fit coefficients are compiled in Tables~\ref{tab:NumFitM1}-\ref{tab:NumFitM4}.
To illustrate the bin statistics, we also show the number of events in each bin, both for 
our total BSM sample (see Section \ref{sec:Recast}) and the SM sample.

\begin{figure*}[hbtp]\begin{center}
 \includegraphics[width=0.32\textwidth]{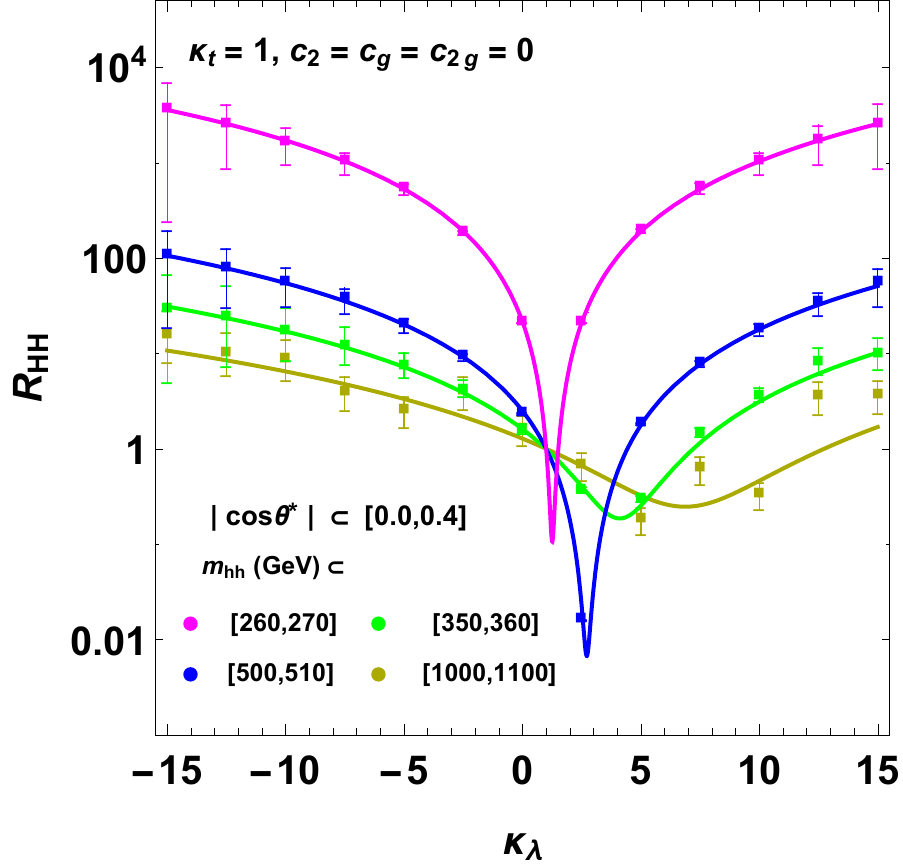}
 \includegraphics[width=0.32\textwidth]{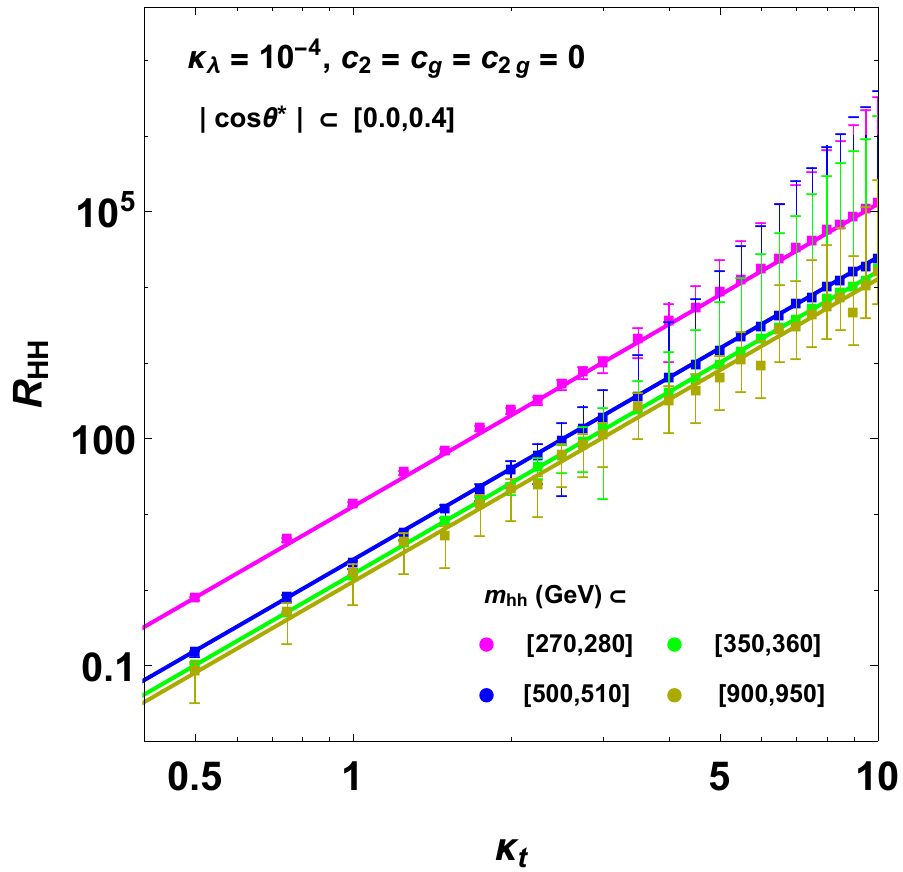}
  \includegraphics[width=0.32\textwidth]{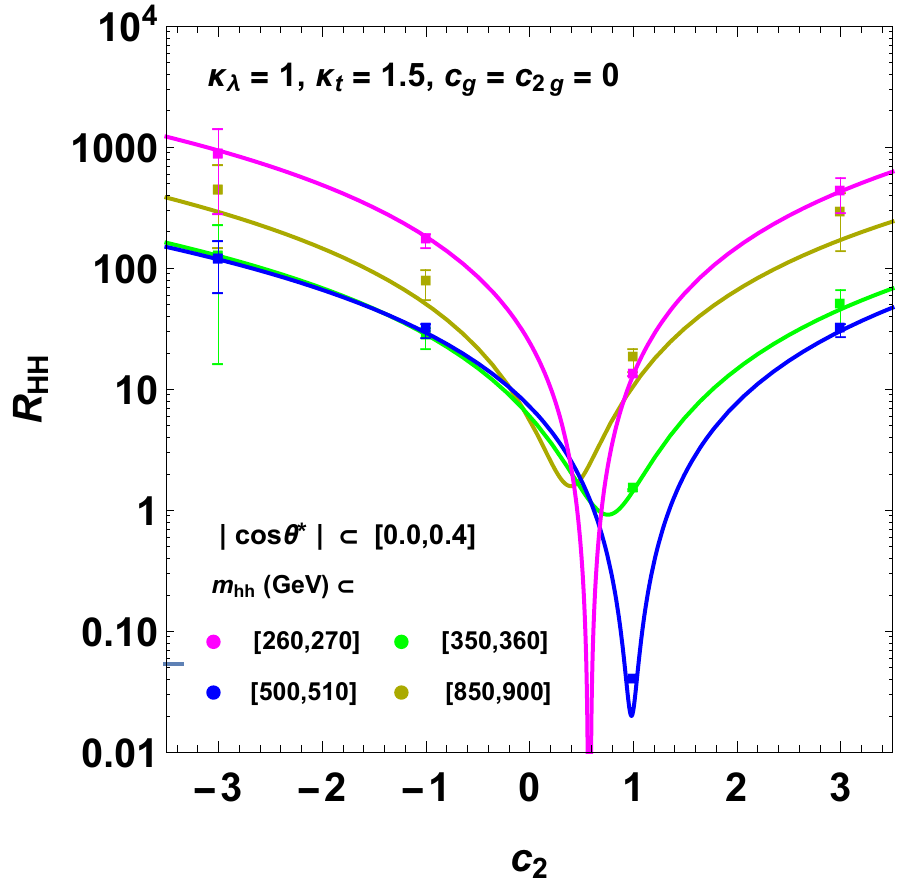}
\caption{\footnotesize{Comparison of $\RHH^j$, obtained from the analytical formula (\ref{eq:cx}), with the same ratio, 
as derived directly via MC, for the most central bin in $\cos \theta^\ast$ (betweem 0 and 0.4) and for four different bins in 
$m_{hh}$, which are between 260\,GeV and 270\,GeV (threshold), 350 GeV and 360 GeV (interference region), 
500 GeV and 510 GeV, and 1000 GeV and 1100 GeV (high mass tails). From left to right we show a scan in 
$k_{\lambda}$, $k_{t}$ and $c_{2}$, respectively, while keeping the other parameters at fixed values 
%\com{rightmost really $\kappa_t=1.5$? Last mhh bin? Axes labels kt plot... errors for large kt? (probably due to population...?!)}. 
The error bars have a pure statistical source related with the generated number of events.}  \label{fig:fitkt}}
\end{center}\end{figure*}

In Figure~\ref{fig:fitkt} we provide some examples of the fit agreement with the dataset points to selected $m_{hh}$ ranges % within $\{260-270,350-360,500-510,1000-1100\}$\,GeV,
considering the bin central in $cos\,\theta^*$. Here, we display three parameter space directions, which are 
$\kappa_\lambda$ for fixed $\kappa_t = 1$, $\kappa_t$ for $\kappa_\lambda =10^{-4}$ and 
$c_2$ for $\kappa_\lambda =1$, $\kappa_t=1.5$. The errors on the data points are purely statistical, while the coloured lines correspond to the fit result, Eq. (\ref{eq:cx}). 
It becomes evident from the results that, even if it is true that for every bin one needs only a minimum of $N_{\rm fit}=15$ different samples to determine all coefficients, due to limited MC statistics the result of such a 
procedure would have a limited reliability. 
Concerning the behaviour of the cross section in different bins, the results confirm the expectation that a change in the 
scalar self coupling $\kappa_\lambda$ affects mostly the threshold region, while the ($D>4$)  
$\bar t t hh$ contact interaction has a large effect on the high energy tail. Moreover, the 
pronounced destructive interference present in the cross section in the vicinity of the SM 
point $\kappa_\lambda=1$ becomes evident via the peaked minimum, visible in the left-most plot.

\begin{figure}[h]
  \centering
  \includegraphics[width=0.32\textwidth]{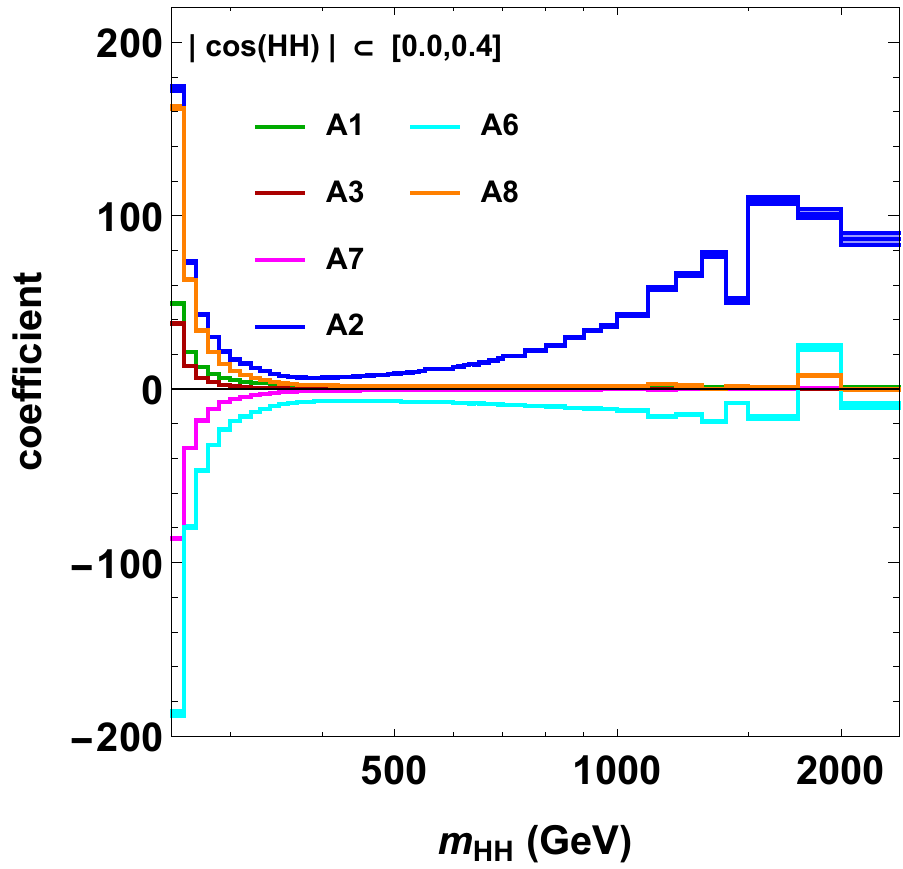}
  \includegraphics[width=0.32\textwidth]{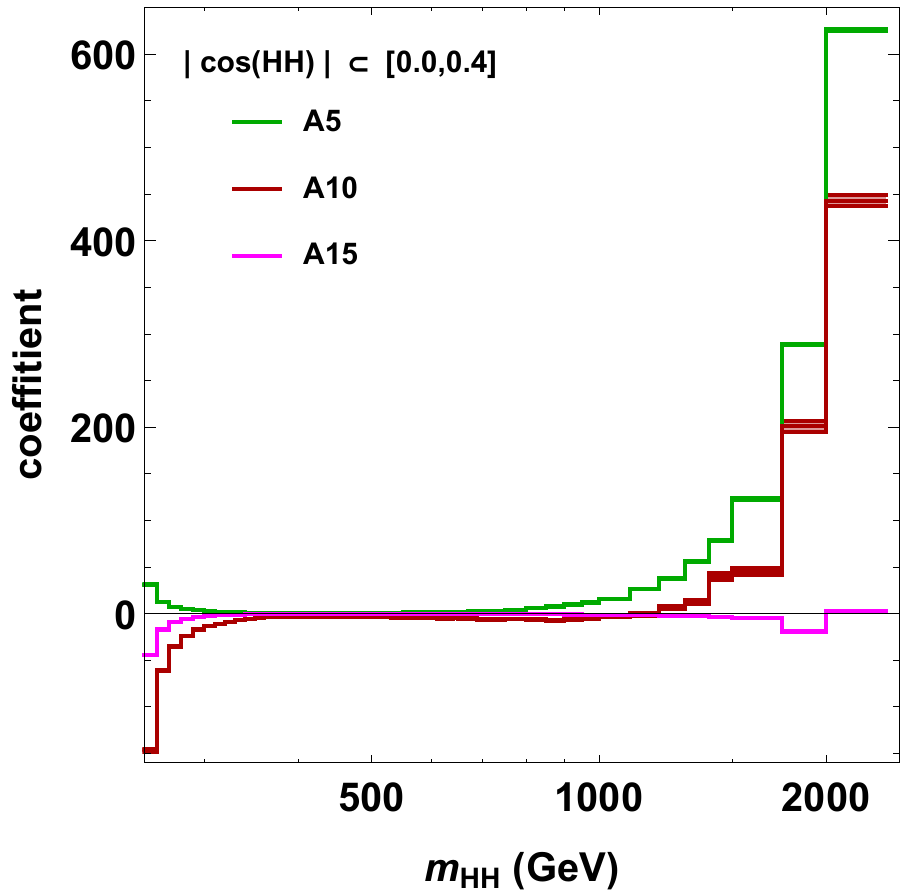}
  \includegraphics[width=0.32\textwidth]{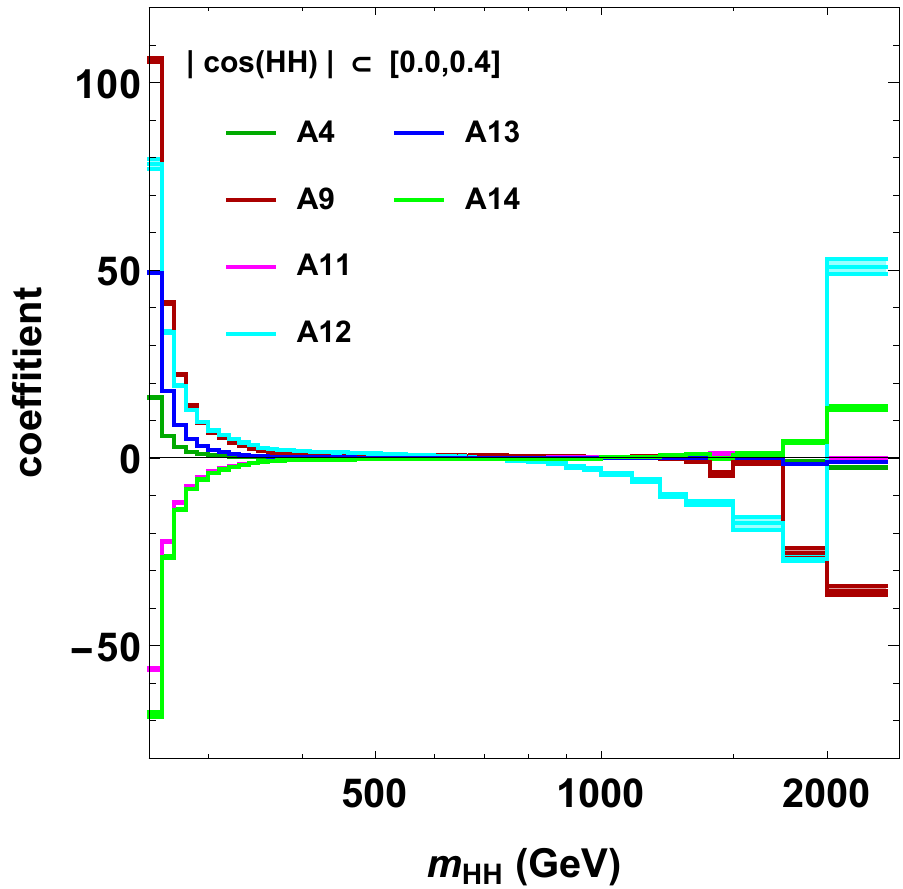}
  \caption{\footnotesize Values of the fit coefficients entering $R_{hh}^j$, differential in $m_{hh}$. We only display the 
results for the central $\cos \theta^\ast$ bin (betweem 0 and 0.4). From left to right, we show the coefficients belonging 
to subspaces (S1,S2), S4, and (S3, S5, S6), respectively (see Table~\ref{tab:datasets}). 
    \label{fig:DiffRhh}
  }
\end{figure}

In Figure~\ref{fig:DiffRhh} we display the value of the coefficients of $\RHH^j$ for the central $\cos \theta^\ast$ bin as a function of $m_{hh}$, grouped in the subspaces given in Table \ref{tab:datasets}. 
As a first observation, we see that at high invariant masses $A_{2,5,10}$ become 
most important, which coincide with the terms that parametrize genuine higher dimensional effects 
(containing exclusively $c_{2,g,2g}$), in agreement with expectations.
We further note that for the subspace that corresponds solely to a variation of SM parameters (S1, with coefficients $A_{1,3,7}$), the absolute value of the coefficients does not surpass 100, and they are peaked towards low $m_{hh}$. 
In particular, the value of $A_7$ (that controls the triangle-box interference
term $\sim \kappa_t^2 \kappa_t \kappa_{\lambda}$) is  always negative for $m_{hh} < 400$\,GeV. 
In subspace S2 (coefficients $A_{2,6,8}$), which features the contributions from the new $\bar t t hh$ contact interaction
(including interference with SM-like contributions), the coefficients are in general larger compared to those in S1, resulting in 
stronger variations with the corresponding parameters close to the $hh$ threshold, and show more pronounced $m_{hh}$ tails, as discussed before. Finally, all the remaining terms contain gluon-Higgs contact interactions and as such also feature 
pronounced contributions at large energies. The Feynman-diagrammatic representation of the different contributions
to the cross section can be studied in Figure 1 of \cite{Carvalho:2015ttv}.

After this general discussion of the behavior of the cross section and the fit coefficients, in the next section we will 
employ Eq. (\ref{eq:cx}) to reinterpret experimental analyses.

\section{Recast of ATLAS and CMS measurements}
\label{sec:Recast}

In this section we establish and validate the reweighting method and estimate the present limits on the Higgs couplings in a multi-dimensional space. The constraints are obtained by a recast of preliminary results from the ATLAS and CMS collaborations~\cite{ATLAS-CONF-2016-004, CMS-PAS-HIG-16-032} in the $HH \rightarrow \gamma\gamma b\bar{b}$ channel, that refer, respectively, to samples of 3.2 fb$^{-1}$ and 2.7 fb$^{-1}$ of 13 TeV data. Results with a slightly better sensitivity were published employing 8 TeV data \cite{Aad:2015xja,Aad:2014yja, Khachatryan:2016sey}, but we decided to use the preliminary results since they are obtained at the same center-of-mass energy that will be used for the next 
generation of LHC results. 
Between the publications of these and the release of this paper, other searches were performed by both collaboration in several channels~\cite{Sirunyan:2017guj,Sirunyan:2017djm,Aad:2015xja} and with more luminosity, were the CMS analyses also include results for beyond the SM (BSM) scans in $\kappa_{\lambda}$ and $\kappa_{t}$. 

The choice of the $HH \rightarrow \gamma\gamma b\bar{b}$ channel results from the fact that this final state is easy to simulate and to reconstruct using a parametric model of the ATLAS and CMS detectors. Moreover, both analyses were performed with a sequential application of selections (in contrast to a multivariate analysis that is a kind of standard now), making them easy to recast.
It is known that this final state is less constrained in regions of the phase-space compared to other 
final states \cite{Aad:2015xja}. Nevertheless, since our goal is the proof of principle of the recast technique, we 
are not hunting for the latest and best limit. 
In the case of more sophisticated (multivariate) analyses, a reinterpretation of the results is more subtle, 
however still possible if samples exist for all relevant kinematic configurations, as provided in \cite{Carvalho:2015ttv}
(see discussion at the end of Section~\ref{sec:models}).

\subsection{Signal simulation and reweighing}

We employ Eq.~\ref{eq:cx} to reweight a set of base events to different points in the parameter space. As base events, we use the 12 benchmarks defined in~\cite{Carvalho:2015ttv,Carvalho:2016rys,deFlorian:2016spz} and for each of them we simulate 100,000 events. Due to the construction of the benchmarks, the resulting ensemble $\mathcal{S}_{\rm BSM}$ of 1,200,000 events populates well the EFT space we are probing. In addition 100,000 SM-like events are generated for comparison but kept out of $\mathcal{S}_{\rm BSM}$. % \com{Now bigger sample?}.

The hard scattering events are generated at LO accuracy using the model described in section \ref{sec:diffit}. Following Ref.~\cite{Carvalho:2016rys} we normalize the events to a total cross section with a k-factor is applied to include corrections up to NNLO (augmented with NNLL re-summation) in QCD ($\sigmaHH^{\rm SM}=33.5\pm 1.4$\,fb ~\cite{deFlorian:2016spz}). %, with an error of ${\cal O}(5 \%)$. 
Note that the normalization only becomes relevant to predict the absolute number of events, as used in Section \ref{sec:models}. In fact, up to NNLO, the 
k-factors are found to be mostly flat in kinematic variables as well as as a function of the EFT parameters
(in the bulk of the parmeterspace)~\cite{deFlorian:2017qfk} 
%\com{Double check subtleties/weaken?! NNLO distributions available now $\to$ check!} 
and this suggests that the distributions are well described by performing the reweighing 
with LO information and employing flat k-factors to correct for QCD corrections at the end is a reasonable approximation. Furthermore, we used the NLO set of the PDF4LHC parton densities~\cite{Butterworth:2015oua, Dulat:2015mca, Harland-Lang:2014zoa, Ball:2014uwa} as well as the parton shower and hadronization infrastructure of the \py\ package~\cite{Sjostrand:2014zea}. Finally, we simulate the response of an LHC-like detector by using \de~\cite{deFavereau:2013fsa}.

To reweight simulated events from $\mathcal{S}_{\rm BSM}$ to any other point in parameter space (such as
the SM point, which we will use for validation), we apply an event-by-event weight 
\begin{equation}
	W_{i} = \frac{\RHH (m_{HH}^i,cos\,\theta^*_i)}{N_{i}} \cdot \frac{\sigmaHH^{\rm SM}\, 
	{\rm Frac}_{\rm SM}^j}{\sigmaHH\ C_{\rm norm}}
	\label{eq:weight}
\end{equation}
%\com{Add explanation that/why we normalize the total result to one (dividing by $\sigmaHH$)}
where $m_{hh}^i$ and $cos\,\theta^*_i$ refer to the ME level variables corresponding to event $i$, and $N_{i}$ is the number of events in bin $j$ that contains $m_{hh}^i$ and $cos\,\theta^*_i$ in the reweighting sample $\mathcal{S}_{\rm BSM}$. Along the same lines, $\RHH (m_{HH}^i,cos\,\theta^*_i)$ is given by $\RHH^j$ (Eq. (\ref{eq:cx})) for this bin. 
It is our choice to normalize the weights in such a way that the resulting BSM distribution is normalized to one if no cut is applied and to the signal efficiency if any selection is applied. In this manner we decouple the shape information from the total cross section information. %, that were derived in \cite{Carvalho:2015ttv} with an independent set of samples.

The normalization coefficient $C_{\rm norm}$ is obtained as
\begin{equation}
C_{\rm norm} = \sum\limits_{j} \RHH^j  \frac{\sigmaHH^{\rm SM}}{\sigmaHH} {\rm Frac}_{\rm SM}^j
\end{equation}
and is used to unitarize the reweighed shape. In the ideal case $C_{\rm norm} = 1$. Nevertheless, fluctuations in the samples used to derive the weights can induce a departure from unitarity. For the SM shape $C_{\rm norm} = 1$, since the bin-by-bin fits are performed fixing the SM point and statistical errors occur only for BSM points where for some samples there are bins with inevitably low statistics (for example in the high $m_{HH}$ bins of threshold-like EFT points). After the renormalization procedure we observe a good agreement within the statistical uncertainties between MC simulation and reweighed samples, as can be seen in the next section.

%and is used to unitarize the reweighted shape. In the ideal case $C_{\rm norm} = 1$. Nevertheless, fluctuations in the samples used to derive the weights can induce a departure from unitarity. For the SM shape $C_{\rm norm} = 1$, since the bin-by-bin fits are performed fixing the SM point and statistical errors occur only for BSM points. After the renormalization procedure we observe a good agreement within the statistical uncertainties between MC simulation and reweighted samples, see below.

\subsection{The analyses}

We consider the number of events $N_{\rm obs}$ observed by each experiment in a signal region and $B$ the number of the expected background events. In the ATLAS documents both values are directly provided in the text, while in the CMS document they are extracted from figures imposing a $M_{\gamma\gamma}$ window  identical to the ATLAS one.  %, read from the reference plot.

\begin{figure}
  \centering
  \includegraphics[width=0.42\textwidth]{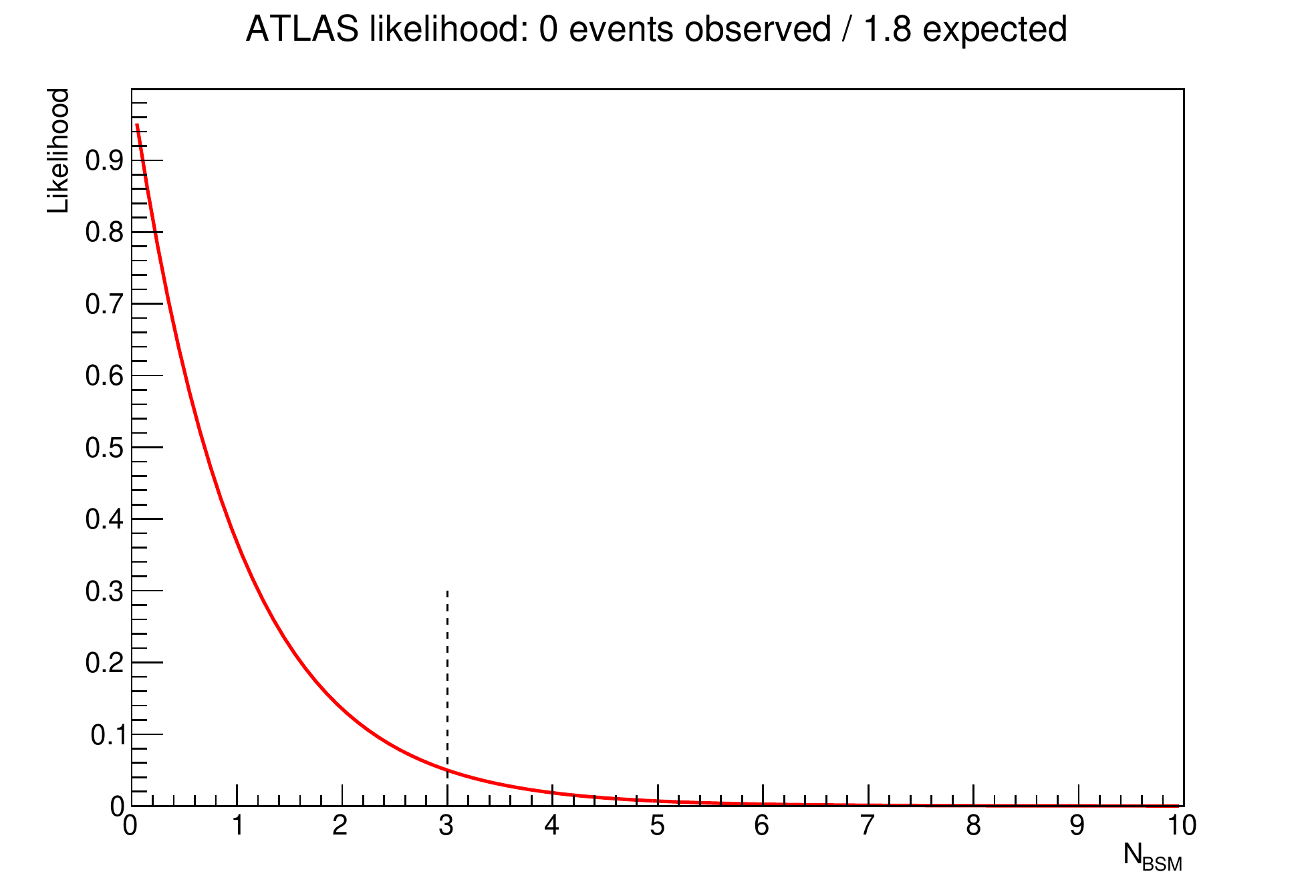}
  \includegraphics[width=0.42\textwidth]{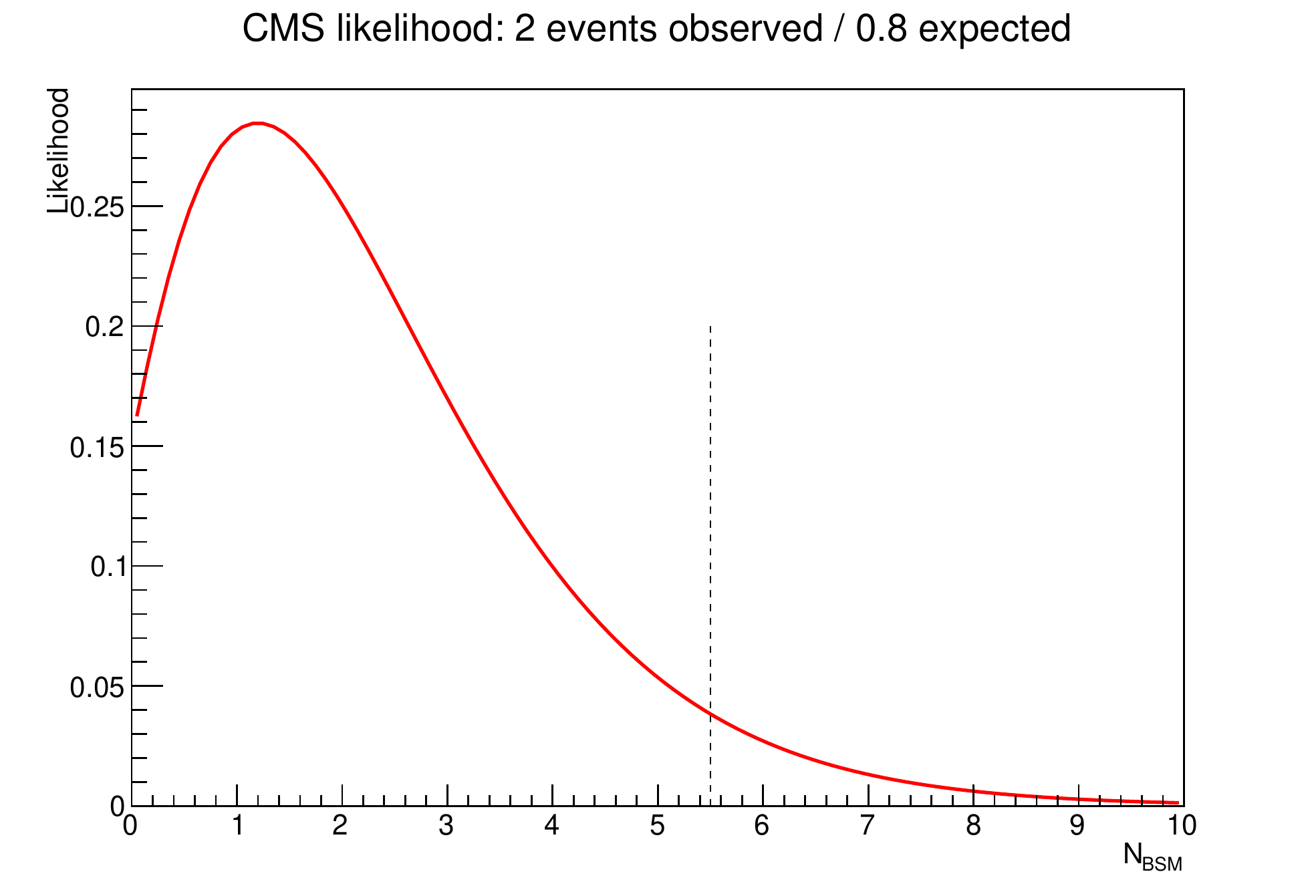}
\caption{\footnotesize ATLAS and CMS likelihoods.   \label{fig:likelihood}
  }
\end{figure}

Both experiments  found that the observed events are statistically compatibles with the expected SM backgrounds. Therefore, we can use this observation to derive exclusion limits on the maximal allowed number of (expected) events in a NP scenario $N_{\rm BSM}$ that could be compatible at 95\% CL with the observed data.   
%\com{Why mention $B$ here? Doesn't $N_{\rm BSM}$ include $B$?}. 
We build a Poissonian statistical likelihood
\begin{equation}
\mathcal{L} = e^{-(B+S)} \frac{(B+S)^{N_{\rm obs}}}{N_{\rm obs}!}\,,
\end{equation}
%\com{Shouldn't it read sth like $e^{-N_{\rm BSM}}N_{\rm BSM}^{\rm N_{\rm obs}}/N_{\rm obs}!$ ?}
where $S$ is the number of Signal events. Bayesian hypothesis with a uniform prior on $S$ allows to invert the likelihood to obtain a posterior as a function of $S$, normalized to unity between $S=[0,\infty]$. The likelihoods are shown in Figure \ref{fig:likelihood}. We integrate then the likelihood from 0 to 95\% to obtain the 2 standard deviation confidence level value 
for $N_{\rm BSM}$. The number of observed, expected and excluded events are provided in Table \ref{tab:likelihood}. In this table we also include the number of events that would be excluded at $L$=100 fb$^{-1}$, projected assuming 
$N_{\rm obs}(\mbox{100 fb}^{-1}) = B(\mbox{100 fb}^{-1}) = L * B/L'$, where $L'$ is the data luminosity currently used.

\begin{table}
\centering
\begin{tabular}{c|c|c|c|c|c|c|c}
  \multicolumn{4}{c}{ATLAS} &\multicolumn{4}{c}{CMS}  \\\hline
$N_{\rm obs}$ & $B$ & $N_{\rm BSM}$ (2.7 fb$^{-1}$) & $N_{\rm BSM}$ (100 fb$^{-1}$)& 
$N_{\rm obs}$ & $B$ & $N_{\rm BSM}$ (3.2 fb$^{-1}$) & $N_{\rm BSM}$ (100 fb$^{-1}$)\\\hline
0 & 1.8 & 3.0 & 16.3 & 2  & 0.8 & 5.45 & 12 \\\hline

\end{tabular}
\caption{\label{tab:likelihood} \footnotesize Upper limits on cross section times luminosity in the analysis window, $N_{\rm BSM}$, from the recast of ATLAS and CMS results.
%\com{Explain table - call expected = B, since do not expect any signal? Should be 0.8 for CMS? Why 100 fbinv CMS value smaller than ATLAS, although opposite for smaller lumi?}
}
\end{table}

To translate the limit on $N_{\rm BSM}$ into the limit on the BSM cross section $\sigma_{\rm BSM}$ we need to know the selection efficiency $\epsilon$ within the signal box and the luminosity $L$: $\sigma_{\rm BSM} = N_{\rm BSM}/(L \epsilon)$. For both analyses we can extract (or estimate)  $\epsilon_{\rm SM}$, \textit{i.e.,} the efficiency for the SM-like $hh$ production, from the information provided in the respective papers. What is not known a priori is the efficiency for a given BSM model with the associated choice of parameters. To estimate this, we employ a mock-up of the ATLAS and CMS analyses with \de. The resulting efficiency $\epsilon_{D}$ is then rescaled (if needed) by a fudge factor $f_{\rm SM} = \epsilon_{\rm SM}/\epsilon_{\rm SM, D}$ to get an estimate of the real efficiency $\epsilon_{\rm BSM} \approx f_{\rm SM} \cdot \epsilon_{D}$.
 This factor takes into account the limited information we benefit from to emulate the analysis and the approximations made in \de\ to describe the detector response, such as the efficiency of detecting photons or tagging b-jets.
Note that this step could be avoided if the experimental collaborations would make public the full information (including the generation level $m_{hh}$ and $|cos\,\theta^*|$ variables) on the
events that pass the signal selection. 
%\com{Check statement... if we know all the selection-cuts, can't we just re-scale the distributions via $R_{hh}$ to obtain
%the efficiencies for BSM points? Besides this, if b-tagging, photon-id, etc depends on kinematics, we are not taking these effects 
%into account for BSM points, right?}

In Table~\ref{tab:recastCuts} we summarize the set of selections of both ATLAS and CMS searches. It is important to notice that the same \de card was used for both experiments. This approximation is valid for photons in the energy range considered in $hh$ production since both experiments are more than 90\% efficient to tag them. The difference in fiducial acceptance is emulated at the level of selections. For jets, both experiments use the anti-k$_T$ algorithm \cite{AKT} with radius parameter 0.4. Their acceptance and calibrations are similar at those energies. The major difference comes from the b-tagging working point (WP). Since both experiments require both jets to be b-tagged we just rescale the typical efficiency of the WP used in \de within our $p_T$ and $\eta$ range (63\% -- see Ref. \cite{DelphesCMS}) to the efficiency of the WP declared in the public notes, {\it i.e.,} 85\% for ATLAS and 78\% for CMS. Finally, for the projections to $100\,$fb$^{-1}$ given below, we assume
the observed number of events to match the expected and no improvements in the analysis, which is a very conservative assumption.

\begin{table}
\centering
\begin{tabular}{c|rr}
 variable & ATLAS & CMS \\\hline
$|\eta_{max}|$ & 2.37 & 2.5 \\
Rejected fiducial region in $|\eta|$ & [1.37, 1.52] & [1.44, 1.57] \\
Leading photon $(p_{T,min})$ & --- &  30 \\
Subleading photon $(p_{T,min})$ & --- & 20 \\
Leading photon $(p_{T,min})/M_{\gamma\gamma}$ & 0.35 & 1/3 \\
Trailing photon $(p_{T,min} )/M_{\gamma\gamma}$ & \multicolumn{2}{c}{\ \ 1/4} \\
$\Delta R$ with any jet  & \multicolumn{2}{c}{\ $> 0.4$} \\\hline
$|\eta_{max, b}|$ & 2.5 & 2.4 \\
%$p_{T,b}^{min}$ & \multicolumn{2}{c}{25 GeV} \\
Leading b-jet $p_{T,min}$ (GeV) & 55  & 25 \\
Trailing b-jet $p_{T,min}$ (GeV) & 35  & 25 \\\hline
$M_{\gamma\gamma}$ window (GeV) & \multicolumn{2}{c}{[122, 128]}\\ 
$M_{b\bar{b}}$ window (GeV) & [95,135] & [80,200]\\
$M_{\gamma\gamma b\bar{b}}$ minimum (GeV) & --- & 350 \\\hline
\end{tabular}
\caption{\footnotesize Selections used in the recast analyses. \label{tab:recastCuts}}
\end{table}

After the reweighting according to the b-tagging properties and the selections of Table~\ref{tab:recastCuts},
we find the efficiencies shown in Table~\ref{tab:effC}. 
It appears that for the ATLAS analysis our efficiencies have to be rescaled by $f_{\rm SM} =1.41$. Beside the usual suspects described above to explain the size of the factor, we suspect that the photon efficiency is slightly better in the real ATLAS analysis than in the \de card, but the exact amount is hard to guess from the details given in the conference note.

In the CMS case the efficiency of the 2 b-tag and 1 b-tag categories are only provided together and no explicit information on how it is distributed between them is given. One may notice that a similar categorization was used at 8\, TeV where the events appeared to be evenly split between the two categories, incidentally \cite{Khachatryan:2016sey}. %\com{I.e., the number of events in both categories was the same at 8 TeV and now at 13 TeV they give 20\% efficiency total, such that we assume there is 10\% of the initial events in each category (so taken together both categories make 20\%). Correct?} 
We apply the same assumption to the 13~TeV analysis and  split the efficiencies evenly between the two categories ($\epsilon_{\rm SM, CMS} \approx 10\%$ per category). 

\begin{table}
\centering
\begin{tabular}{c|c|c||c|c|c}
  \multicolumn{3}{c}{ATLAS} &\multicolumn{3}{c}{CMS}  \\
$\epsilon_{\rm SM, D}$ & $\epsilon_{\rm SM}$ & $f_{\rm SM}$ & $\epsilon_{\rm SM, D}$ & $\epsilon_{\rm SM}$ & $f_{\rm SM}$   \\\hline
7.1\% & 10\%  & 1.41 & 10.8\% & $\approx 10\%$ & 1 \\\hline
\end{tabular}
\caption{\label{tab:effC} \footnotesize Results from the ATLAS and CMS-like recast. See text for details.}
\end{table}

\subsection{Validation of the analytical reweighing}

\begin{figure}
  \centering
  \includegraphics[width=0.32\textwidth]{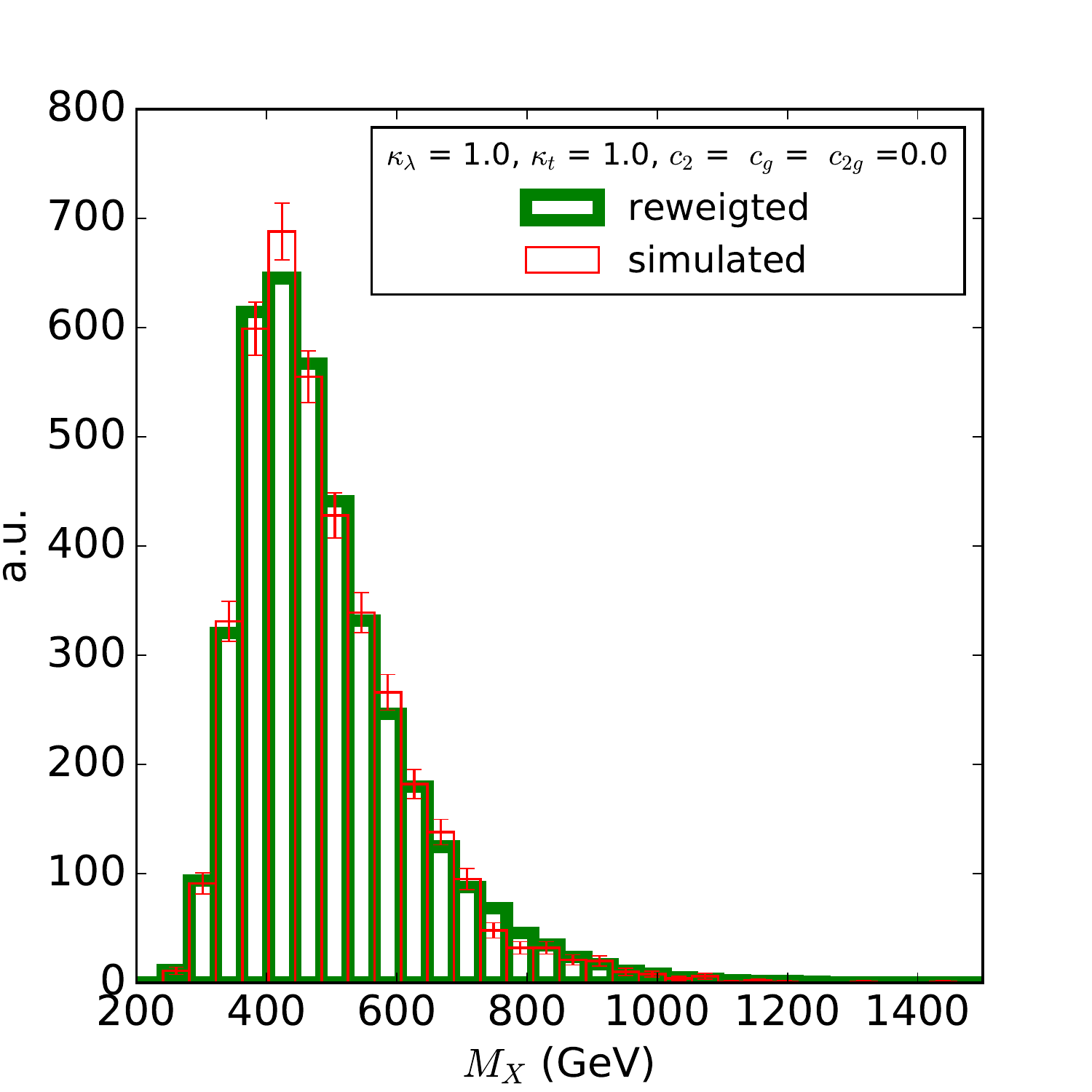}
  \includegraphics[width=0.32\textwidth]{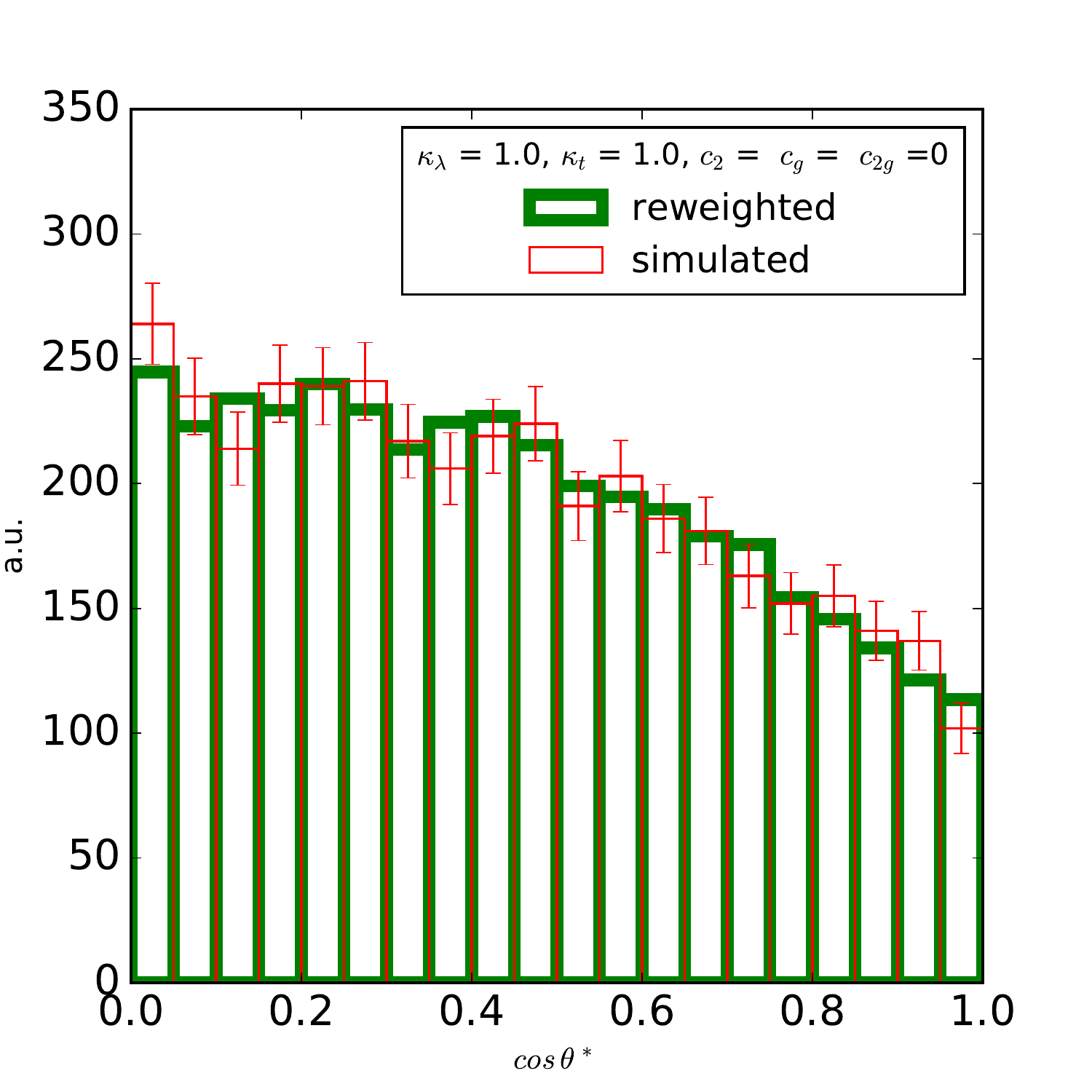}
  \includegraphics[width=0.32\textwidth]{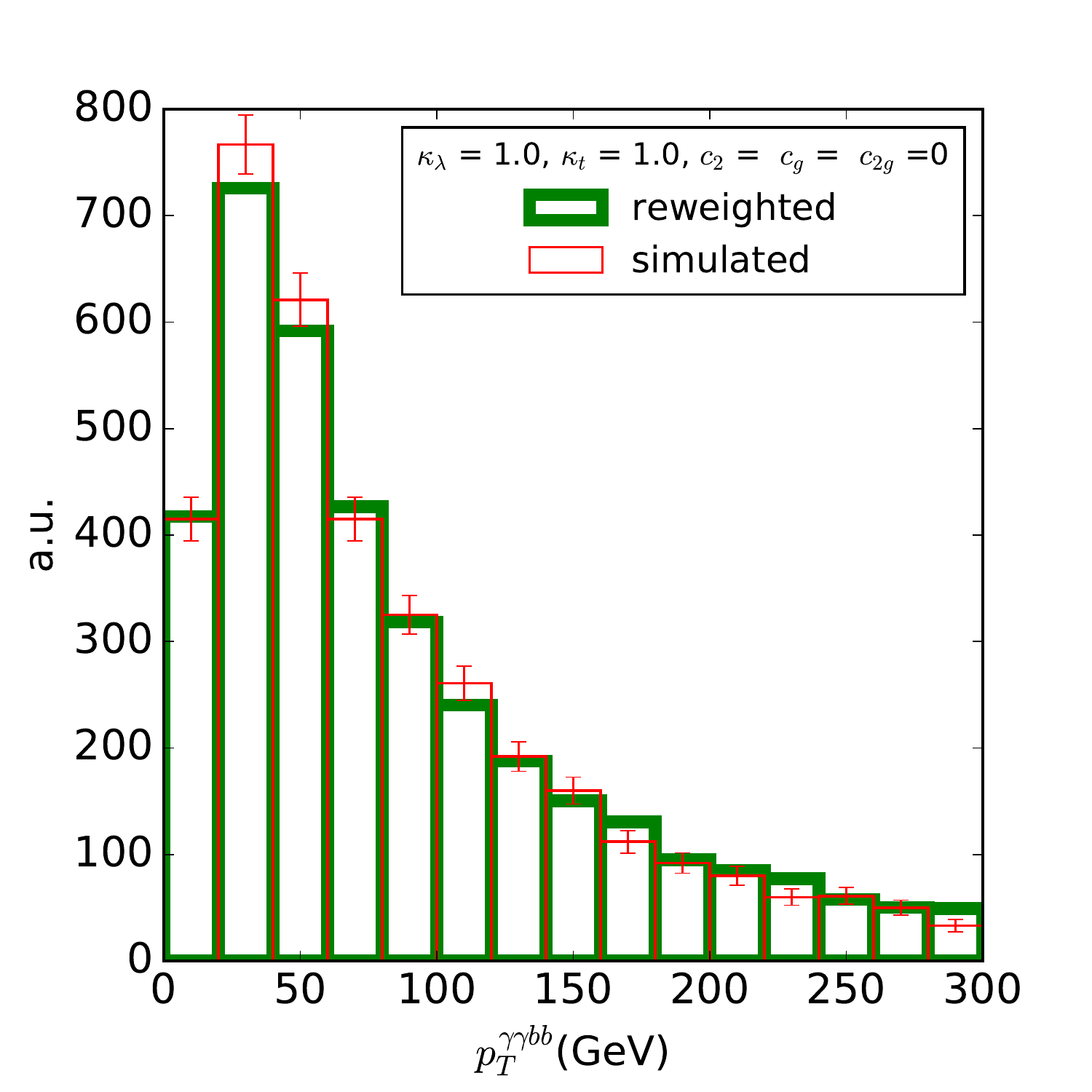}
  \caption{\footnotesize Reconstructed variables after ATLAS-like selection for the di-Higgs system. The histograms display the signal efficiency times 100,000 events. 
    \label{fig:RecastSM}
  }
\end{figure}

In Figure \ref{fig:RecastSM} we show the comparison of the actual simulation and the reweighting for an SM-like signal for 3 main reconstructed kinematic variables describing the hh system: $m_X \equiv m_{\gamma \gamma b \bar b} - m_{b \bar b} - m_{\gamma \gamma } + 250\, {\rm GeV}$, \mbox{cos $\theta^{*}_{\rm \gamma \gamma b \bar b}$} and $p_{\rm T, \gamma \gamma b \bar b}$. The first variable was shown to be the best estimate of $m_{hh}$ in Ref. \cite{CMS-PAS-HIG-16-032}. The statistical errors per bin are shown only for the plain simulation, since the shapes derived by reweighting are constructed from 12 times as much bare events and thus expected to have less statistical fluctuations. In general we find a very good agreement between the distributions. Comparisons for different benchmarks provide similar picture and are given in Appendix~\ref{app:shapes}. 
Moreover, in Figure~\ref{fig:RecastBSM} we show reweighted distributions in $m_X$, after ATLAS-like selections, for BSM points not within the set of (simulated) benchmarks. Beyond points featuring non-vanishing contact interactions (cg=-c2g=1 and c2=1), we also consider the maximum box-triangle interference (kl=2.4), thereby exploring three different kinematic regimes. In all the cases we see smooth shapes. 

\begin{figure}
  \centering
  \includegraphics[width=0.32\textwidth]{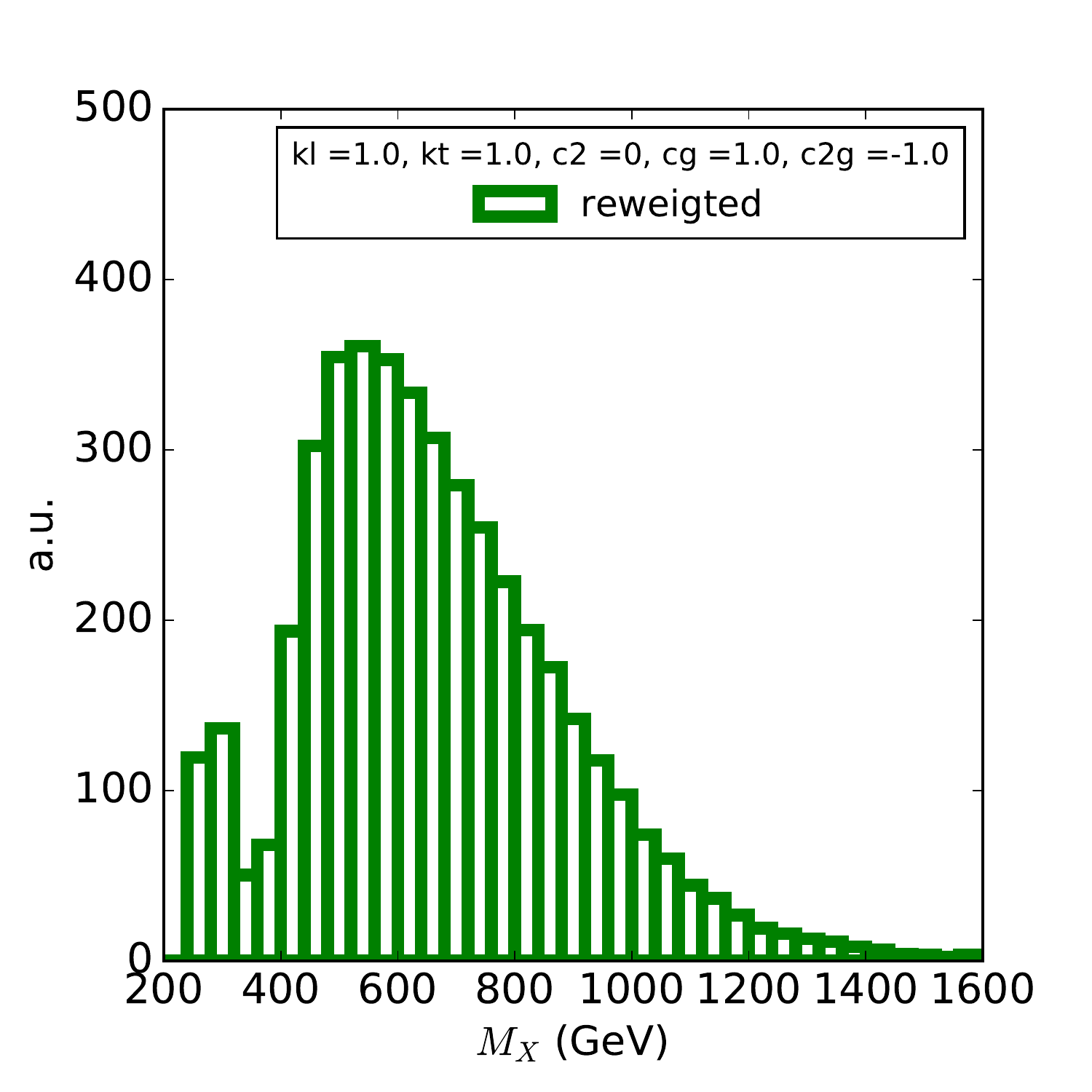}
  \includegraphics[width=0.32\textwidth]{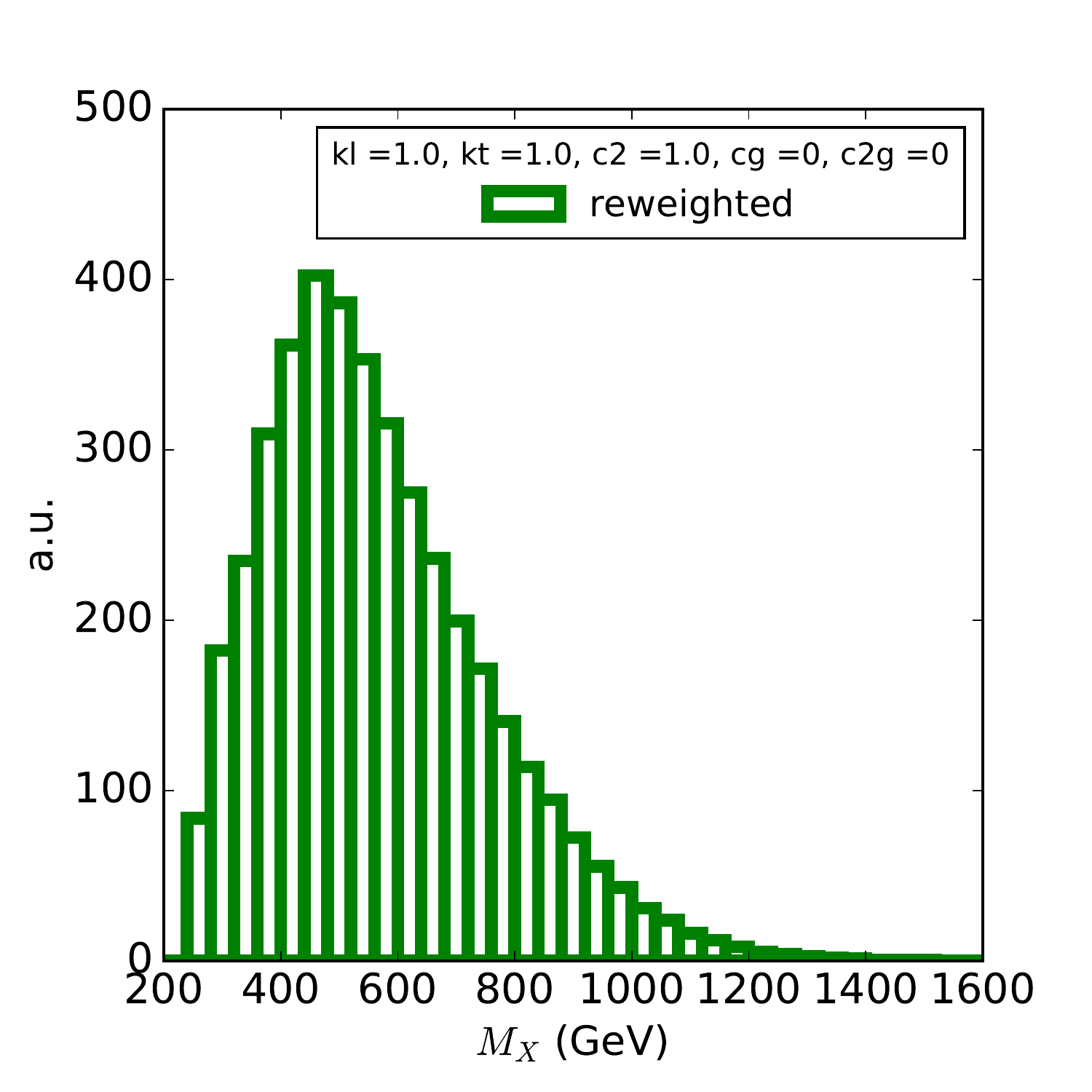}
  \includegraphics[width=0.32\textwidth]{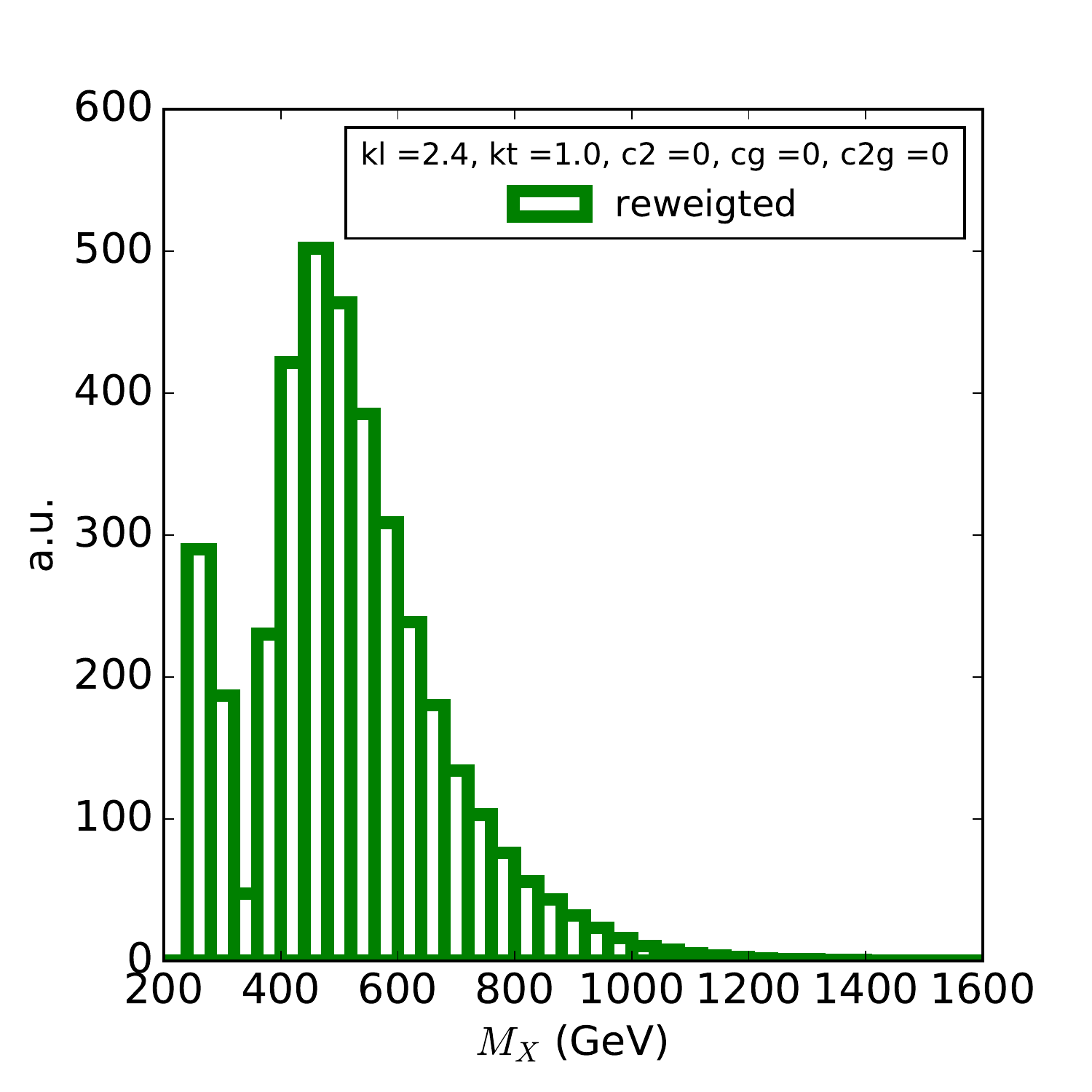}
  \caption{\footnotesize Reconstructed variables after ATLAS-like selection for the $m_X$ variable for three BSM points with different kinematic properties. The histograms display the signal efficiency times 100,000 events. 
    \label{fig:RecastBSM}
  }
\end{figure}

Finally, the efficiencies of different benchmark points obtained from the MC simulation are given
in Table \ref{tab:effBench} ($\epsilon_{MC}$), together with the 
reweighted efficiencies as obtained after applying the renormalization 
via $C_{norm}$. The agreement of both efficiencies is remarkable, specially in the points where EFT is linear 
(where the parameter space scan used for the fit is more dense, see table~\ref{tab:datasets}). 
%\com{Besides (a bit) BM 3:
%what could happen? Put other model plots?}.

\begin{table}
\centering
\begin{tabular}{c|ccccc|c|cc||cc}
 & & & & & & & \multicolumn{2}{c|}{ATLAS-like} & \multicolumn{2}{|c}{CMS-like} \\
$N$ &$\kappa_\lambda$ & $\kappa_t$ & $c_2$ & $c_g$ & $c_{2g}$& $C_{\rm norm}$ & $\epsilon_{\rm MC}$  &  $\epsilon_{\rm rew}$ & $\epsilon_{\rm MC}$ & $\epsilon_{\rm rew}$  \\\hline
0	& 1.0 & 1.0 & 0.0 & 0.0 & 0.0 &	1.0	&	10.0	&	10.0	&	11.46	&	11.47	\\\hline\hline
1	& 7.5 & 1.0 & -1.0 & 0.0 & 0.0 &	0.94	&	10.6	&	10.6	&	10.83	&	10.79	\\
2	& 1.0 & 1.0 & 0.5 & -0.8 & 0.6 &	0.71	&	10.3	&	10.4	&	11.46	&	11.26	\\
3	& 1.0 & 1.5 & -1.5 & 0.0 & -0.8 &	0.96	&	9.93	&	10.5	&	10.89	&	11.68	\\
4	& -3.5 & 1.0 & -3.0 & 0.0 & 0.0 &	0.98	&	9.42	&	9.37	&	9.89	&	9.72	\\
5	& 1.0 & 1.0 & 0.0 & 0.8 & -1.0 &	0.88	&	11.6	&	11.5	&	12.47	&	12.50	\\
6	& 2.4 & 1.0 & 0.0 & 0.2 & -0.2 &	0.96	&	8.47	&	8.73	&	7.27	&	7.74	\\
7	& 5.0 & 1.0 & 0.0 & 0.2 & -0.2 &	1.01	&	6.24	&	6.17	&	3.30	&	3.23	\\
8	& 15.0 & 1.0 & 0.0 & -1.0 & 1.0 &	0.92	&	10.2	&	10.3	&	10.64	&	10.74	\\
9	& 1.0 & 1.0 & 1.0 & -0.6 & 0.6 &	0.86	&	11.6	&	11.6	&	12.92	&	12.79	\\
10	& 10.0 & 1.5 & -1.0 & 0.0 & 0.0 &	1.01	&	6.67	&	6.60	&	3.48	&	3.48	\\
11	& 2.4 & 1.0 & 0.0 & 1.0 & -1.0 &	0.95	&	8.16	&	8.26	&	6.72	&	6.81	\\
12	& 15.0 & 1.0 & 1.0 & 0.0 & 0.0 &	1.0	&	7.80	&	7.52	&	6.23	&	6.17	\\
 \hline
\end{tabular}
\caption{\footnotesize The renormalization factor ($C_{\rm norm}$) for benchmark point $N$ as well as MC and reweighted sample efficiencies.\label{tab:effBench}}
\end{table}

\section{Model dependent interpretations}
\label{sec:models}

We illustrate the method described in Section \ref{sec:Recast} reinterpreting EFT bounds, applying them to concrete NP
setups. We consider the explicit models collected in Table \ref{tab:models}. 
%After mapping them to the EFT description, we use the events in ${\cal S}_{\rm BSM}$ that pass the signal-region selections and make use of Eq. \ref{eq:cx} to reweigh the events according to the different distributions of the cross section scanning the EFT parameter space to predict the number of events in signal region, as described in Section \ref{sec:Recast}. 
%\com{Do we really need these tables,or just the number of events per bin?} 
%This information will then be used to exclude regions of the parameter space of the models, as described below. 
%, together with the reconstruction level variables \com{specify?!}

\begin{table}[h!]
\begin{center}
\begin{tabular}{c|c c }
Model & NP integrated out & Ref. \\
\hline
1 & real scalar singlet with explicit $Z_2$ breaking &  \cite{Dawson:2017vgm,deBlas:2014mba}\\
2 & real scalar singlet with spontaneous $Z_2$ breaking & \cite{Dawson:2017vgm} \\
3 & real scalar triplet & \cite{Dawson:2017vgm,deBlas:2014mba} \\
4 & complex scalar triplet & \cite{Dawson:2017vgm,deBlas:2014mba} \\
5 & quartet scalar with $Y=1/2$ & \cite{Dawson:2017vgm,deBlas:2014mba} \\
6 & quartet scalar with $Y=3/2$ & \cite{Dawson:2017vgm,deBlas:2014mba} \\
\hline
7 & 2HDM (addtl. scalars heavy + $Z_2$) & \cite{Belusca-Maito:2016dqe} \\
\hline
8 & vector-like quark: $T$ (singlet top partner) & \cite{delAguila:2000rc} \\
9 & vector-like lepton: $E$ (flavor universal singlet) & \cite{delAguila:2008pw} \\
\hline
10 & MCHM$_5$ & \cite{Giudice:2007fh,Contino:2013kra,Grober:2010yv,Contino:2010rs} \\
11 & MCHM$_4$ & \cite{Giudice:2007fh,Contino:2013kra,Grober:2010yv,Contino:2010rs} \\
\end{tabular}
\end{center}
\caption{\label{tab:models} Explicit models considered in this work.}
\end{table}

In Table \ref{tab:EffCoup} we provide the relations between the effective couplings
for the various models under consideration, arising after integrating out the NP at tree level, 
where we express dependent couplings in terms 
of independent ones, the latter given in the second column. The relation with the fundamental parameters of
the models is given in Table~\ref{tab:paras}, which contains equivalent information.
Note that the fact that we integrate out NP at tree level might appear problematic, 
given that the effective couplings in Table~\ref{tab:EffCoup} enter Higgs-pair production only at one-loop and,
in principle, potential loop induced contact interactions with gluons ($c_{g,2g}$) could enter the process 
at tree-level, thus lifting a loop-suppression in integrating out NP. 
However, since the scalars considered in models 1-7 are 
assumed to be color neutral, no such effective interactions are in fact induced at the one-loop level. 
The same is true for integrating out vector-like leptons (model 9). The vector-like quarks (model 8), 
on the other hand, do not couple to the Higgs boson without involving further SM fermions, thus
also not generating Higgs-gluon contact interactions at one-loop. % \com{Mixing? Check cancellation (1301.5856),
%then would have no effect at all... how in hh?}. 

\begin{table}[th!]
\begin{center}
\begin{tabular}{c|c| c c c}
Model & Free Parameters & $\kappa_\lambda$ & $\kappa_t$ & $c_2$ \\
\hline
1 & $\kappa_t, \kappa_\lambda$ & $\kappa_\lambda$ & $\kappa_t$ & $ \kappa_t -1$ \\
2 & $\kappa_t$ & $3 \kappa_t - 2$ & $\kappa_t$ & $ \kappa_t -1$ \\
3 & $\kappa_\lambda, c_2$ & $ \kappa_\lambda$ & $1$ & $c_2$ \\
4 & $\kappa_t, \kappa_\lambda$ & $ \kappa_\lambda$ & $\kappa_t$ & $ 2 \kappa_t -2$ \\
5 & $\kappa_\lambda$ & $ \kappa_\lambda$ & $1$ & $0$ \\
6 & $\kappa_\lambda$ & $ \kappa_\lambda$ & $1$ & $0$ \\
\hline
\hline
7 & $\kappa_t, \kappa_\lambda$ & $ \kappa_\lambda$ & $\kappa_t$ & $3/2 (\kappa_t-1)$ \\
\hline
\hline
8 & $\kappa_t$ & $1$ & $\kappa_t$ & $3/2 (\kappa_t-1)$ \\
9 & $\kappa_t$ & $\kappa_t$ & $\kappa_t$ & $0$ \\
\hline
\hline
10 & $\kappa_t$ & $ \kappa_t$ & $\kappa_t$ & $\kappa_t(\kappa_t+\sqrt{\kappa_t^2+8})/4-1$ \\
11 & $\kappa_t$ & $ \kappa_t$ & $\kappa_t$ & $(\kappa_t^2-1)/2$ \\
\end{tabular}
\end{center}
\caption{\label{tab:EffCoup} Correlations between effective couplings in various explicit models. While the total {\it number} of free parameters
is invariant, when there was freedom {\it which} parameters to treat as free, we always chose $\kappa_t$, expressing the other
couplings in terms of the latter.
%\com{\small Expand MCHMs around $\kappa_t=1$? Check everything... in particula VLL}
}
\end{table}

\begin{table}[th!]
\centering
\begin{tabular}{c|c| c c c}
Model & Fund. Parameters & $\kappa_\lambda$ & $\kappa_t$ & $c_2$ \\
\hline
1 & $\alpha, m_2, \lambda_\alpha$ & $ 1 - \frac 3 2 t_\alpha^2 + t_\alpha^2\, (\lambda_\alpha - t_\alpha\frac{m_2}{v})/\lambda_{\rm SM} $ &
  $ 1 - \frac{t_\alpha^2}{2} $ & $-\frac{t_\alpha^2}{2}$ \\
2 & $\alpha$  &  $ 1 - \frac 3 2 t_\alpha^2 $ & $ 1 - \frac{t_\alpha^2}{2} $ & $ -\frac{t_\alpha^2}{2} $\\
3 & $\beta, m_{H^+}, m_H$ & $ 1 + 4 s_\beta^2\, (3 + \frac{m_{H^+}^2}{v^2 \lambda_{\rm SM}})\frac{m_{H^+}^4 }{m_H^4} $ & 
  $ 1 $ & $- 2 s_\beta^2 \frac{m_{H^+}^4}{m_H^4}$ \\
4 & $\beta, m_A, m_H$ & $ 1 + 2 s_\beta^2\, (3 + \frac{4 m_A^2}{v^2 \lambda_{\rm SM}})\frac{m_A^4}{m_H^4} $ & 
$  1 - 2 s_\beta^2 \frac{m_A^4}{m_H^4} $ & $-4 s_\beta^2 \frac{m_A^4}{m_H^4} $ \\ 
5 & $\beta, m_A, m_H$ & $ 1 + \frac{24}{7} t_\beta^2\frac{m_A^4}{m_H^2 v^2 \lambda_{\rm SM}} $ & $1$ & 0 \\ 
6 & $\beta, m_A, m_H$ & $  1 + \frac{8}{3} t_\beta^2 \frac{m_A^4}{m_H^2 v^2 \lambda_{\rm SM}} $ & $1$ & 0 \\
\hline
\hline
7 & $\beta,Z_6,m_H$ & $ 1 - \frac{3 Z_6^2}{2\lambda_{\rm SM}} \frac{v^2}{m_H^2}$ & $1 -  \frac{Z_6}{t_\beta} \frac{v^2}{m_H^2}$ & $-\frac{3 Z_6}{2t_\beta} \frac{v^2}{m_H^2}$ \\
\hline
\hline
8 & $\lambda_{Tt},M_T$ & $1$ & $ 1- V_{tb}\frac{|\lambda_{Tt}|^2 v^2}{2M_T^2}    $ & $- 3 V_{tb} \frac{|\lambda_{Tt}|^2 v^2}{4M_T^2} $ \\
9 & $\lambda_{E\ell},M_E$ & $ 1 + \frac{|\lambda_{E\ell}|^2v^2}{4M_E^2}  $ & $ 1 + \frac{|\lambda_{E\ell}|^2 v^2}{4M_E^2} $ & $0$ \\
\hline
\hline
10 & $\xi$ & $ \frac{(1 - 2 \xi)}{\sqrt{1 - \xi}}$ & $\frac{1 - 2 \xi}{\sqrt{1 - \xi}}$ & $-2\xi$ \\
11 & $\xi$ & $\sqrt{1 - \xi}$ & $\sqrt{1 - \xi}$ & $-\frac{\xi}{2}$ \\
\end{tabular}
\caption{\label{tab:paras} Effective couplings in terms of physical parameters of various explicit models.
Here, $\alpha$ is the mixing angle between the two scalars, while $\beta = \arccos(v_1/v)$ is the arccosine 
of the ratio of the vev of the (first) doublet and the electroweak vev $v\approx246\,$GeV, and we defined 
$s_x \equiv \sin x$ ($t_x \equiv \tan x$). A common mass $m_H$ is assumed for the heavy scalars, besides 
for $H^+$ and $A$ in models 3-6, with masses $m_{H^+}, m_A$. Beyond that, $m_2$ is the coefficient 
of the triple-singlet coupling and $\lambda_\alpha$ that of the bi-quadratic scalar term, 
while $Z_6$ multiplies $|H_1|^2H_1^\dagger H_2$ in the 2HDM. Moreover, $M_T$ and $M_E$
are the masses of the heavy vector-like quark and lepton, respectively, and 
$\lambda_{Tt},\lambda_{E\ell}$ are the coefficients of their (Yukawa-type) couplings with the 
SM fermions, mediated by the Higgs. Finally, $\xi \equiv v^2/f^2$ 
parametrizes the composite Higgs non-linearity, with $f$ the Pseudo-Goldstone decay constant.
See references given in Table \ref{tab:models} for more details. 
%\com{$\lambda$ vs. $\lambda_{\rm SM}$?; VL T $\to$ change to flavor universal?
%Replace $Z_6 v^2/m_H^2$ with $-c_{\beta-\alpha}$? VL quarks fine (prepared
%mail to answer Maxime)}
} 
\end{table}

The situation becomes more subtle for the composite Higgs setups (models 10 and 11). Here,
the potential effect of loop-induced Higgs-gluon contact interactions due to integrating out 
fermionic resonances appearing in the models (that could become relevant in parts of the parameter 
space) cancels with additional corrections to the Yukawa couplings generated by the very same same fields. 
Thus, considering only the Higgs-non-linearities that lead to the anomalous couplings as given in Table~\ref{tab:EffCoup}
(see also Table \ref{tab:paras}) leads effectively to a correct description. In summary, the effective couplings
given in Table \ref{tab:EffCoup} provide an appropriate description of all models at hand to leading approximation,
given that the NP is heavy such that the EFT framework is valid (see, {\it e.g.}, \cite{Contino:2016jqw}).

Using the map from the model parameters to the anomalous couplings in Tables~\ref{tab:EffCoup} and \ref{tab:paras}, 
we determine the sensitivity of the LHC analyses to the independent model parameters
(which differs between the various models due to the different correlations). The reweighted differential information 
is used to determine the expected number of events that populate the signal region in each analysis and a parameter point is 
excluded if it exceeds the upper limits derived in Section~\ref{sec:Recast}. Given the low amount of data in 
the analyses considered, together with the inherent difficulty of the channel, we do not expect to provide competitive 
limits on the models at the current stage, but rather present these results as an academic exercise 
to demonstrate the application of our tool in the future.\footnote{Note that 
at the current stage we do {\it not} include the effect of the anomalous couplings
on the Higgs decays ($\kappa_t$ is entering $h\to \gamma \gamma$ at the one-loop level
and has an indirect effect on the branchings by changing $h\to gg$) as well as
further operators generated in the models at hand that can modify the Higgs boson branching ratios (in particular $h \to bb$). 
As the Higgs boson is a narrow particle those effects do not change the hh kinematics and the effect is secondary with respect with the scope of this paper.  
%These effects will be included in v2 of the paper. %\com{Give good reason!!}
} 

\begin{figure}
  \centering
  \includegraphics[width=0.47\textwidth]{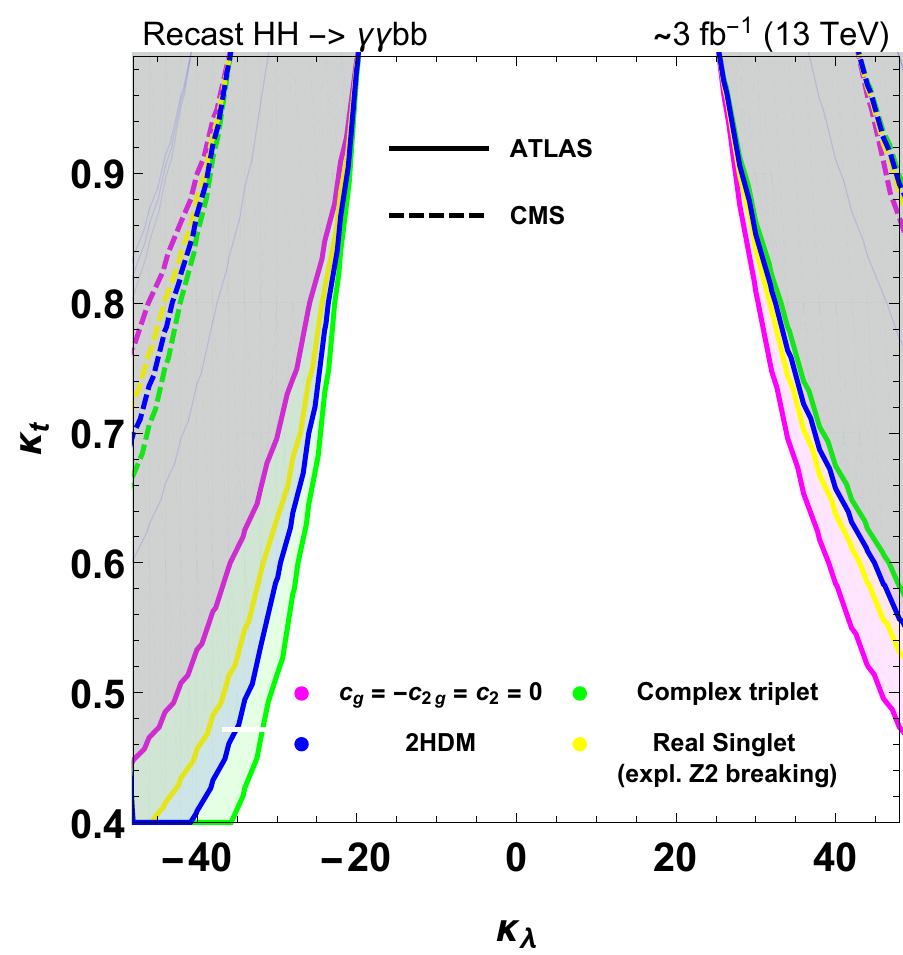}
  \includegraphics[width=0.47\textwidth]{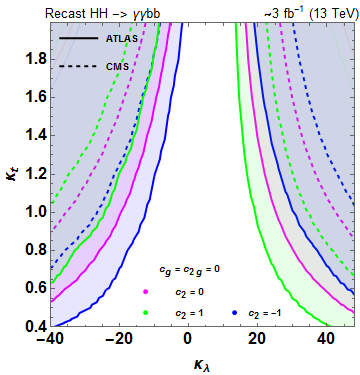}
  \caption{\footnotesize ATLAS (continuous lines) and CMS (dashed lines) 95\% exclusion bounds in 
the $\kappa_\lambda - \kappa_t$ plane, reinterpreted in terms of different models (left) and in the pure EFT
framework for different values of $c_2$ (right). See text for details.
    \label{fig:toyklkt}
  }
\end{figure}

We begin with the limits obtained for models with additional scalars in the $\kappa_\lambda - \kappa_t$ plane, given in 
Figure~\ref{fig:toyklkt}. In the left plot, the cases of a real scalar singlet with explicit $Z_2$ breaking, a
complex triplet scalar, and the 2HDM are compared with the scenario where only these two effective couplings are allowed to vary and the others are set to zero\footnote{That has been considered by the CMS collaboration }. 
Note that in each of the models considered, only  $\kappa_t$ between 0 and 1 are permitted
\footnote{While a large depletion seems only viable in the presence of cancellations with other NP contributions 
to single Higgs boson production and decay, here we focus mostly on $hh$ production and only
take into account rigorous theoretical constraints at this stage, leaving a combined phenomenological analysis 
for future work. }. % \com{I added this... you agree with the comment?}
In these examples, the presence of $c_2$ (for $\kappa_t \neq 1$) leads to mild variations in the final sensitivity in this plane. 
The impact of $c_2$ on the bounds is further quantified in the right plot, where we show the effect of setting this coefficient to $\pm1$, which shifts 
the contours left or right - in agreement with the tendencies observable in the left plot.

\begin{figure}[h]
  \centering
  \includegraphics[width=0.32\textwidth]{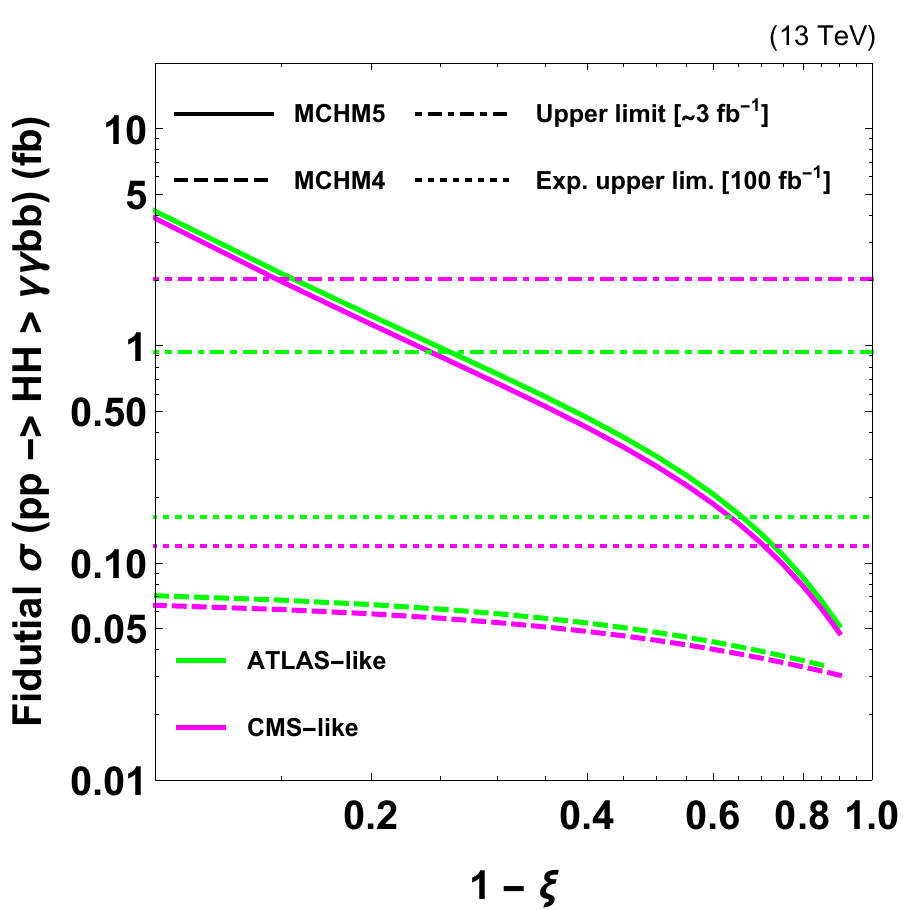}
  \includegraphics[width=0.32\textwidth]{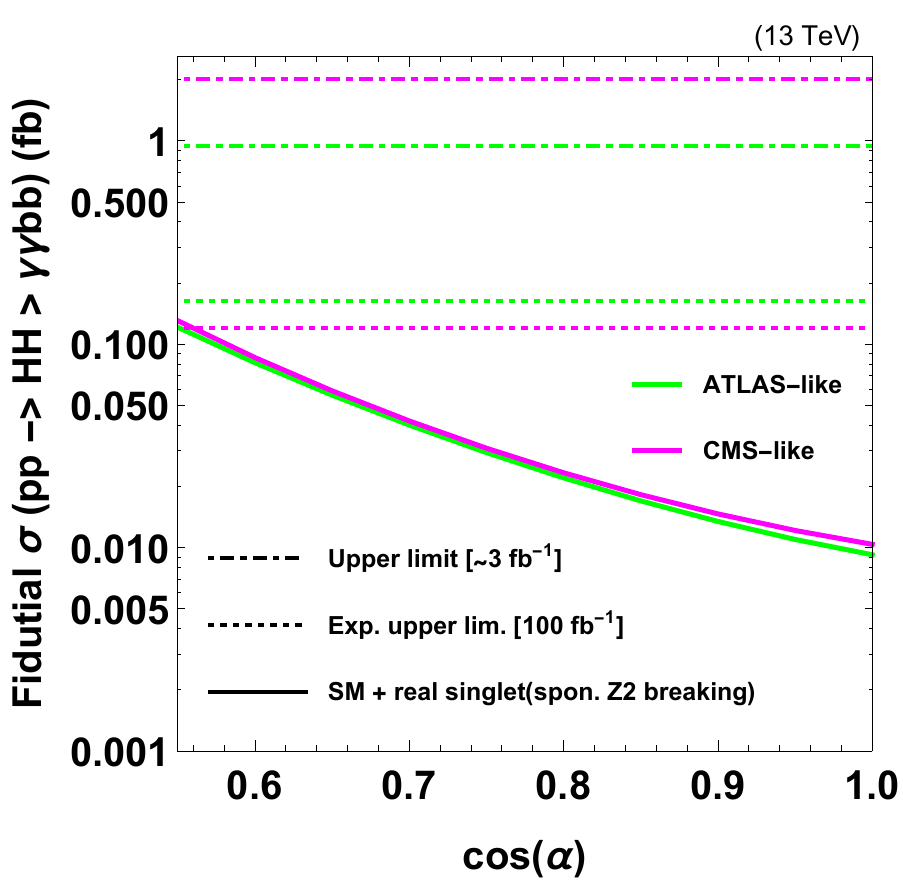}
  \includegraphics[width=0.3\textwidth]{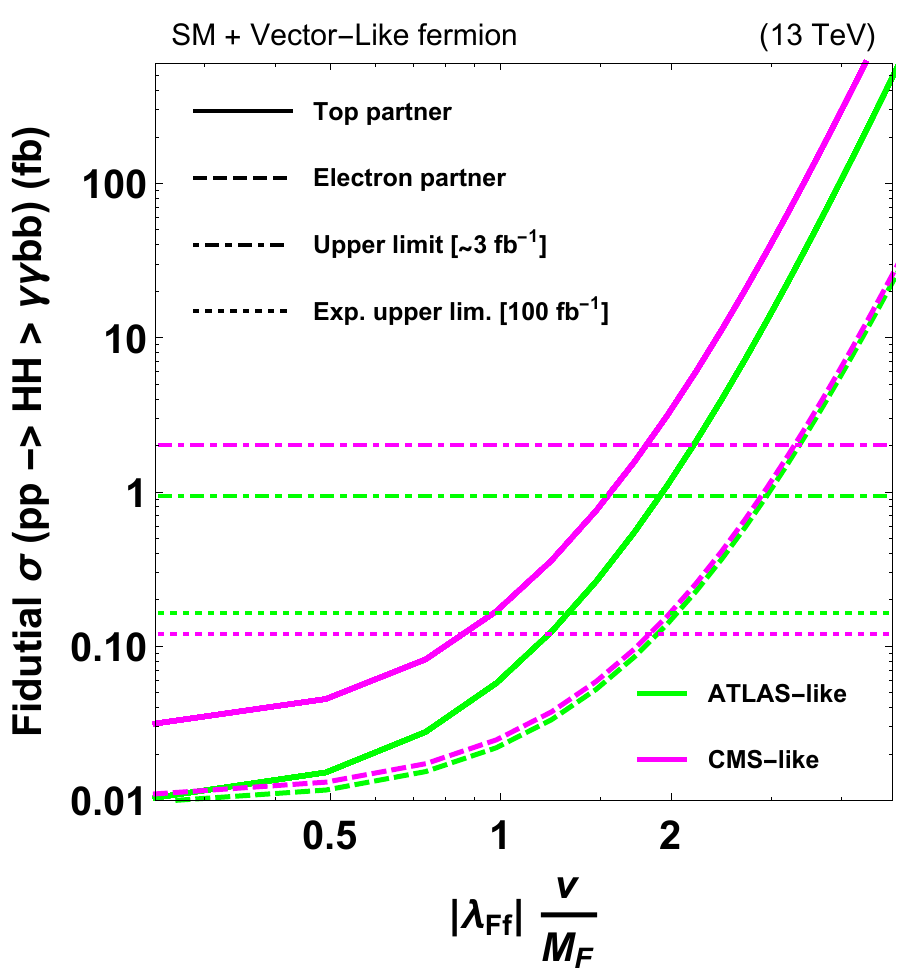}
  \caption{\footnotesize Predictions of the one-parameter
models in Table~\ref{tab:models} and 95\% CL exclusion bounds. See text for details.
% \com{Axis labels? Electron $\to$ Lepton? For $\cos(\alpha) \to 0$ should explode... also for $\xi \to 1$ (cut?) $\to$ 
%Limits... . In light of poor constraints, shouldn't we go for HL LHC projections???}
    \label{fig:Recast1DModels}
  }
\end{figure}

Moving to the models which can be described with one parameter (combination), we show in Figure~\ref{fig:Recast1DModels} the fiducial cross section as a function of the free (fundamental) parameter for the MCHM (left), real singlet scalar with spontaneous $\mathbb{Z}_2$ breaking (center), and singlet vector-like fermion (right) models. 
For comparison, we also provide current exclusion limits from ATLAS and CMS as well as projections to $100\,$fb$^{-1}$, that
will exclude the part of the parameter space corresponding to a larger fiducial cross section.
%\com{Shouldn't we go for 3 inv ab? HH is in fact a high lumi project...}

For both MCHM scenarios, the signal efficiency is observed to be rather flat as the kinematics are close to the SM-like case. % \com{$\to$ linear behavior... not fully evident from plots/formulas... c2 should change kinematics...}.
The sensitivity thus is mostly determined via a simple rescaling of the total rate. The current analyses are sensitive to values of $\xi\sim0.9$ in the case of the MCHM$_5$ and values of $\xi\sim0.2$ can be probed with $100\,$fb$^{-1}$, while no bound is obtained for the MCHM$_4$. We observe a similar pattern in the case of the singlet, where the kinematics are not greatly affected and no bound is obtained for mixing angles $\cos \alpha \gtrsim 0.5$. % \com{Comment on
%other bounds for alpha? Rho parameter constraints for triplet?}. 

In the case of vector-like fermions, the free parameter scales with the Yukawa mixing over the mass of the new states
and a decoupling is exhibited with the latter approaching infinity, as expected. % \com{E and T should approach same limit...}. 
As shown in Table~\ref{tab:EffCoup}, the vector-like quark model does not modify $\kappa_\lambda$. Combining this with the sensitivity information of Figure~\ref{fig:toyklkt}, it is not surprising that (expected) limits are rather weak, stemming mostly from a non-vanishing $c_2$ and 
residing in the region of large couplings ($\lambda_{Tt}\gtrsim 1$) and/or small masses
($M_T \lesssim 1$\,TeV). The vector-like leptons, on the other hand, do modify $\kappa_\lambda$ and also identically affect $\kappa_t$. Since $c_2$ is unchanged, they correspond exactly to the benchmark scenario considered in the experimental analyses (with the additional constraint $\kappa_t \equiv \kappa_\lambda$) and the limits are also rather weak. 

\begin{figure}
  \centering
  \includegraphics[width=0.32\textwidth]{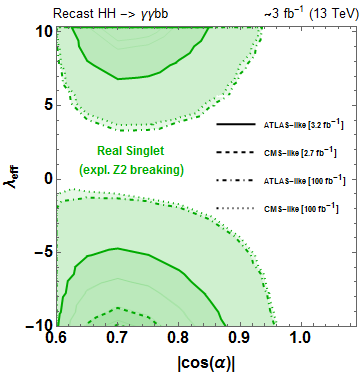}
  \includegraphics[width=0.32\textwidth]{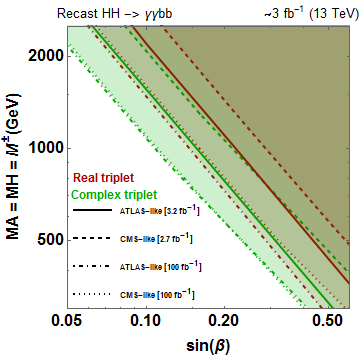}
  \includegraphics[width=0.31\textwidth]{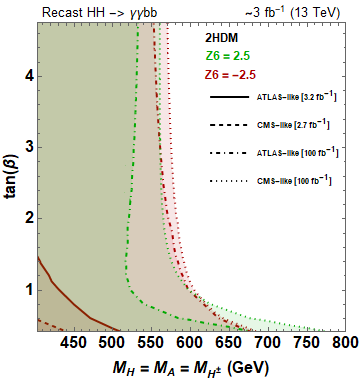}
  \caption{\footnotesize ATLAS (continuous lines) and CMS (dashed lines) 95\% exclusion bounds (as well as projected constraints) 
for different explicit models given in Table~\ref{tab:models}, as derived in the EFT framework. 
%{\bf Those are toy exclusions, most of the points needs to be run. 
%\com{Singlet region should close for cos to zero... mention? Cut off triplet plot for low masses (EFT not valid): 500\,GeV? and high mass (perturbativity): 1 TeV? Could we plot alignment parameter on upper x-axis? 
%2HDM regions open again for low tanb? Check behavior. Check for effects of other operators!}} 
    \label{fig:Recast2DModels}	
  }
\end{figure}

Finally, Figure~\ref{fig:Recast2DModels} displays the limits obtained for the models with more than one free parameter, this time in terms of the different
'fundamental' model parameters (and including the real scalar triplet): the real singlet with explicit $\mathbb{Z}_2$ symmetry breaking (left), the real/complex triplet (center) and the 2HDM (right). Where possible, we reduce the parameter space to two degrees of freedom:
In the triplet models we set the heavy scalar masses to be equal while for the real singlet we can define an effective trilinear coupling
\begin{align}
    \lambda_{\text{eff}} = \lambda_\alpha-t_\alpha\frac{m_2}{v} \,.
\end{align}
In the 2HDM, it is not trivial to reduce the $(t_\beta, Z_6, m_H)$ parameter space. However, $Z_6$ is related to the usual alignment parameter, $\cos_{\beta-\alpha}$ and the neutral scalar masses by~\cite{Belusca-Maito:2016dqe}
\begin{align}
    v^2Z_6 = -\cos_{\beta-\alpha}\sin_{\beta-\alpha}(m_H^2-m_h^2),
\end{align}
%Global fits to Higgs signal strengths 
%\textbf{\small [CITE, also for alignment relation above... Falkowski 2HEFT?]} 
%constrain this alignment parameter to be not too far from zero. We choose to fix it to $\cos_{\beta-\alpha}=0.2$ to obtain a relation between $Z_6$ and $m_H$. 
which we use to fix a relation between the latter, showing predictions for two fixed values of $Z_6=\pm 2.5$.\footnote{Fixing a relation between the alignment parameter and the heavy scalar mass (keeping $Z_6$ finite) 
leads to a proper decoupling behaviour, with $m_H \to \infty$.}

The current constraints on the real singlet lie in regions of large $\lambda_{\text{eff}}$ of order 5--10 and of sizable 
mixing with the Higgs. The bounds will however improve significantly with increasing luminosity. For the triplets, the analyses are sensitive to large mixings for masses of a few hundred GeV while increasing the heavy scalar mass leads
to an increased sensitivity, including also smaller mixing angles $\beta$. % \com{Relate to precision EW? 1410.7703}. 
This subtle behaviour is due to the fact that 
$m_H^2$ is proportional to the size of scalar quartic interactions (times the electroweak vev squared), such that the coupling strength increases
with the scalar mass. We thus cut off the parameter space at $m_H^2 \sim 4 \pi v^2 \sim 1$\,TeV, in order to remain at a reliable (perturbative)
behaviour of the theory. %\com{Check Appendix B of \cite{Dawson:2017vgm}... there are clearly some O(1) factors,
%and cancellations might happen (what about explicit mass $M^2$?), but I guess the (naive) limit above is fine...} \com{Models 5,6?}
Finally, in the case of the 2HDM, a sensitivity to scalar masses in the TeV region is only possible for small $\tan\beta$, which amplifies the corrections to $\kappa_t$ and $c_2$. %\com{Comment further, including relation with alignment parameter?}
%\com{Comment on 1504.05596, 1703.10614?}

\section{Mapping between smooth scan and shape benchmarks}
\label{sec:map}

Most of the times the experimental searches are not easy to recast as soon as shape analysis or multivariative analysis is used. In this case we may ask ourselves if at some level of confidence we can use results obtained for shape benchmarks for a general study of the parameter space. 
%level of confidence we can use the results for a general interpretation of the results presented before, expressed in terms of shape benchmarks. 
In this chapter we test this premise on the scans presented in the last chapter. 
We verify how the same criteria used to define the shape benchmarks as representatives of clusters of similar shapes from a large parameter space scan can help to predict the closest experimental limit (provided there is a list of results for the shape benchmarks), and how this limit can approximate the ``real`` limit .

As a quick reminder, in Ref.~\cite{Carvalho:2015ttv} a two sample  Test Statistic (TS) was defined to order the degree of similarity between two samples. The log-likelihood ratio function of the hypothetical case in which the two samples under test share the same parent distribution is the product over the bins of the probability to observe $n_{i,1}$ and $n_{i,2}$ event counts in bin $i$ from the two samples $S_1$ and $S_2$ and can be written as:    
\begin{equation}
TS =  -2 \sum_{i=1}^{N_{bins}} \left[ log(n_{i,1}!) + log(n_{i,2}!) -2log\left(  \frac{n_{i,1}+n_{i,2}}{2}! \right) \right]\,.
\label{eq:TS}
\end{equation} 
This quantity is constructed in a manner that it is ``$\chi^{2}$ distributed``~\cite{bakercousins,Wilks:1938dza} and therefore can be directly used as an ordering parameter to decide between pairs of test samples which of them are the most likely to be  compatible with the same parental distribution.  In other words, the values $TS_{ij}$ and $TS_{kl}$ obtained respectively by testing the compatibility of samples $ij$ and $kl$ are suitable to determine if samples $S_i$ and $S_j$ are more similar to each other than are  samples $S_k$ and $S_l$: this is the case if $TS_{ij}>TS_{kl}$. %\com{change a bit this last sentence, it is copy and paste of the clustering paper.}
 
One solution for constructing analyses optimal in continuous scans is, for example, the use the $TS$ as quantity to decide, given a test sample to which shape benchmark this one is most similar to and check if this prediction corresponds to the closest experimental limit. At this point we should emphasise that we are after the {\it closest} experimental limit. The {\it real} experimental limit, that would be directly derived for the test point, would be a bit different from the one from the {\it closest} shape benchmark. 

\begin{figure}[h]
  \centering
  \includegraphics[width=0.42\textwidth]{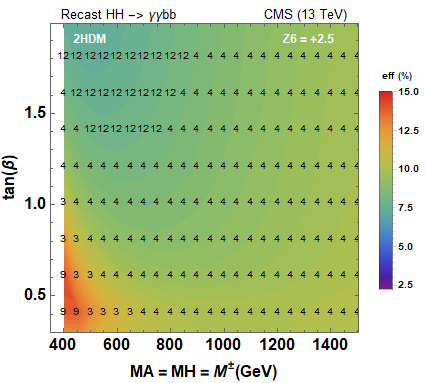}
  \includegraphics[width=0.42\textwidth]{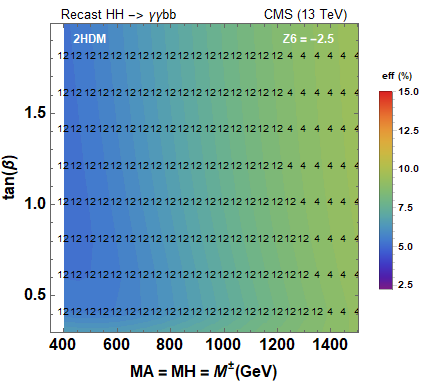}
  \caption{\footnotesize Efficiency maps for the signal region of the CMS-like analysis to the 2HDM benchmarks we consider. The markers superimposed correspond to the closest shape benchmark. Details in the text. 
    \label{fig:Map2HDM}
  }
\end{figure}

\begin{figure}[h]
  \centering
  \includegraphics[width=0.32\textwidth]{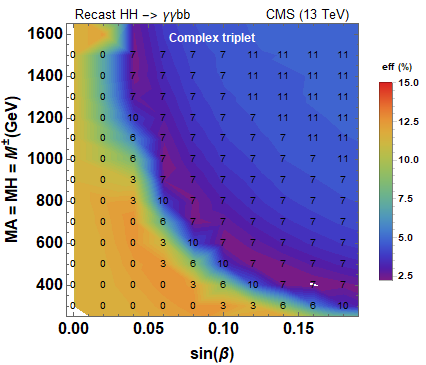}
  \includegraphics[width=0.33\textwidth]{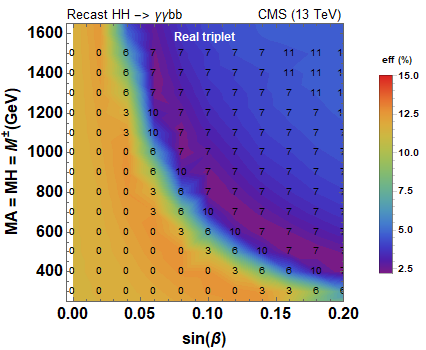}
  \includegraphics[width=0.32\textwidth]{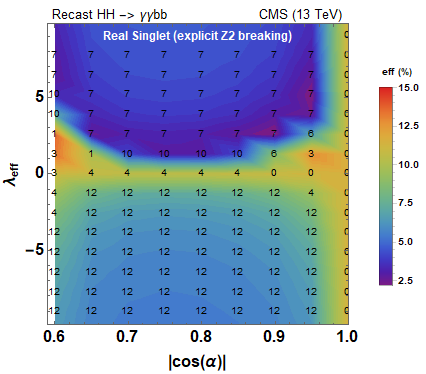}
  \caption{\footnotesize Efficiency maps for the signal region of the CMS-like analysis to the triplet and singlet benchmarks we consider. The markers superimposed correspond to the closest shape benchmark. Details in the text.
    \label{fig:MapSinglet}
  }
\end{figure}

Since we are considering cut-and-count examples the "experimental limits" can be well identified with the signal efficiency in the signal region. As the CMS-like analysis contains a cut in the $m_{\gamma\gamma b\overline{b}}$ variable the signal efficiency is more sensitive to the BSM physics, therefore we will only do this mapping exercise for the CMS-like case.     
In figures~\ref{fig:Map2HDM} and~\ref{fig:MapSinglet} we show the efficiency maps for the cases of the two 2HDM scans and the triplet and singlet extension cases. Of course the quality of color interpolation on the figures depend on the density of points inspected. To each inspected point a marker is superimposed, symbolizing the closest shape benchmark according to eq.~\ref{eq:TS}. The first fact to notice is that as suggested in~\cite{Carvalho:2015ttv}, the regions belonging to the same benchmark tend to enclose fully connected regions of parameter space, that we will call {\it islands}. We also notice that within those fully connected regions the closest benchmark indeed has a signal efficiency that is the closest to the true one. This conclusion is not so precise near the boundaries between islands.

This approach has multiple applications: If used by the theory side it is an approximation of  reality, that can be employed to obtain a first estimate of the constraints on a specific model  without the need to expensive MC simulations or recast. This usage would hold for the case of including further BSM effects such as the inclusion of additional EFT operators (for example, the chromomagnetic term~\cite{Maltoni:2016yxb}) or the departure from the EFT framework towards explicit inclusion of particles in the loops (for example~\cite{Dawson:2015oha,Cacciapaglia:2017gzh}).   
On the experimental side usage, we can imagine a situation where it is advantageous for the discovery potential of a hypothetical BSM signal to construct different selections (or multivariate variables) for the different shape benchmarks. In this case it would be unclear how to extend the interpretation of the search to a smooth scan to decide which set of cuts for use to a point that is not a shape benchmark, and the $TS$-test offers a solution for keeping the best search sensitivity also for the case of smooth parameter space scans. 

\section{Implementation in Rosetta}

Although the numerical values of the coefficients that can be used to construct the signal predictions are given in appendix~\ref{sec:appCoef}, for  convenience, the input and algorithm to perform sample reweighing are implemented as a module of the {\sc Rosetta} package~\cite{Falkowski:2015wza}. {\sc Rosetta} provides a framework for EFT basis translations along with a suite of modules ranging from the calculation of Higgs branching fractions via an interface to {\sc eHDECAY}~\cite{Contino:2014aaa} to the compatibility of a given parameter space point to electroweak precision data and Higgs signal strength measurements. All modules can be exploited in a basis-independent way once a basis definition has be implemented in the package. The package may be found on \url{http://rosetta.hepforge.org/}.

The results of this paper are implemented as in extension of the existing \texttt{dihiggs} module~\cite{Brooijmans:2016vro}. In this module we had implemented the example of usage of the elements of the machinery that do not require expensive Monte Carlo simulations, namely the mapping of parameter space points into shape benchmarks done in the last section. From this implementation it is straightforward 
for the user to extend it to the case where the calculation of event weights is of interest.
After downloading the package, one can invoke the \texttt{dihiggs} interface by calling the command line executable,
\begin{verbatim}
>> bin/rosetta dihiggs [OPTIONS]
\end{verbatim}
The package is designed to receive SLHA formatted parameter cards specifying the coefficients of an existing basis implementation. The required translations are then performed to obtain the anomalous Higgs couplings parameters relevant for pair production and decay. Alternatively, one can directly specify the values of the anomalous coupling parameters  via the self-explanatory \texttt{--kl}, \texttt{--kt}, \texttt{--c2}, \texttt{--cg} and \texttt{--c2g} optional arguments. The input parameter card has therefore been made optional, to be specified by the \texttt{--param\_card} option.

Along with computing the inclusive Higgs pair production cross section, the module will also compute the correct Higgs branching fractions for that point in parameter space using either {\sc eHDECAY} or an internal interpolating function. Using the results of this paper, we have also added the functionality to return the closest benchmark according to the test defined in formula~\ref{eq:TS}. Finally, the computation of the higgs-pair production process is promoted to differential level, via an internal function.
Inside \url{interfaces/dihiggs/AnalyticalReweighter.py} the user can find a function \texttt{weight(variables)}, where a matrix of weights is calculated from vectors containing the generation level variables $m_{HH}$ and $\cos\theta^*$ according to formula~\ref{eq:weight}. Support for implementing the same algorithm in C++ fashion may be given by directly emailing us.

\section{Conclusions}

%\com{Clean Preamble!}

Understanding the properties of the Higgs potential is of utmost importance
to eventually answer one of the key questions in particle physics, namely what is the origin of EWSB.
Examining the production of Higgs pairs is a crucial experiment to achieve this
goal, with a focus both on the total cross section, as well as on distributions, which contain
valuable information on the nature of potential NP.

In this article we presented semi-analytic  results for the distributions of the Higgs-pair production cross section in a well defined and maximally general parametrization of nature, as it appears at low energies, employing EFT.
This allows us to express potential deviations from the SM in a consistent way as coefficients of effective 
operators and constraints on these operators will provide us a guidance on how nature could look at shortest distances.

Furthermore, we employed the formula to recast exclusion bounds, derived assuming certain benchmark
points in parameter space with a given kinematics, to points with modified kinematic distributions.
The presented method is crucial to cover the full EFT parameter space, taking correctly
into account the efficiencies of signal selections.

Finally, the results presented are also useful to confront explicit models, mapped to an EFT, 
efficiently with constraints from the LHC, providing a bridge between (explicit) theories
and data. We demonstrated this procedure, using recent ATLAS and CMS results,
for various NP setups, like models with additional scalars, including 2HDM, vector-like fermions, and minimal composite
Higgs models, delivering also a dictionary between their explicit parameters and 
effective couplings after electroweak symmetry breaking.

%It turns out that \com{...} feature the best prospects for competitive constraints on its parameters...
%\com{Singlet with expl Z2break unique regarding hh: dependence on lambda, other models could be fully constrained without
%$hh$?!}.

\section{Acknowledgements}
FG and AC are grateful for the hospitality of the CERN theory division during the completion of this work.
A.C is grateful to Andre Tinoco Mendes, Luca Cadamuro, Giacomo Ortona, Olivier Bondu, Konstantin Androsov, Andrey Pozdnyakov, Rafael Teixeira de Lima, Martino Dall'Osso and Tommaso Dorigo for the fruitful discussions and encouragement to pursue this work.  K. M. is supported in part by the Belgian Federal Science Policy Office through the Interuniversity Attraction Pole P7/37 and by the European Union’s Horizon 2020 research and innovation programme under the Marie Skłodowska-Curie grant agreement No. 707983.

%make public tools that allow the LHC experiments to provide a detailed interpretation of the non-resonant di-Higgs production from finite the computing resources.  Our goal is to facilitate a global combination of the HH results with other processes in specific BSM models.
 
\begin{appendices}
%\clearpage\mbox{}\clearpage
\clearpage
\appendix
\section{Numeric tables with the coefficients}
\label{sec:appCoef}
\begin{figure}[h]
  \centering
  \includegraphics[width=0.42\textwidth]{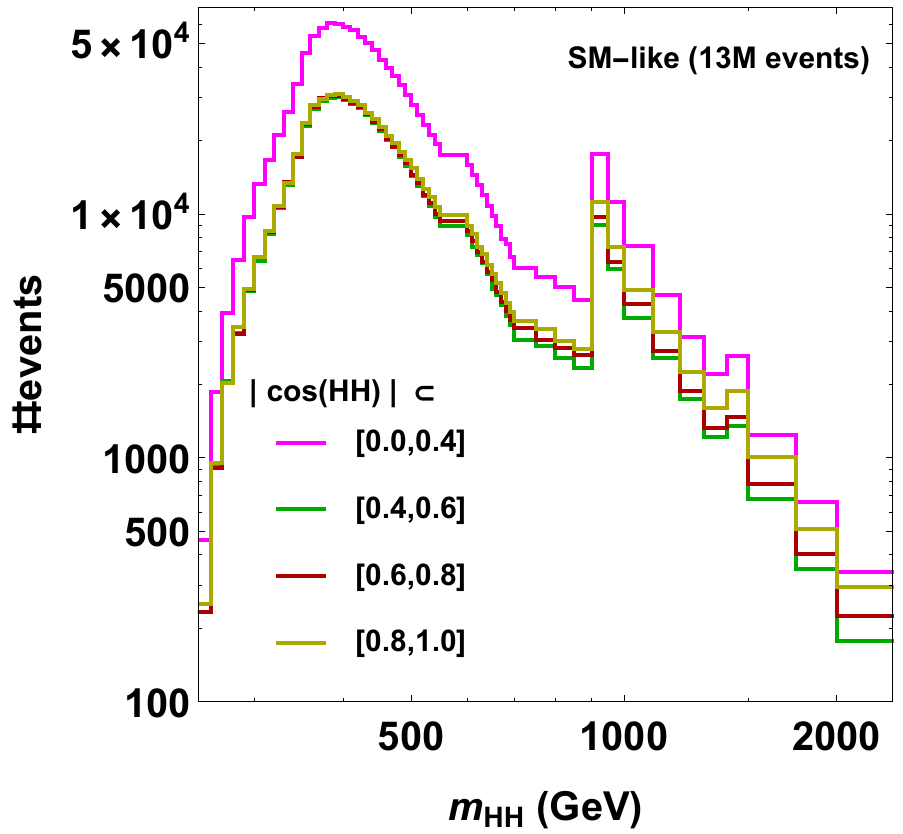}
  \includegraphics[width=0.42\textwidth]{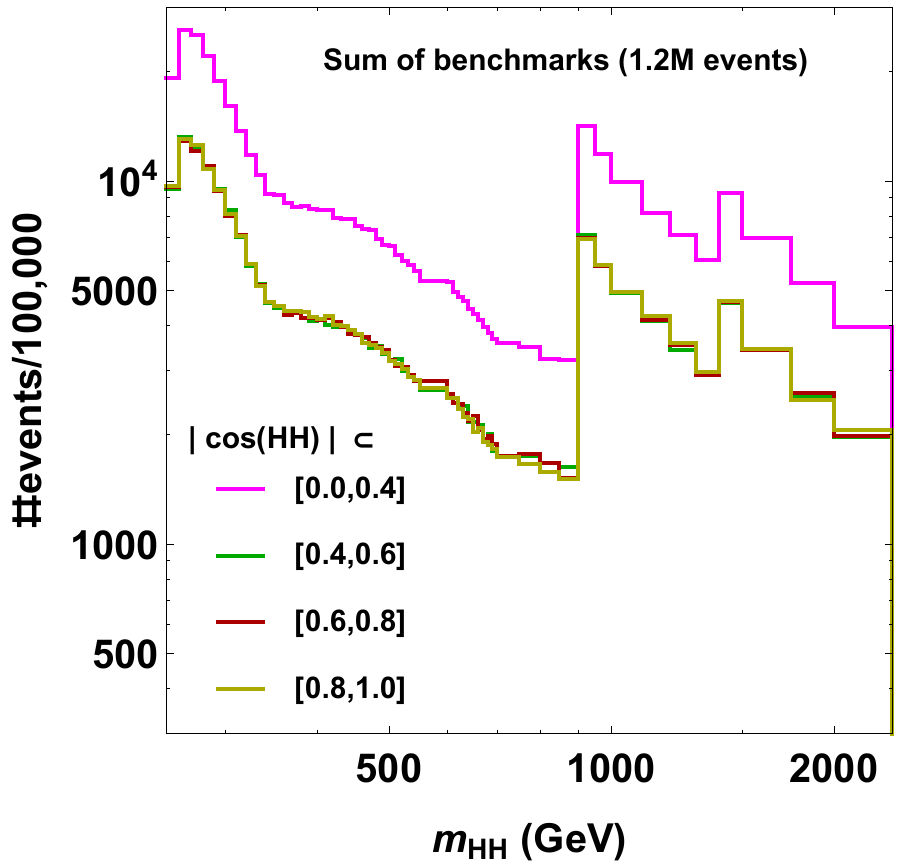}  
  \caption{\footnotesize The bin-by-bin statistic of events in a 3M SM sample and on a 1.2M reweighting sample constructed by the plain sum of shape benchmarks. 
    \label{fig:Nenvents}
  }
\end{figure}

\begin{figure}[h]
  \centering
  \includegraphics[width=0.32\textwidth]{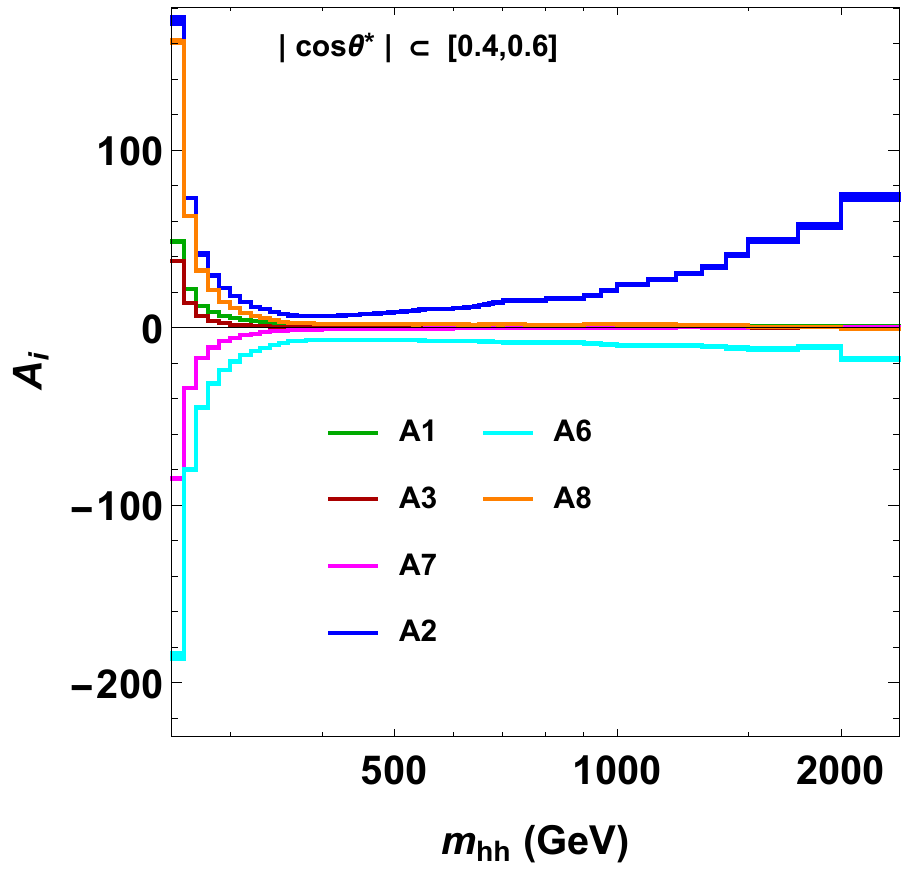}
  \includegraphics[width=0.32\textwidth]{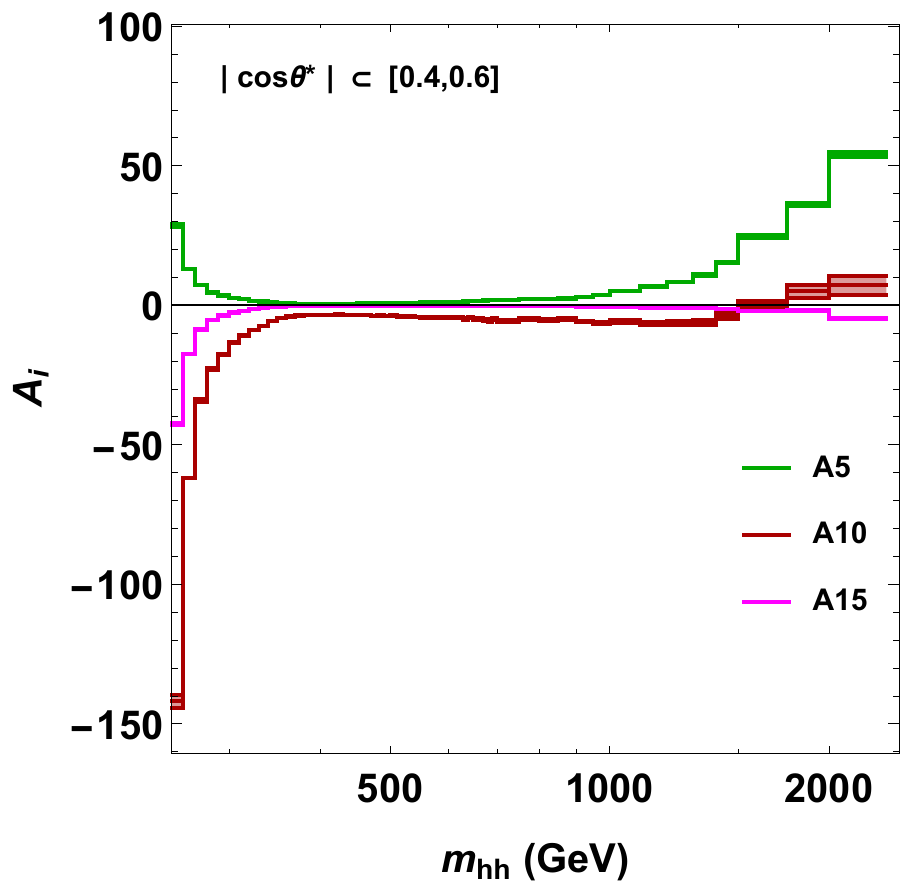}
  \includegraphics[width=0.32\textwidth]{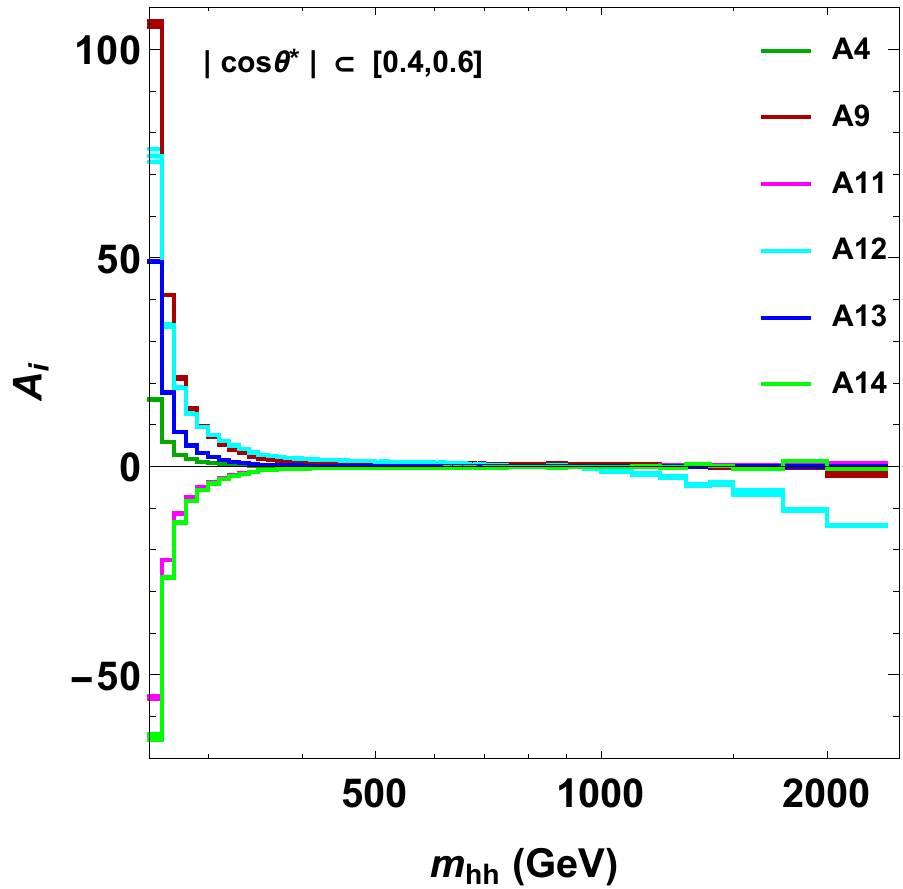}
  \includegraphics[width=0.32\textwidth]{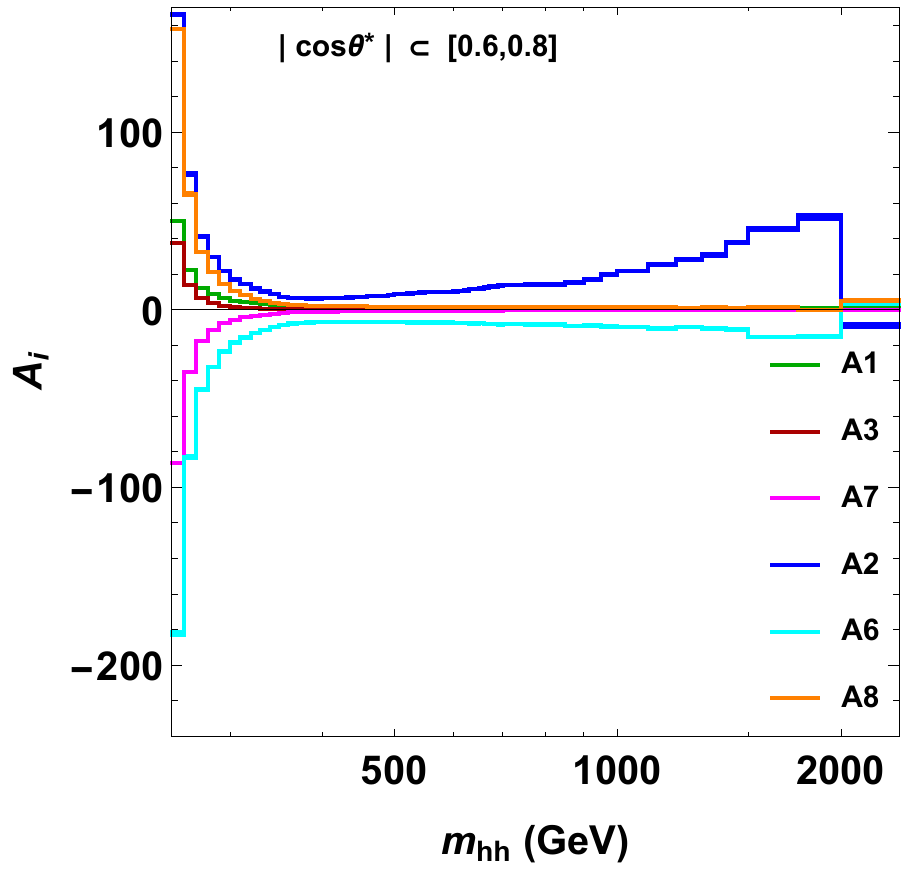}
  \includegraphics[width=0.32\textwidth]{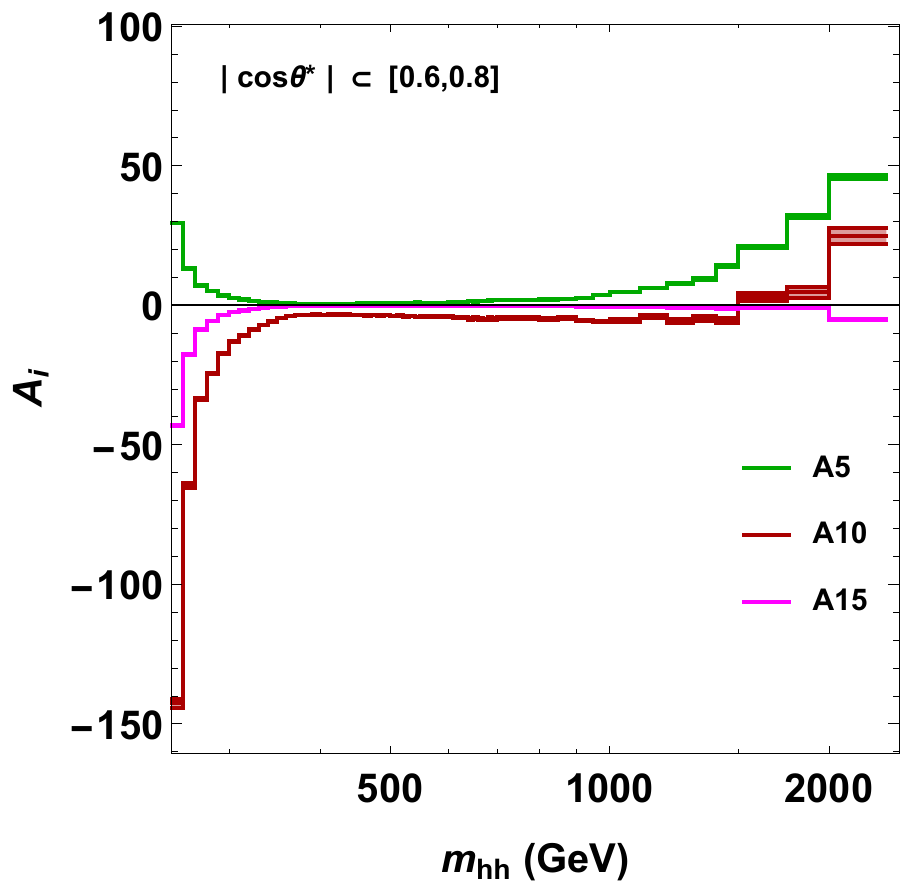}
  \includegraphics[width=0.32\textwidth]{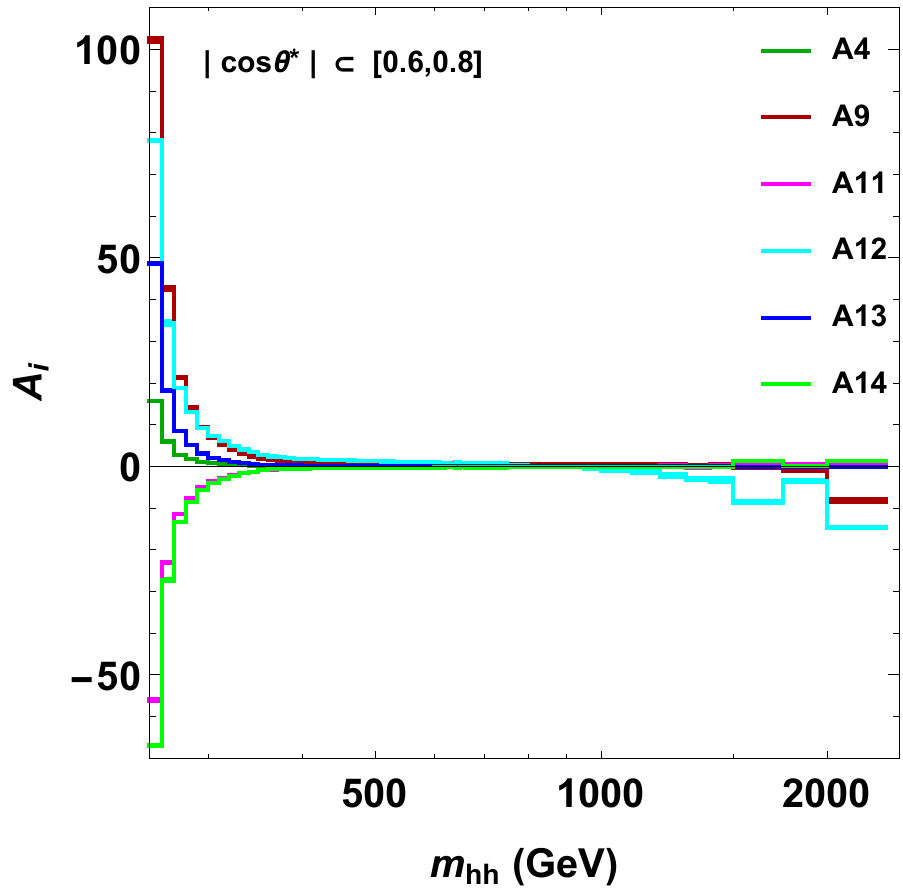}
  \includegraphics[width=0.32\textwidth]{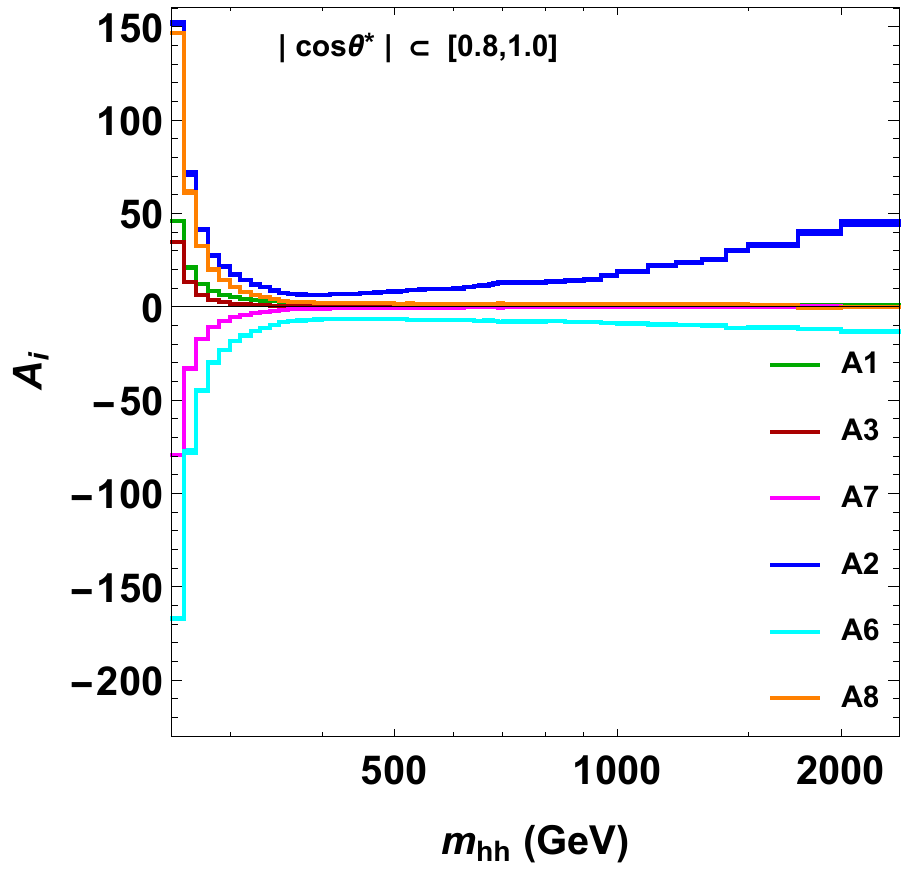}
  \includegraphics[width=0.32\textwidth]{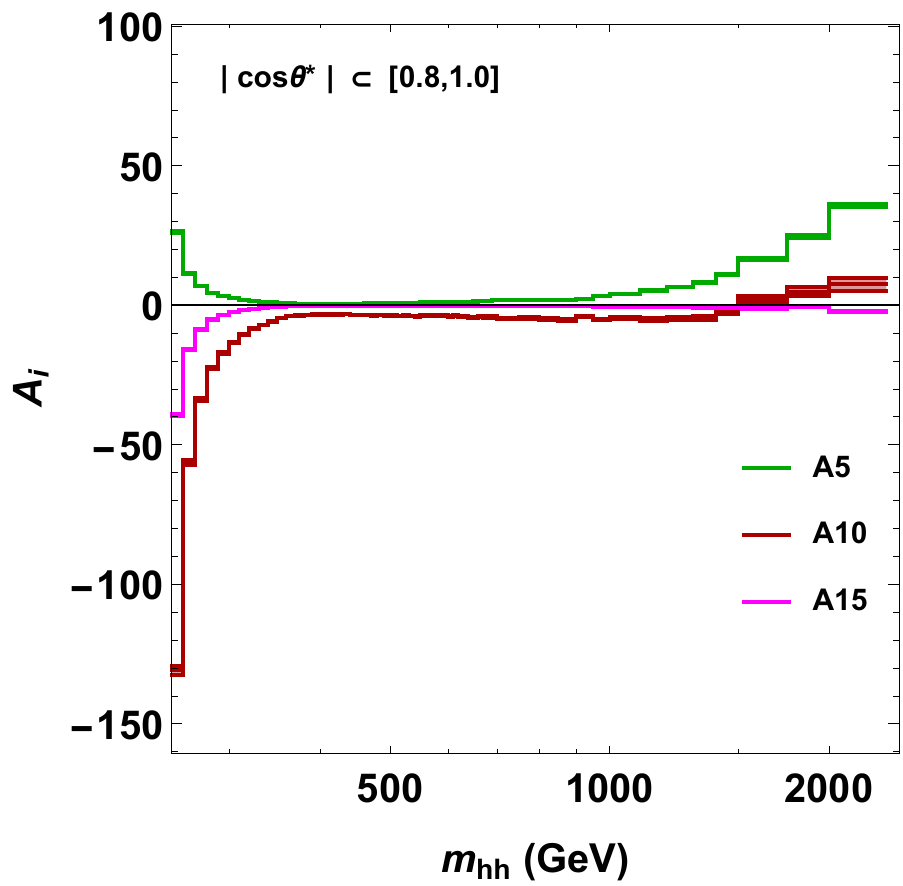}
  \includegraphics[width=0.32\textwidth]{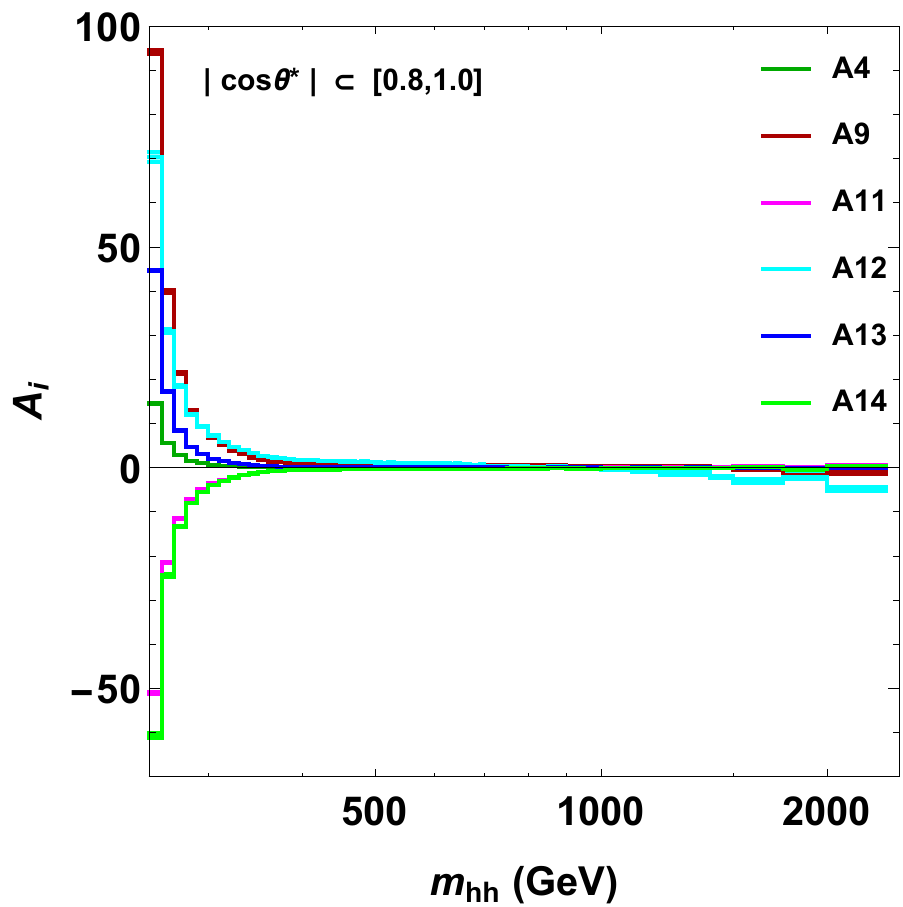}
  \caption{\footnotesize Values of the fit coefficients entering $R_{\rm HH}^j$, differential in $m_{HH}$. We only display the results for the central $\cos \theta^\ast$ bin (betweem 0 and 0.4). From left to right we show respectivelly the coefficients relative to subspace S1 and S2, S4 and finally S3, S5 and S6 (see table~\ref{tab:datasets}). 
    \label{fig:DiffRhh09}
  }
\end{figure}

Tables \ref{tab:NumFitM1} to \ref{tab:NumFitM4} contains the numeric values of the formula~\ref{eq:fit}. The first columns of the tables stand for the number of the bin, followed by the mean $m_{hh}$ and $cos\,\theta^*$ of that bin. It is also displayed the number of events found in that bin for a 13,000,000 events SM sample and for the 1,200,000 events of the reweighting sample $\mathcal{S}_{\rm BSM}$ used in this paper. The same informations can be found visually in figures~\ref{fig:Nenvents} (for the number of events/bin) and~\ref{fig:DiffRhh09} (for the values of the coefficients on the not central $cos\,\theta^*$ bins). 

\begin{sidewaystable}[t]
\centering
{\tiny
\begin{tabular}{c|cc|cc|ccccccccccccccc}
\hline
$j$ & $m_{hh}$ & $cos\,\theta^*$ & $N_{SM}^j$ & $N_{Sum}^j$  
& A1  & A2 & A3 & A4 & A5  & A6 & A7 & A8 & A9  & A10 & A11 & A12 & A13  & A14 & A15 \\ 
\hline
1 & 255 & 0.2 & 461 & 19239 & 4.96E+01 & 1.73E+02 & 3.78E+01 & 1.61E+01 & 3.07E+01 & -1.87E+02 & -8.64E+01 & 1.62E+02 & 1.06E+02 & -1.47E+02 & -5.62E+01 & 7.83E+01 & 4.93E+01 & -6.84E+01 & -4.45E+01  \\
2 & 255 & 0.5 & 232 & 9483 & 4.86E+01 & 1.73E+02 & 3.75E+01 & 1.60E+01 & 2.87E+01 & -1.85E+02 & -8.51E+01 & 1.61E+02 & 1.06E+02 & -1.42E+02 & -5.54E+01 & 7.44E+01 & 4.90E+01 & -6.50E+01 & -4.27E+01  \\
3 & 255 & 0.7 & 233 & 9604 & 5.00E+01 & 1.66E+02 & 3.75E+01 & 1.57E+01 & 2.94E+01 & -1.82E+02 & -8.65E+01 & 1.58E+02 & 1.02E+02 & -1.43E+02 & -5.61E+01 & 7.81E+01 & 4.86E+01 & -6.69E+01 & -4.31E+01  \\
4 & 255 & 0.9 & 252 & 9642 & 4.57E+01 & 1.52E+02 & 3.46E+01 & 1.45E+01 & 2.62E+01 & -1.67E+02 & -7.94E+01 & 1.46E+02 & 9.41E+01 & -1.31E+02 & -5.10E+01 & 7.03E+01 & 4.47E+01 & -6.08E+01 & -3.93E+01  \\
5 & 265 & 0.2 & 1862 & 26019 & 2.15E+01 & 7.32E+01 & 1.35E+01 & 5.82E+00 & 1.26E+01 & -7.96E+01 & -3.40E+01 & 6.31E+01 & 4.13E+01 & -6.13E+01 & -2.23E+01 & 3.35E+01 & 1.78E+01 & -2.64E+01 & -1.72E+01  \\
6 & 265 & 0.5 & 935 & 13160 & 2.19E+01 & 7.27E+01 & 1.35E+01 & 5.78E+00 & 1.30E+01 & -8.00E+01 & -3.44E+01 & 6.28E+01 & 4.10E+01 & -6.20E+01 & -2.25E+01 & 3.38E+01 & 1.77E+01 & -2.67E+01 & -1.74E+01  \\
7 & 265 & 0.7 & 904 & 12882 & 2.23E+01 & 7.64E+01 & 1.39E+01 & 5.91E+00 & 1.31E+01 & -8.28E+01 & -3.52E+01 & 6.53E+01 & 4.27E+01 & -6.45E+01 & -2.30E+01 & 3.44E+01 & 1.81E+01 & -2.72E+01 & -1.77E+01  \\
8 & 265 & 0.9 & 947 & 13002 & 2.09E+01 & 7.11E+01 & 1.32E+01 & 5.61E+00 & 1.14E+01 & -7.76E+01 & -3.31E+01 & 6.15E+01 & 4.00E+01 & -5.64E+01 & -2.14E+01 & 3.10E+01 & 1.72E+01 & -2.44E+01 & -1.58E+01  \\
9 & 275 & 0.2 & 3897 & 25137 & 1.26E+01 & 4.31E+01 & 6.59E+00 & 2.83E+00 & 7.20E+00 & -4.68E+01 & -1.82E+01 & 3.39E+01 & 2.22E+01 & -3.54E+01 & -1.19E+01 & 1.92E+01 & 8.65E+00 & -1.38E+01 & -9.04E+00  \\
10 & 275 & 0.5 & 2058 & 12280 & 1.21E+01 & 4.14E+01 & 6.27E+00 & 2.71E+00 & 7.20E+00 & -4.51E+01 & -1.74E+01 & 3.22E+01 & 2.11E+01 & -3.41E+01 & -1.14E+01 & 1.89E+01 & 8.25E+00 & -1.35E+01 & -8.83E+00  \\
11 & 275 & 0.7 & 2023 & 12088 & 1.22E+01 & 4.11E+01 & 6.36E+00 & 2.74E+00 & 6.99E+00 & -4.49E+01 & -1.75E+01 & 3.25E+01 & 2.12E+01 & -3.37E+01 & -1.15E+01 & 1.87E+01 & 8.35E+00 & -1.34E+01 & -8.74E+00  \\
12 & 275 & 0.9 & 2024 & 12509 & 1.22E+01 & 4.12E+01 & 6.35E+00 & 2.76E+00 & 6.87E+00 & -4.49E+01 & -1.75E+01 & 3.25E+01 & 2.14E+01 & -3.38E+01 & -1.15E+01 & 1.85E+01 & 8.38E+00 & -1.33E+01 & -8.71E+00  \\
13 & 285 & 0.2 & 6460 & 21999 & 8.67E+00 & 2.99E+01 & 3.80E+00 & 1.63E+00 & 4.65E+00 & -3.21E+01 & -1.15E+01 & 2.13E+01 & 1.39E+01 & -2.38E+01 & -7.51E+00 & 1.27E+01 & 4.98E+00 & -8.40E+00 & -5.50E+00  \\
14 & 285 & 0.5 & 3245 & 10960 & 8.53E+00 & 2.94E+01 & 3.76E+00 & 1.62E+00 & 4.56E+00 & -3.17E+01 & -1.13E+01 & 2.10E+01 & 1.38E+01 & -2.29E+01 & -7.39E+00 & 1.26E+01 & 4.94E+00 & -8.29E+00 & -5.42E+00  \\
15 & 285 & 0.7 & 3203 & 10987 & 8.69E+00 & 2.99E+01 & 3.81E+00 & 1.64E+00 & 4.89E+00 & -3.21E+01 & -1.15E+01 & 2.14E+01 & 1.41E+01 & -2.46E+01 & -7.53E+00 & 1.31E+01 & 5.00E+00 & -8.64E+00 & -5.65E+00  \\
16 & 285 & 0.9 & 3439 & 10778 & 8.28E+00 & 2.74E+01 & 3.57E+00 & 1.53E+00 & 4.32E+00 & -2.99E+01 & -1.09E+01 & 1.99E+01 & 1.29E+01 & -2.24E+01 & -7.12E+00 & 1.21E+01 & 4.69E+00 & -7.95E+00 & -5.18E+00  \\
17 & 295 & 0.2 & 9651 & 18825 & 6.37E+00 & 2.17E+01 & 2.35E+00 & 1.01E+00 & 3.47E+00 & -2.34E+01 & -7.72E+00 & 1.43E+01 & 9.34E+00 & -1.72E+01 & -5.04E+00 & 9.53E+00 & 3.08E+00 & -5.73E+00 & -3.72E+00  \\
18 & 295 & 0.5 & 4808 & 9502 & 6.43E+00 & 2.23E+01 & 2.36E+00 & 1.01E+00 & 3.39E+00 & -2.40E+01 & -7.80E+00 & 1.45E+01 & 9.47E+00 & -1.76E+01 & -5.10E+00 & 9.40E+00 & 3.09E+00 & -5.68E+00 & -3.70E+00  \\
19 & 295 & 0.7 & 4868 & 9392 & 6.37E+00 & 2.18E+01 & 2.34E+00 & 9.97E-01 & 3.43E+00 & -2.36E+01 & -7.70E+00 & 1.43E+01 & 9.33E+00 & -1.72E+01 & -5.01E+00 & 9.24E+00 & 3.05E+00 & -5.60E+00 & -3.67E+00  \\
20 & 295 & 0.9 & 4920 & 9420 & 6.29E+00 & 2.14E+01 & 2.30E+00 & 9.87E-01 & 3.32E+00 & -2.32E+01 & -7.59E+00 & 1.41E+01 & 9.22E+00 & -1.72E+01 & -4.98E+00 & 9.26E+00 & 3.02E+00 & -5.57E+00 & -3.63E+00  \\
21 & 305 & 0.2 & 13170 & 16088 & 5.07E+00 & 1.71E+01 & 1.58E+00 & 6.62E-01 & 2.52E+00 & -1.86E+01 & -5.65E+00 & 1.04E+01 & 6.75E+00 & -1.31E+01 & -3.65E+00 & 7.24E+00 & 2.05E+00 & -4.02E+00 & -2.58E+00  \\
22 & 305 & 0.5 & 6387 & 8299 & 5.17E+00 & 1.77E+01 & 1.64E+00 & 6.83E-01 & 2.59E+00 & -1.91E+01 & -5.81E+00 & 1.08E+01 & 7.00E+00 & -1.34E+01 & -3.74E+00 & 7.38E+00 & 2.12E+00 & -4.13E+00 & -2.66E+00  \\
23 & 305 & 0.7 & 6631 & 8002 & 5.05E+00 & 1.69E+01 & 1.57E+00 & 6.64E-01 & 2.48E+00 & -1.85E+01 & -5.63E+00 & 1.04E+01 & 6.79E+00 & -1.29E+01 & -3.66E+00 & 7.19E+00 & 2.05E+00 & -3.97E+00 & -2.56E+00  \\
24 & 305 & 0.9 & 6612 & 8109 & 5.06E+00 & 1.70E+01 & 1.58E+00 & 6.49E-01 & 2.55E+00 & -1.85E+01 & -5.63E+00 & 1.04E+01 & 6.67E+00 & -1.33E+01 & -3.62E+00 & 7.23E+00 & 2.02E+00 & -3.99E+00 & -2.57E+00  \\
25 & 315 & 0.2 & 16525 & 13745 & 4.28E+00 & 1.46E+01 & 1.15E+00 & 4.69E-01 & 2.04E+00 & -1.58E+01 & -4.43E+00 & 8.22E+00 & 5.25E+00 & -1.10E+01 & -2.83E+00 & 6.03E+00 & 1.47E+00 & -3.10E+00 & -1.96E+00  \\
26 & 315 & 0.5 & 8236 & 7014 & 4.32E+00 & 1.44E+01 & 1.16E+00 & 4.70E-01 & 2.05E+00 & -1.57E+01 & -4.49E+00 & 8.16E+00 & 5.19E+00 & -1.08E+01 & -2.85E+00 & 6.00E+00 & 1.48E+00 & -3.12E+00 & -1.98E+00  \\
27 & 315 & 0.7 & 8413 & 7076 & 4.23E+00 & 1.44E+01 & 1.13E+00 & 4.55E-01 & 2.04E+00 & -1.56E+01 & -4.36E+00 & 8.06E+00 & 5.10E+00 & -1.08E+01 & -2.78E+00 & 5.94E+00 & 1.43E+00 & -3.03E+00 & -1.92E+00  \\
28 & 315 & 0.9 & 8510 & 7038 & 4.21E+00 & 1.43E+01 & 1.12E+00 & 4.54E-01 & 1.94E+00 & -1.55E+01 & -4.33E+00 & 8.02E+00 & 5.10E+00 & -1.05E+01 & -2.75E+00 & 5.82E+00 & 1.43E+00 & -3.00E+00 & -1.89E+00  \\
29 & 325 & 0.2 & 21078 & 11797 & 3.65E+00 & 1.22E+01 & 8.32E-01 & 3.23E-01 & 1.65E+00 & -1.33E+01 & -3.48E+00 & 6.38E+00 & 3.98E+00 & -9.04E+00 & -2.17E+00 & 4.94E+00 & 1.04E+00 & -2.37E+00 & -1.47E+00  \\
30 & 325 & 0.5 & 10542 & 5837 & 3.64E+00 & 1.17E+01 & 8.31E-01 & 3.22E-01 & 1.59E+00 & -1.31E+01 & -3.47E+00 & 6.26E+00 & 3.90E+00 & -8.86E+00 & -2.16E+00 & 4.95E+00 & 1.03E+00 & -2.37E+00 & -1.46E+00  \\
31 & 325 & 0.7 & 10614 & 5908 & 3.62E+00 & 1.19E+01 & 8.19E-01 & 3.19E-01 & 1.58E+00 & -1.31E+01 & -3.44E+00 & 6.25E+00 & 3.89E+00 & -8.85E+00 & -2.15E+00 & 4.84E+00 & 1.02E+00 & -2.33E+00 & -1.45E+00  \\
32 & 325 & 0.9 & 10752 & 5900 & 3.60E+00 & 1.18E+01 & 8.07E-01 & 3.10E-01 & 1.48E+00 & -1.30E+01 & -3.40E+00 & 6.19E+00 & 3.82E+00 & -8.42E+00 & -2.12E+00 & 4.68E+00 & 1.00E+00 & -2.23E+00 & -1.38E+00  \\
33 & 335 & 0.2 & 26047 & 10369 & 3.19E+00 & 1.04E+01 & 6.24E-01 & 2.31E-01 & 1.27E+00 & -1.15E+01 & -2.82E+00 & 5.10E+00 & 3.10E+00 & -7.31E+00 & -1.72E+00 & 4.05E+00 & 7.60E-01 & -1.80E+00 & -1.09E+00  \\
34 & 335 & 0.5 & 13112 & 5183 & 3.18E+00 & 1.04E+01 & 6.17E-01 & 2.28E-01 & 1.27E+00 & -1.15E+01 & -2.80E+00 & 5.08E+00 & 3.08E+00 & -7.37E+00 & -1.70E+00 & 4.00E+00 & 7.50E-01 & -1.79E+00 & -1.09E+00  \\
35 & 335 & 0.7 & 13521 & 5144 & 3.14E+00 & 1.02E+01 & 5.99E-01 & 2.17E-01 & 1.21E+00 & -1.13E+01 & -2.73E+00 & 4.95E+00 & 2.97E+00 & -7.05E+00 & -1.66E+00 & 3.94E+00 & 7.22E-01 & -1.73E+00 & -1.04E+00  \\
36 & 335 & 0.9 & 13340 & 5134 & 3.16E+00 & 1.03E+01 & 6.09E-01 & 2.25E-01 & 1.26E+00 & -1.14E+01 & -2.77E+00 & 5.01E+00 & 3.04E+00 & -7.19E+00 & -1.68E+00 & 4.01E+00 & 7.40E-01 & -1.77E+00 & -1.07E+00  \\
37 & 345 & 0.2 & 34128 & 9211 & 2.80E+00 & 8.81E+00 & 4.56E-01 & 1.53E-01 & 9.71E-01 & -9.93E+00 & -2.26E+00 & 4.01E+00 & 2.32E+00 & -5.88E+00 & -1.31E+00 & 3.32E+00 & 5.28E-01 & -1.34E+00 & -7.74E-01  \\
38 & 345 & 0.5 & 17291 & 4591 & 2.78E+00 & 8.60E+00 & 4.47E-01 & 1.52E-01 & 9.44E-01 & -9.78E+00 & -2.23E+00 & 3.92E+00 & 2.28E+00 & -5.75E+00 & -1.30E+00 & 3.26E+00 & 5.20E-01 & -1.31E+00 & -7.64E-01  \\
39 & 345 & 0.7 & 17129 & 4630 & 2.79E+00 & 8.84E+00 & 4.51E-01 & 1.53E-01 & 9.82E-01 & -9.93E+00 & -2.24E+00 & 3.99E+00 & 2.33E+00 & -5.89E+00 & -1.31E+00 & 3.29E+00 & 5.26E-01 & -1.33E+00 & -7.74E-01  \\
40 & 345 & 0.9 & 17533 & 4638 & 2.77E+00 & 8.61E+00 & 4.44E-01 & 1.50E-01 & 9.50E-01 & -9.77E+00 & -2.22E+00 & 3.91E+00 & 2.26E+00 & -5.83E+00 & -1.29E+00 & 3.29E+00 & 5.15E-01 & -1.33E+00 & -7.70E-01  \\
41 & 355 & 0.2 & 45693 & 9105 & 2.50E+00 & 7.43E+00 & 3.37E-01 & 1.00E-01 & 7.25E-01 & -8.60E+00 & -1.83E+00 & 3.16E+00 & 1.72E+00 & -4.68E+00 & -9.97E-01 & 2.67E+00 & 3.68E-01 & -9.87E-01 & -5.40E-01  \\
42 & 355 & 0.5 & 22904 & 4473 & 2.49E+00 & 7.44E+00 & 3.37E-01 & 1.01E-01 & 6.78E-01 & -8.61E+00 & -1.83E+00 & 3.17E+00 & 1.73E+00 & -4.57E+00 & -1.00E+00 & 2.67E+00 & 3.68E-01 & -9.81E-01 & -5.35E-01  \\
43 & 355 & 0.7 & 23414 & 4521 & 2.48E+00 & 7.23E+00 & 3.31E-01 & 9.89E-02 & 7.09E-01 & -8.47E+00 & -1.81E+00 & 3.10E+00 & 1.69E+00 & -4.54E+00 & -9.89E-01 & 2.66E+00 & 3.61E-01 & -9.73E-01 & -5.36E-01  \\
44 & 355 & 0.9 & 23458 & 4508 & 2.46E+00 & 7.22E+00 & 3.26E-01 & 9.78E-02 & 7.04E-01 & -8.42E+00 & -1.79E+00 & 3.07E+00 & 1.68E+00 & -4.53E+00 & -9.78E-01 & 2.65E+00 & 3.56E-01 & -9.61E-01 & -5.27E-01  \\
45 & 365 & 0.2 & 53582 & 8685 & 2.31E+00 & 6.82E+00 & 2.72E-01 & 7.55E-02 & 5.88E-01 & -7.93E+00 & -1.58E+00 & 2.73E+00 & 1.43E+00 & -4.01E+00 & -8.21E-01 & 2.32E+00 & 2.86E-01 & -8.02E-01 & -4.22E-01  \\
46 & 365 & 0.5 & 26937 & 4279 & 2.30E+00 & 6.73E+00 & 2.71E-01 & 7.41E-02 & 6.14E-01 & -7.87E+00 & -1.57E+00 & 2.70E+00 & 1.40E+00 & -4.07E+00 & -8.15E-01 & 2.38E+00 & 2.82E-01 & -8.21E-01 & -4.31E-01  \\
47 & 365 & 0.7 & 27395 & 4265 & 2.29E+00 & 6.64E+00 & 2.66E-01 & 7.35E-02 & 5.80E-01 & -7.78E+00 & -1.55E+00 & 2.66E+00 & 1.39E+00 & -3.89E+00 & -8.08E-01 & 2.30E+00 & 2.78E-01 & -7.93E-01 & -4.16E-01  \\
48 & 365 & 0.9 & 27758 & 4378 & 2.28E+00 & 6.54E+00 & 2.62E-01 & 7.16E-02 & 5.69E-01 & -7.71E+00 & -1.54E+00 & 2.62E+00 & 1.35E+00 & -3.91E+00 & -7.93E-01 & 2.28E+00 & 2.72E-01 & -7.80E-01 & -4.10E-01  \\
49 & 375 & 0.2 & 57750 & 8481 & 2.19E+00 & 6.64E+00 & 2.35E-01 & 6.04E-02 & 5.44E-01 & -7.58E+00 & -1.42E+00 & 2.50E+00 & 1.26E+00 & -3.77E+00 & -6.98E-01 & 2.10E+00 & 2.36E-01 & -7.04E-01 & -3.62E-01  \\
50 & 375 & 0.5 & 29067 & 4355 & 2.19E+00 & 6.56E+00 & 2.34E-01 & 6.21E-02 & 5.21E-01 & -7.53E+00 & -1.42E+00 & 2.48E+00 & 1.27E+00 & -3.68E+00 & -7.15E-01 & 2.09E+00 & 2.39E-01 & -7.02E-01 & -3.64E-01  \\
51 & 375 & 0.7 & 29727 & 4311 & 2.17E+00 & 6.46E+00 & 2.27E-01 & 5.92E-02 & 5.17E-01 & -7.46E+00 & -1.40E+00 & 2.43E+00 & 1.22E+00 & -3.66E+00 & -6.91E-01 & 2.09E+00 & 2.29E-01 & -6.91E-01 & -3.58E-01  \\
52 & 375 & 0.9 & 29382 & 4370 & 2.18E+00 & 6.48E+00 & 2.31E-01 & 6.09E-02 & 5.36E-01 & -7.49E+00 & -1.41E+00 & 2.45E+00 & 1.24E+00 & -3.78E+00 & -7.07E-01 & 2.16E+00 & 2.34E-01 & -7.22E-01 & -3.74E-01  \\
53 & 385 & 0.2 & 60252 & 8527 & 2.08E+00 & 6.46E+00 & 2.02E-01 & 5.14E-02 & 5.17E-01 & -7.27E+00 & -1.28E+00 & 2.29E+00 & 1.13E+00 & -3.63E+00 & -6.09E-01 & 1.97E+00 & 1.99E-01 & -6.39E-01 & -3.31E-01  \\
54 & 385 & 0.5 & 29918 & 4342 & 2.09E+00 & 6.49E+00 & 2.04E-01 & 5.12E-02 & 4.88E-01 & -7.30E+00 & -1.29E+00 & 2.30E+00 & 1.13E+00 & -3.64E+00 & -6.12E-01 & 1.96E+00 & 1.99E-01 & -6.43E-01 & -3.30E-01  \\
55 & 385 & 0.7 & 30410 & 4184 & 2.08E+00 & 6.35E+00 & 2.01E-01 & 5.16E-02 & 4.84E-01 & -7.21E+00 & -1.28E+00 & 2.26E+00 & 1.12E+00 & -3.56E+00 & -6.12E-01 & 1.96E+00 & 1.98E-01 & -6.34E-01 & -3.28E-01  \\
56 & 385 & 0.9 & 30534 & 4336 & 2.07E+00 & 6.27E+00 & 2.01E-01 & 4.97E-02 & 4.98E-01 & -7.15E+00 & -1.27E+00 & 2.24E+00 & 1.10E+00 & -3.49E+00 & -6.00E-01 & 1.93E+00 & 1.96E-01 & -6.27E-01 & -3.18E-01  \\
57 & 395 & 0.2 & 59804 & 8361 & 2.01E+00 & 6.50E+00 & 1.83E-01 & 4.41E-02 & 5.09E-01 & -7.14E+00 & -1.19E+00 & 2.18E+00 & 1.05E+00 & -3.51E+00 & -5.41E-01 & 1.86E+00 & 1.74E-01 & -5.94E-01 & -3.03E-01  \\
58 & 395 & 0.5 & 30041 & 4104 & 2.01E+00 & 6.44E+00 & 1.81E-01 & 4.59E-02 & 4.82E-01 & -7.09E+00 & -1.19E+00 & 2.16E+00 & 1.05E+00 & -3.46E+00 & -5.53E-01 & 1.80E+00 & 1.76E-01 & -5.74E-01 & -2.97E-01  \\
59 & 395 & 0.7 & 30417 & 4176 & 2.00E+00 & 6.29E+00 & 1.78E-01 & 4.40E-02 & 4.94E-01 & -7.00E+00 & -1.17E+00 & 2.11E+00 & 1.02E+00 & -3.33E+00 & -5.38E-01 & 1.80E+00 & 1.71E-01 & -5.65E-01 & -2.93E-01  \\
60 & 395 & 0.9 & 30800 & 4206 & 1.99E+00 & 6.29E+00 & 1.75E-01 & 4.43E-02 & 4.79E-01 & -6.98E+00 & -1.17E+00 & 2.10E+00 & 1.02E+00 & -3.34E+00 & -5.34E-01 & 1.77E+00 & 1.70E-01 & -5.56E-01 & -2.90E-01  \\
\hline
\end{tabular}}
\caption{\footnotesize \label{tab:NumFitM1}
Coefficients of equation~\ref{eq:cx}. There $j$ is the bin number and $m_{hh}$ and $cos\,\theta^*$ the central values in the bin. $N_{SM}^j$ ( $N_{Sum}^j$ ) is the number of events in that bin for SM sample (sum of benchmarks) out of 3M (1.2M) events. 
}
\end{sidewaystable}
%%%%%%%%%%%%%%%%%%%%%%%%%%%%%%%%%%%%%%%%%%%%%%%%%%%%%%%%%%%%%%%%%%%%%%%%%%%%%%%%%%%%%%%%%%
%%%%%%%%%%%%%%%%%%%%%%%%%%%%%%%%%%%%%%%%%%%%%%%%%%%%%%%%%%%%%%%%%%%%%%%%%%%%%%%%%%%%%%%%%%
%%%%%%%%%%%%%%%%%%%%%%%%%%%%%%%%%%%%%%%%%%%%%%%%%%%%%%%%%%%%%%%%%%%%%%%%%%%%%%%%%%%%%%%%%%
\begin{sidewaystable}[t]
\centering
{\tiny
\begin{tabular}{c|cc|cc|ccccccccccccccc}
\hline
$j$ & $m_{hh}$ & $cos\,\theta^*$ & $N_{SM}^j$ & $N_{Sum}^j$  
& A1  & A2 & A3 & A4 & A5  & A6 & A7 & A8 & A9  & A10 & A11 & A12 & A13  & A14 & A15 \\ 
\hline
61 & 405 & 0.2 & 58279 & 8303 & 1.95E+00 & 6.59E+00 & 1.65E-01 & 4.02E-02 & 4.95E-01 & -7.02E+00 & -1.11E+00 & 2.09E+00 & 9.82E-01 & -3.51E+00 & -4.88E-01 & 1.71E+00 & 1.55E-01 & -5.42E-01 & -2.85E-01  \\
62 & 405 & 0.5 & 29217 & 4146 & 1.95E+00 & 6.55E+00 & 1.66E-01 & 4.08E-02 & 4.84E-01 & -7.03E+00 & -1.12E+00 & 2.09E+00 & 9.81E-01 & -3.46E+00 & -4.94E-01 & 1.73E+00 & 1.57E-01 & -5.48E-01 & -2.87E-01  \\
63 & 405 & 0.7 & 29553 & 4133 & 1.94E+00 & 6.51E+00 & 1.63E-01 & 4.10E-02 & 5.06E-01 & -6.98E+00 & -1.10E+00 & 2.06E+00 & 1.00E+00 & -3.45E+00 & -4.94E-01 & 1.78E+00 & 1.56E-01 & -5.57E-01 & -2.96E-01  \\
64 & 405 & 0.9 & 30212 & 4148 & 1.93E+00 & 6.29E+00 & 1.60E-01 & 3.92E-02 & 4.81E-01 & -6.85E+00 & -1.09E+00 & 2.00E+00 & 9.41E-01 & -3.38E+00 & -4.76E-01 & 1.71E+00 & 1.50E-01 & -5.37E-01 & -2.81E-01  \\
65 & 415 & 0.2 & 56376 & 8307 & 1.89E+00 & 6.67E+00 & 1.51E-01 & 3.77E-02 & 4.91E-01 & -6.93E+00 & -1.04E+00 & 2.01E+00 & 9.40E-01 & -3.55E+00 & -4.44E-01 & 1.64E+00 & 1.41E-01 & -5.16E-01 & -2.75E-01  \\
66 & 415 & 0.5 & 28125 & 3992 & 1.90E+00 & 6.64E+00 & 1.52E-01 & 3.63E-02 & 5.18E-01 & -6.95E+00 & -1.05E+00 & 2.01E+00 & 9.21E-01 & -3.54E+00 & -4.38E-01 & 1.59E+00 & 1.39E-01 & -5.11E-01 & -2.73E-01  \\
67 & 415 & 0.7 & 28448 & 4191 & 1.89E+00 & 6.57E+00 & 1.48E-01 & 3.72E-02 & 4.88E-01 & -6.88E+00 & -1.03E+00 & 1.98E+00 & 9.21E-01 & -3.41E+00 & -4.44E-01 & 1.62E+00 & 1.39E-01 & -4.95E-01 & -2.65E-01  \\
68 & 415 & 0.9 & 28836 & 4239 & 1.88E+00 & 6.53E+00 & 1.46E-01 & 3.44E-02 & 4.99E-01 & -6.88E+00 & -1.03E+00 & 1.96E+00 & 8.99E-01 & -3.39E+00 & -4.31E-01 & 1.63E+00 & 1.34E-01 & -5.04E-01 & -2.64E-01  \\
69 & 425 & 0.2 & 53180 & 7880 & 1.85E+00 & 6.81E+00 & 1.39E-01 & 3.46E-02 & 4.83E-01 & -6.89E+00 & -9.85E-01 & 1.95E+00 & 8.91E-01 & -3.44E+00 & -4.00E-01 & 1.56E+00 & 1.27E-01 & -4.92E-01 & -2.68E-01  \\
70 & 425 & 0.5 & 27049 & 3953 & 1.84E+00 & 6.70E+00 & 1.37E-01 & 3.18E-02 & 5.02E-01 & -6.83E+00 & -9.77E-01 & 1.93E+00 & 8.69E-01 & -3.40E+00 & -3.85E-01 & 1.52E+00 & 1.21E-01 & -4.77E-01 & -2.55E-01  \\
71 & 425 & 0.7 & 27261 & 4078 & 1.83E+00 & 6.74E+00 & 1.34E-01 & 3.31E-02 & 4.98E-01 & -6.86E+00 & -9.68E-01 & 1.91E+00 & 8.76E-01 & -3.43E+00 & -3.93E-01 & 1.55E+00 & 1.22E-01 & -4.77E-01 & -2.59E-01  \\
72 & 425 & 0.9 & 27770 & 3992 & 1.84E+00 & 6.58E+00 & 1.35E-01 & 3.38E-02 & 4.73E-01 & -6.80E+00 & -9.70E-01 & 1.89E+00 & 8.65E-01 & -3.39E+00 & -4.03E-01 & 1.53E+00 & 1.24E-01 & -4.67E-01 & -2.53E-01  \\
73 & 435 & 0.2 & 50181 & 7864 & 1.81E+00 & 6.99E+00 & 1.29E-01 & 3.20E-02 & 5.24E-01 & -6.89E+00 & -9.34E-01 & 1.90E+00 & 8.51E-01 & -3.44E+00 & -3.60E-01 & 1.48E+00 & 1.15E-01 & -4.66E-01 & -2.62E-01  \\
74 & 435 & 0.5 & 25410 & 3949 & 1.80E+00 & 6.94E+00 & 1.28E-01 & 3.30E-02 & 5.03E-01 & -6.84E+00 & -9.30E-01 & 1.89E+00 & 8.65E-01 & -3.54E+00 & -3.74E-01 & 1.53E+00 & 1.17E-01 & -4.76E-01 & -2.66E-01  \\
75 & 435 & 0.7 & 25549 & 3945 & 1.80E+00 & 6.88E+00 & 1.27E-01 & 3.19E-02 & 4.98E-01 & -6.84E+00 & -9.29E-01 & 1.88E+00 & 8.43E-01 & -3.32E+00 & -3.70E-01 & 1.53E+00 & 1.15E-01 & -4.77E-01 & -2.65E-01  \\
76 & 435 & 0.9 & 25962 & 3981 & 1.79E+00 & 6.70E+00 & 1.23E-01 & 3.10E-02 & 5.02E-01 & -6.72E+00 & -9.13E-01 & 1.82E+00 & 8.37E-01 & -3.26E+00 & -3.60E-01 & 1.51E+00 & 1.11E-01 & -4.63E-01 & -2.58E-01  \\
77 & 445 & 0.2 & 46676 & 7830 & 1.77E+00 & 7.25E+00 & 1.21E-01 & 2.93E-02 & 5.44E-01 & -6.89E+00 & -8.90E-01 & 1.88E+00 & 8.25E-01 & -3.53E+00 & -3.25E-01 & 1.40E+00 & 1.05E-01 & -4.48E-01 & -2.57E-01  \\
78 & 445 & 0.5 & 23501 & 3816 & 1.76E+00 & 7.21E+00 & 1.19E-01 & 3.15E-02 & 5.23E-01 & -6.88E+00 & -8.82E-01 & 1.85E+00 & 8.50E-01 & -3.55E+00 & -3.37E-01 & 1.40E+00 & 1.07E-01 & -4.46E-01 & -2.60E-01  \\
79 & 445 & 0.7 & 23756 & 3781 & 1.76E+00 & 7.05E+00 & 1.18E-01 & 3.15E-02 & 5.31E-01 & -6.83E+00 & -8.83E-01 & 1.83E+00 & 8.19E-01 & -3.52E+00 & -3.48E-01 & 1.49E+00 & 1.08E-01 & -4.59E-01 & -2.64E-01  \\
80 & 445 & 0.9 & 24417 & 3888 & 1.75E+00 & 6.88E+00 & 1.14E-01 & 2.91E-02 & 4.99E-01 & -6.73E+00 & -8.67E-01 & 1.77E+00 & 7.96E-01 & -3.50E+00 & -3.35E-01 & 1.45E+00 & 1.01E-01 & -4.34E-01 & -2.46E-01  \\
81 & 455 & 0.2 & 42819 & 7511 & 1.74E+00 & 7.53E+00 & 1.14E-01 & 3.16E-02 & 5.87E-01 & -6.95E+00 & -8.55E-01 & 1.86E+00 & 8.38E-01 & -3.70E+00 & -3.26E-01 & 1.44E+00 & 1.05E-01 & -4.58E-01 & -2.74E-01  \\
82 & 455 & 0.5 & 21866 & 3773 & 1.73E+00 & 7.40E+00 & 1.09E-01 & 2.89E-02 & 5.63E-01 & -6.84E+00 & -8.37E-01 & 1.81E+00 & 8.16E-01 & -3.50E+00 & -3.05E-01 & 1.33E+00 & 9.68E-02 & -4.15E-01 & -2.51E-01  \\
83 & 455 & 0.7 & 22275 & 3770 & 1.74E+00 & 7.30E+00 & 1.10E-01 & 2.87E-02 & 5.99E-01 & -6.87E+00 & -8.46E-01 & 1.80E+00 & 8.17E-01 & -3.52E+00 & -3.20E-01 & 1.47E+00 & 9.85E-02 & -4.50E-01 & -2.65E-01  \\
84 & 455 & 0.9 & 22753 & 3793 & 1.72E+00 & 7.10E+00 & 1.05E-01 & 2.95E-02 & 4.97E-01 & -6.74E+00 & -8.24E-01 & 1.72E+00 & 7.83E-01 & -3.52E+00 & -3.13E-01 & 1.35E+00 & 9.50E-02 & -4.04E-01 & -2.44E-01  \\
85 & 465 & 0.2 & 39481 & 7354 & 1.71E+00 & 7.78E+00 & 1.07E-01 & 2.87E-02 & 6.14E-01 & -7.01E+00 & -8.19E-01 & 1.82E+00 & 8.13E-01 & -3.69E+00 & -2.90E-01 & 1.37E+00 & 9.40E-02 & -4.32E-01 & -2.66E-01  \\
86 & 465 & 0.5 & 20095 & 3678 & 1.70E+00 & 7.61E+00 & 1.04E-01 & 2.70E-02 & 5.67E-01 & -6.88E+00 & -8.08E-01 & 1.79E+00 & 7.65E-01 & -3.52E+00 & -2.76E-01 & 1.28E+00 & 8.90E-02 & -4.09E-01 & -2.49E-01  \\
87 & 465 & 0.7 & 20239 & 3716 & 1.69E+00 & 7.62E+00 & 1.02E-01 & 2.75E-02 & 5.98E-01 & -6.88E+00 & -7.97E-01 & 1.78E+00 & 7.88E-01 & -3.68E+00 & -2.72E-01 & 1.31E+00 & 8.86E-02 & -4.10E-01 & -2.56E-01  \\
88 & 465 & 0.9 & 20845 & 3561 & 1.70E+00 & 7.37E+00 & 1.02E-01 & 2.71E-02 & 5.78E-01 & -6.82E+00 & -8.04E-01 & 1.73E+00 & 7.63E-01 & -3.53E+00 & -2.91E-01 & 1.34E+00 & 8.85E-02 & -4.15E-01 & -2.58E-01  \\
89 & 475 & 0.2 & 36626 & 7301 & 1.68E+00 & 8.04E+00 & 9.92E-02 & 2.77E-02 & 6.41E-01 & -7.01E+00 & -7.77E-01 & 1.80E+00 & 7.93E-01 & -3.85E+00 & -2.58E-01 & 1.30E+00 & 8.59E-02 & -4.25E-01 & -2.73E-01  \\
90 & 475 & 0.5 & 18719 & 3471 & 1.66E+00 & 7.79E+00 & 9.50E-02 & 2.77E-02 & 6.23E-01 & -6.86E+00 & -7.56E-01 & 1.73E+00 & 7.59E-01 & -3.83E+00 & -2.59E-01 & 1.26E+00 & 8.41E-02 & -4.10E-01 & -2.63E-01  \\
91 & 475 & 0.7 & 18720 & 3577 & 1.67E+00 & 7.74E+00 & 9.74E-02 & 2.71E-02 & 6.18E-01 & -6.89E+00 & -7.69E-01 & 1.75E+00 & 7.61E-01 & -3.56E+00 & -2.53E-01 & 1.25E+00 & 8.37E-02 & -4.03E-01 & -2.60E-01  \\
92 & 475 & 0.9 & 19475 & 3528 & 1.66E+00 & 7.56E+00 & 9.27E-02 & 2.57E-02 & 6.43E-01 & -6.80E+00 & -7.53E-01 & 1.68E+00 & 7.52E-01 & -3.66E+00 & -2.54E-01 & 1.29E+00 & 8.07E-02 & -3.95E-01 & -2.56E-01  \\
93 & 485 & 0.2 & 33567 & 6891 & 1.66E+00 & 8.20E+00 & 9.44E-02 & 2.72E-02 & 6.87E-01 & -7.00E+00 & -7.51E-01 & 1.76E+00 & 7.65E-01 & -3.89E+00 & -2.38E-01 & 1.27E+00 & 8.05E-02 & -4.22E-01 & -2.81E-01  \\
94 & 485 & 0.5 & 17086 & 3503 & 1.65E+00 & 8.14E+00 & 9.01E-02 & 2.51E-02 & 6.31E-01 & -6.96E+00 & -7.35E-01 & 1.72E+00 & 7.57E-01 & -3.67E+00 & -2.29E-01 & 1.15E+00 & 7.46E-02 & -3.83E-01 & -2.55E-01  \\
95 & 485 & 0.7 & 17444 & 3473 & 1.65E+00 & 7.88E+00 & 9.22E-02 & 2.71E-02 & 6.45E-01 & -6.87E+00 & -7.39E-01 & 1.72E+00 & 7.39E-01 & -3.69E+00 & -2.44E-01 & 1.25E+00 & 8.03E-02 & -4.05E-01 & -2.67E-01  \\
96 & 485 & 0.9 & 17889 & 3453 & 1.65E+00 & 7.74E+00 & 8.88E-02 & 2.57E-02 & 6.32E-01 & -6.83E+00 & -7.35E-01 & 1.67E+00 & 7.25E-01 & -3.60E+00 & -2.54E-01 & 1.31E+00 & 7.70E-02 & -4.05E-01 & -2.64E-01  \\
97 & 495 & 0.2 & 30583 & 6650 & 1.63E+00 & 8.50E+00 & 8.81E-02 & 2.71E-02 & 7.01E-01 & -7.07E+00 & -7.20E-01 & 1.74E+00 & 7.43E-01 & -4.01E+00 & -2.28E-01 & 1.15E+00 & 7.63E-02 & -3.89E-01 & -2.75E-01  \\
98 & 495 & 0.5 & 15630 & 3332 & 1.62E+00 & 8.42E+00 & 8.51E-02 & 2.55E-02 & 7.09E-01 & -7.03E+00 & -7.04E-01 & 1.71E+00 & 7.38E-01 & -3.91E+00 & -2.15E-01 & 1.18E+00 & 7.27E-02 & -3.83E-01 & -2.64E-01  \\
99 & 495 & 0.7 & 16167 & 3422 & 1.62E+00 & 8.12E+00 & 8.43E-02 & 2.67E-02 & 6.75E-01 & -6.93E+00 & -7.05E-01 & 1.67E+00 & 7.46E-01 & -3.58E+00 & -2.27E-01 & 1.19E+00 & 7.45E-02 & -3.76E-01 & -2.64E-01  \\
100 & 495 & 0.9 & 16559 & 3360 & 1.61E+00 & 7.97E+00 & 8.24E-02 & 2.45E-02 & 6.04E-01 & -6.84E+00 & -6.96E-01 & 1.64E+00 & 6.93E-01 & -3.66E+00 & -2.19E-01 & 1.19E+00 & 6.94E-02 & -3.64E-01 & -2.50E-01  \\
101 & 505 & 0.2 & 27997 & 6594 & 1.61E+00 & 8.88E+00 & 8.44E-02 & 2.49E-02 & 7.67E-01 & -7.14E+00 & -6.93E-01 & 1.75E+00 & 7.18E-01 & -3.98E+00 & -1.91E-01 & 1.13E+00 & 6.90E-02 & -3.88E-01 & -2.79E-01  \\
102 & 505 & 0.5 & 14226 & 3210 & 1.60E+00 & 8.68E+00 & 8.12E-02 & 2.36E-02 & 7.12E-01 & -7.08E+00 & -6.82E-01 & 1.69E+00 & 7.27E-01 & -3.85E+00 & -1.82E-01 & 1.00E+00 & 6.42E-02 & -3.53E-01 & -2.62E-01  \\
103 & 505 & 0.7 & 14414 & 3214 & 1.60E+00 & 8.65E+00 & 8.08E-02 & 2.64E-02 & 7.03E-01 & -7.07E+00 & -6.84E-01 & 1.70E+00 & 7.28E-01 & -3.92E+00 & -2.16E-01 & 1.20E+00 & 7.09E-02 & -3.83E-01 & -2.74E-01  \\
104 & 505 & 0.9 & 15322 & 3179 & 1.59E+00 & 8.07E+00 & 7.66E-02 & 2.36E-02 & 6.92E-01 & -6.89E+00 & -6.69E-01 & 1.59E+00 & 6.72E-01 & -3.81E+00 & -2.00E-01 & 1.12E+00 & 6.34E-02 & -3.52E-01 & -2.58E-01  \\
105 & 515 & 0.2 & 25283 & 6251 & 1.60E+00 & 9.27E+00 & 8.12E-02 & 2.63E-02 & 8.25E-01 & -7.29E+00 & -6.78E-01 & 1.76E+00 & 7.08E-01 & -4.13E+00 & -1.91E-01 & 1.09E+00 & 6.81E-02 & -3.94E-01 & -2.97E-01  \\
106 & 515 & 0.5 & 12999 & 3224 & 1.59E+00 & 8.94E+00 & 7.86E-02 & 2.34E-02 & 8.22E-01 & -7.11E+00 & -6.67E-01 & 1.72E+00 & 7.04E-01 & -4.01E+00 & -1.85E-01 & 1.09E+00 & 6.41E-02 & -3.72E-01 & -2.81E-01  \\
107 & 515 & 0.7 & 13375 & 3082 & 1.58E+00 & 8.80E+00 & 7.52E-02 & 2.24E-02 & 7.30E-01 & -7.07E+00 & -6.55E-01 & 1.64E+00 & 6.82E-01 & -3.78E+00 & -1.73E-01 & 1.11E+00 & 5.93E-02 & -3.55E-01 & -2.58E-01  \\
108 & 515 & 0.9 & 13918 & 3122 & 1.58E+00 & 8.34E+00 & 7.29E-02 & 2.33E-02 & 7.45E-01 & -6.89E+00 & -6.52E-01 & 1.58E+00 & 6.53E-01 & -3.87E+00 & -1.98E-01 & 1.04E+00 & 6.14E-02 & -3.37E-01 & -2.59E-01  \\
109 & 525 & 0.2 & 23150 & 6002 & 1.57E+00 & 9.56E+00 & 7.63E-02 & 2.35E-02 & 8.11E-01 & -7.29E+00 & -6.48E-01 & 1.73E+00 & 7.18E-01 & -4.11E+00 & -1.62E-01 & 1.01E+00 & 6.09E-02 & -3.58E-01 & -2.77E-01  \\
110 & 525 & 0.5 & 11867 & 2985 & 1.57E+00 & 9.19E+00 & 7.43E-02 & 2.29E-02 & 8.29E-01 & -7.11E+00 & -6.43E-01 & 1.69E+00 & 6.33E-01 & -4.01E+00 & -1.68E-01 & 1.01E+00 & 5.93E-02 & -3.64E-01 & -2.77E-01  \\
111 & 525 & 0.7 & 11919 & 3060 & 1.58E+00 & 9.18E+00 & 7.47E-02 & 2.38E-02 & 8.38E-01 & -7.20E+00 & -6.51E-01 & 1.68E+00 & 6.78E-01 & -4.00E+00 & -1.78E-01 & 1.08E+00 & 6.15E-02 & -3.69E-01 & -2.85E-01  \\
112 & 525 & 0.9 & 12678 & 3024 & 1.56E+00 & 8.79E+00 & 7.06E-02 & 2.03E-02 & 7.84E-01 & -7.04E+00 & -6.32E-01 & 1.59E+00 & 6.51E-01 & -3.78E+00 & -1.67E-01 & 1.07E+00 & 5.40E-02 & -3.38E-01 & -2.57E-01  \\
113 & 535 & 0.2 & 21003 & 5873 & 1.56E+00 & 9.97E+00 & 7.24E-02 & 2.46E-02 & 9.18E-01 & -7.37E+00 & -6.29E-01 & 1.74E+00 & 7.09E-01 & -4.29E+00 & -1.48E-01 & 9.94E-01 & 5.85E-02 & -3.71E-01 & -3.05E-01  \\
114 & 535 & 0.5 & 10788 & 2875 & 1.56E+00 & 9.50E+00 & 7.28E-02 & 2.41E-02 & 8.73E-01 & -7.20E+00 & -6.34E-01 & 1.66E+00 & 6.58E-01 & -4.16E+00 & -1.56E-01 & 9.26E-01 & 5.82E-02 & -3.59E-01 & -2.96E-01  \\
115 & 535 & 0.7 & 11040 & 2912 & 1.56E+00 & 9.31E+00 & 7.09E-02 & 2.49E-02 & 8.52E-01 & -7.20E+00 & -6.28E-01 & 1.66E+00 & 6.70E-01 & -3.94E+00 & -1.63E-01 & 1.02E+00 & 5.88E-02 & -3.59E-01 & -2.94E-01  \\
116 & 535 & 0.9 & 11742 & 2872 & 1.54E+00 & 8.95E+00 & 6.58E-02 & 2.17E-02 & 8.09E-01 & -7.07E+00 & -6.10E-01 & 1.57E+00 & 6.72E-01 & -3.87E+00 & -1.49E-01 & 9.72E-01 & 5.17E-02 & -3.26E-01 & -2.68E-01  \\
117 & 545 & 0.2 & 19327 & 5631 & 1.55E+00 & 1.03E+01 & 7.02E-02 & 2.47E-02 & 9.57E-01 & -7.44E+00 & -6.18E-01 & 1.73E+00 & 7.20E-01 & -4.36E+00 & -1.46E-01 & 9.50E-01 & 5.64E-02 & -3.65E-01 & -3.10E-01  \\
118 & 545 & 0.5 & 9660 & 2819 & 1.54E+00 & 1.01E+01 & 6.80E-02 & 2.34E-02 & 8.68E-01 & -7.38E+00 & -6.06E-01 & 1.69E+00 & 6.43E-01 & -4.26E+00 & -1.41E-01 & 8.97E-01 & 5.34E-02 & -3.33E-01 & -2.80E-01  \\
119 & 545 & 0.7 & 10006 & 2814 & 1.54E+00 & 9.64E+00 & 6.67E-02 & 2.13E-02 & 9.98E-01 & -7.21E+00 & -6.03E-01 & 1.62E+00 & 6.75E-01 & -4.28E+00 & -1.48E-01 & 1.02E+00 & 5.24E-02 & -3.55E-01 & -2.95E-01  \\
120 & 545 & 0.9 & 10606 & 2817 & 1.53E+00 & 9.20E+00 & 6.24E-02 & 2.15E-02 & 8.68E-01 & -7.10E+00 & -5.89E-01 & 1.55E+00 & 6.24E-01 & -4.00E+00 & -1.40E-01 & 9.53E-01 & 4.91E-02 & -3.16E-01 & -2.74E-01  \\
\hline
\end{tabular}}
\caption{\footnotesize \label{tab:NumFitM2}
Coefficients of equation~\ref{eq:cx}. There $j$ is the bin number and $m_{hh}$ and $cos\,\theta^*$ the central values in the bin. $N_{SM}^j$ ( $N_{Sum}^j$ ) is the number of events in that bin for SM sample (sum of benchmarks) out of 3M (1.2M) events. 
 }
\end{sidewaystable}
%%%%%%%%%%%%%%%%%%%%%%%%%%%%%%%%%%%%%%%%%%%%%%%%%%%%%%%%%%%%%%%%%%%%%%%%%%%%%%%%%%%%%%%%%%
%%%%%%%%%%%%%%%%%%%%%%%%%%%%%%%%%%%%%%%%%%%%%%%%%%%%%%%%%%%%%%%%%%%%%%%%%%%%%%%%%%%%%%%%%%
%%%%%%%%%%%%%%%%%%%%%%%%%%%%%%%%%%%%%%%%%%%%%%%%%%%%%%%%%%%%%%%%%%%%%%%%%%%%%%%%%%%%%%%%%%
\begin{sidewaystable}[t]
\centering
{\tiny
\begin{tabular}{c|cc|cc|ccccccccccccccc}
\hline
$j$ & $m_{hh}$ & $cos\,\theta^*$ & $N_{SM}^j$ & $N_{Sum}^j$  
& A1  & A2 & A3 & A4 & A5  & A6 & A7 & A8 & A9  & A10 & A11 & A12 & A13  & A14 & A15 \\ 
\hline
121 & 555 & 0.2 & 17431 & 5293 & 1.53E+00 & 1.06E+01 & 6.71E-02 & 2.41E-02 & 9.93E-01 & -7.53E+00 & -5.95E-01 & 1.72E+00 & 6.88E-01 & -4.61E+00 & -1.22E-01 & 7.89E-01 & 5.29E-02 & -3.38E-01 & -3.10E-01  \\
122 & 555 & 0.5 & 8936 & 2644 & 1.52E+00 & 1.04E+01 & 6.48E-02 & 2.38E-02 & 1.02E+00 & -7.48E+00 & -5.87E-01 & 1.67E+00 & 7.05E-01 & -4.35E+00 & -1.18E-01 & 8.55E-01 & 5.04E-02 & -3.37E-01 & -3.09E-01  \\
123 & 555 & 0.7 & 9328 & 2799 & 1.51E+00 & 9.88E+00 & 6.08E-02 & 2.43E-02 & 8.84E-01 & -7.27E+00 & -5.75E-01 & 1.57E+00 & 6.48E-01 & -3.98E+00 & -1.43E-01 & 9.36E-01 & 5.16E-02 & -3.37E-01 & -2.93E-01  \\
124 & 555 & 0.9 & 9834 & 2680 & 1.51E+00 & 9.28E+00 & 5.80E-02 & 2.09E-02 & 9.79E-01 & -7.04E+00 & -5.67E-01 & 1.48E+00 & 6.08E-01 & -3.85E+00 & -1.43E-01 & 9.15E-01 & 4.67E-02 & -3.14E-01 & -2.84E-01  \\
125 & 565 & 0.2 & 15754 & 5247 & 1.51E+00 & 1.10E+01 & 6.33E-02 & 2.38E-02 & 1.07E+00 & -7.57E+00 & -5.76E-01 & 1.72E+00 & 6.82E-01 & -4.65E+00 & -1.05E-01 & 8.20E-01 & 4.84E-02 & -3.42E-01 & -3.22E-01  \\
126 & 565 & 0.5 & 8163 & 2570 & 1.50E+00 & 1.09E+01 & 6.12E-02 & 2.48E-02 & 1.05E+00 & -7.52E+00 & -5.64E-01 & 1.70E+00 & 6.72E-01 & -4.56E+00 & -1.23E-01 & 7.31E-01 & 5.08E-02 & -3.33E-01 & -3.25E-01  \\
127 & 565 & 0.7 & 8475 & 2591 & 1.50E+00 & 1.02E+01 & 5.91E-02 & 2.19E-02 & 9.23E-01 & -7.32E+00 & -5.63E-01 & 1.57E+00 & 5.92E-01 & -4.14E+00 & -1.25E-01 & 8.35E-01 & 4.51E-02 & -3.02E-01 & -2.82E-01  \\
128 & 565 & 0.9 & 8926 & 2521 & 1.49E+00 & 9.65E+00 & 5.48E-02 & 1.77E-02 & 9.59E-01 & -7.18E+00 & -5.46E-01 & 1.48E+00 & 5.70E-01 & -3.98E+00 & -1.18E-01 & 8.91E-01 & 3.90E-02 & -2.88E-01 & -2.68E-01  \\
129 & 575 & 0.2 & 14425 & 4932 & 1.51E+00 & 1.13E+01 & 6.29E-02 & 2.49E-02 & 1.13E+00 & -7.61E+00 & -5.72E-01 & 1.71E+00 & 6.31E-01 & -4.68E+00 & -1.17E-01 & 7.71E-01 & 5.00E-02 & -3.38E-01 & -3.34E-01  \\
130 & 575 & 0.5 & 7293 & 2435 & 1.50E+00 & 1.08E+01 & 6.15E-02 & 2.50E-02 & 1.08E+00 & -7.51E+00 & -5.66E-01 & 1.66E+00 & 6.25E-01 & -4.33E+00 & -1.30E-01 & 8.36E-01 & 5.05E-02 & -3.41E-01 & -3.18E-01  \\
131 & 575 & 0.7 & 7724 & 2447 & 1.49E+00 & 1.06E+01 & 5.58E-02 & 2.14E-02 & 1.14E+00 & -7.44E+00 & -5.47E-01 & 1.57E+00 & 5.71E-01 & -4.47E+00 & -9.68E-02 & 7.42E-01 & 3.99E-02 & -3.02E-01 & -3.14E-01  \\
132 & 575 & 0.9 & 8263 & 2514 & 1.48E+00 & 9.63E+00 & 5.31E-02 & 2.22E-02 & 1.02E+00 & -7.16E+00 & -5.32E-01 & 1.47E+00 & 5.56E-01 & -3.84E+00 & -1.20E-01 & 9.35E-01 & 4.39E-02 & -3.25E-01 & -3.05E-01  \\
133 & 585 & 0.2 & 13097 & 4793 & 1.49E+00 & 1.17E+01 & 5.81E-02 & 2.40E-02 & 1.24E+00 & -7.66E+00 & -5.49E-01 & 1.70E+00 & 6.45E-01 & -4.87E+00 & -1.01E-01 & 7.38E-01 & 4.56E-02 & -3.36E-01 & -3.44E-01  \\
134 & 585 & 0.5 & 6752 & 2365 & 1.48E+00 & 1.13E+01 & 5.67E-02 & 2.03E-02 & 1.14E+00 & -7.63E+00 & -5.32E-01 & 1.64E+00 & 6.12E-01 & -4.50E+00 & -7.31E-02 & 6.76E-01 & 3.94E-02 & -3.06E-01 & -3.04E-01  \\
135 & 585 & 0.7 & 6994 & 2431 & 1.48E+00 & 1.09E+01 & 5.24E-02 & 1.82E-02 & 1.21E+00 & -7.46E+00 & -5.28E-01 & 1.57E+00 & 6.15E-01 & -4.38E+00 & -7.73E-02 & 6.89E-01 & 3.29E-02 & -2.85E-01 & -3.05E-01  \\
136 & 585 & 0.9 & 7315 & 2349 & 1.46E+00 & 1.04E+01 & 4.92E-02 & 2.17E-02 & 1.18E+00 & -7.31E+00 & -5.14E-01 & 1.51E+00 & 5.77E-01 & -4.13E+00 & -1.10E-01 & 8.55E-01 & 3.96E-02 & -3.15E-01 & -3.16E-01  \\
137 & 595 & 0.2 & 11949 & 4652 & 1.47E+00 & 1.21E+01 & 5.55E-02 & 2.30E-02 & 1.37E+00 & -7.78E+00 & -5.28E-01 & 1.70E+00 & 5.97E-01 & -5.07E+00 & -7.11E-02 & 6.46E-01 & 4.09E-02 & -3.24E-01 & -3.57E-01  \\
138 & 595 & 0.5 & 6266 & 2399 & 1.47E+00 & 1.15E+01 & 5.21E-02 & 2.29E-02 & 1.26E+00 & -7.55E+00 & -5.23E-01 & 1.61E+00 & 5.39E-01 & -4.96E+00 & -7.91E-02 & 7.21E-01 & 3.92E-02 & -3.19E-01 & -3.41E-01  \\
139 & 595 & 0.7 & 6356 & 2287 & 1.47E+00 & 1.16E+01 & 5.06E-02 & 1.72E-02 & 1.21E+00 & -7.77E+00 & -5.17E-01 & 1.61E+00 & 6.62E-01 & -4.40E+00 & -6.27E-02 & 5.85E-01 & 2.91E-02 & -2.66E-01 & -2.98E-01  \\
140 & 595 & 0.9 & 6860 & 2229 & 1.46E+00 & 1.04E+01 & 4.72E-02 & 1.88E-02 & 1.25E+00 & -7.35E+00 & -5.07E-01 & 1.46E+00 & 5.73E-01 & -3.98E+00 & -9.44E-02 & 6.98E-01 & 3.31E-02 & -2.78E-01 & -3.11E-01  \\
141 & 605 & 0.2 & 10709 & 4428 & 1.48E+00 & 1.28E+01 & 5.66E-02 & 2.19E-02 & 1.45E+00 & -8.02E+00 & -5.40E-01 & 1.76E+00 & 6.54E-01 & -4.87E+00 & -7.05E-02 & 5.46E-01 & 3.88E-02 & -3.19E-01 & -3.63E-01  \\
142 & 605 & 0.5 & 5642 & 2169 & 1.45E+00 & 1.18E+01 & 5.03E-02 & 2.13E-02 & 1.33E+00 & -7.56E+00 & -5.02E-01 & 1.60E+00 & 6.07E-01 & -4.45E+00 & -5.30E-02 & 6.19E-01 & 3.52E-02 & -3.00E-01 & -3.43E-01  \\
143 & 605 & 0.7 & 5791 & 2178 & 1.45E+00 & 1.17E+01 & 4.84E-02 & 2.33E-02 & 1.36E+00 & -7.68E+00 & -5.02E-01 & 1.59E+00 & 6.16E-01 & -5.01E+00 & -9.17E-02 & 8.45E-01 & 3.89E-02 & -3.51E-01 & -3.63E-01  \\
144 & 605 & 0.9 & 6135 & 2211 & 1.46E+00 & 1.11E+01 & 4.86E-02 & 2.38E-02 & 1.20E+00 & -7.59E+00 & -5.08E-01 & 1.49E+00 & 6.21E-01 & -4.40E+00 & -9.10E-02 & 8.26E-01 & 3.78E-02 & -3.04E-01 & -3.35E-01  \\
145 & 615 & 0.2 & 9790 & 4281 & 1.46E+00 & 1.31E+01 & 5.11E-02 & 2.42E-02 & 1.48E+00 & -7.96E+00 & -5.07E-01 & 1.72E+00 & 6.50E-01 & -5.26E+00 & -4.91E-02 & 5.54E-01 & 3.73E-02 & -3.28E-01 & -3.77E-01  \\
146 & 615 & 0.5 & 4916 & 2161 & 1.44E+00 & 1.27E+01 & 4.85E-02 & 2.31E-02 & 1.53E+00 & -7.84E+00 & -4.86E-01 & 1.65E+00 & 6.20E-01 & -5.02E+00 & -5.46E-02 & 5.66E-01 & 3.78E-02 & -3.26E-01 & -3.77E-01  \\
147 & 615 & 0.7 & 5332 & 2255 & 1.42E+00 & 1.18E+01 & 4.26E-02 & 2.18E-02 & 1.32E+00 & -7.60E+00 & -4.64E-01 & 1.51E+00 & 5.63E-01 & -4.66E+00 & -6.89E-02 & 5.66E-01 & 3.29E-02 & -2.71E-01 & -3.33E-01  \\
148 & 615 & 0.9 & 5649 & 2037 & 1.43E+00 & 1.13E+01 & 4.39E-02 & 1.99E-02 & 1.32E+00 & -7.60E+00 & -4.78E-01 & 1.47E+00 & 4.68E-01 & -4.41E+00 & -6.24E-02 & 6.00E-01 & 3.06E-02 & -2.64E-01 & -3.27E-01  \\
149 & 625 & 0.2 & 8863 & 4128 & 1.44E+00 & 1.37E+01 & 4.95E-02 & 2.22E-02 & 1.63E+00 & -8.10E+00 & -4.91E-01 & 1.71E+00 & 6.33E-01 & -5.34E+00 & -4.42E-02 & 4.71E-01 & 3.54E-02 & -3.08E-01 & -3.80E-01  \\
150 & 625 & 0.5 & 4656 & 2129 & 1.43E+00 & 1.28E+01 & 4.48E-02 & 2.45E-02 & 1.53E+00 & -7.85E+00 & -4.75E-01 & 1.63E+00 & 5.89E-01 & -5.05E+00 & -5.95E-02 & 5.67E-01 & 3.42E-02 & -3.02E-01 & -3.83E-01  \\
151 & 625 & 0.7 & 4759 & 2053 & 1.44E+00 & 1.26E+01 & 4.77E-02 & 2.17E-02 & 1.55E+00 & -8.01E+00 & -4.90E-01 & 1.56E+00 & 5.74E-01 & -4.96E+00 & -6.23E-02 & 6.95E-01 & 3.43E-02 & -3.12E-01 & -3.74E-01  \\
152 & 625 & 0.9 & 5211 & 2089 & 1.42E+00 & 1.13E+01 & 4.11E-02 & 1.65E-02 & 1.40E+00 & -7.56E+00 & -4.59E-01 & 1.45E+00 & 4.46E-01 & -4.14E+00 & -6.08E-02 & 5.80E-01 & 2.74E-02 & -2.52E-01 & -3.10E-01  \\
153 & 635 & 0.2 & 7896 & 3959 & 1.45E+00 & 1.44E+01 & 5.19E-02 & 2.34E-02 & 1.67E+00 & -8.33E+00 & -5.04E-01 & 1.78E+00 & 6.19E-01 & -5.35E+00 & -5.67E-02 & 4.37E-01 & 3.79E-02 & -3.08E-01 & -3.90E-01  \\
154 & 635 & 0.5 & 4220 & 1939 & 1.44E+00 & 1.36E+01 & 4.69E-02 & 2.27E-02 & 1.60E+00 & -8.04E+00 & -4.83E-01 & 1.69E+00 & 6.45E-01 & -5.29E+00 & -4.83E-02 & 3.12E-01 & 3.29E-02 & -2.70E-01 & -3.75E-01  \\
155 & 635 & 0.7 & 4356 & 1945 & 1.44E+00 & 1.31E+01 & 4.35E-02 & 2.24E-02 & 1.55E+00 & -8.18E+00 & -4.80E-01 & 1.59E+00 & 6.00E-01 & -5.13E+00 & -5.89E-02 & 6.41E-01 & 3.02E-02 & -3.08E-01 & -3.73E-01  \\
156 & 635 & 0.9 & 4743 & 1911 & 1.43E+00 & 1.17E+01 & 4.19E-02 & 1.49E-02 & 1.42E+00 & -7.56E+00 & -4.68E-01 & 1.45E+00 & 4.57E-01 & -4.24E+00 & -4.78E-02 & 5.20E-01 & 2.28E-02 & -2.36E-01 & -3.05E-01  \\
157 & 645 & 0.2 & 7540 & 3806 & 1.42E+00 & 1.41E+01 & 4.54E-02 & 2.12E-02 & 1.80E+00 & -8.25E+00 & -4.69E-01 & 1.66E+00 & 6.08E-01 & -5.03E+00 & -8.73E-03 & 1.68E-01 & 2.65E-02 & -2.84E-01 & -4.06E-01  \\
158 & 645 & 0.5 & 3796 & 2001 & 1.42E+00 & 1.42E+01 & 4.34E-02 & 2.34E-02 & 1.84E+00 & -8.33E+00 & -4.63E-01 & 1.72E+00 & 5.74E-01 & -5.82E+00 & -1.78E-02 & 4.85E-01 & 2.82E-02 & -3.19E-01 & -4.14E-01  \\
159 & 645 & 0.7 & 4006 & 1980 & 1.41E+00 & 1.31E+01 & 3.98E-02 & 1.71E-02 & 1.69E+00 & -8.02E+00 & -4.48E-01 & 1.49E+00 & 5.69E-01 & -4.42E+00 & -3.03E-02 & 5.18E-01 & 2.28E-02 & -2.87E-01 & -3.69E-01  \\
160 & 645 & 0.9 & 4319 & 1872 & 1.40E+00 & 1.23E+01 & 3.60E-02 & 2.03E-02 & 1.47E+00 & -7.91E+00 & -4.32E-01 & 1.43E+00 & 4.38E-01 & -4.37E+00 & -5.54E-02 & 3.44E-01 & 2.52E-02 & -2.33E-01 & -3.45E-01  \\
161 & 655 & 0.2 & 6638 & 3673 & 1.43E+00 & 1.51E+01 & 4.67E-02 & 2.80E-02 & 1.94E+00 & -8.42E+00 & -4.73E-01 & 1.75E+00 & 6.28E-01 & -5.23E+00 & -4.88E-02 & 4.28E-01 & 3.97E-02 & -3.38E-01 & -4.53E-01  \\
162 & 655 & 0.5 & 3507 & 1804 & 1.42E+00 & 1.42E+01 & 4.19E-02 & 3.02E-02 & 1.75E+00 & -8.15E+00 & -4.62E-01 & 1.60E+00 & 6.40E-01 & -4.96E+00 & -7.58E-02 & 4.21E-01 & 4.02E-02 & -3.18E-01 & -4.35E-01  \\
163 & 655 & 0.7 & 3668 & 1888 & 1.46E+00 & 1.36E+01 & 4.33E-02 & 1.45E-02 & 1.82E+00 & -8.24E+00 & -5.06E-01 & 1.63E+00 & 4.69E-01 & -5.07E+00 & -8.88E-02 & 4.32E-01 & 2.08E-02 & -2.81E-01 & -3.69E-01  \\
164 & 655 & 0.9 & 3969 & 1825 & 1.48E+00 & 1.27E+01 & 4.68E-02 & 1.65E-02 & 1.72E+00 & -8.17E+00 & -5.30E-01 & 1.61E+00 & 4.84E-01 & -4.28E+00 & -9.67E-02 & 6.98E-01 & 2.53E-02 & -2.72E-01 & -3.67E-01  \\
165 & 665 & 0.2 & 6000 & 3577 & 1.41E+00 & 1.55E+01 & 4.37E-02 & 2.53E-02 & 2.07E+00 & -8.53E+00 & -4.57E-01 & 1.69E+00 & 5.24E-01 & -5.26E+00 & -1.88E-02 & 5.19E-02 & 3.07E-02 & -2.79E-01 & -4.47E-01  \\
166 & 665 & 0.5 & 3031 & 1740 & 1.43E+00 & 1.52E+01 & 3.69E-02 & 2.52E-02 & 1.93E+00 & -8.46E+00 & -4.71E-01 & 1.66E+00 & 5.07E-01 & -5.78E+00 & -4.99E-02 & 3.75E-01 & 2.45E-02 & -2.84E-01 & -4.50E-01  \\
167 & 665 & 0.7 & 3389 & 1743 & 1.39E+00 & 1.39E+01 & 3.63E-02 & 2.04E-02 & 1.84E+00 & -8.09E+00 & -4.27E-01 & 1.50E+00 & 5.23E-01 & -4.69E+00 & -4.85E-02 & 4.89E-01 & 2.67E-02 & -3.10E-01 & -4.05E-01  \\
168 & 665 & 0.9 & 3643 & 1730 & 1.38E+00 & 1.29E+01 & 3.02E-02 & 2.49E-02 & 1.79E+00 & -8.04E+00 & -4.09E-01 & 1.38E+00 & 5.80E-01 & -4.84E+00 & -6.70E-02 & 3.99E-01 & 2.79E-02 & -2.62E-01 & -3.99E-01  \\
169 & 675 & 0.2 & 5498 & 3478 & 1.41E+00 & 1.63E+01 & 4.32E-02 & 2.58E-02 & 2.26E+00 & -8.77E+00 & -4.50E-01 & 1.71E+00 & 6.15E-01 & -5.98E+00 & 1.29E-02 & 1.17E-01 & 2.82E-02 & -2.97E-01 & -4.86E-01  \\
170 & 675 & 0.5 & 2857 & 1747 & 1.39E+00 & 1.53E+01 & 3.69E-02 & 2.23E-02 & 2.14E+00 & -8.42E+00 & -4.23E-01 & 1.59E+00 & 5.29E-01 & -5.00E+00 & -1.17E-02 & 9.29E-02 & 2.27E-02 & -2.79E-01 & -4.59E-01  \\
171 & 675 & 0.7 & 3037 & 1762 & 1.38E+00 & 1.44E+01 & 3.59E-02 & 1.41E-02 & 1.88E+00 & -8.22E+00 & -4.20E-01 & 1.50E+00 & 4.35E-01 & -4.72E+00 & 1.18E-02 & 2.06E-01 & 1.11E-02 & -2.05E-01 & -3.52E-01  \\
172 & 675 & 0.9 & 3351 & 1660 & 1.37E+00 & 1.29E+01 & 2.87E-02 & 1.90E-02 & 1.83E+00 & -7.99E+00 & -3.97E-01 & 1.31E+00 & 5.25E-01 & -4.63E+00 & 2.69E-02 & 2.81E-01 & 1.08E-02 & -1.88E-01 & -3.78E-01  \\
173 & 685 & 0.2 & 5004 & 3230 & 1.41E+00 & 1.66E+01 & 4.18E-02 & 2.28E-02 & 2.44E+00 & -8.76E+00 & -4.51E-01 & 1.77E+00 & 5.82E-01 & -5.46E+00 & -3.42E-03 & 3.93E-02 & 2.54E-02 & -2.64E-01 & -4.70E-01  \\
174 & 685 & 0.5 & 2567 & 1673 & 1.36E+00 & 1.64E+01 & 2.74E-02 & 2.05E-02 & 2.19E+00 & -8.50E+00 & -3.87E-01 & 1.55E+00 & 4.96E-01 & -5.25E+00 & 2.97E-02 & -5.78E-02 & 1.26E-02 & -1.99E-01 & -4.22E-01  \\
175 & 685 & 0.7 & 2825 & 1669 & 1.37E+00 & 1.44E+01 & 3.43E-02 & 2.69E-02 & 2.06E+00 & -8.20E+00 & -4.09E-01 & 1.47E+00 & 4.28E-01 & -5.02E+00 & -1.70E-02 & 1.21E-01 & 2.66E-02 & -2.49E-01 & -4.47E-01  \\
176 & 685 & 0.9 & 3009 & 1573 & 1.32E+00 & 1.34E+01 & 2.12E-02 & 2.24E-02 & 1.85E+00 & -7.83E+00 & -3.44E-01 & 1.22E+00 & 4.58E-01 & -4.95E+00 & 5.60E-02 & 8.23E-02 & 8.12E-03 & -1.77E-01 & -3.95E-01  \\
177 & 695 & 0.2 & 4437 & 3211 & 1.39E+00 & 1.78E+01 & 4.01E-02 & 2.34E-02 & 2.54E+00 & -8.98E+00 & -4.34E-01 & 1.77E+00 & 6.15E-01 & -6.45E+00 & 3.90E-02 & -2.07E-03 & 2.09E-02 & -2.78E-01 & -4.94E-01  \\
178 & 695 & 0.5 & 2323 & 1628 & 1.34E+00 & 1.63E+01 & 2.28E-02 & 2.92E-02 & 2.37E+00 & -8.48E+00 & -3.63E-01 & 1.48E+00 & 6.05E-01 & -5.11E+00 & 1.27E-01 & 3.28E-01 & 1.60E-03 & -2.71E-01 & -4.67E-01  \\
179 & 695 & 0.7 & 2631 & 1517 & 1.36E+00 & 1.52E+01 & 2.72E-02 & 1.89E-02 & 2.09E+00 & -8.99E+00 & -3.85E-01 & 1.31E+00 & 5.15E-01 & -4.63E+00 & 1.65E-02 & 3.30E-02 & 1.00E-02 & -1.76E-01 & -3.93E-01  \\
180 & 695 & 0.9 & 2795 & 1508 & 1.39E+00 & 1.38E+01 & 2.80E-02 & 1.83E-02 & 1.82E+00 & -8.25E+00 & -4.18E-01 & 1.39E+00 & 5.41E-01 & -5.19E+00 & -2.79E-02 & 2.79E-01 & 1.17E-02 & -1.69E-01 & -3.61E-01  \\
\hline
\end{tabular}}
\caption{\footnotesize \label{tab:NumFitM3}
Coefficients of equation~\ref{eq:cx}. There $j$ is the bin number and $m_{hh}$ and $cos\,\theta^*$ the central values in the bin. $N_{SM}^j$ ( $N_{Sum}^j$ ) is the number of events in that bin for SM sample (sum of benchmarks) out of 3M (1.2M) events. 
 }
\end{sidewaystable}
%%%%%%%%%%%%%%%%%%%%%%%%%%%%%%%%%%%%%%%%%%%%%%%%%%%%%%%%%%%%%%%%%%%%%%%%%%%%%%%%%%%%%%%%%%
%%%%%%%%%%%%%%%%%%%%%%%%%%%%%%%%%%%%%%%%%%%%%%%%%%%%%%%%%%%%%%%%%%%%%%%%%%%%%%%%%%%%%%%%%%
%%%%%%%%%%%%%%%%%%%%%%%%%%%%%%%%%%%%%%%%%%%%%%%%%%%%%%%%%%%%%%%%%%%%%%%%%%%%%%%%%%%%%%%%%%
\begin{sidewaystable}[t]
\centering
{\tiny
\begin{tabular}{c|cc|cc|ccccccccccccccc}
\hline
$j$ & $m_{hh}$ & $cos\,\theta^*$ & $N_{SM}^j$ & $N_{Sum}^j$  
& A1  & A2 & A3 & A4 & A5  & A6 & A7 & A8 & A9  & A10 & A11 & A12 & A13  & A14 & A15 \\ 
\hline
181 & 725 & 0.2 & 17502 & 14116 & 1.38E+00 & 1.89E+01 & 3.89E-02 & 2.79E-02 & 3.04E+00 & -9.10E+00 & -4.14E-01 & 1.74E+00 & 6.51E-01 & -6.46E+00 & 3.20E-02 & -2.13E-01 & 2.65E-02 & -2.89E-01 & -5.68E-01  \\
182 & 725 & 0.5 & 9025 & 7065 & 1.38E+00 & 1.81E+01 & 3.67E-02 & 2.51E-02 & 2.92E+00 & -9.09E+00 & -4.20E-01 & 1.66E+00 & 4.68E-01 & -5.96E+00 & 1.07E-04 & -7.26E-02 & 2.29E-02 & -2.54E-01 & -5.35E-01  \\
183 & 725 & 0.7 & 9716 & 6968 & 1.36E+00 & 1.70E+01 & 3.30E-02 & 2.43E-02 & 2.64E+00 & -8.81E+00 & -3.96E-01 & 1.54E+00 & 4.38E-01 & -5.51E+00 & 2.04E-03 & 2.75E-02 & 2.17E-02 & -2.54E-01 & -4.98E-01  \\
184 & 725 & 0.9 & 11138 & 6927 & 1.35E+00 & 1.45E+01 & 2.76E-02 & 2.01E-02 & 2.32E+00 & -8.23E+00 & -3.74E-01 & 1.34E+00 & 3.80E-01 & -4.23E+00 & -3.04E-03 & 8.47E-02 & 1.68E-02 & -2.00E-01 & -4.24E-01  \\
185 & 775 & 0.2 & 11119 & 11818 & 1.35E+00 & 2.21E+01 & 3.37E-02 & 2.52E-02 & 4.00E+00 & -9.76E+00 & -3.87E-01 & 1.82E+00 & 4.59E-01 & -5.99E+00 & 6.72E-02 & -7.34E-01 & 1.47E-02 & -2.38E-01 & -6.42E-01  \\
186 & 775 & 0.5 & 5932 & 5860 & 1.35E+00 & 2.09E+01 & 3.13E-02 & 2.52E-02 & 3.73E+00 & -9.51E+00 & -3.81E-01 & 1.69E+00 & 3.65E-01 & -6.48E+00 & 5.64E-02 & -6.03E-01 & 1.46E-02 & -2.20E-01 & -6.05E-01  \\
187 & 775 & 0.7 & 6339 & 5820 & 1.35E+00 & 1.97E+01 & 2.92E-02 & 2.26E-02 & 3.53E+00 & -9.51E+00 & -3.75E-01 & 1.61E+00 & 4.87E-01 & -5.88E+00 & 3.67E-02 & -4.22E-01 & 1.49E-02 & -2.08E-01 & -5.61E-01  \\
188 & 775 & 0.9 & 7332 & 5854 & 1.33E+00 & 1.68E+01 & 2.49E-02 & 2.06E-02 & 3.25E+00 & -8.78E+00 & -3.50E-01 & 1.37E+00 & 3.87E-01 & -5.18E+00 & 9.66E-03 & 8.89E-02 & 1.56E-02 & -2.24E-01 & -5.21E-01  \\
189 & 825 & 0.2 & 7360 & 9903 & 1.32E+00 & 2.50E+01 & 2.69E-02 & 3.22E-02 & 5.44E+00 & -1.01E+01 & -3.45E-01 & 1.78E+00 & 3.83E-01 & -6.48E+00 & 4.43E-02 & -9.91E-01 & 2.38E-02 & -2.76E-01 & -7.94E-01  \\
190 & 825 & 0.5 & 3736 & 4896 & 1.29E+00 & 2.43E+01 & 1.85E-02 & 3.30E-02 & 5.08E+00 & -1.01E+01 & -3.11E-01 & 1.65E+00 & 5.13E-01 & -5.89E+00 & 9.75E-02 & -1.24E+00 & 8.74E-03 & -1.42E-01 & -7.18E-01  \\
191 & 825 & 0.7 & 4244 & 4945 & 1.31E+00 & 2.18E+01 & 2.30E-02 & 2.14E-02 & 4.60E+00 & -9.87E+00 & -3.29E-01 & 1.54E+00 & 3.60E-01 & -5.51E+00 & 1.24E-01 & -8.78E-01 & 2.19E-03 & -1.39E-01 & -6.49E-01  \\
192 & 825 & 0.9 & 4851 & 4925 & 1.24E+00 & 1.87E+01 & 1.17E-02 & 1.94E-02 & 4.01E+00 & -9.04E+00 & -2.53E-01 & 1.14E+00 & 2.83E-01 & -4.62E+00 & 8.91E-02 & -4.38E-01 & 3.46E-03 & -1.57E-01 & -5.51E-01  \\
193 & 875 & 0.2 & 4654 & 8172 & 1.30E+00 & 2.98E+01 & 2.85E-02 & 3.46E-02 & 7.42E+00 & -1.10E+01 & -3.32E-01 & 1.92E+00 & 4.42E-01 & -7.19E+00 & 1.09E-01 & -1.45E+00 & 1.69E-02 & -3.06E-01 & -9.82E-01  \\
194 & 875 & 0.5 & 2565 & 4106 & 1.28E+00 & 2.70E+01 & 1.64E-02 & 3.04E-02 & 6.75E+00 & -1.02E+01 & -2.93E-01 & 1.60E+00 & 4.63E-01 & -6.46E+00 & 1.10E-01 & -1.97E+00 & 6.73E-03 & -5.93E-02 & -8.21E-01  \\
195 & 875 & 0.7 & 2724 & 4133 & 1.28E+00 & 2.55E+01 & 1.70E-02 & 2.91E-02 & 6.09E+00 & -1.05E+01 & -2.97E-01 & 1.49E+00 & 3.51E-01 & -4.27E+00 & 4.37E-02 & -1.31E+00 & 1.11E-02 & -1.07E-01 & -7.38E-01  \\
196 & 875 & 0.9 & 3276 & 4231 & 1.22E+00 & 2.18E+01 & 1.07E-02 & 2.53E-02 & 5.17E+00 & -9.72E+00 & -2.29E-01 & 1.19E+00 & 3.80E-01 & -5.19E+00 & 4.52E-02 & -7.93E-01 & 1.27E-02 & -1.87E-01 & -6.85E-01  \\
197 & 925 & 0.2 & 3129 & 7083 & 1.28E+00 & 3.36E+01 & 2.19E-02 & 3.67E-02 & 9.56E+00 & -1.14E+01 & -3.00E-01 & 1.88E+00 & 3.07E-01 & -6.12E+00 & 1.39E-01 & -2.38E+00 & 1.11E-02 & -1.93E-01 & -1.10E+00  \\
198 & 925 & 0.5 & 1742 & 3407 & 1.03E+00 & 3.04E+01 & -9.08E-03 & 3.96E-02 & 8.28E+00 & -1.04E+01 & -2.21E-02 & 1.14E+00 & 1.75E-01 & -6.56E+00 & 2.48E-01 & -2.62E+00 & 8.58E-04 & -2.60E-01 & -1.06E+00  \\
199 & 925 & 0.7 & 1868 & 3521 & 9.69E-01 & 2.82E+01 & -2.12E-02 & 2.81E-02 & 7.86E+00 & -9.75E+00 & 5.17E-02 & 8.69E-01 & 4.32E-01 & -5.70E+00 & 3.20E-01 & -2.23E+00 & -2.01E-02 & -1.69E-01 & -9.18E-01  \\
200 & 925 & 0.9 & 2250 & 3568 & 1.21E+00 & 2.36E+01 & 7.54E-03 & 2.14E-02 & 6.44E+00 & -9.92E+00 & -2.17E-01 & 1.09E+00 & 1.47E-01 & -4.96E+00 & 1.19E-01 & -1.35E+00 & -4.62E-03 & -5.32E-02 & -7.23E-01  \\
201 & 975 & 0.2 & 2201 & 6044 & 1.21E+00 & 3.64E+01 & 8.15E-03 & 4.38E-02 & 1.18E+01 & -1.15E+01 & -2.23E-01 & 1.72E+00 & 1.17E-01 & -5.30E+00 & 1.92E-01 & -3.03E+00 & 9.64E-03 & -1.83E-01 & -1.28E+00  \\
202 & 975 & 0.5 & 1209 & 2953 & 1.16E+00 & 3.38E+01 & -8.08E-04 & -9.26E-03 & 1.08E+01 & -1.07E+01 & -1.60E-01 & 1.39E+00 & 4.27E-01 & -6.40E+00 & 4.63E-01 & -4.60E+00 & -7.89E-02 & 4.13E-01 & -8.24E-01  \\
203 & 975 & 0.7 & 1321 & 2907 & 1.13E+00 & 3.06E+01 & -5.96E-03 & 6.21E-02 & 9.35E+00 & -1.05E+01 & -1.27E-01 & 1.41E+00 & 2.12E-01 & -4.80E+00 & -3.67E-01 & -3.08E+00 & 8.40E-02 & -1.93E-01 & -1.02E+00  \\
204 & 975 & 0.9 & 1600 & 2972 & 1.27E+00 & 2.53E+01 & 1.28E-02 & 3.33E-02 & 8.14E+00 & -1.04E+01 & -2.81E-01 & 1.39E+00 & 2.82E-01 & -4.59E+00 & -5.28E-02 & -1.40E+00 & 2.92E-02 & -1.15E-01 & -8.53E-01  \\
205 & 1050 & 0.2 & 2616 & 9223 & 1.23E+00 & 4.27E+01 & 8.65E-03 & 2.65E-02 & 1.58E+01 & -1.23E+01 & -2.43E-01 & 1.77E+00 & 1.47E-02 & -3.06E+00 & 2.06E-01 & -4.39E+00 & -1.35E-02 & 1.13E-01 & -1.25E+00  \\
206 & 1050 & 0.5 & 1346 & 4610 & 1.05E+00 & 4.08E+01 & -1.64E-02 & 3.47E-02 & 1.53E+01 & -1.16E+01 & -3.81E-02 & 1.36E+00 & -3.61E-01 & -3.88E+00 & 2.52E-01 & -4.24E+00 & -1.42E-02 & -2.12E-03 & -1.31E+00  \\
207 & 1050 & 0.7 & 1465 & 4631 & 1.13E+00 & 3.78E+01 & -3.49E-03 & 3.25E-02 & 1.40E+01 & -1.14E+01 & -1.29E-01 & 1.16E+00 & 3.90E-01 & -5.35E+00 & 2.47E-01 & -3.33E+00 & -3.52E-03 & -8.93E-02 & -1.22E+00  \\
208 & 1050 & 0.9 & 1872 & 4663 & 1.25E+00 & 3.00E+01 & 1.24E-02 & 2.44E-02 & 1.09E+01 & -1.15E+01 & -2.66E-01 & 1.48E+00 & 3.08E-02 & -2.40E+00 & 1.00E-01 & -2.14E+00 & -5.21E-03 & -5.26E-02 & -9.68E-01  \\
209 & 1150 & 0.2 & 1238 & 6964 & 1.30E+00 & 5.79E+01 & 1.66E-02 & 3.80E-02 & 2.62E+01 & -1.56E+01 & -3.21E-01 & 2.87E+00 & 2.23E-01 & -1.17E+00 & 2.03E-01 & -6.08E+00 & -8.01E-03 & 1.19E-01 & -1.89E+00  \\
210 & 1150 & 0.5 & 679 & 3433 & 7.80E-01 & 4.90E+01 & -3.94E-02 & 3.43E-03 & 2.46E+01 & -1.23E+01 & 2.55E-01 & 5.29E-01 & -1.90E-01 & -3.89E-02 & 2.46E-01 & -6.28E+00 & -4.77E-02 & -6.06E-01 & -1.96E+00  \\
211 & 1150 & 0.7 & 776 & 3427 & 1.23E+00 & 4.55E+01 & 5.35E-03 & -2.54E-02 & 2.07E+01 & -1.53E+01 & -2.34E-01 & 1.41E+00 & -2.54E-01 & 2.91E+00 & 4.69E-01 & -8.56E+00 & -1.24E-01 & 1.26E+00 & -9.31E-01  \\
212 & 1150 & 0.9 & 1005 & 3429 & 1.07E+00 & 3.30E+01 & -9.35E-03 & -9.55E-04 & 1.66E+01 & -1.11E+01 & -6.33E-02 & 7.15E-01 & -4.01E-01 & 2.26E+00 & 2.72E-01 & -3.05E+00 & -3.97E-02 & 3.07E-02 & -1.13E+00  \\
213 & 1250 & 0.2 & 660 & 5212 & 1.12E+00 & 6.59E+01 & -1.48E-02 & 8.33E-03 & 3.78E+01 & -1.46E+01 & -1.07E-01 & 1.96E+00 & 1.00E-01 & 6.37E+00 & 6.15E-01 & -1.00E+01 & -1.19E-01 & 7.35E-01 & -2.11E+00  \\
214 & 1250 & 0.5 & 348 & 2542 & 9.88E-01 & 5.70E+01 & -2.71E-02 & 1.63E-02 & 3.60E+01 & -1.13E+01 & 3.77E-02 & 1.01E-01 & -4.31E-02 & 4.96E+00 & 5.27E-01 & -1.05E+01 & -7.59E-02 & 1.14E+00 & -1.78E+00  \\
215 & 1250 & 0.7 & 402 & 2605 & 1.07E+00 & 5.20E+01 & -1.49E-02 & -1.00E-01 & 3.18E+01 & -1.50E+01 & -5.17E-02 & 5.38E-02 & -7.12E-01 & 4.58E+00 & 4.33E-01 & -3.61E+00 & -1.01E-01 & 1.41E-01 & -1.07E+00  \\
216 & 1250 & 0.9 & 509 & 2484 & 9.29E-01 & 3.98E+01 & -2.38E-02 & -1.23E-01 & 2.45E+01 & -1.20E+01 & 9.32E-02 & -9.24E-01 & -1.64E+00 & 4.84E+00 & 7.63E-02 & -2.08E+00 & -9.30E-02 & -3.59E-01 & -7.70E-01  \\
217 & 1350 & 0.2 & 340 & 3962 & 1.06E+00 & 7.74E+01 & -1.60E-02 & 6.22E-02 & 5.58E+01 & -1.90E+01 & -4.32E-02 & 1.99E-02 & -5.64E-01 & 1.23E+01 & 3.22E-01 & -1.21E+01 & -2.89E-02 & 9.78E-01 & -2.74E+00  \\
218 & 1350 & 0.5 & 177 & 1971 & 9.60E-01 & 7.36E+01 & -3.41E-02 & 6.42E-03 & 5.40E+01 & -1.77E+01 & 7.30E-02 & -9.79E-01 & -1.68E+00 & 7.04E+00 & 5.99E-01 & -1.43E+01 & -4.18E-02 & -7.98E-01 & -4.75E+00  \\
219 & 1350 & 0.7 & 225 & 1980 & 1.28E+00 & -9.14E+00 & -1.32E-02 & -1.02E-01 & 4.60E+01 & 3.10E+00 & -2.66E-01 & 5.16E+00 & -8.23E+00 & 2.48E+01 & 4.87E-01 & -1.48E+01 & -3.00E-01 & 1.28E+00 & -5.08E+00  \\
220 & 1350 & 0.9 & 294 & 2051 & 1.04E+00 & 4.49E+01 & -1.75E-02 & -1.37E-01 & 3.56E+01 & -1.35E+01 & -2.02E-02 & -1.19E-01 & -1.10E+00 & 7.45E+00 & 4.52E-01 & -4.86E+00 & -3.58E-02 & 3.75E-01 & -2.22E+00  \\
221 & 1450 & 0.2 & 204 & 2906 & 9.32E-01 & 5.12E+01 & -3.31E-02 & -2.27E-02 & 7.82E+01 & -8.01E+00 & 9.96E-02 & 1.67E+00 & -4.38E+00 & 3.96E+01 & 1.16E+00 & -1.21E+01 & -1.51E-01 & -1.64E-01 & -3.66E+00  \\
222 & 1450 & 0.5 & 94 & 1529 & 7.30E-01 & 3.51E+01 & -5.11E-02 & -2.65E-01 & 6.85E+01 & -9.81E+00 & 3.20E-01 & -7.77E-01 & -8.86E+00 & 4.50E+01 & 7.59E-01 & -1.64E+01 & -1.13E-01 & 3.22E-01 & -1.47E+00  \\
223 & 1450 & 0.7 & 100 & 1540 & 7.68E-01 & 1.42E+01 & -5.65E-02 & 1.20E-01 & 6.05E+01 & -2.38E+00 & 2.87E-01 & -8.15E-01 & -1.07E+01 & 4.69E+01 & -9.15E-01 & -1.59E+01 & -5.20E-01 & 1.01E-01 & -5.79E+00  \\
224 & 1450 & 0.9 & 175 & 1447 & 1.15E+00 & 3.34E+00 & -3.01E-02 & -7.15E-01 & 1.89E+01 & -3.30E+00 & -1.16E-01 & 2.56E-01 & -1.11E+01 & 4.71E+01 & -3.49E-01 & 7.61E+00 & -1.90E-01 & 2.17E+00 & 4.48E+00  \\
225 & 1625 & 0.2 & 206 & 4797 & 9.84E-01 & 1.09E+02 & -2.74E-02 & 1.20E-01 & 1.23E+02 & -1.63E+01 & 4.29E-02 & 8.68E-01 & -9.43E-01 & 4.52E+01 & 4.38E-01 & -1.74E+01 & 8.41E-03 & 7.33E-01 & -4.74E+00  \\
226 & 1625 & 0,5 & 87 & 2412 & 9,25E-01 & 1,27E+02 & -5,48E-02 & -1,62E+00 & 1,22E+02 & -3,05E+01 & 1,29E-01 & -3,63E+00 & -1,50E+01 & 1,13E+02 & -6,01E-01 & -7,97E+00 & 2,39E-01 & -2,32E+00 & 8,77E+00  \\
227 & 1625 & 0,7 & 132 & 2386 & 1,13E+00 & 8,98E+01 & -3,89E-02 & -7,53E-03 & 1,04E+02 & -1,90E+01 & -9,08E-02 & 9,30E-01 & -7,16E-01 & 1,71E+01 & 1,71E+00 & -2,34E+01 & 5,03E-02 & -8,86E-02 & -8,09E+00  \\
228 & 1625 & 0,9 & 156 & 2441 & 8,79E-01 & 3,37E+01 & -4,86E-02 & -3,89E-01 & 8,18E+01 & -1,10E+01 & 1,69E-01 & -4,41E-01 & -6,06E+00 & 4,02E+01 & -8,82E-01 & -5,48E+00 & -3,20E-01 & -2,80E-01 & -1,02E+00  \\
229 & 1875 & 0,2 & 47 & 2609 & 6,91E-01 & 1,01E+02 & -8,02E-02 & -8,66E-01 & 2,88E+02 & 2,39E+01 & 3,89E-01 & 8,01E+00 & -2,54E+01 & 2,01E+02 & -1,61E+00 & -2,71E+01 & -1,56E+00 & 4,27E+00 & -1,94E+01  \\
230 & 1875 & 0,5 & 34 & 1290 & 7,49E-01 & 8,85E+00 & -7,73E-02 & -1,88E+00 & 1,91E+02 & 5,16E-01 & 3,28E-01 & -1,36E+00 & -3,51E+01 & 1,86E+02 & 1,59E+00 & -2,97E+01 & 4,36E-01 & -8,83E+00 & 1,47E+01  \\
231 & 1875 & 0,7 & 40 & 1327 & 1,07E+00 & 2,13E+01 & -6,59E-02 & 3,95E-01 & 1,74E+02 & -9,48E+00 & -6,28E-03 & -4,25E-01 & -9,85E+00 & 9,96E+01 & 1,29E+00 & -3,47E+01 & -4,76E-01 & 5,48E+00 & -1,07E+01  \\
232 & 1875 & 0,9 & 52 & 1349 & 8,15E-01 & 6,65E+00 & -6,94E-02 & 2,68E-01 & 1,23E+02 & -4,80E+00 & 2,54E-01 & -1,83E+00 & -9,44E+00 & 6,09E+01 & 1,35E+00 & -2,59E+01 & -3,56E-01 & 2,41E+00 & -8,88E+00  \\
233 & 26000 & 0,2 & 33 & 3477 & 9,97E-01 & 8,66E+01 & -7,32E-02 & -2,56E+00 & 6,26E+02 & -9,41E+00 & 7,58E-02 & -1,79E-01 & -3,54E+01 & 4,43E+02 & -4,77E-01 & 5,10E+01 & -1,12E+00 & 1,34E+01 & 2,00E+00  \\
234 & 26000 & 0,5 & 12 & 1726 & 6,56E-01 & 1,58E+01 & -1,82E-01 & -3,78E+00 & 8,20E+02 & 6,27E+00 & 5,26E-01 & -1,70E-01 & -6,35E+01 & 5,81E+02 & 1,00E+01 & -1,06E+02 & -1,51E+00 & 8,47E-01 & -3,83E+01  \\
235 & 26000 & 0,7 & 19 & 1703 & 9,20E-01 & 4,73E+01 & -8,76E-02 & 1,32E+00 & 5,35E+02 & -7,60E-01 & 1,67E-01 & -2,21E+00 & -1,69E+00 & 6,06E+02 & 1,56E+00 & -9,73E+01 & 4,43E+00 & 2,72E+01 & 1,49E+01  \\
236 & 26000 & 0,9 & 23 & 1652 & 1,07E+00 & 3,51E-01 & -7,49E-02 & -2,41E+00 & 4,12E+02 & -2,89E-01 & 2,39E-03 & 2,43E-01 & -4,54E+01 & 2,61E+02 & 4,37E+00 & -6,92E+01 & 8,71E-01 & 1,34E+00 & 1,60E+01  \\
\hline
\end{tabular}}
\caption{\footnotesize \label{tab:NumFitM4}
Coefficients of equation~\ref{eq:cx}. There $j$ is the bin number and $m_{hh}$ and $cos\,\theta^*$ the central values in the bin. $N_{SM}^j$ ( $N_{Sum}^j$ ) is the number of events in that bin for SM sample (sum of benchmarks) out of 3M (1.2M) events. 
}
\end{sidewaystable}

\section{Distributions validation for shape benchmarks}
\label{app:shapes}

On figures~\ref{fig:RecastBSM1} to~\ref{fig:RecastBSM3} we show the kinematic distributions for reconstructed variables after ATLAS like selections, those are respectively to $m_X$, $\cos\theta^*$, $p_T^{\gamma\gamma b\bar{b}}$ and $p_T (\gamma\gamma)$. 

\begin{figure}
  \centering
  \includegraphics[width=0.32\textwidth]{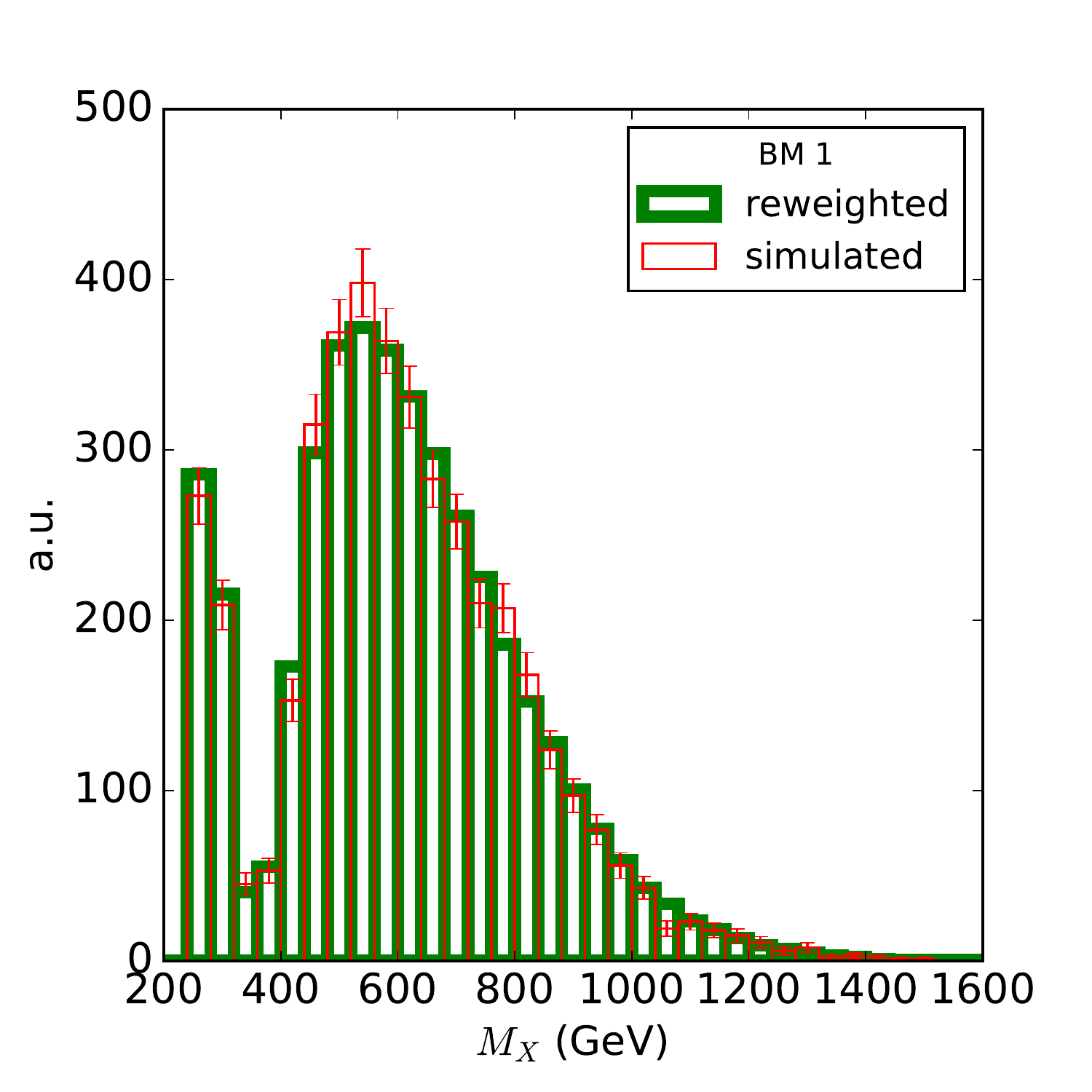}
  \includegraphics[width=0.32\textwidth]{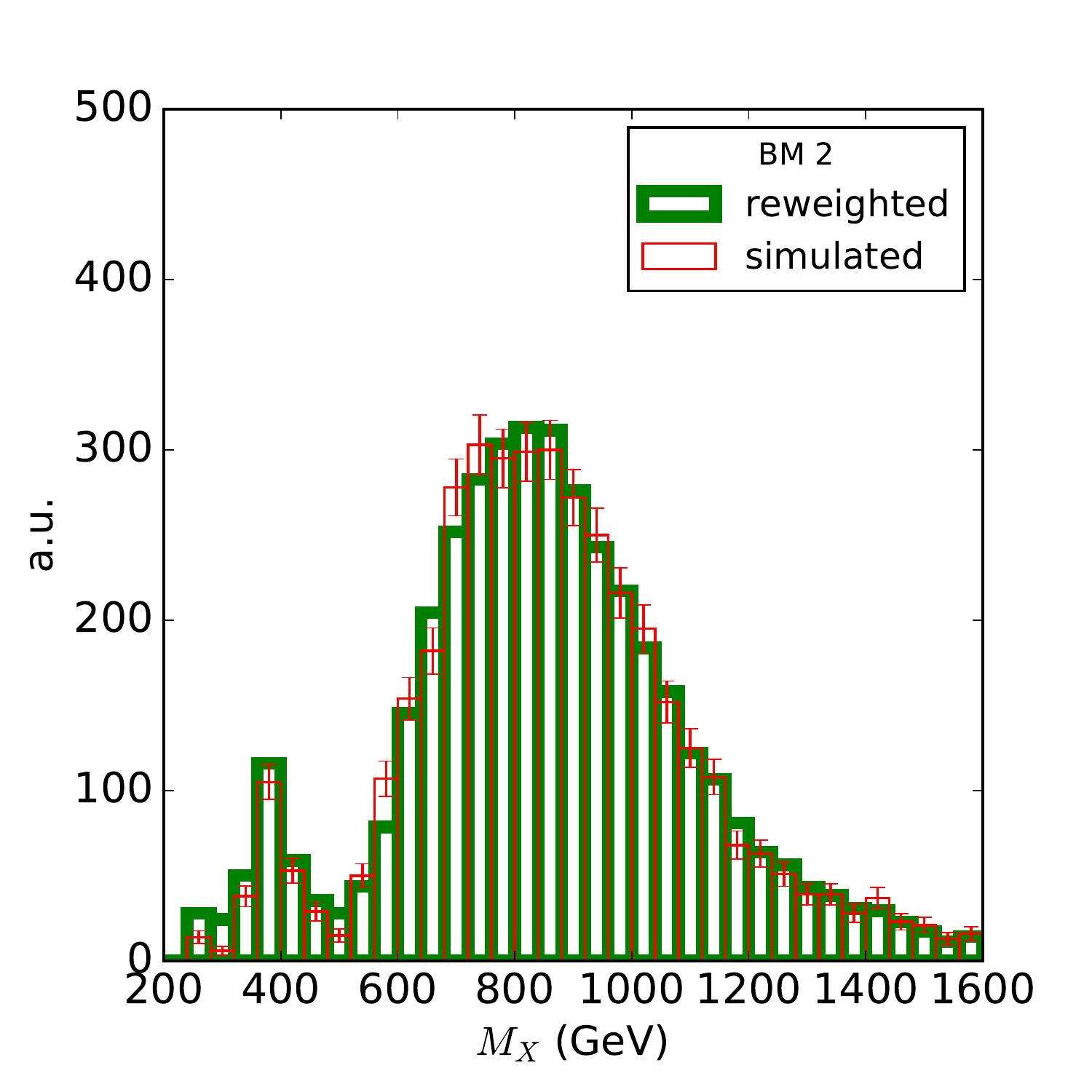}
  \includegraphics[width=0.32\textwidth]{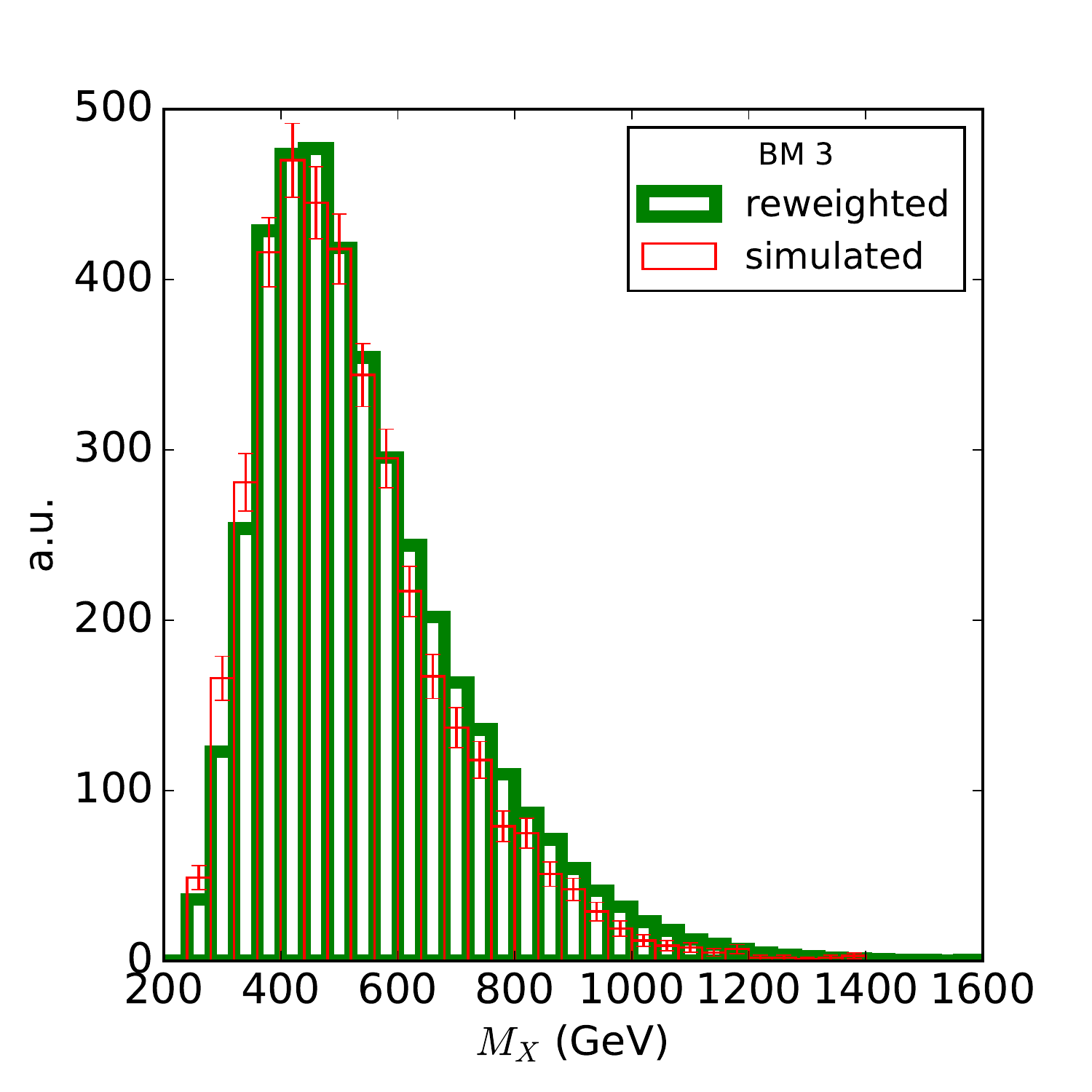}
  \includegraphics[width=0.32\textwidth]{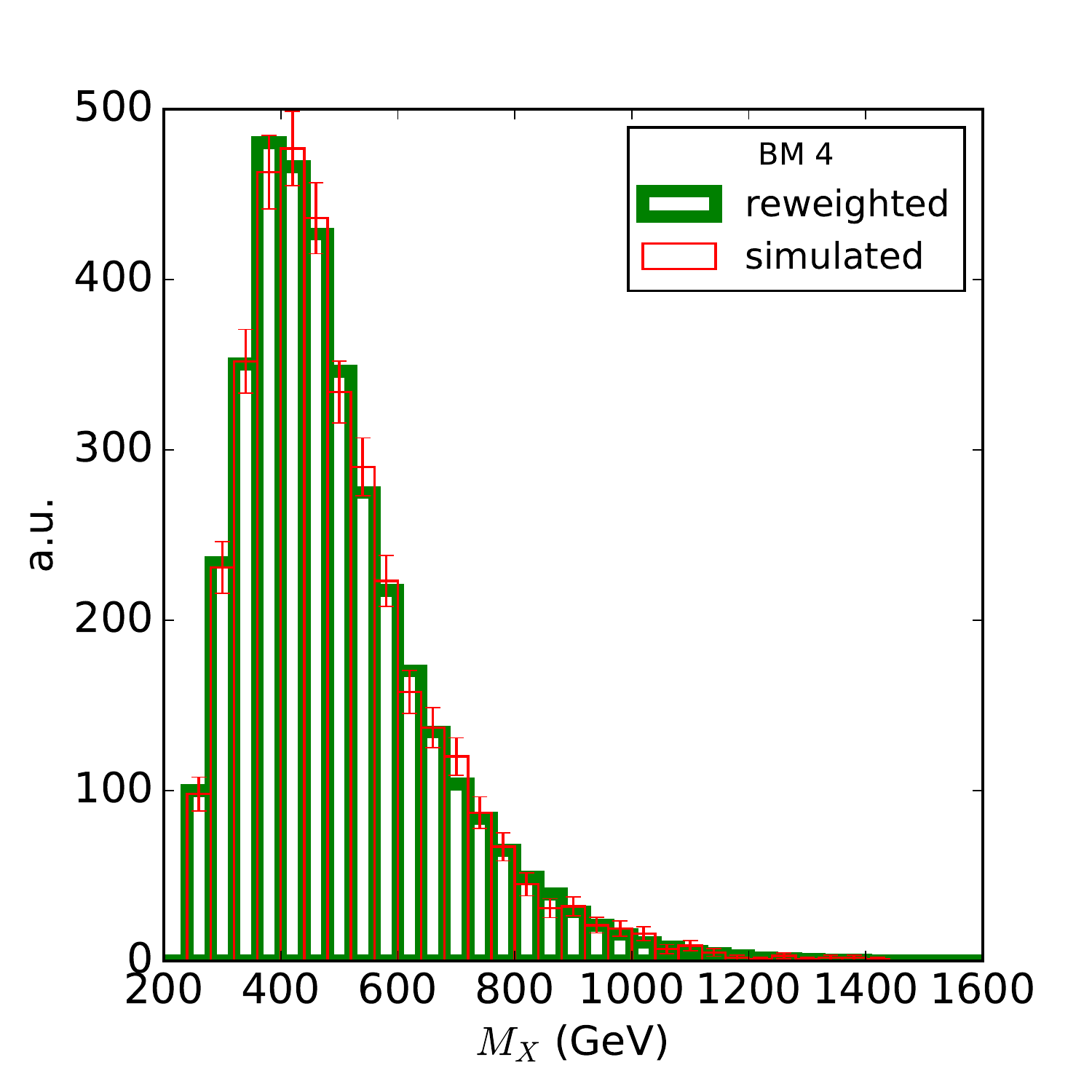}
  \includegraphics[width=0.32\textwidth]{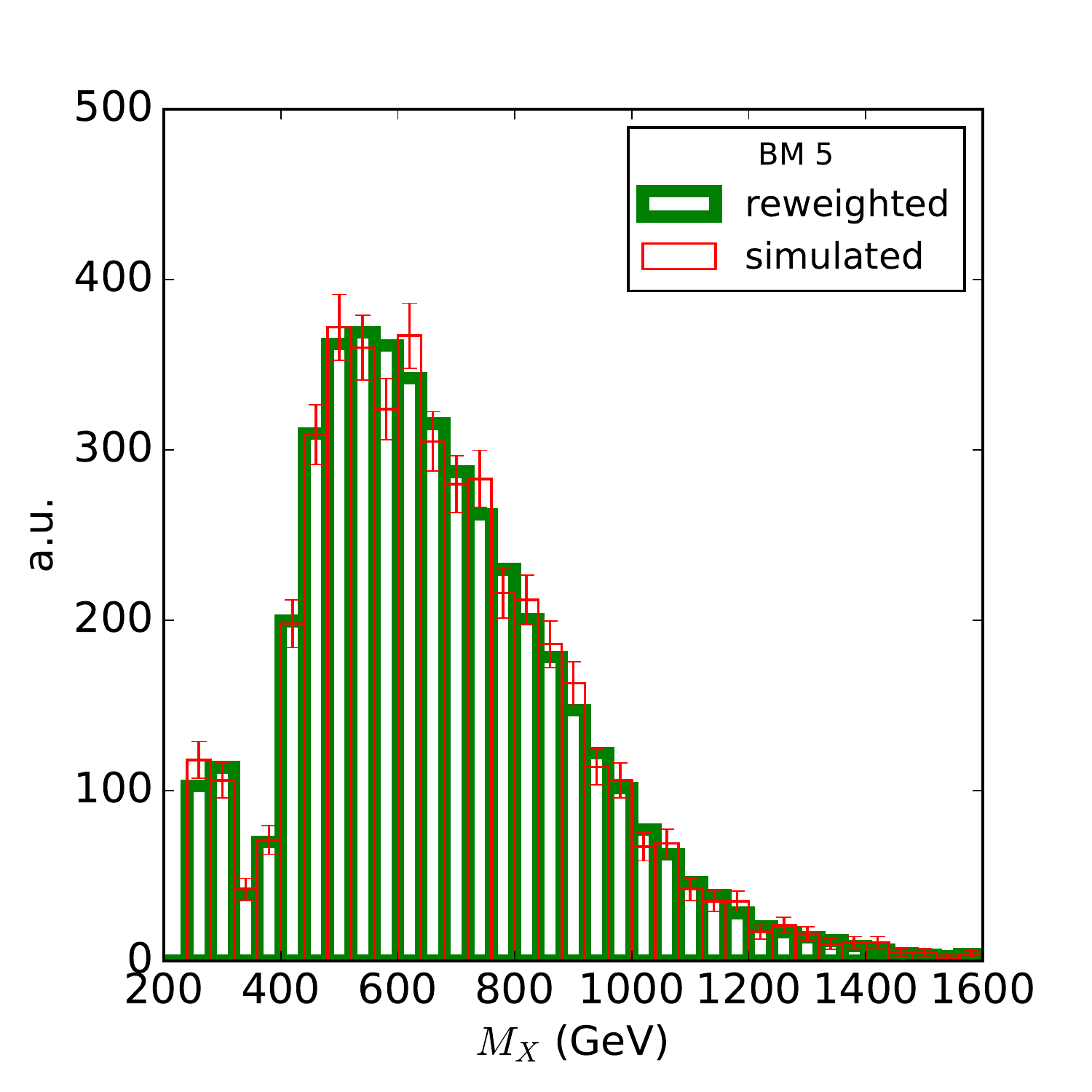}
  \includegraphics[width=0.32\textwidth]{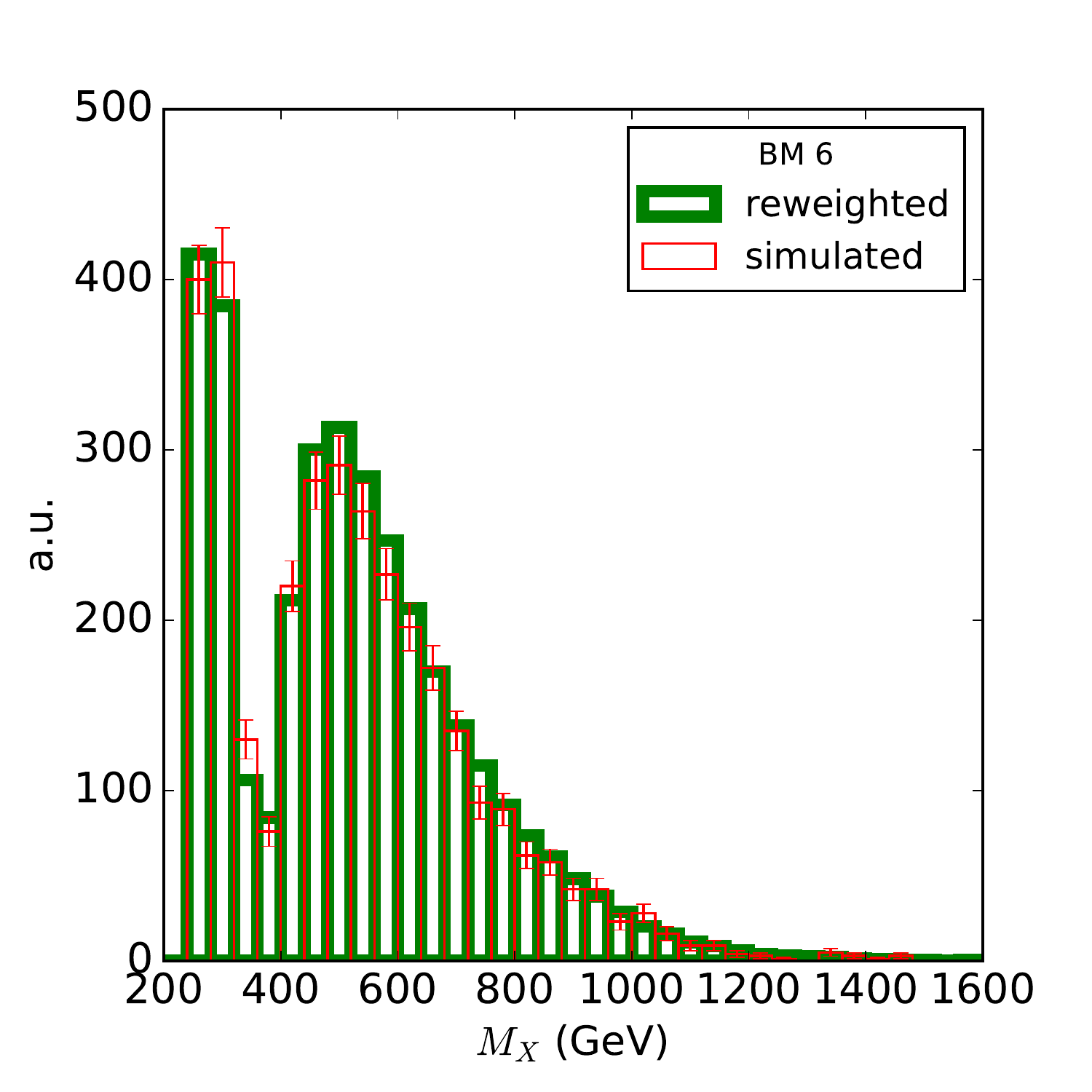}
  \includegraphics[width=0.32\textwidth]{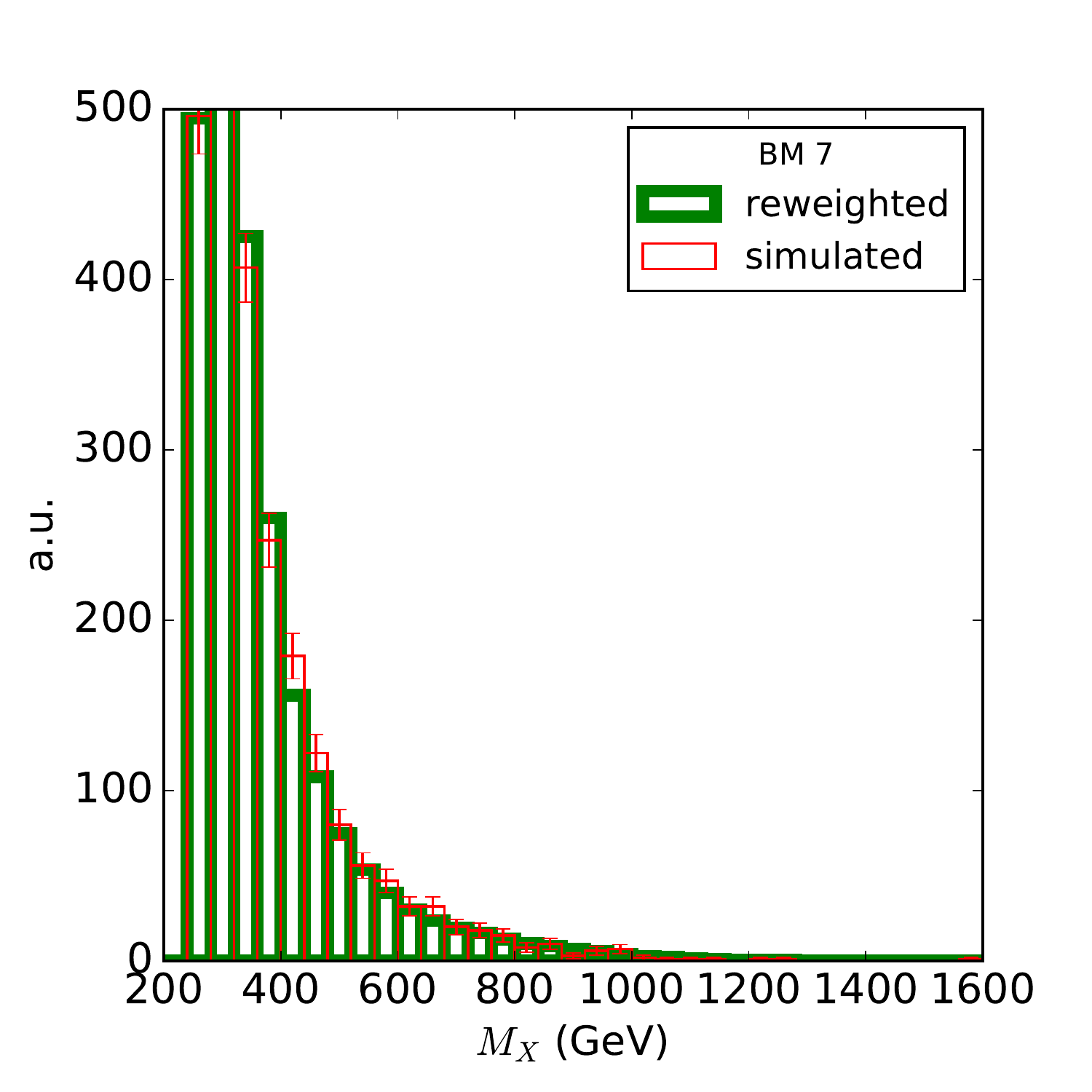}
  \includegraphics[width=0.32\textwidth]{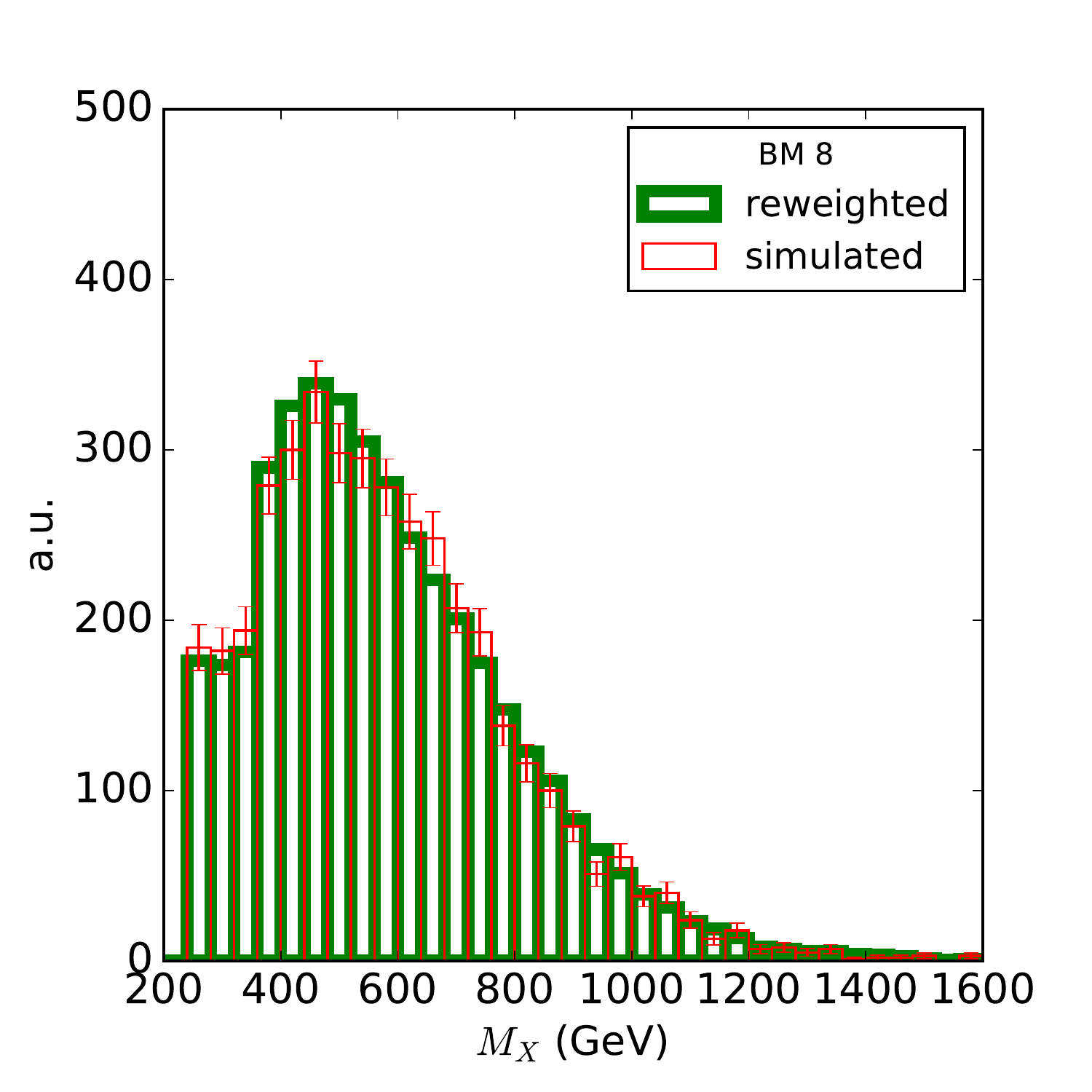}
  \includegraphics[width=0.32\textwidth]{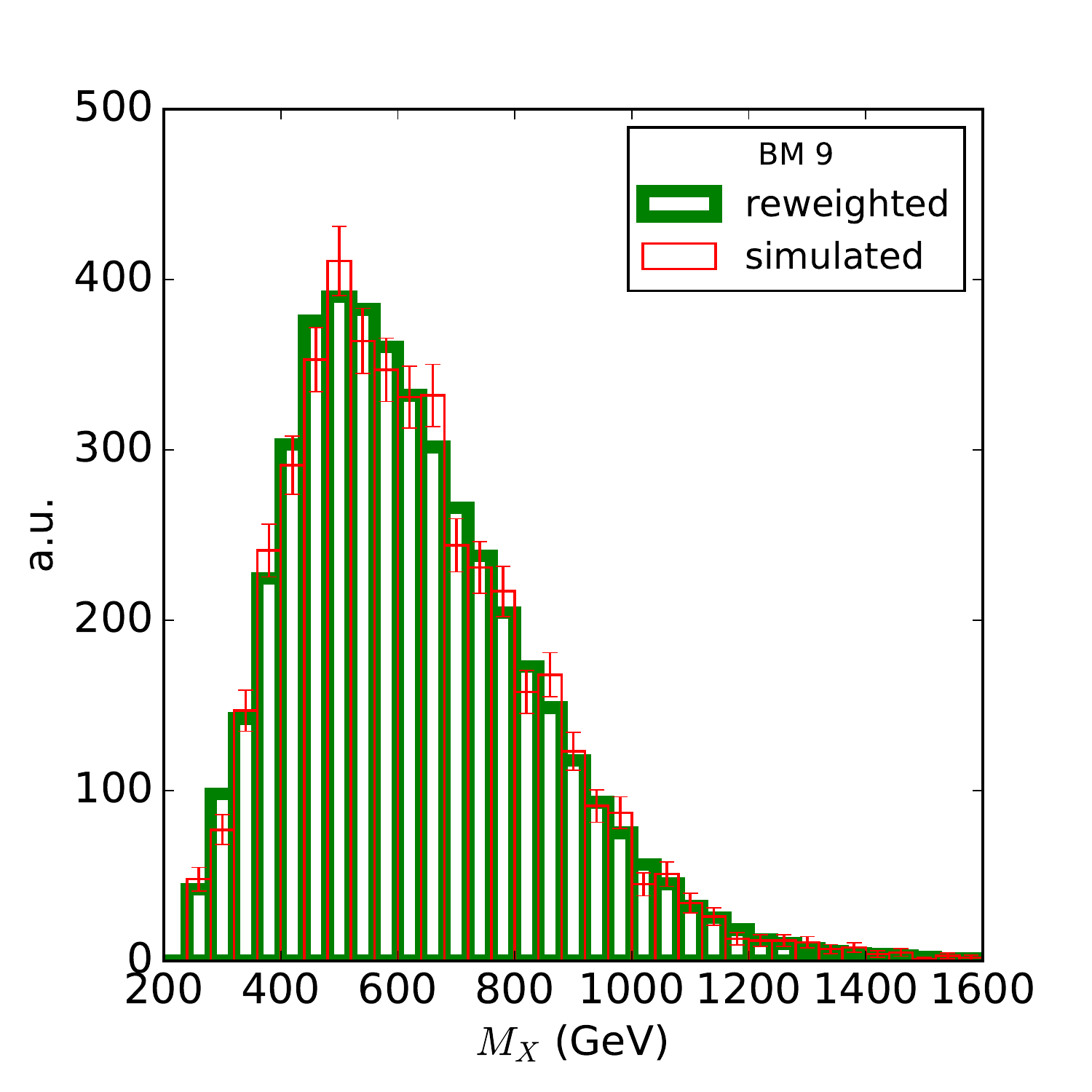}
  \includegraphics[width=0.32\textwidth]{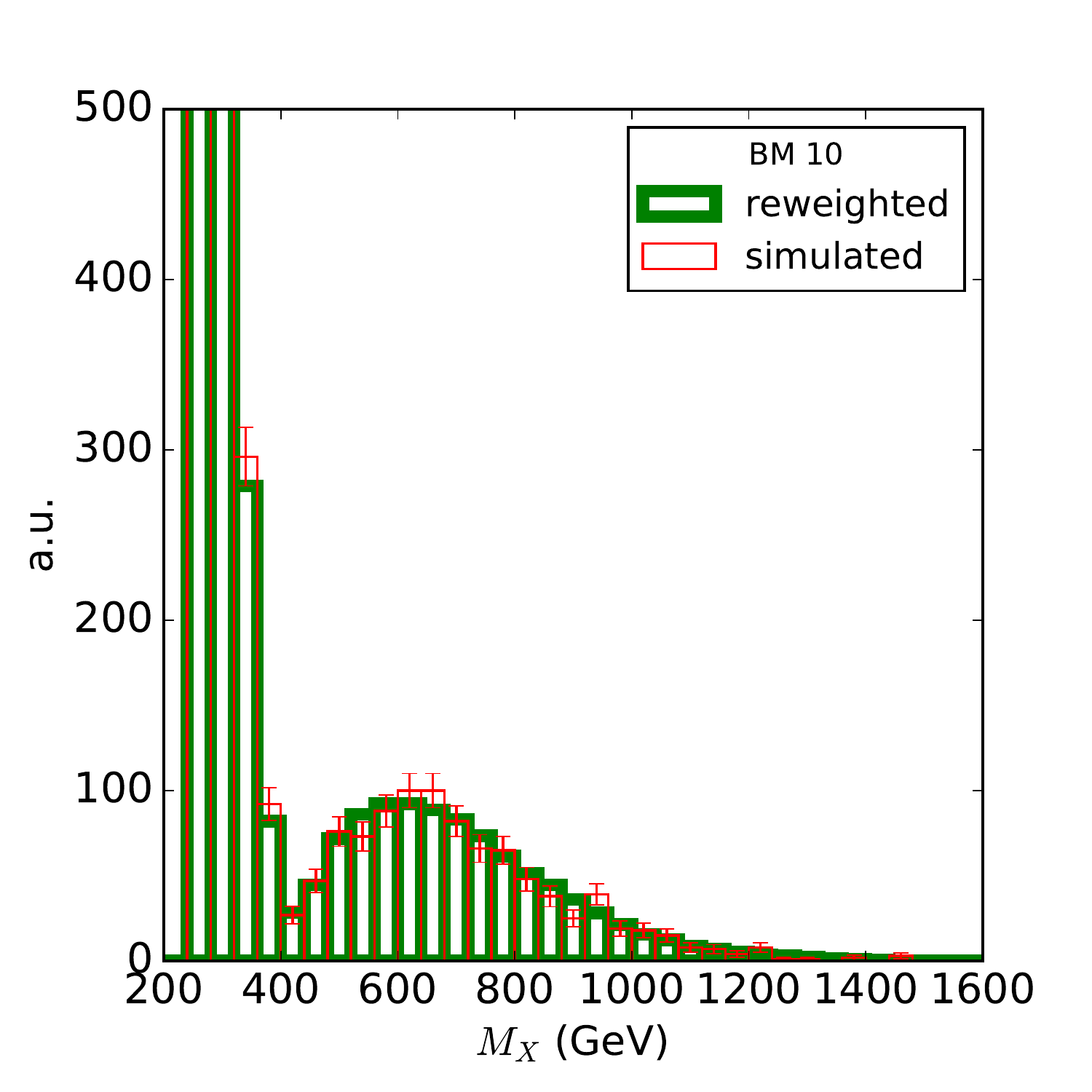}
  \includegraphics[width=0.32\textwidth]{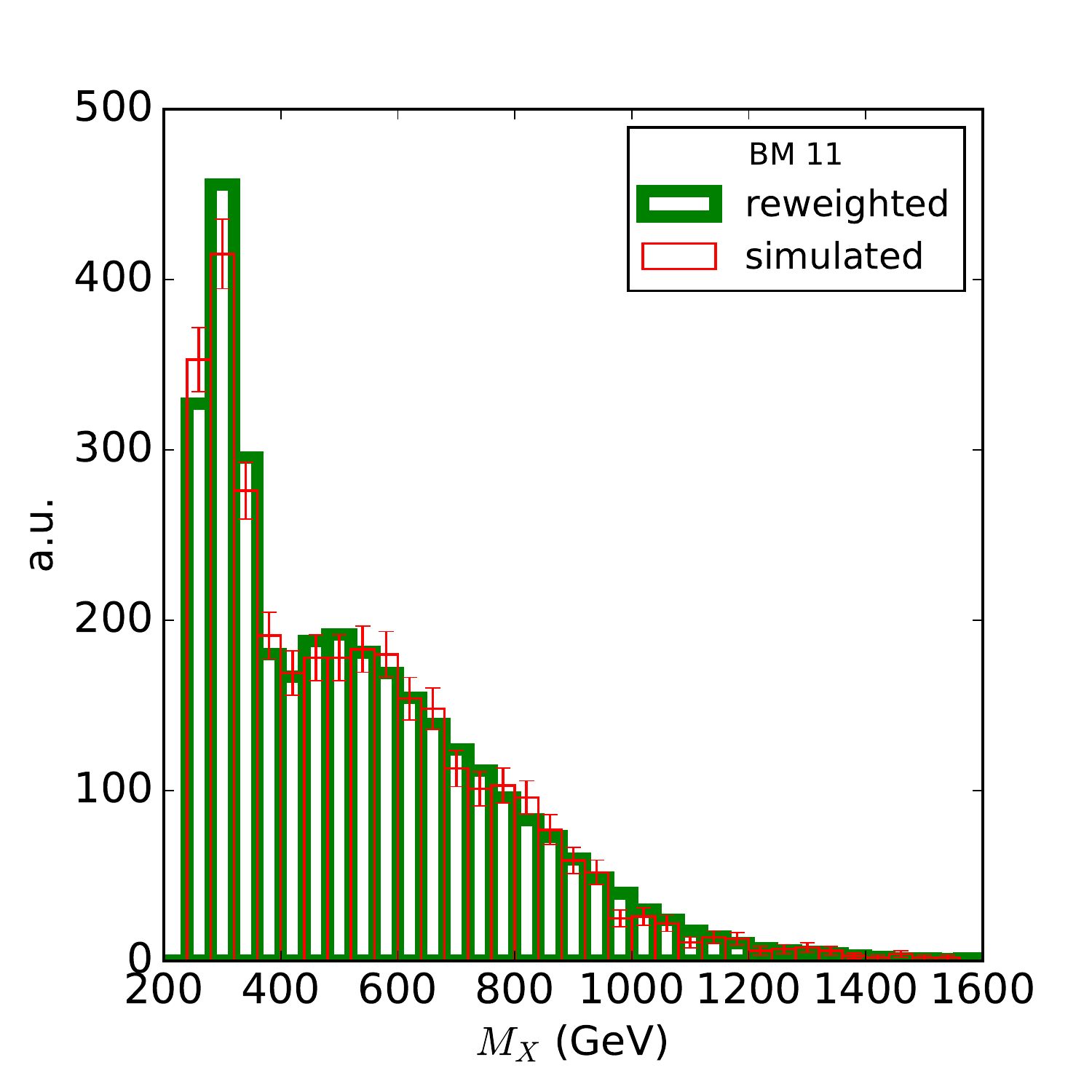}
  \includegraphics[width=0.32\textwidth]{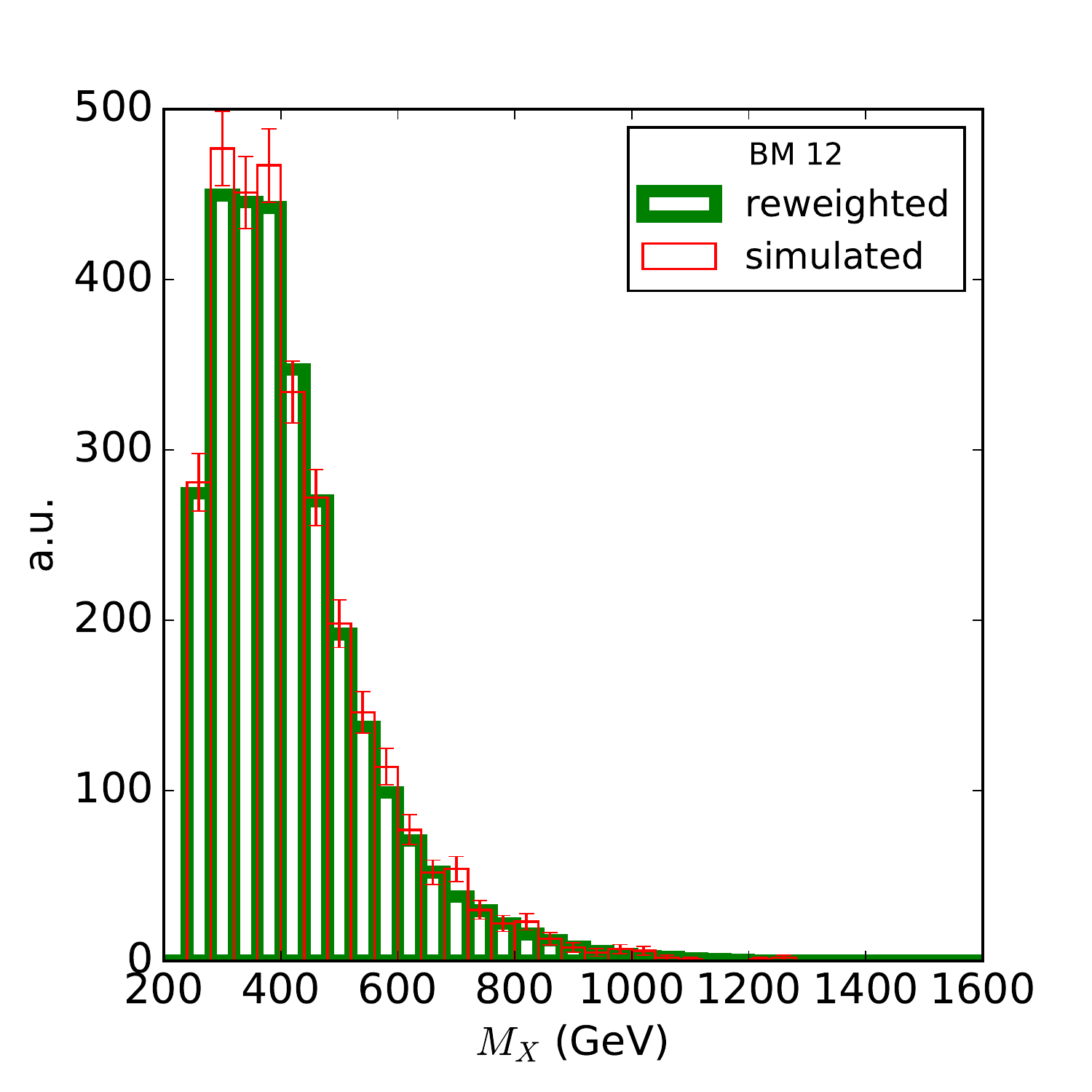}      %
  \caption{\footnotesize The reconstructed reduced mass after ATLAS-like selection.  The histograms are normalized be signal efficiency times 100,000 events.
    \label{fig:RecastBSM2}
  }
\end{figure}

\begin{figure}
  \centering
  \includegraphics[width=0.32\textwidth]{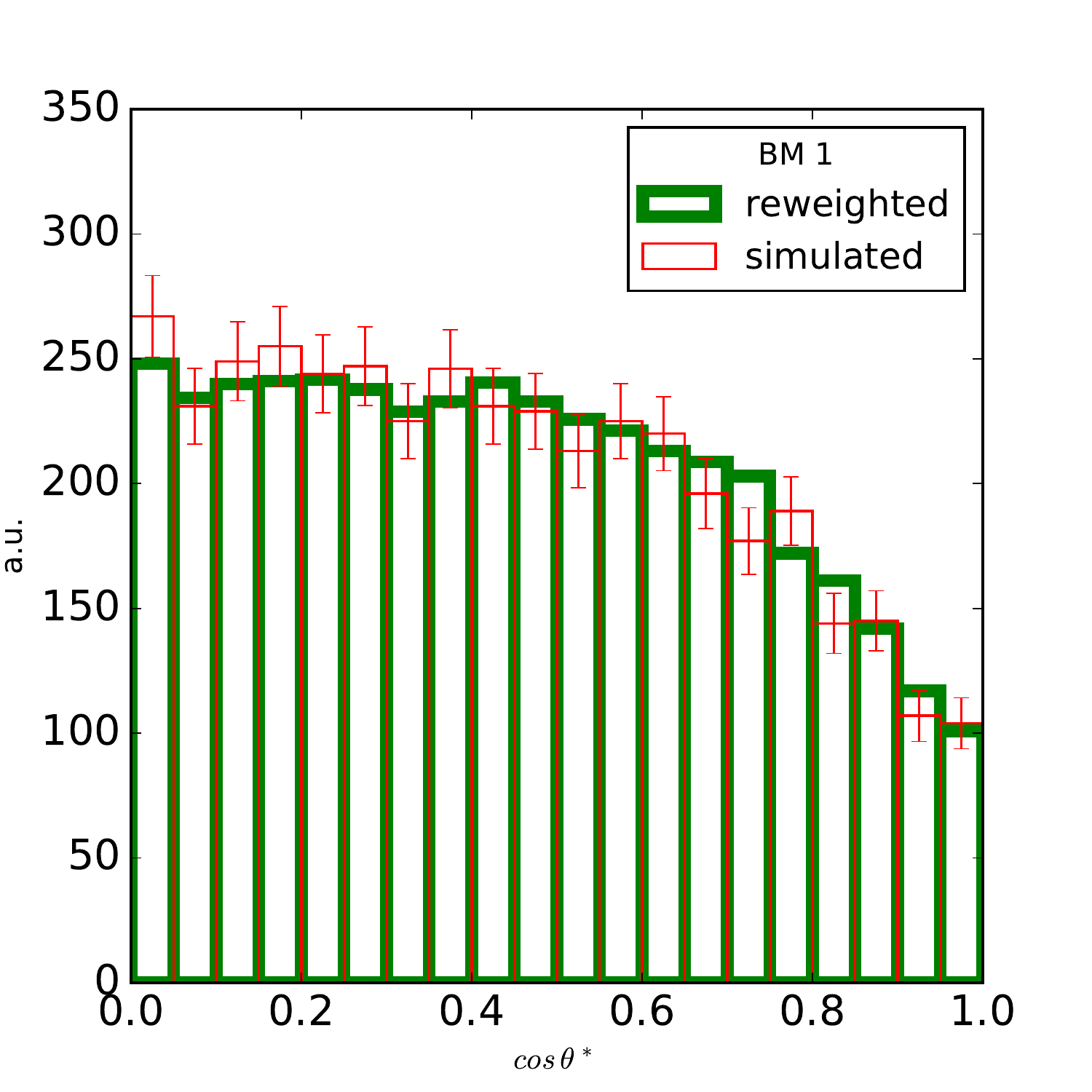}
  \includegraphics[width=0.32\textwidth]{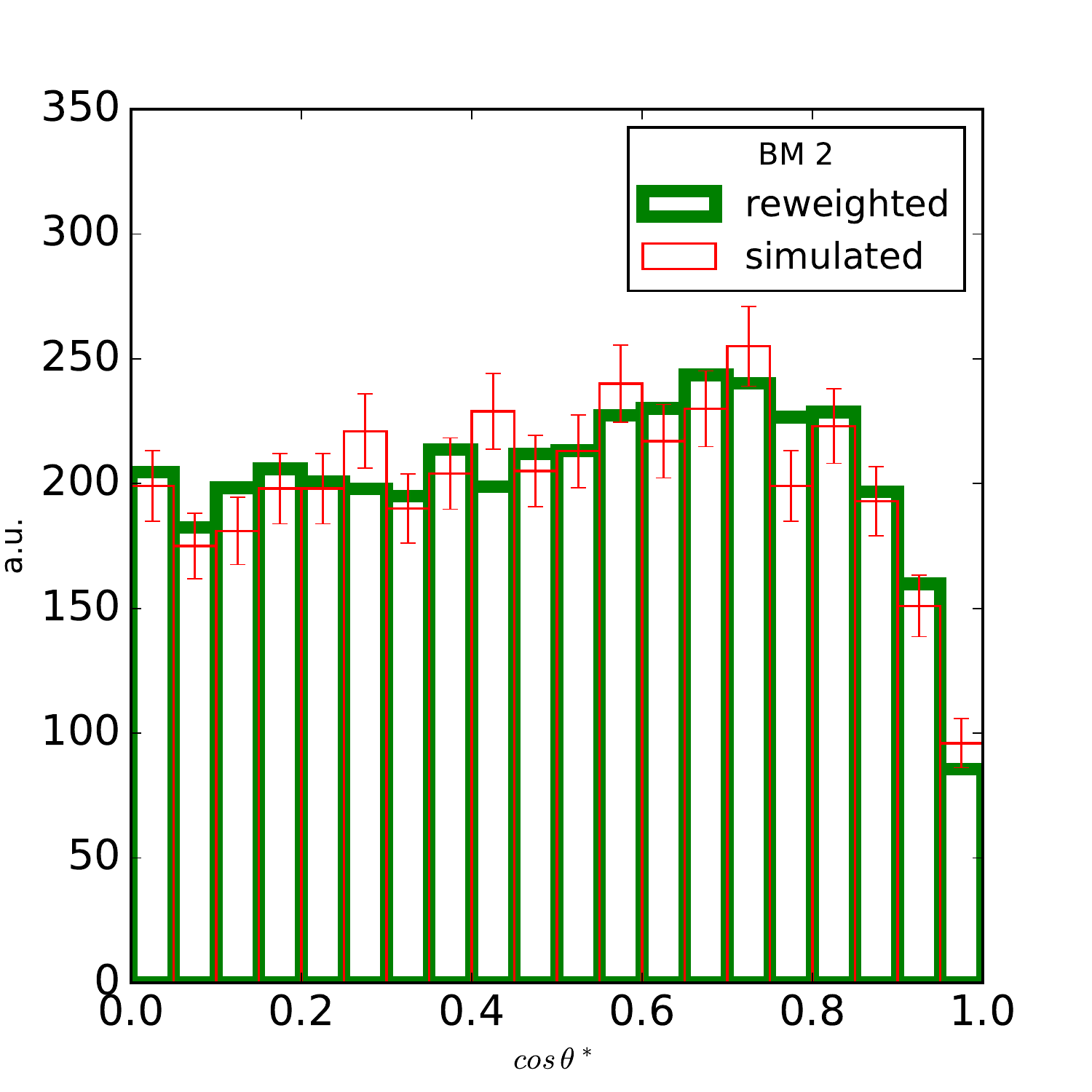}
  \includegraphics[width=0.32\textwidth]{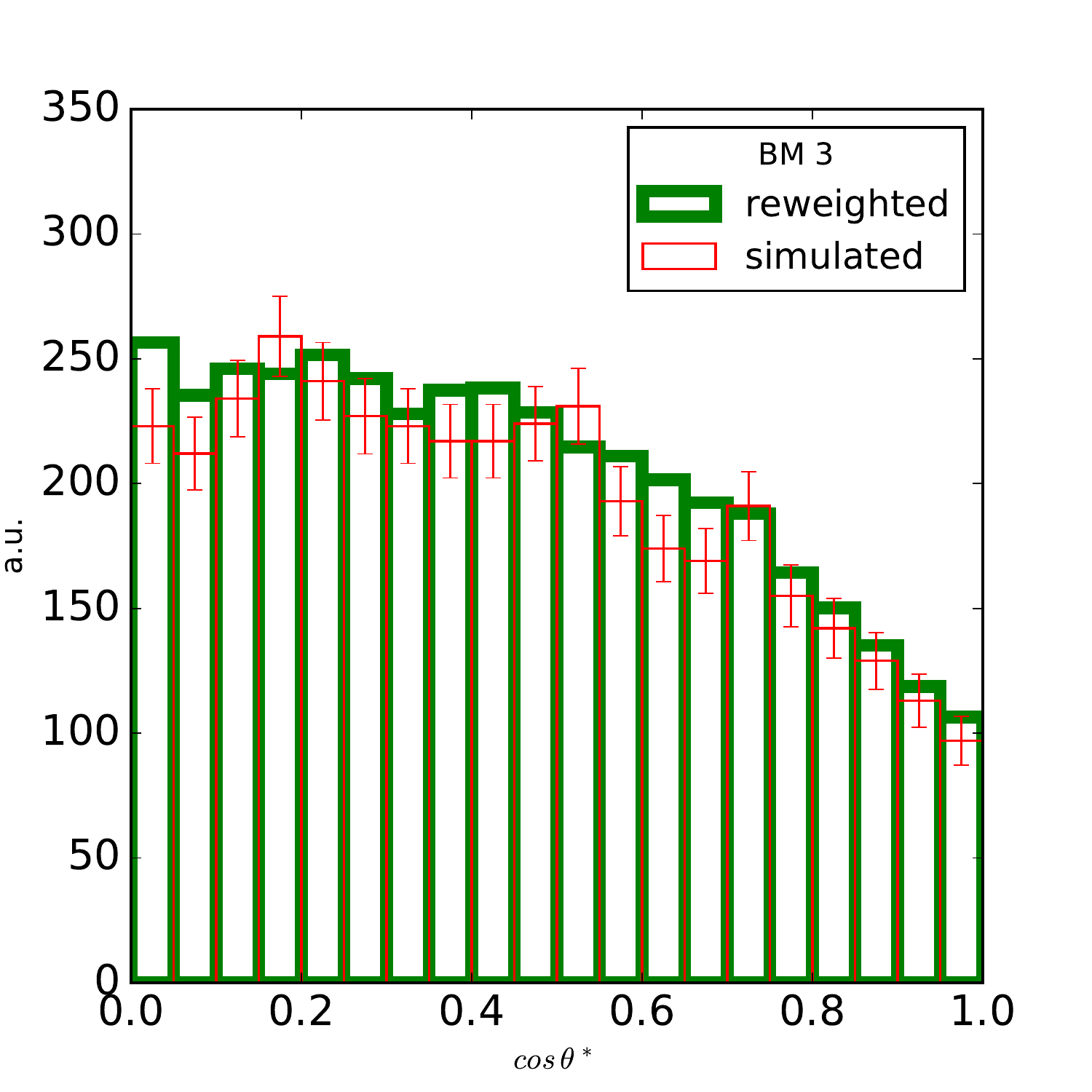}
  \includegraphics[width=0.32\textwidth]{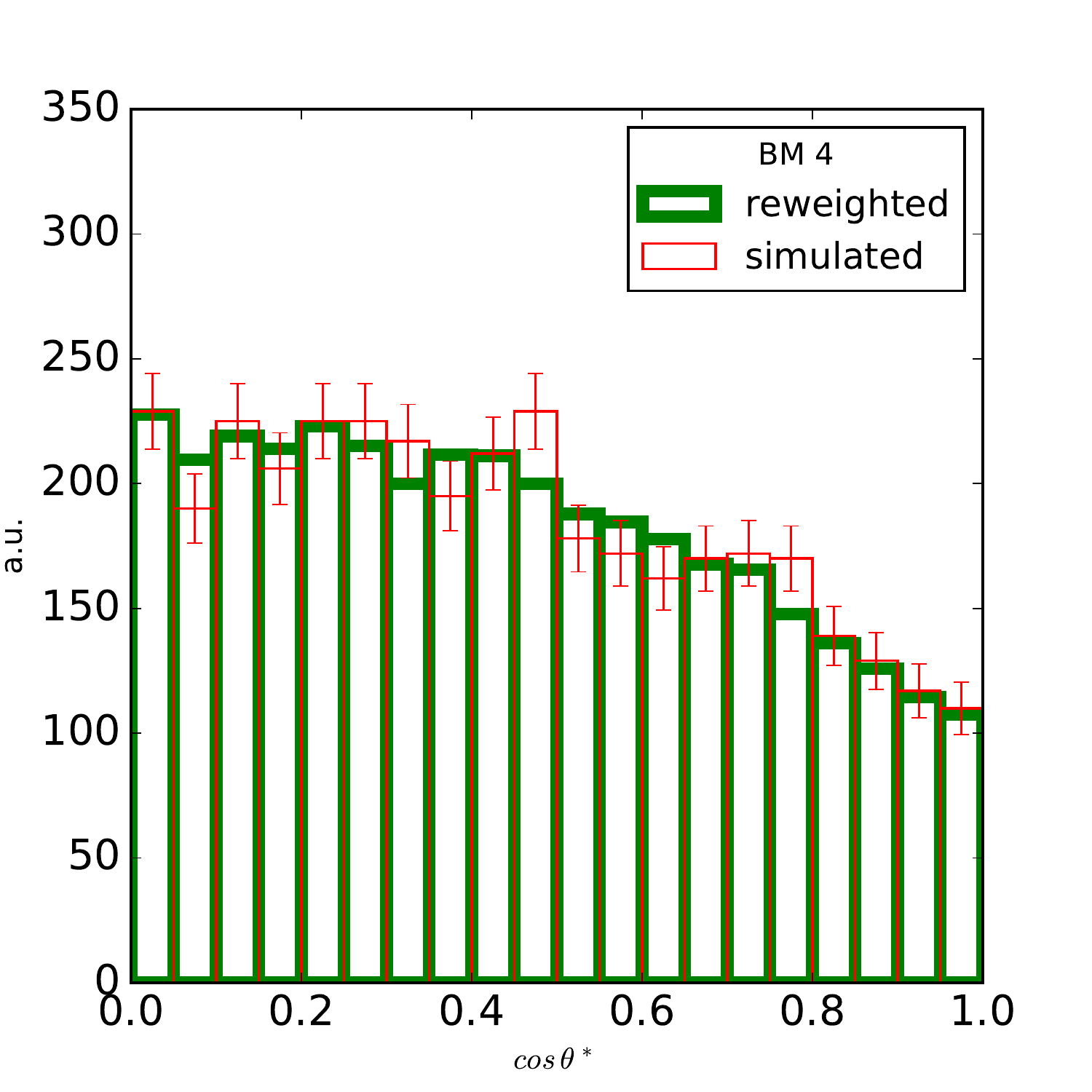}
  \includegraphics[width=0.32\textwidth]{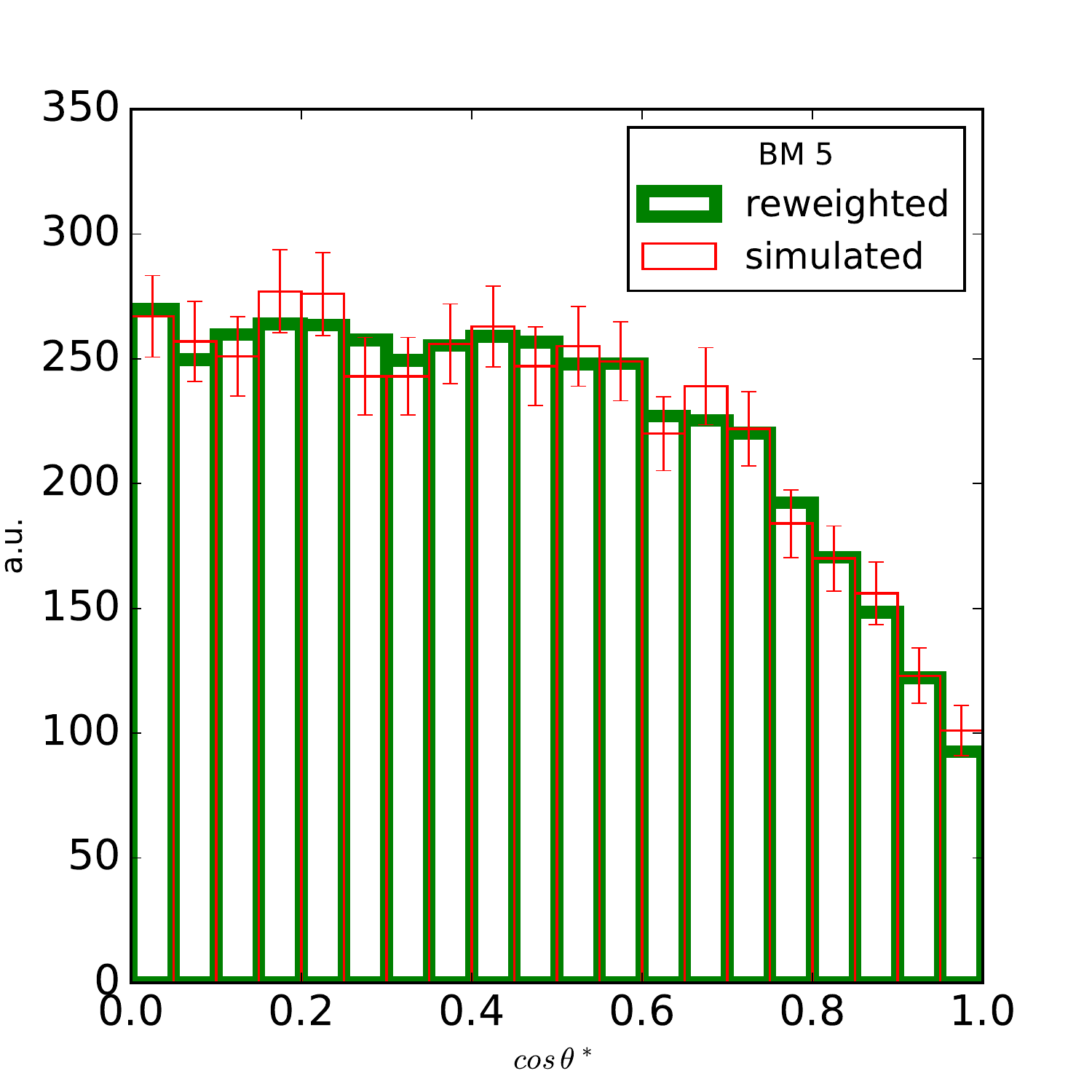}
  \includegraphics[width=0.32\textwidth]{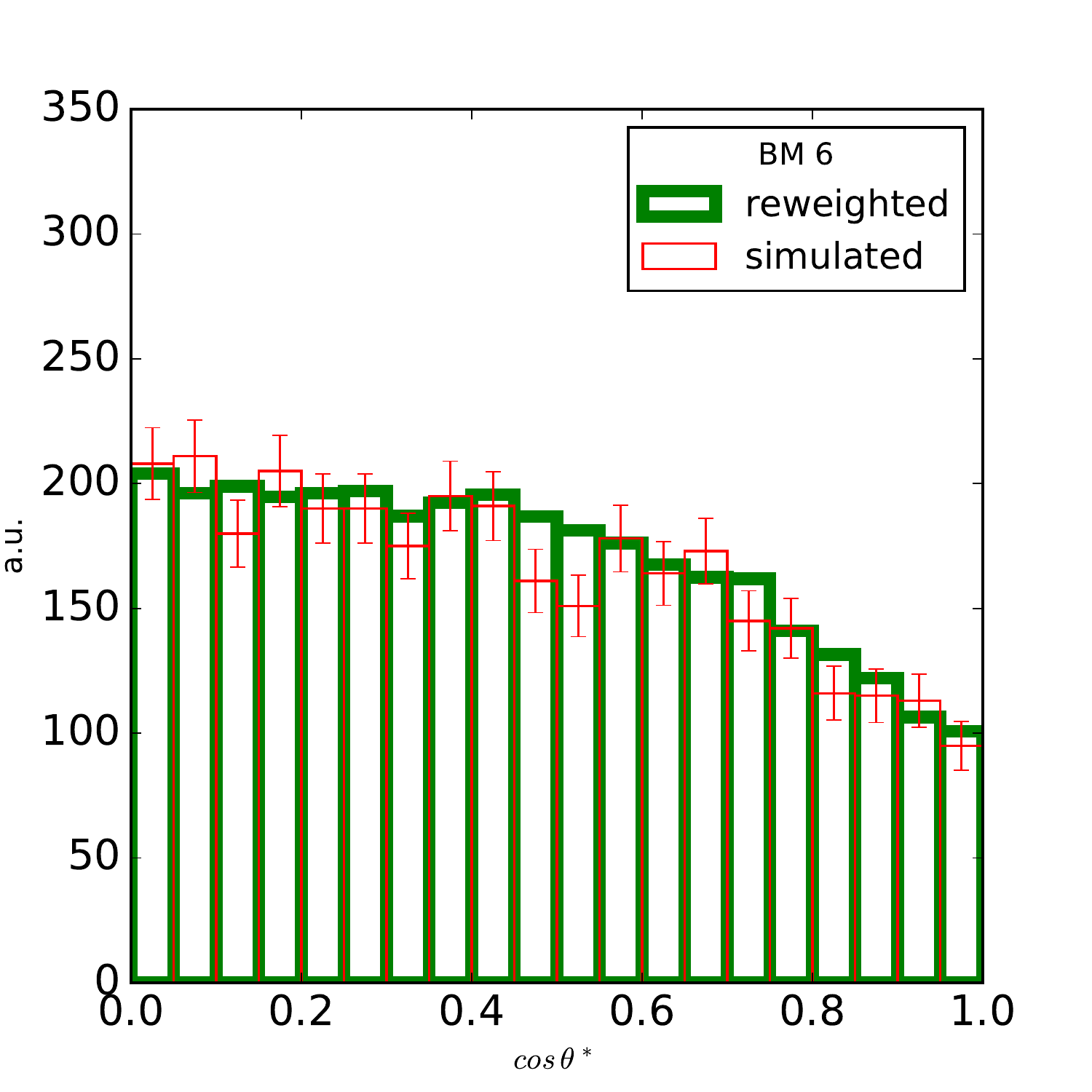}
  \includegraphics[width=0.32\textwidth]{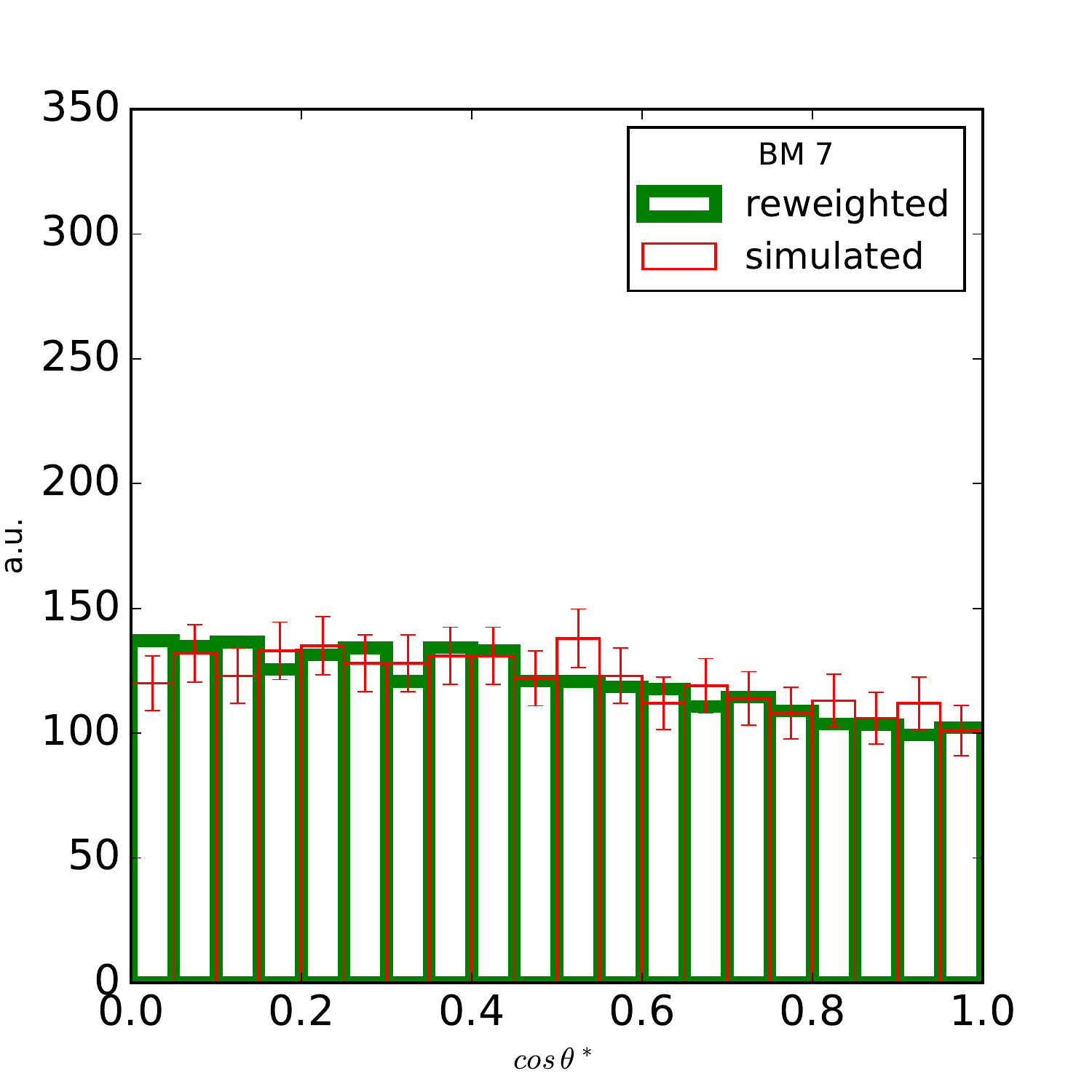}
  \includegraphics[width=0.32\textwidth]{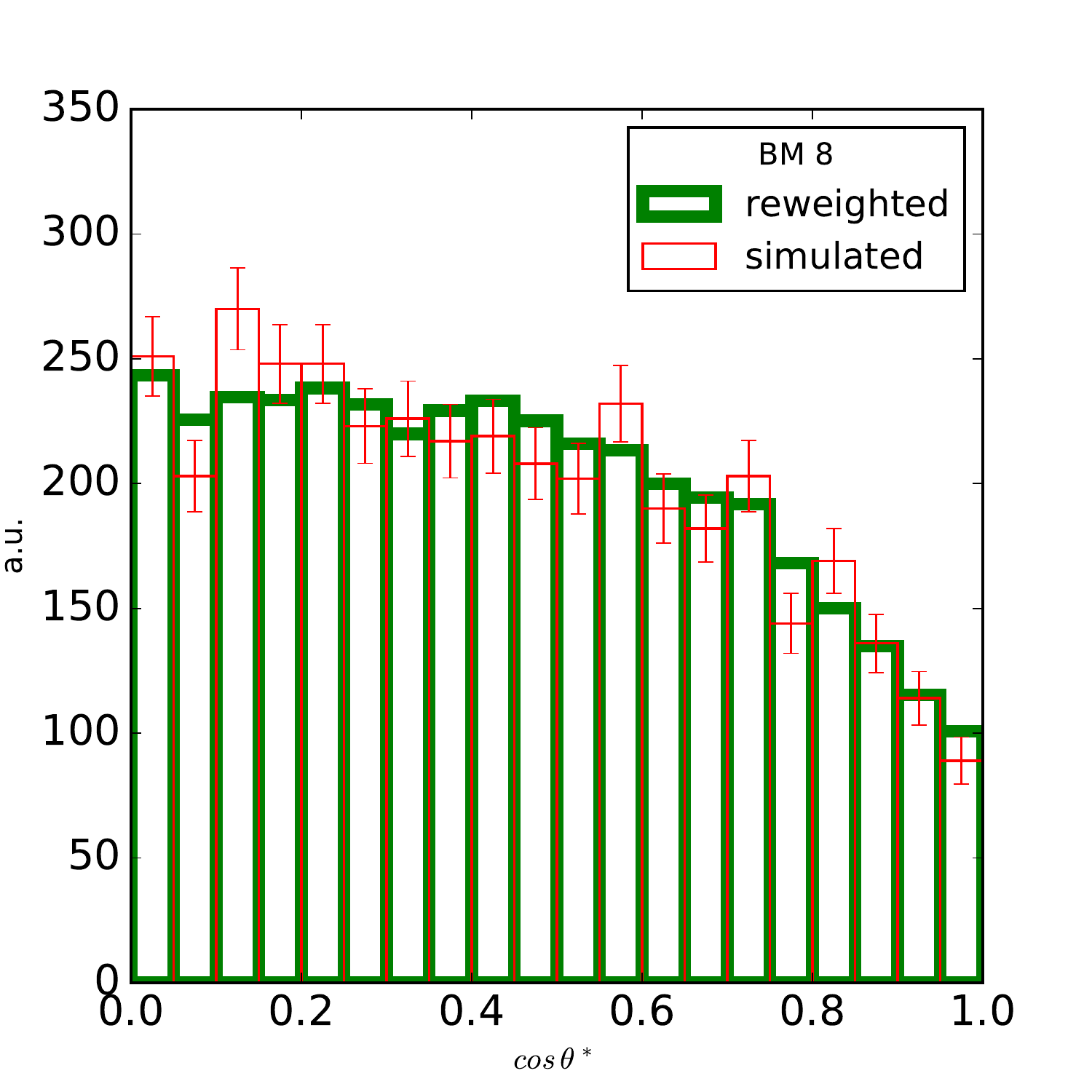}
  \includegraphics[width=0.32\textwidth]{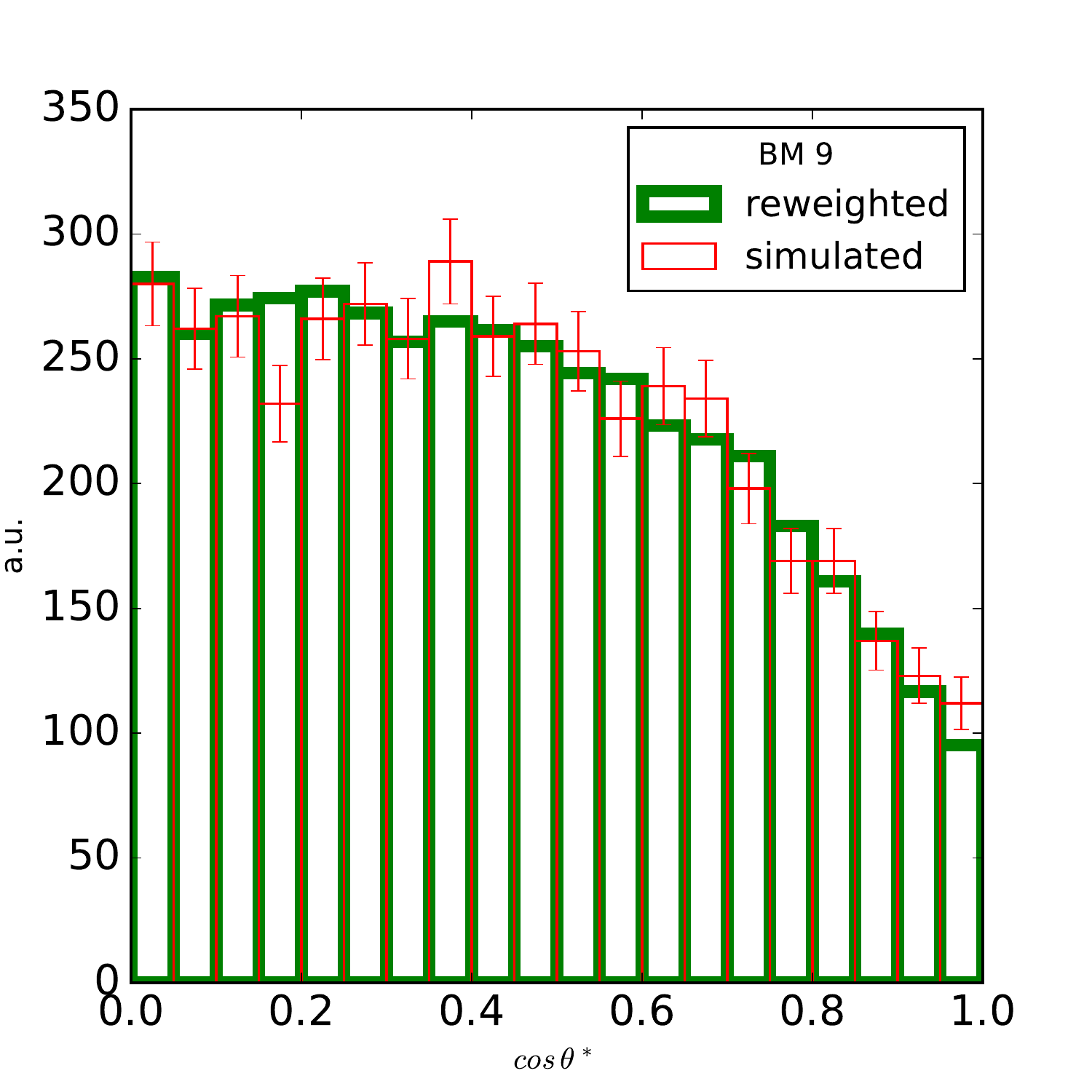}
  \includegraphics[width=0.32\textwidth]{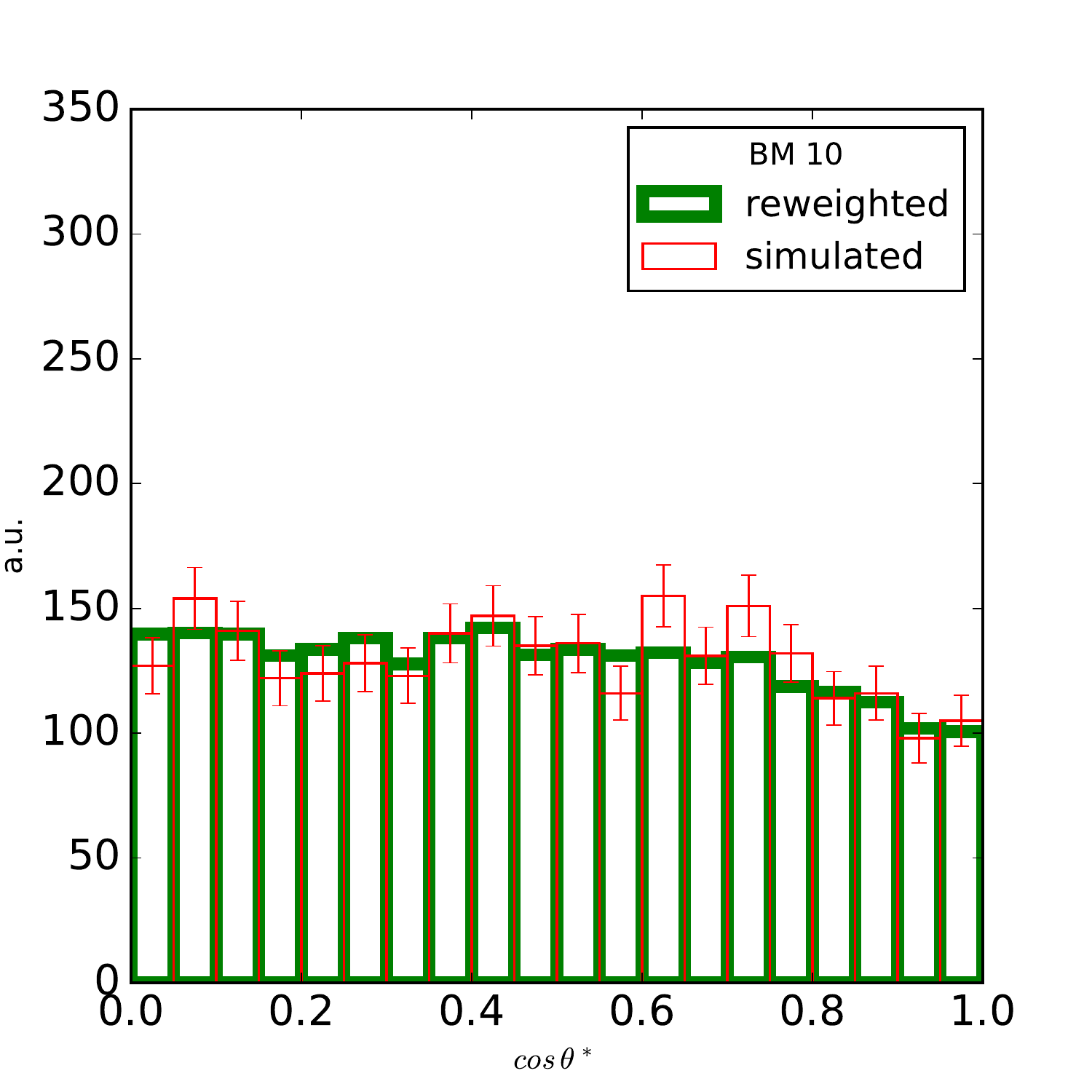}
  \includegraphics[width=0.32\textwidth]{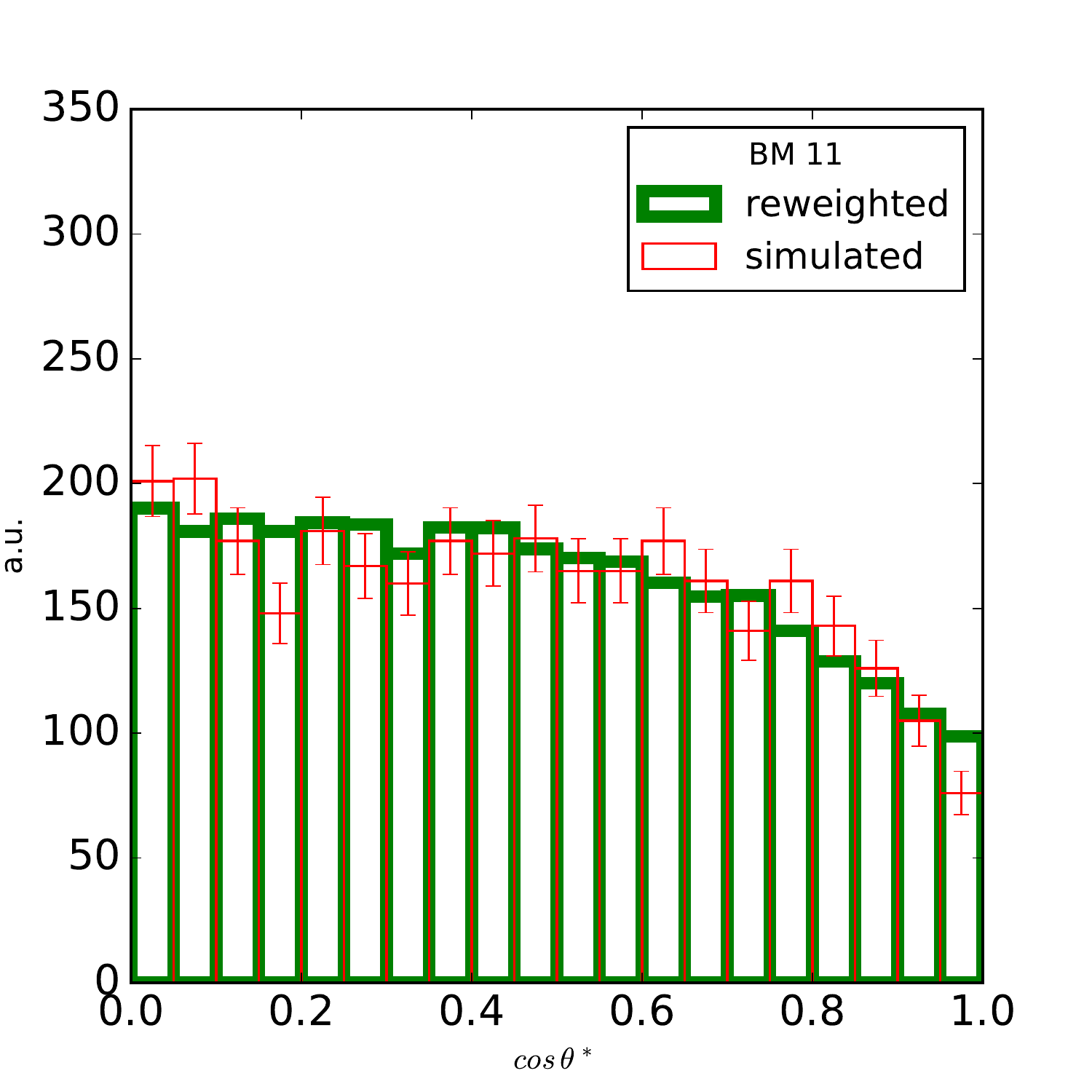}
  \includegraphics[width=0.32\textwidth]{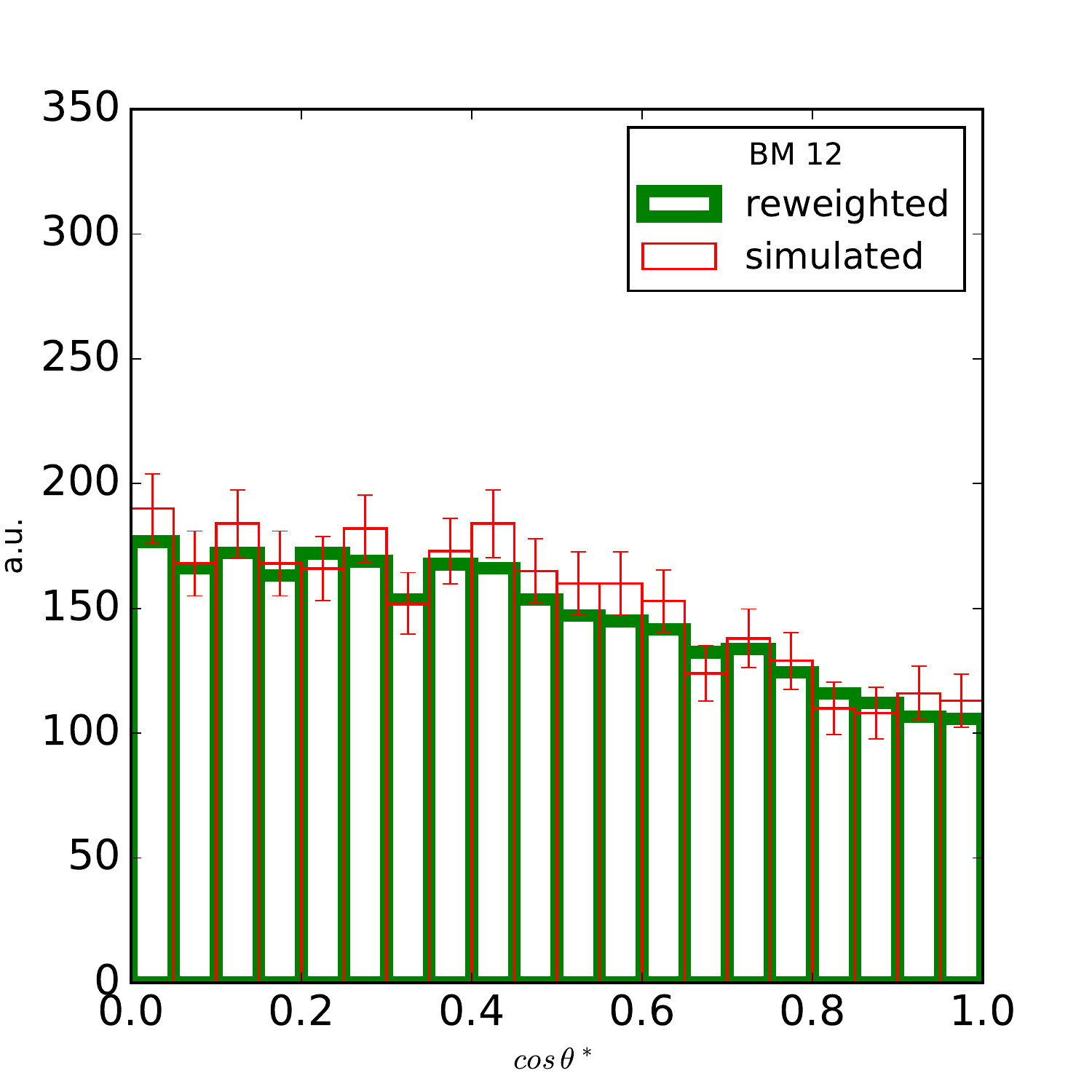}      %
  \caption{\footnotesize The reconstructed $\cos\theta^*$ after ATLAS-like selection.  The histograms are normalized be signal efficiency times 100,000 events.
    \label{fig:RecastBSM1}
  }
\end{figure}

\begin{figure}
  \centering
  \includegraphics[width=0.32\textwidth]{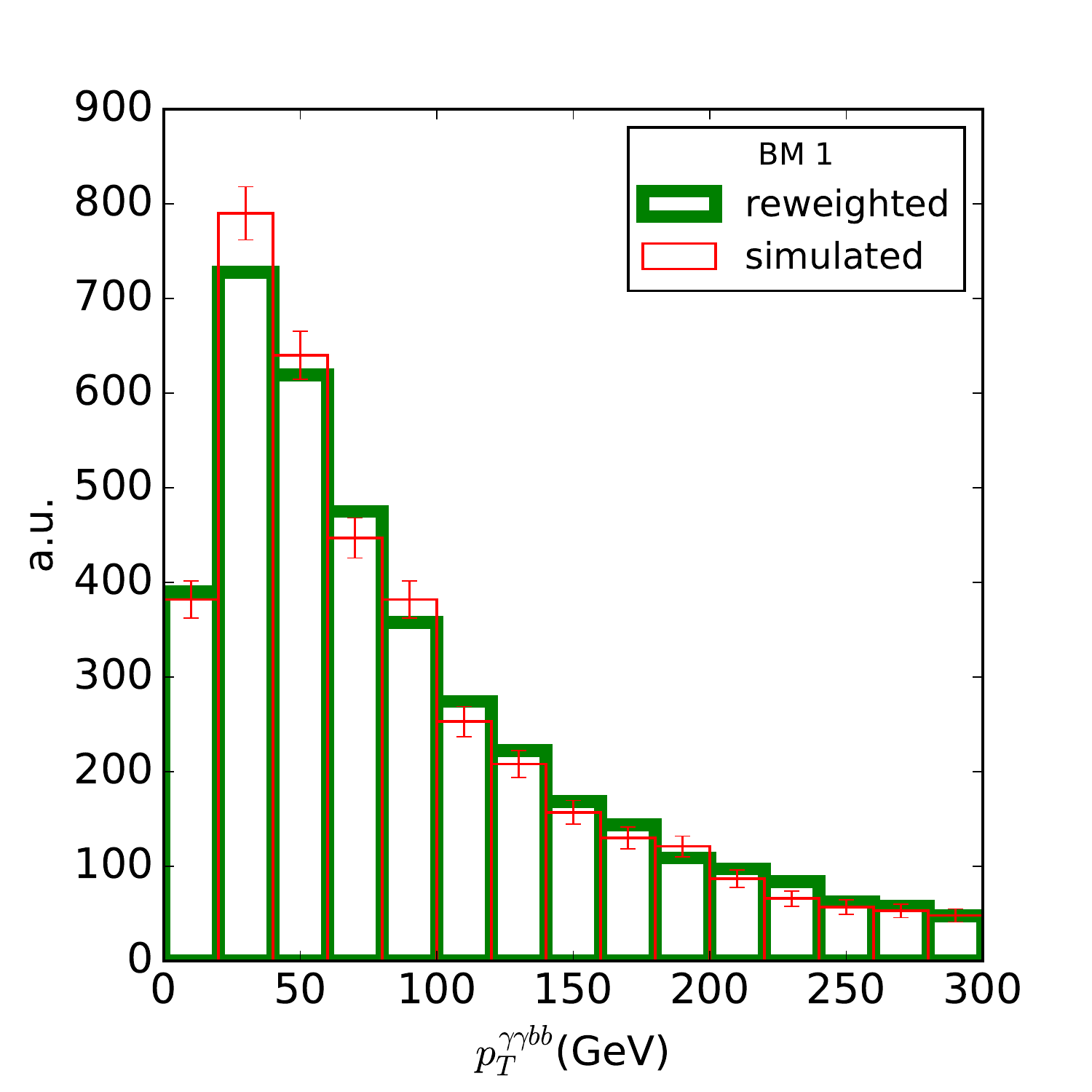}
  \includegraphics[width=0.32\textwidth]{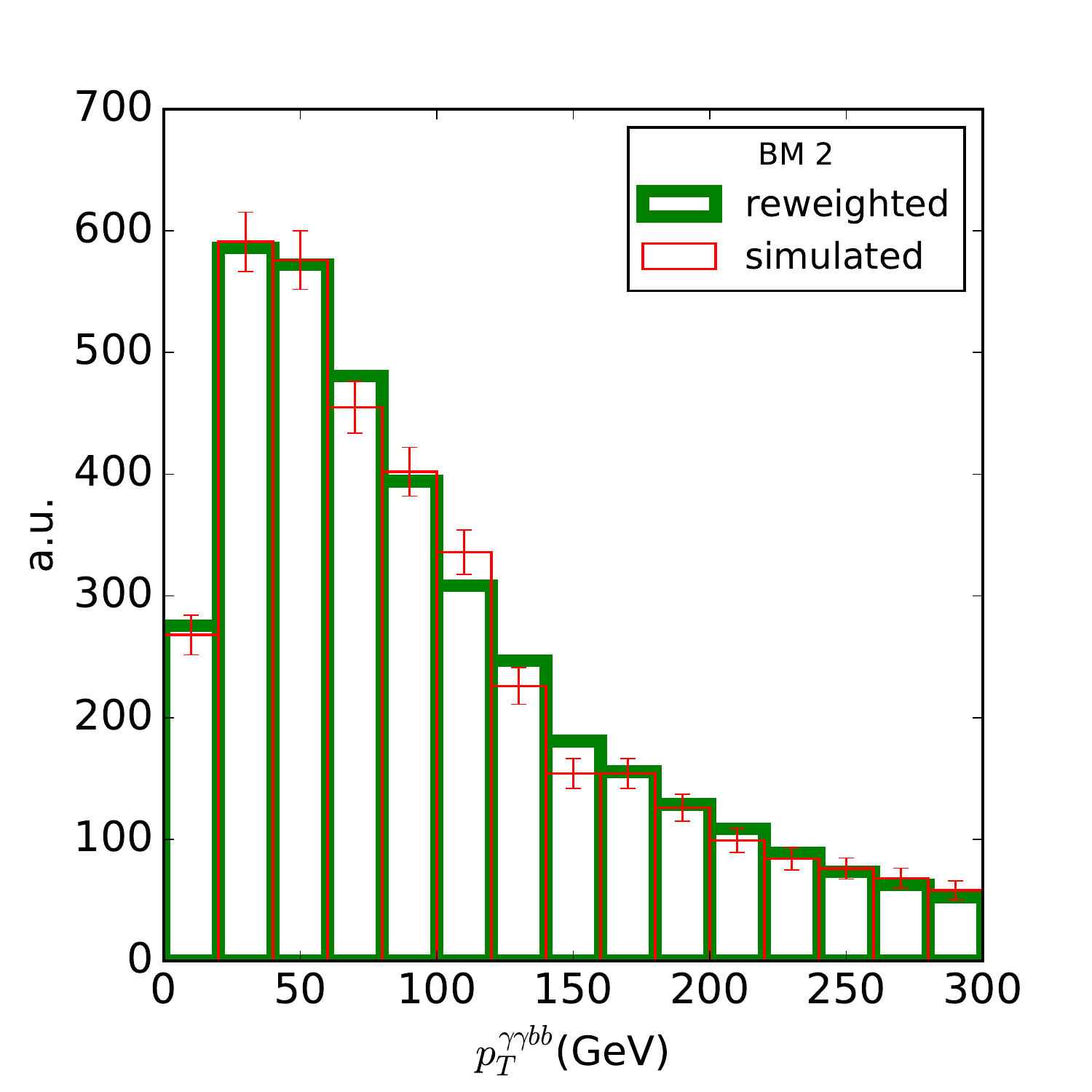}
  \includegraphics[width=0.32\textwidth]{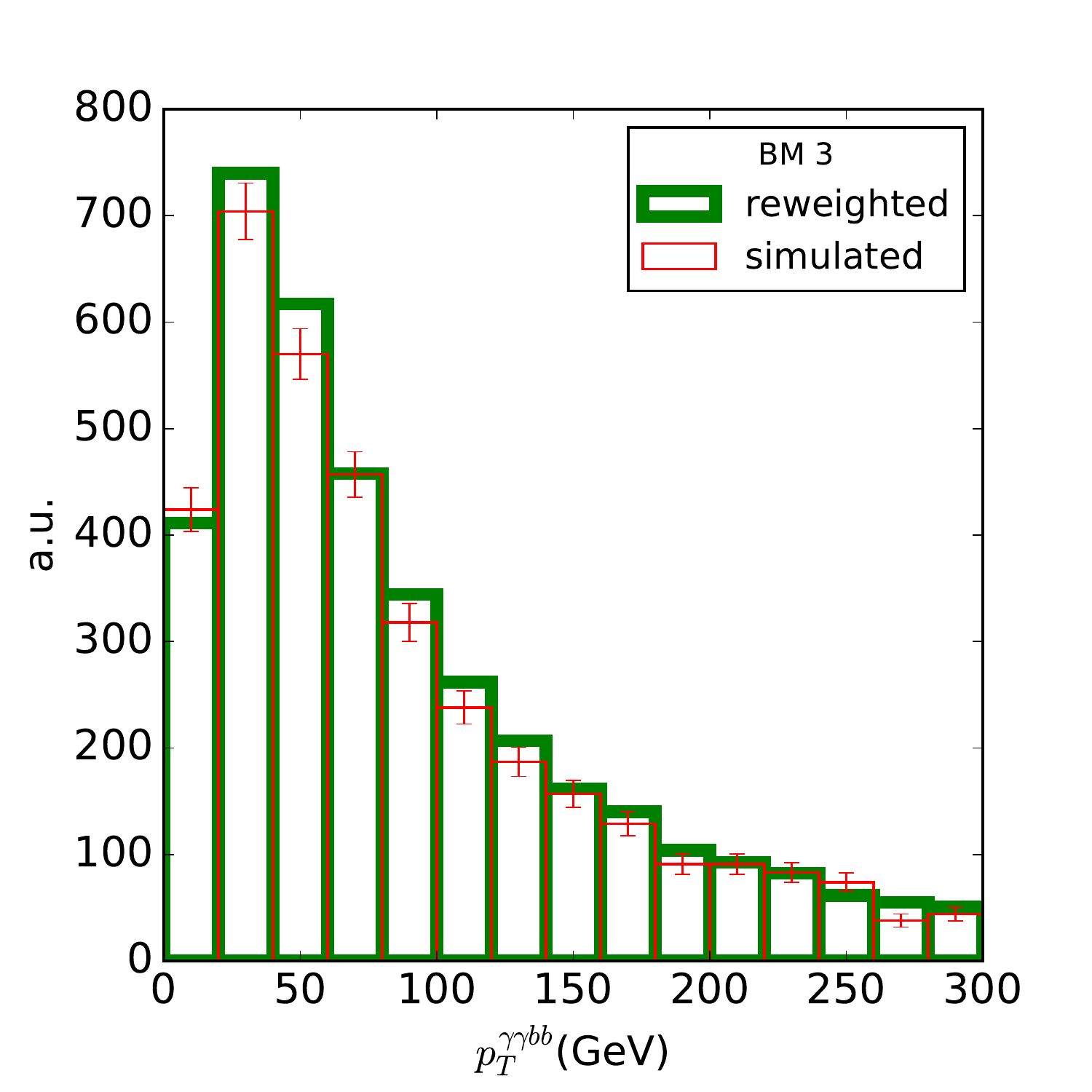}
  \includegraphics[width=0.32\textwidth]{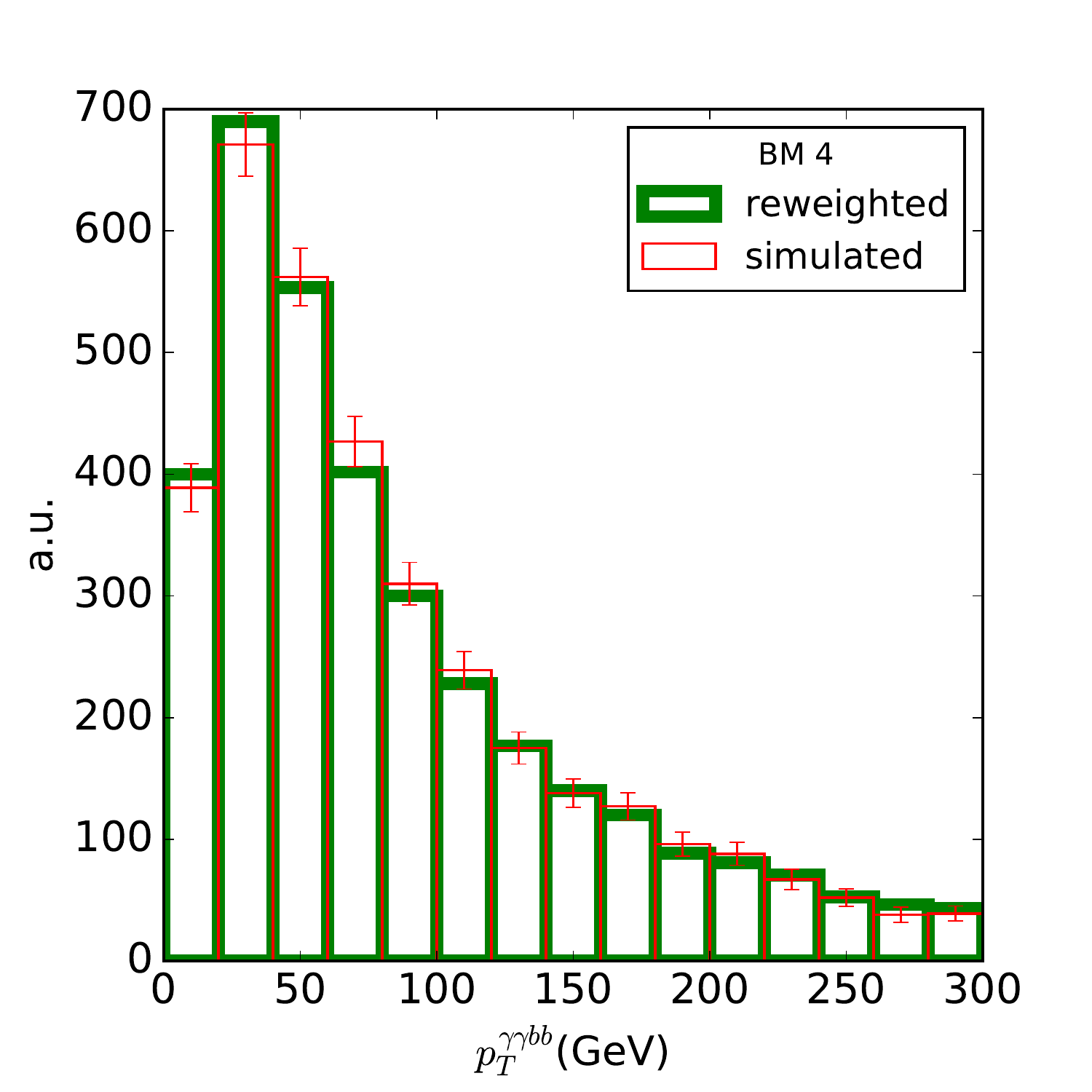}
  \includegraphics[width=0.32\textwidth]{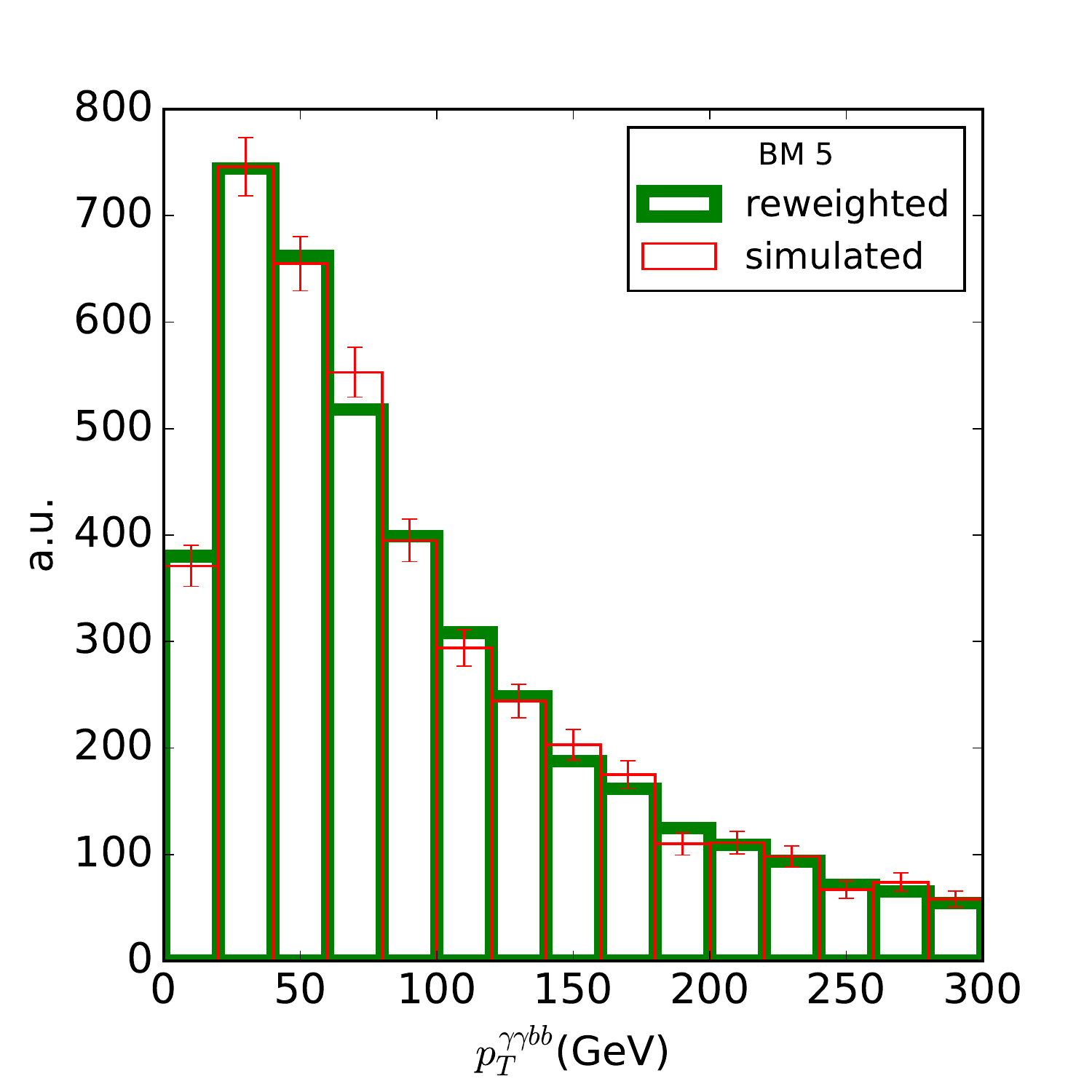}
  \includegraphics[width=0.32\textwidth]{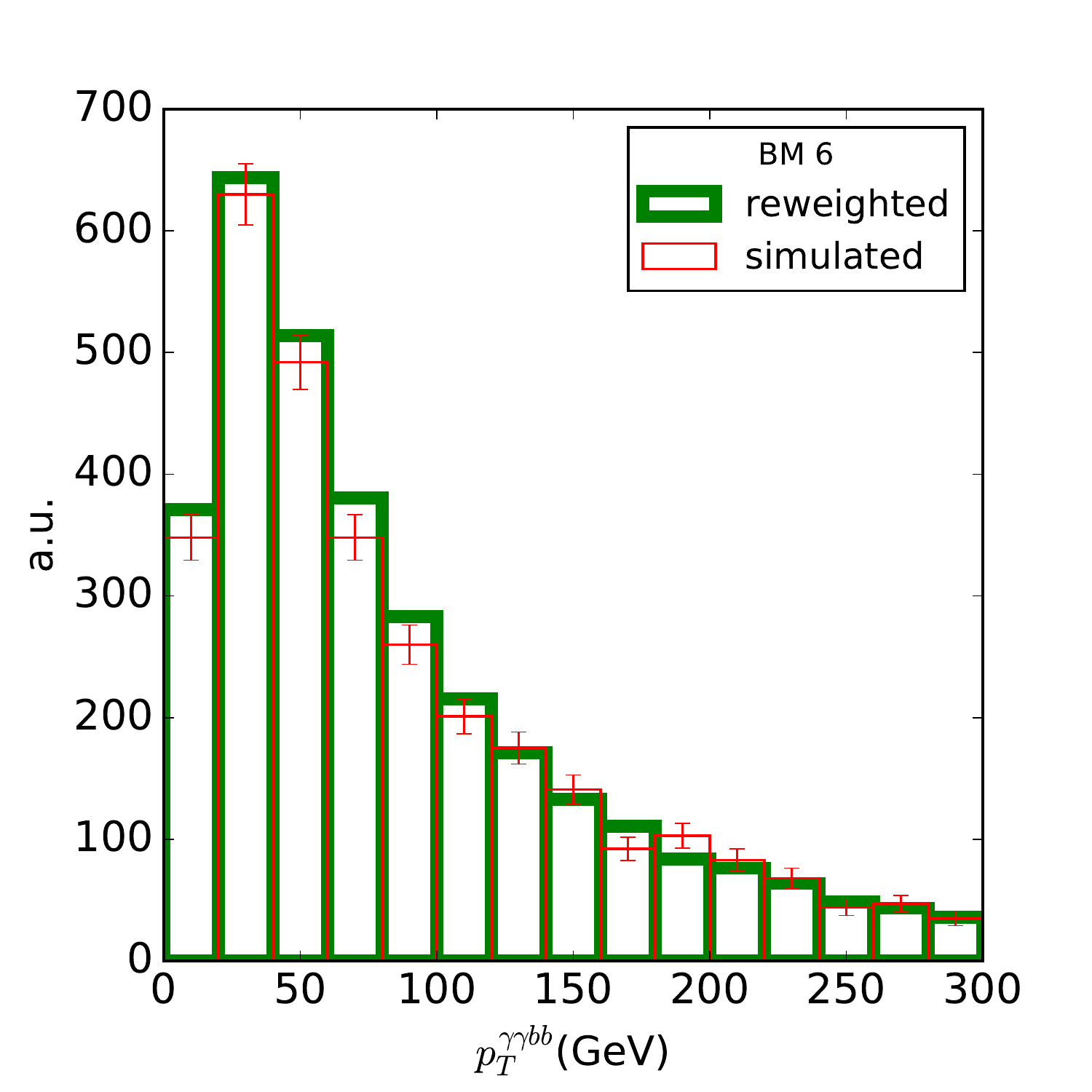}
  \includegraphics[width=0.32\textwidth]{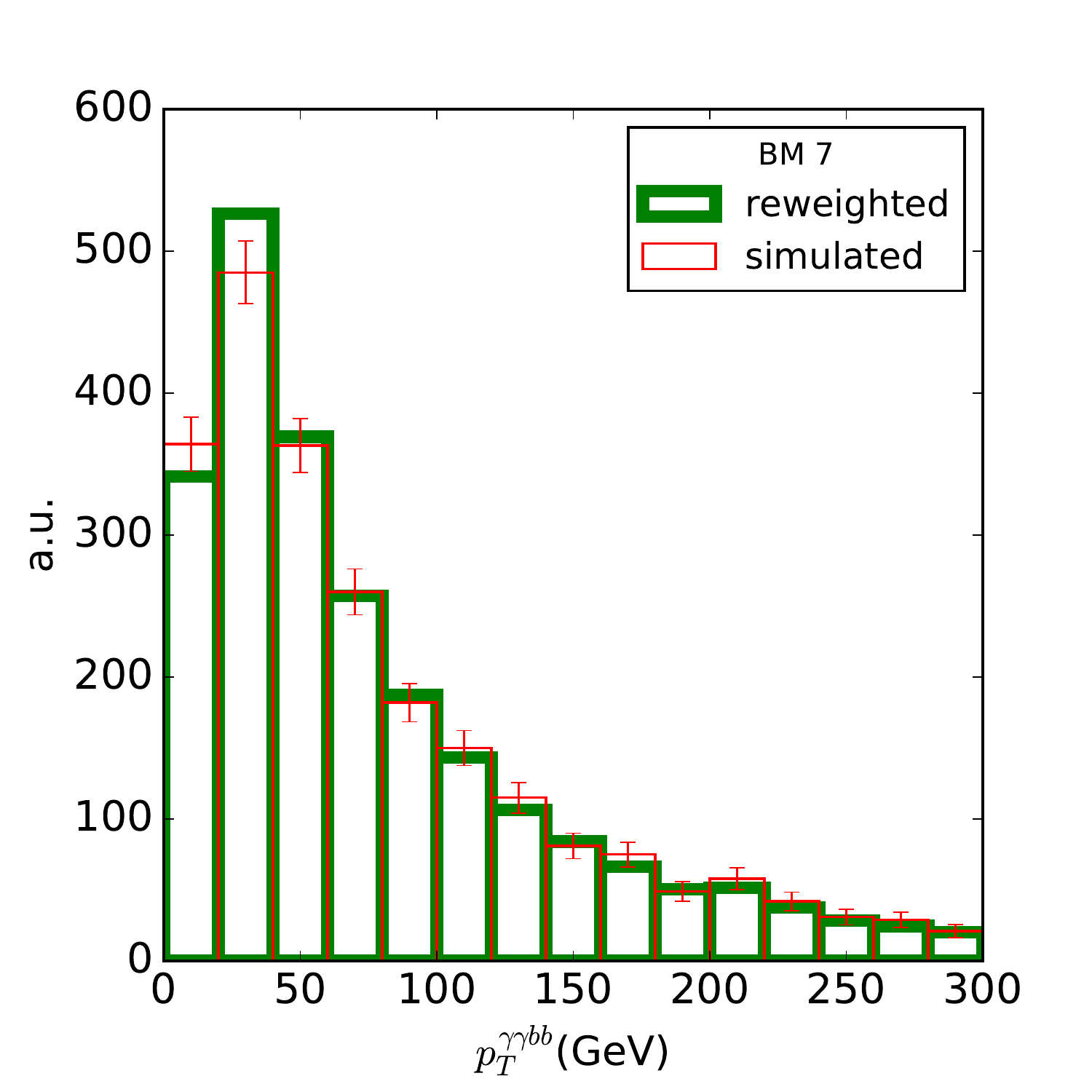}
  \includegraphics[width=0.32\textwidth]{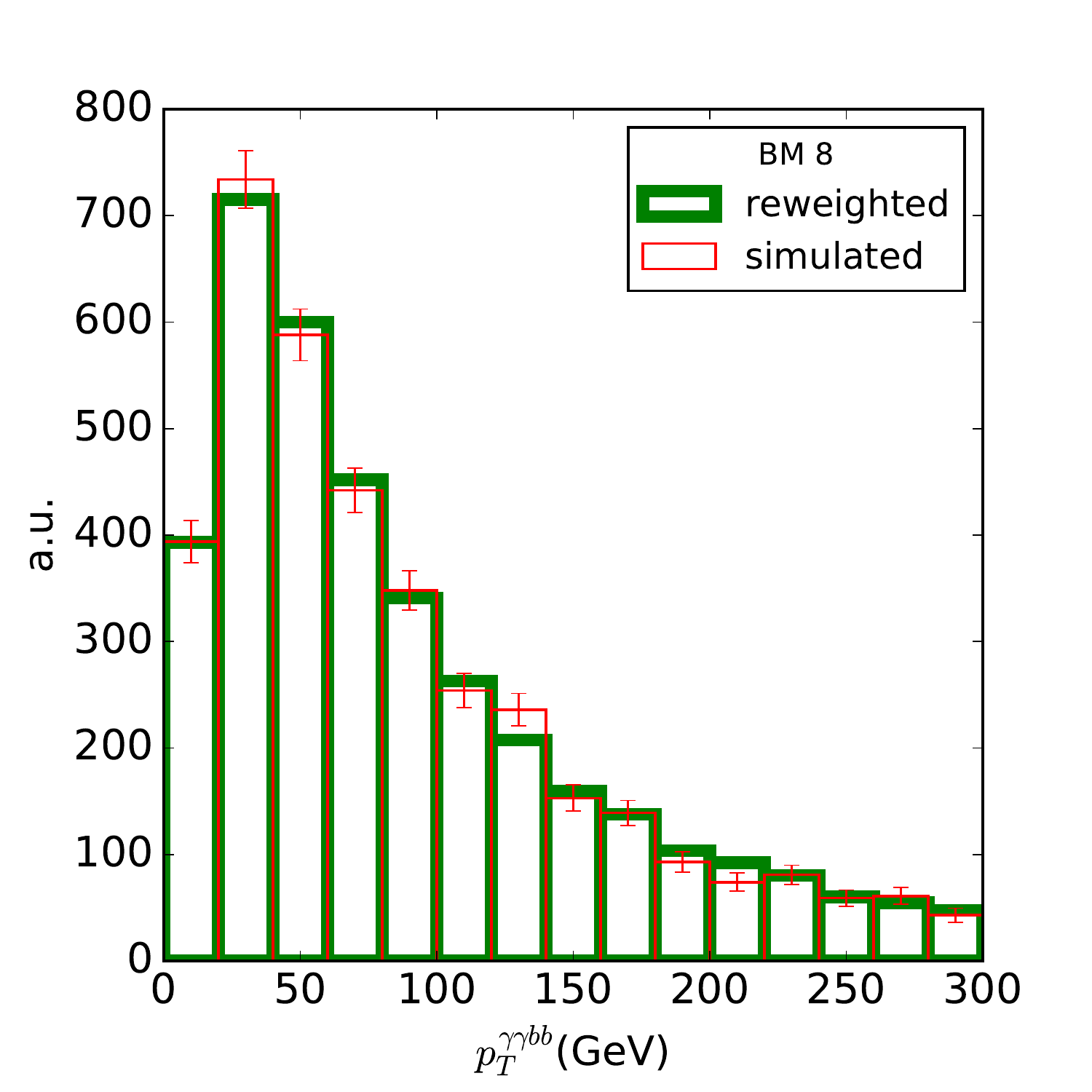}
  \includegraphics[width=0.32\textwidth]{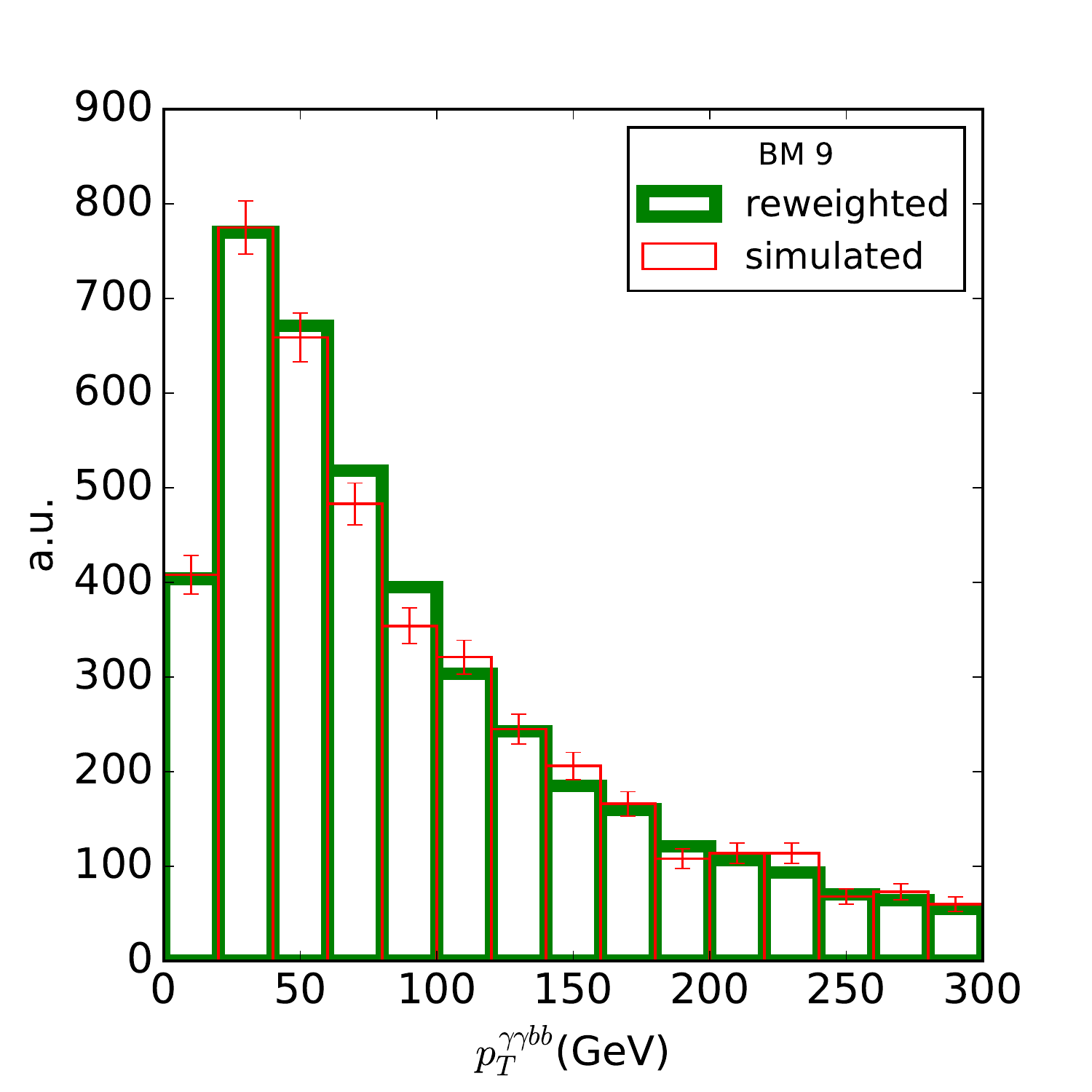}
  \includegraphics[width=0.32\textwidth]{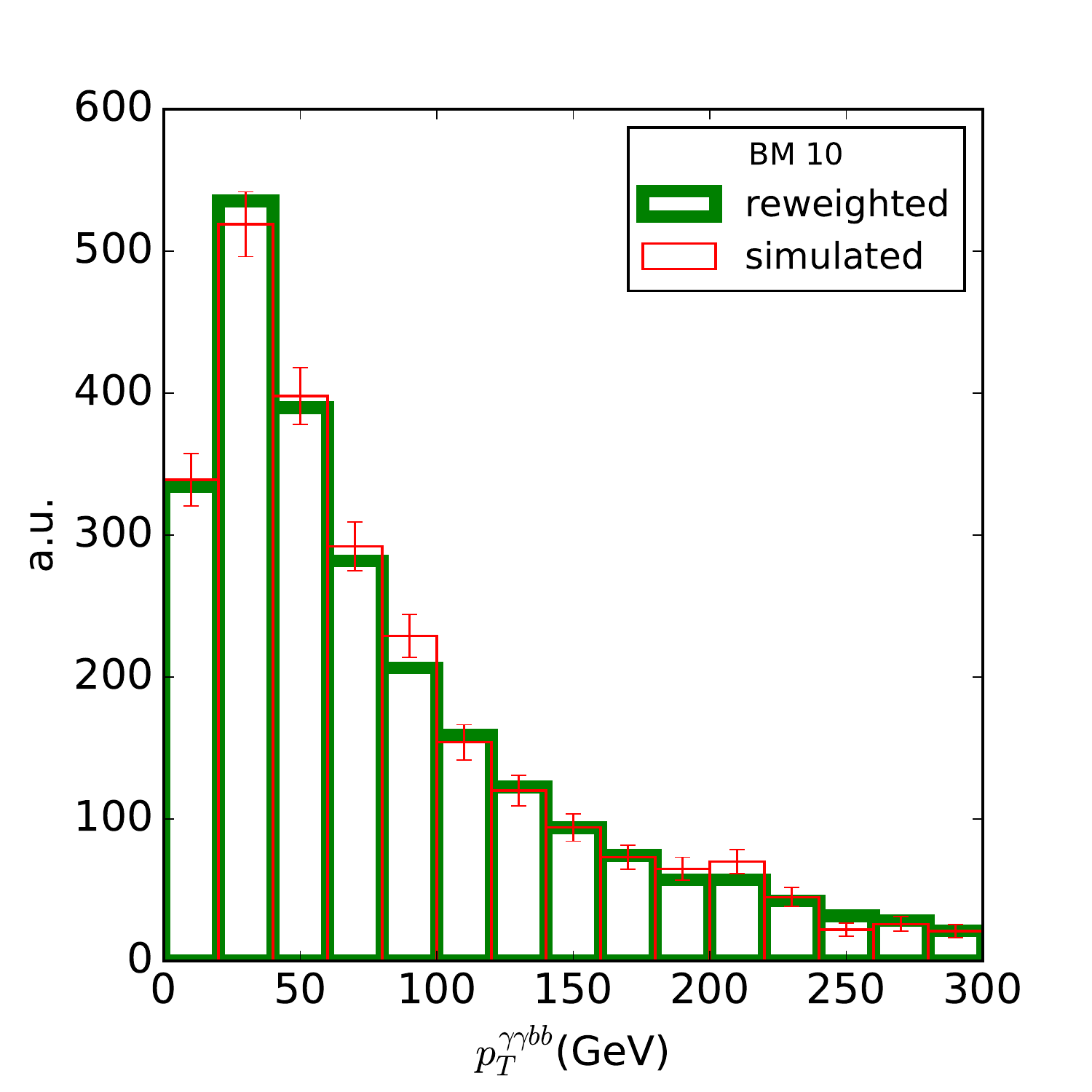}
  \includegraphics[width=0.32\textwidth]{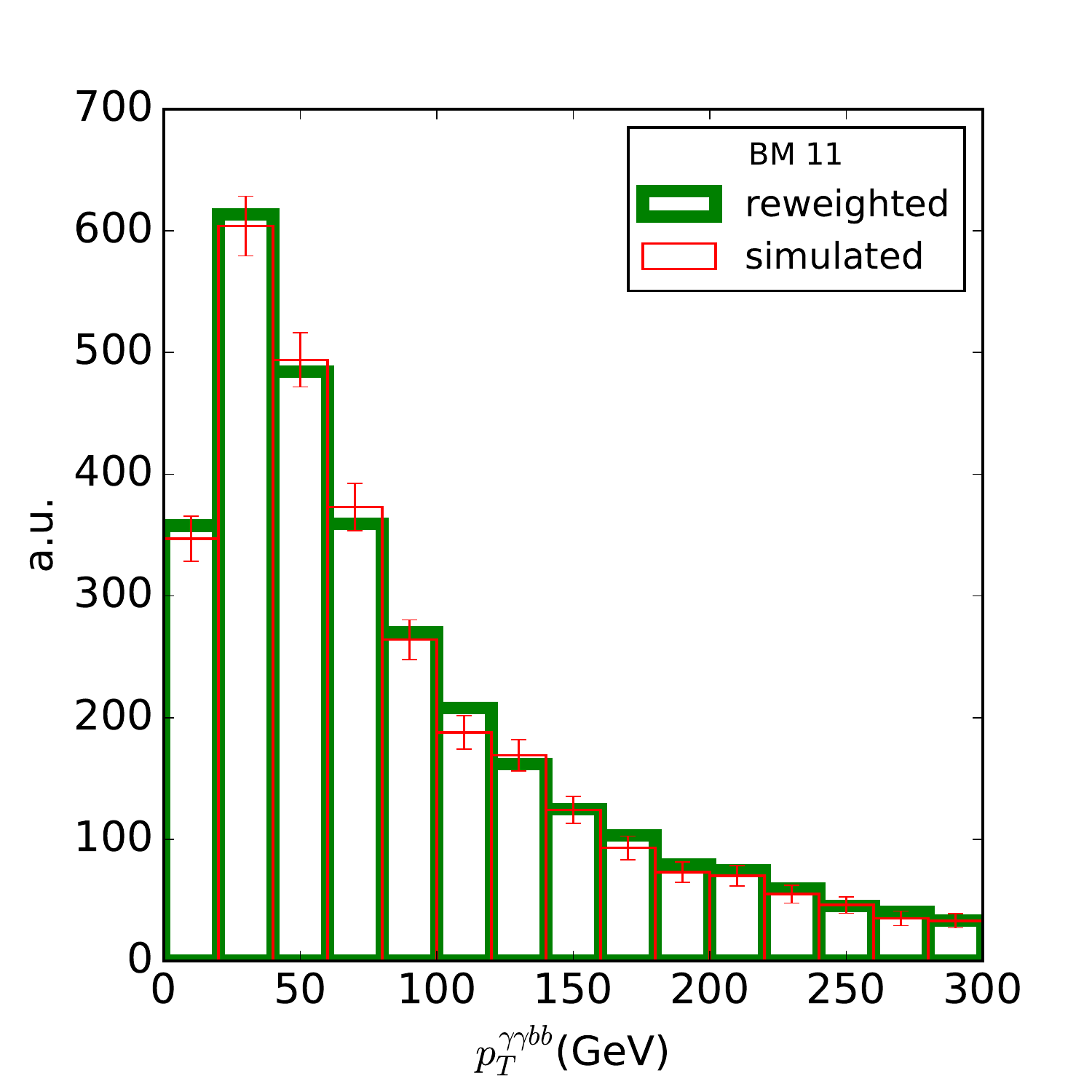}
  \includegraphics[width=0.32\textwidth]{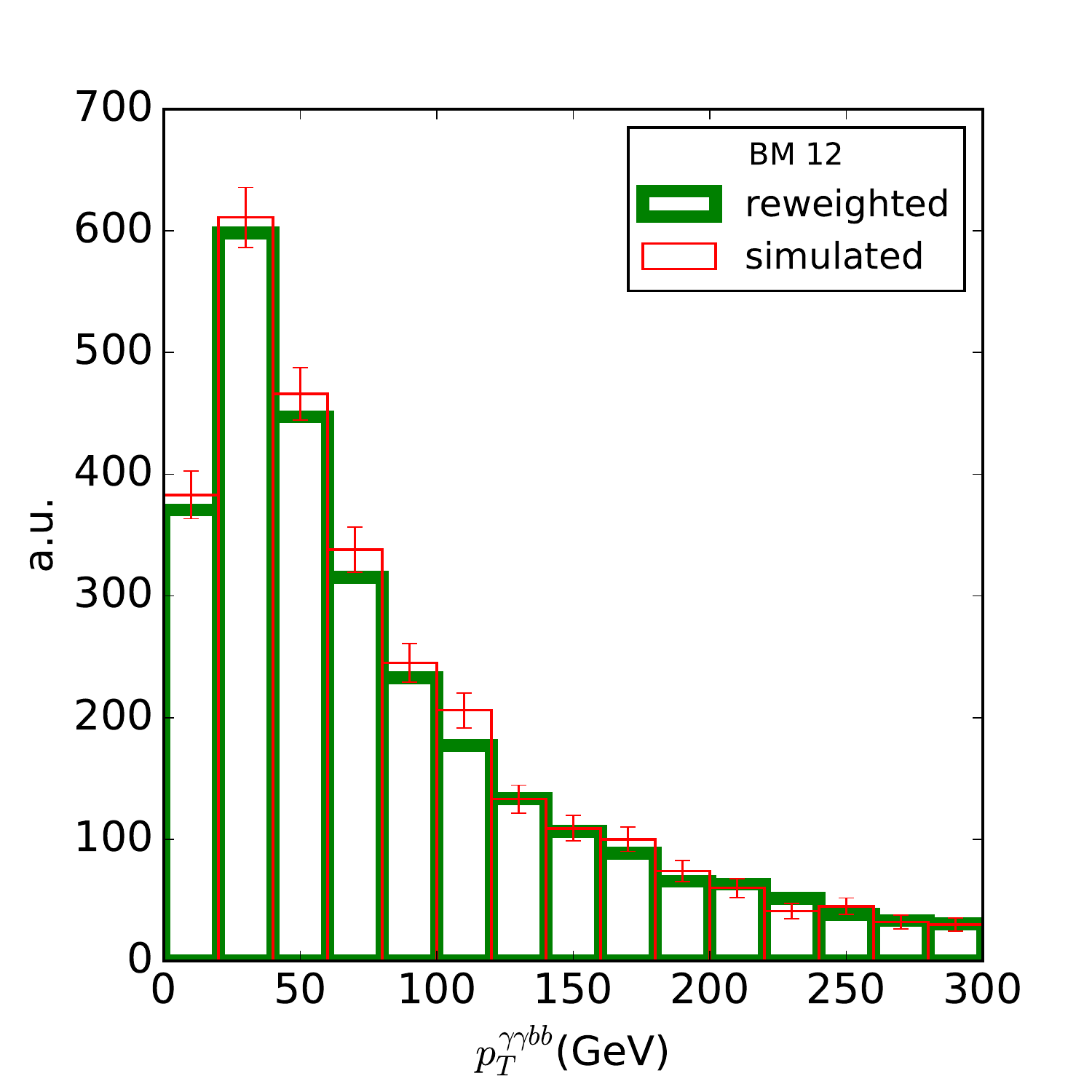}      %
  \caption{\footnotesize The reconstructed $p_T^{\gamma\gamma b\bar{b}}$ after ATLAS-like selection.  The histograms are normalized be signal efficiency times 100,000 events. 
    \label{fig:RecastBSM2}
  }
\end{figure}

\begin{figure}
  \centering
  \includegraphics[width=0.32\textwidth]{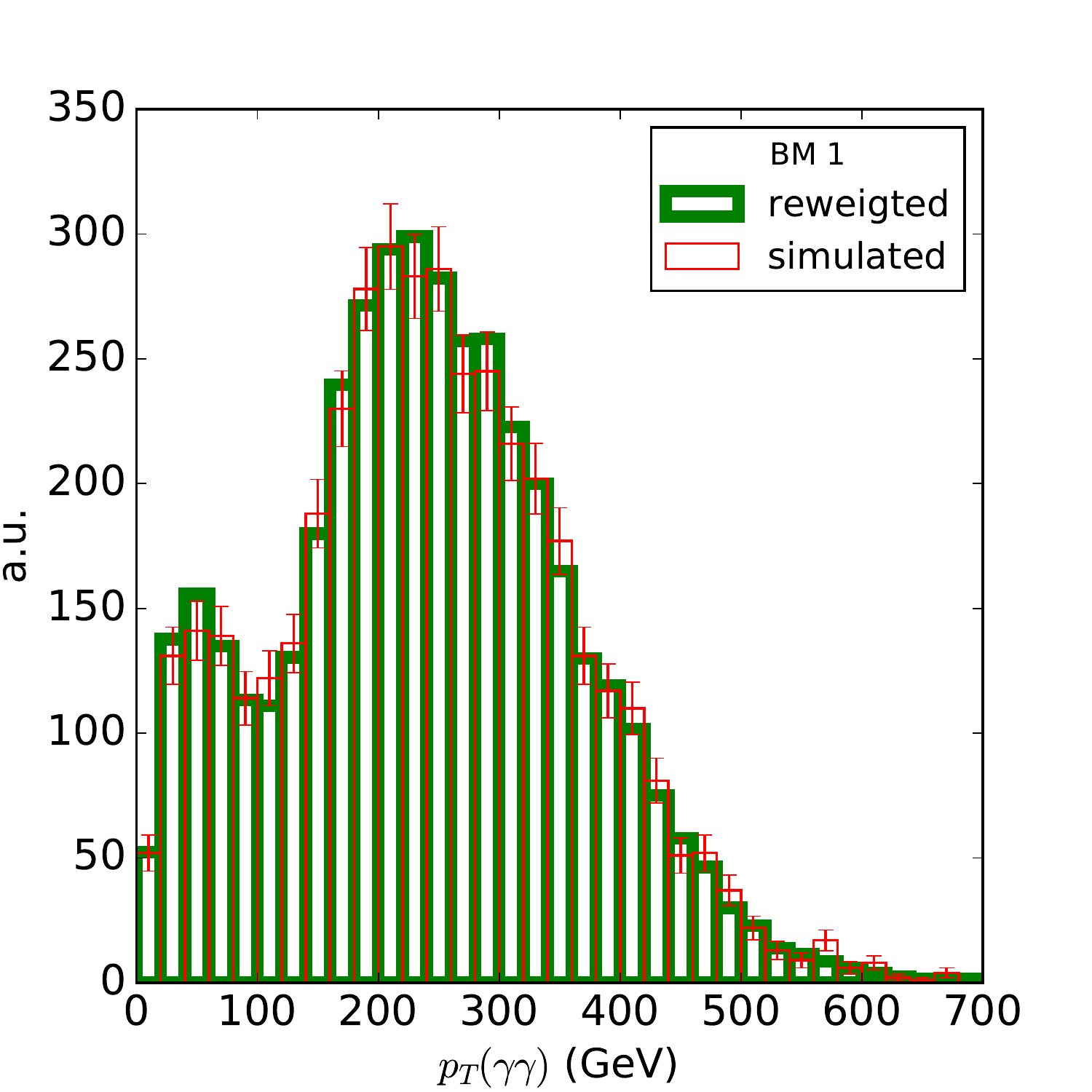}
  \includegraphics[width=0.32\textwidth]{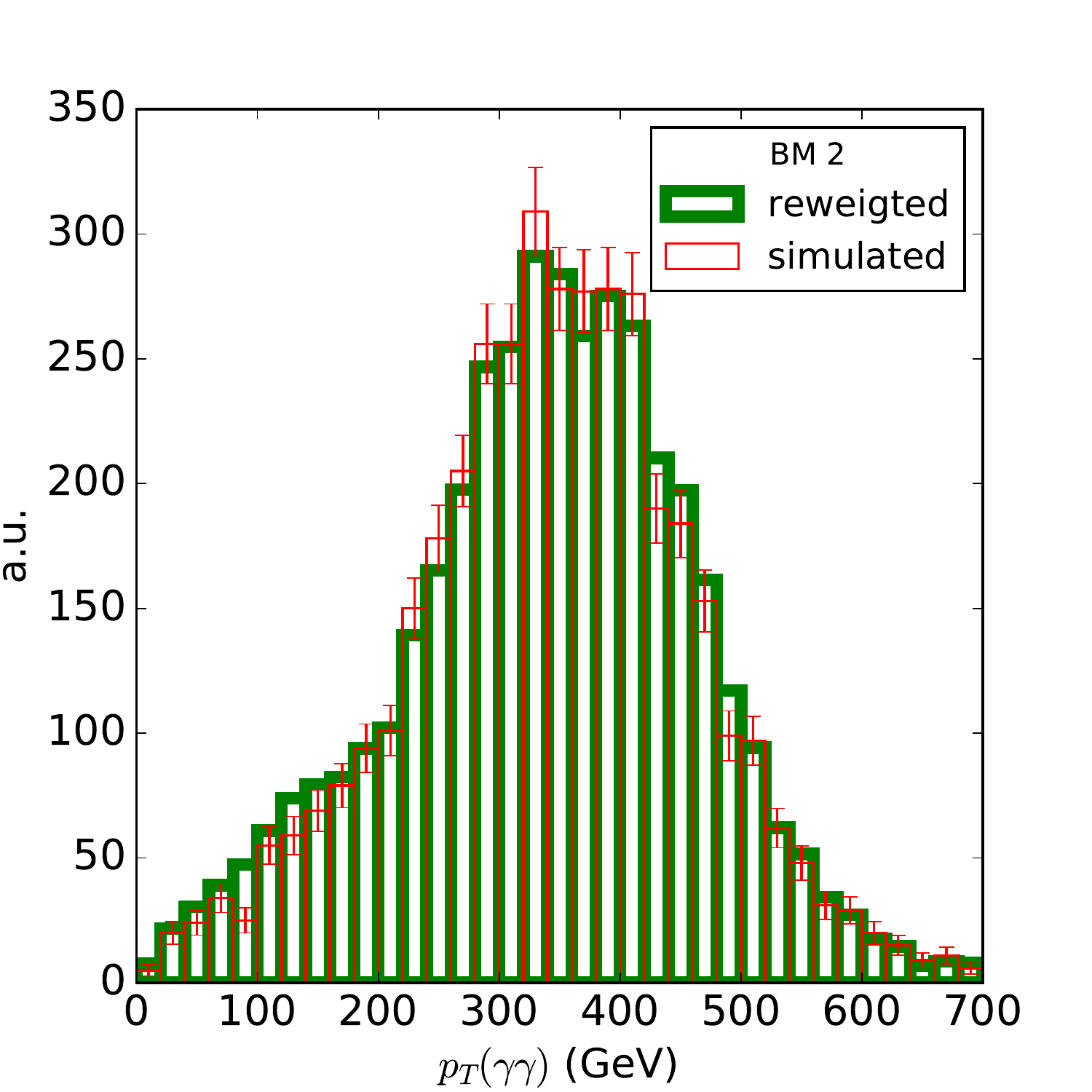}
  \includegraphics[width=0.32\textwidth]{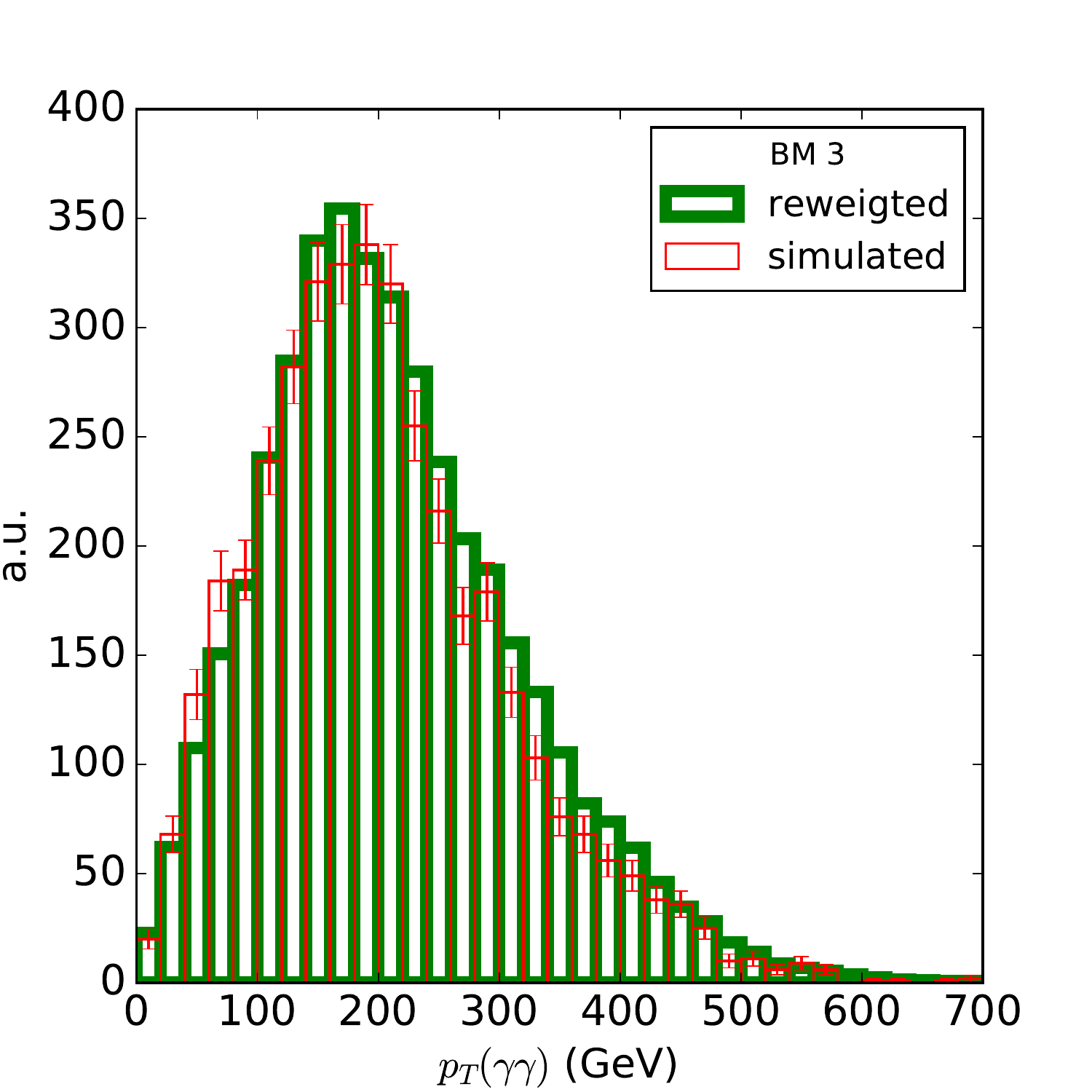}
  \includegraphics[width=0.32\textwidth]{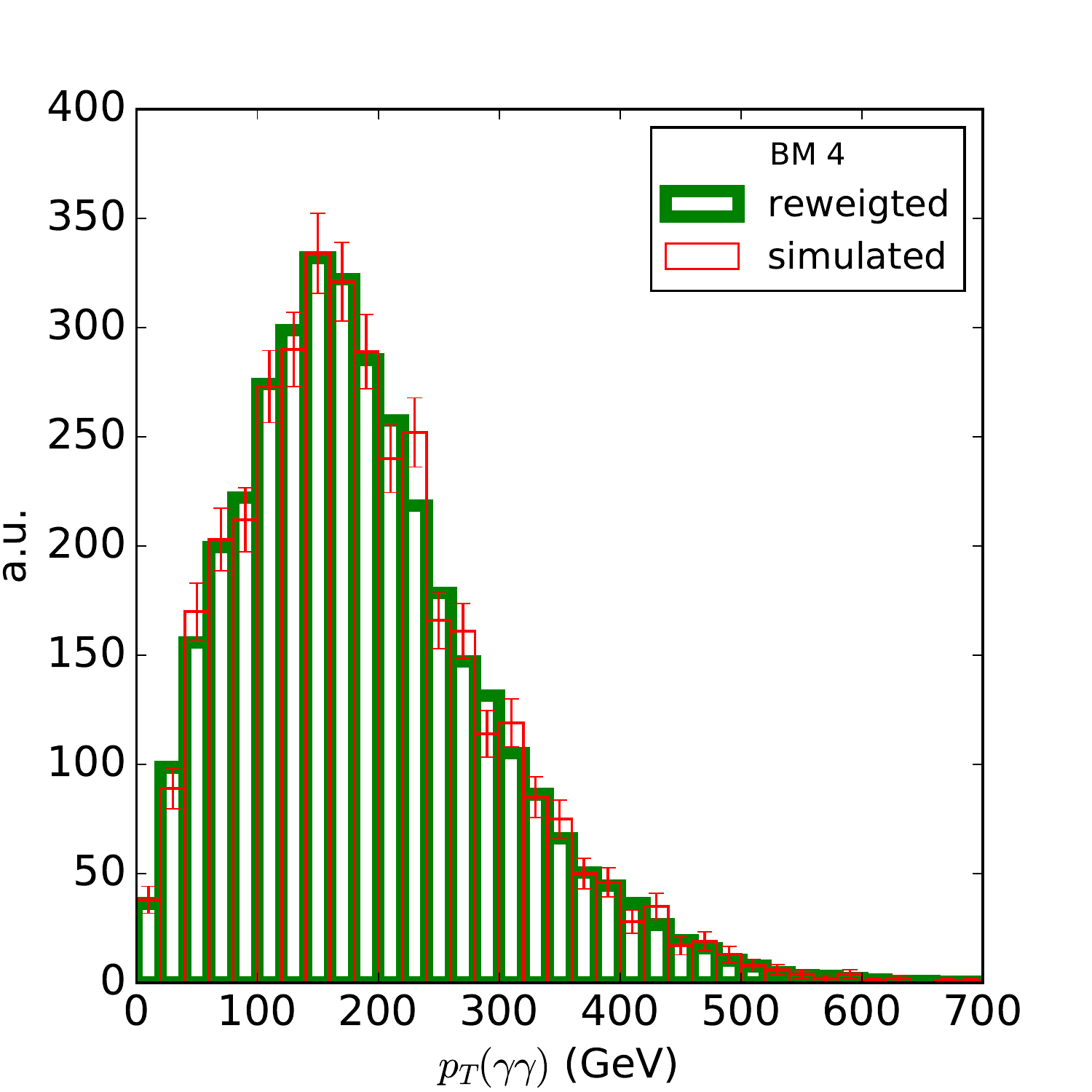}
  \includegraphics[width=0.32\textwidth]{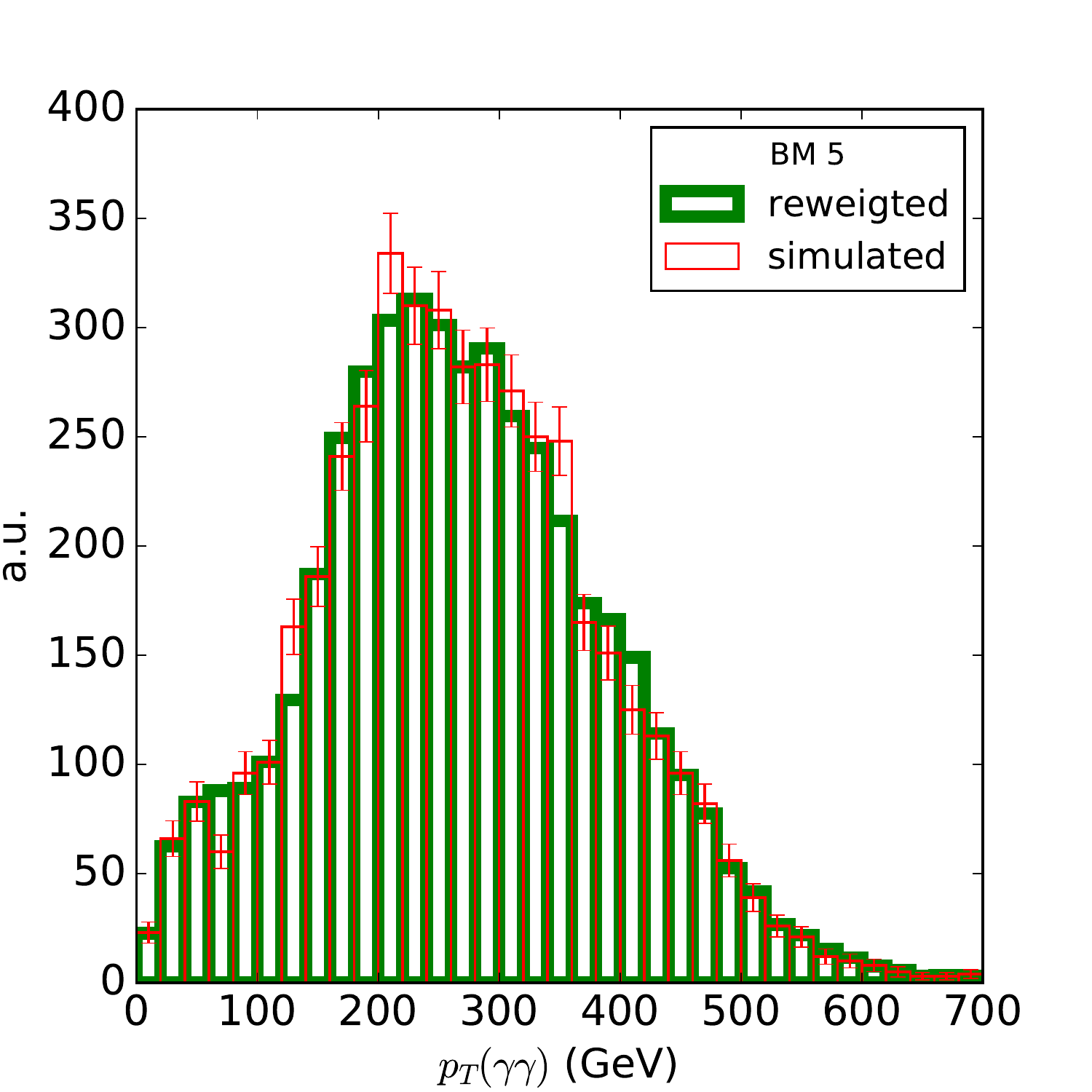}
  \includegraphics[width=0.32\textwidth]{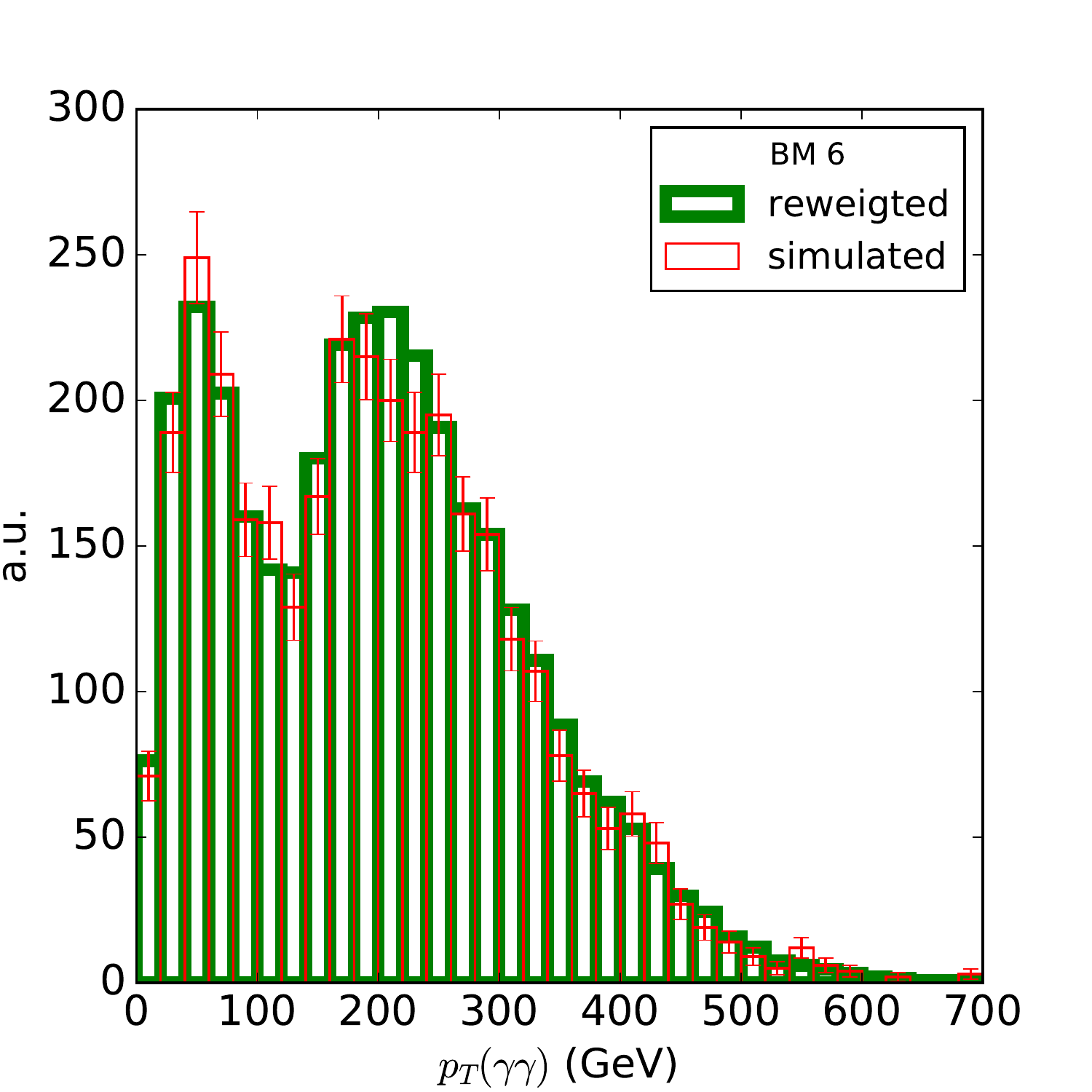}
  \includegraphics[width=0.32\textwidth]{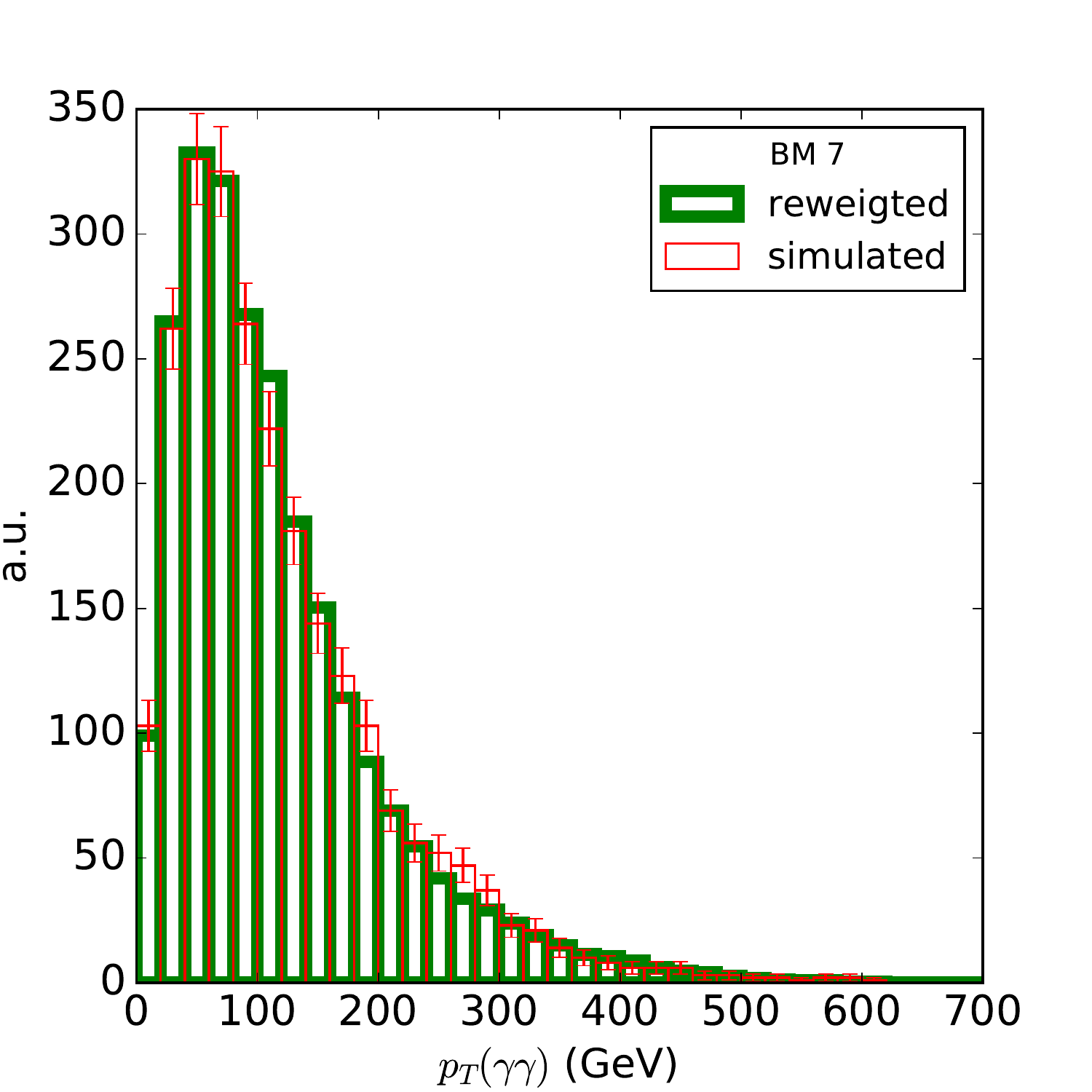}
  \includegraphics[width=0.32\textwidth]{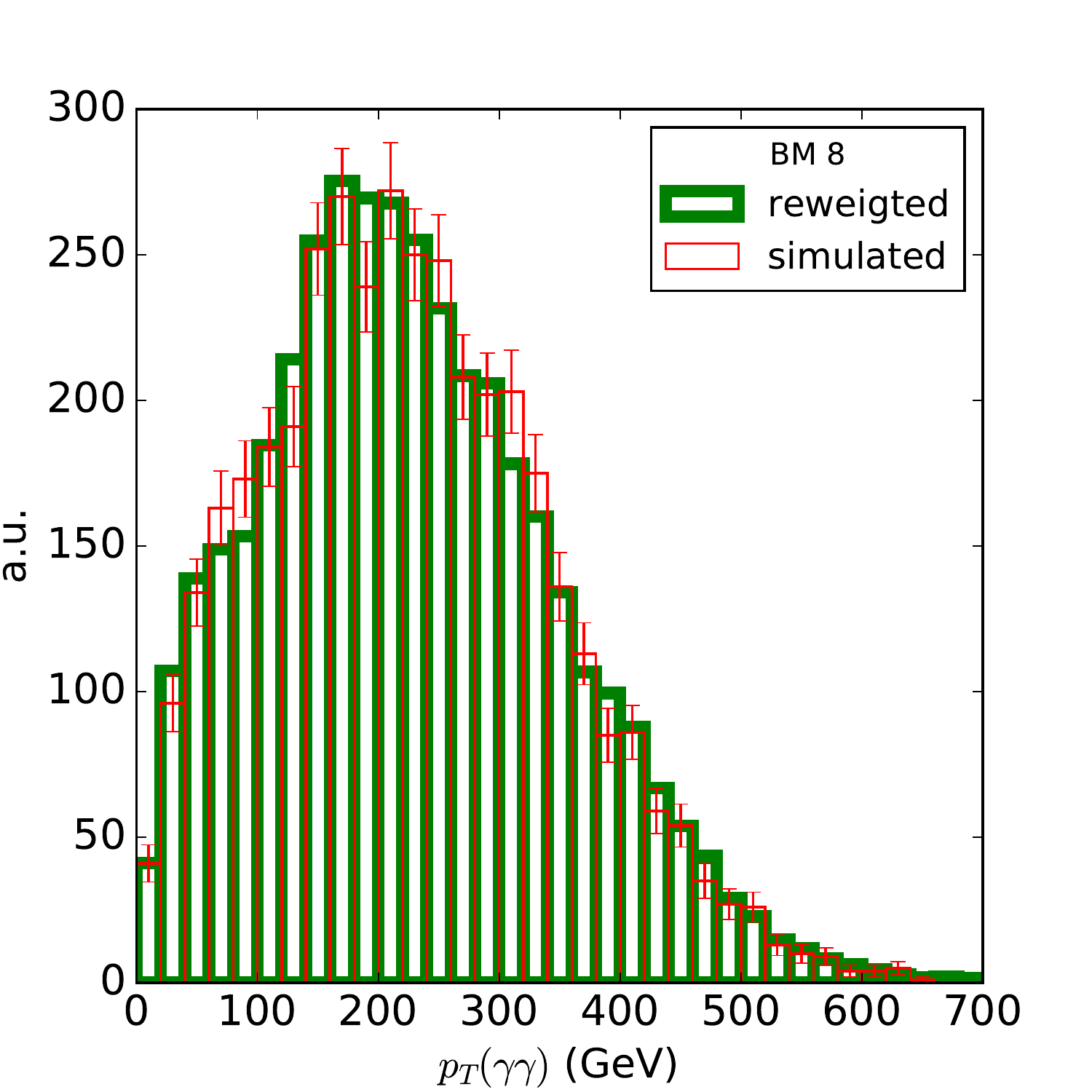}
  \includegraphics[width=0.32\textwidth]{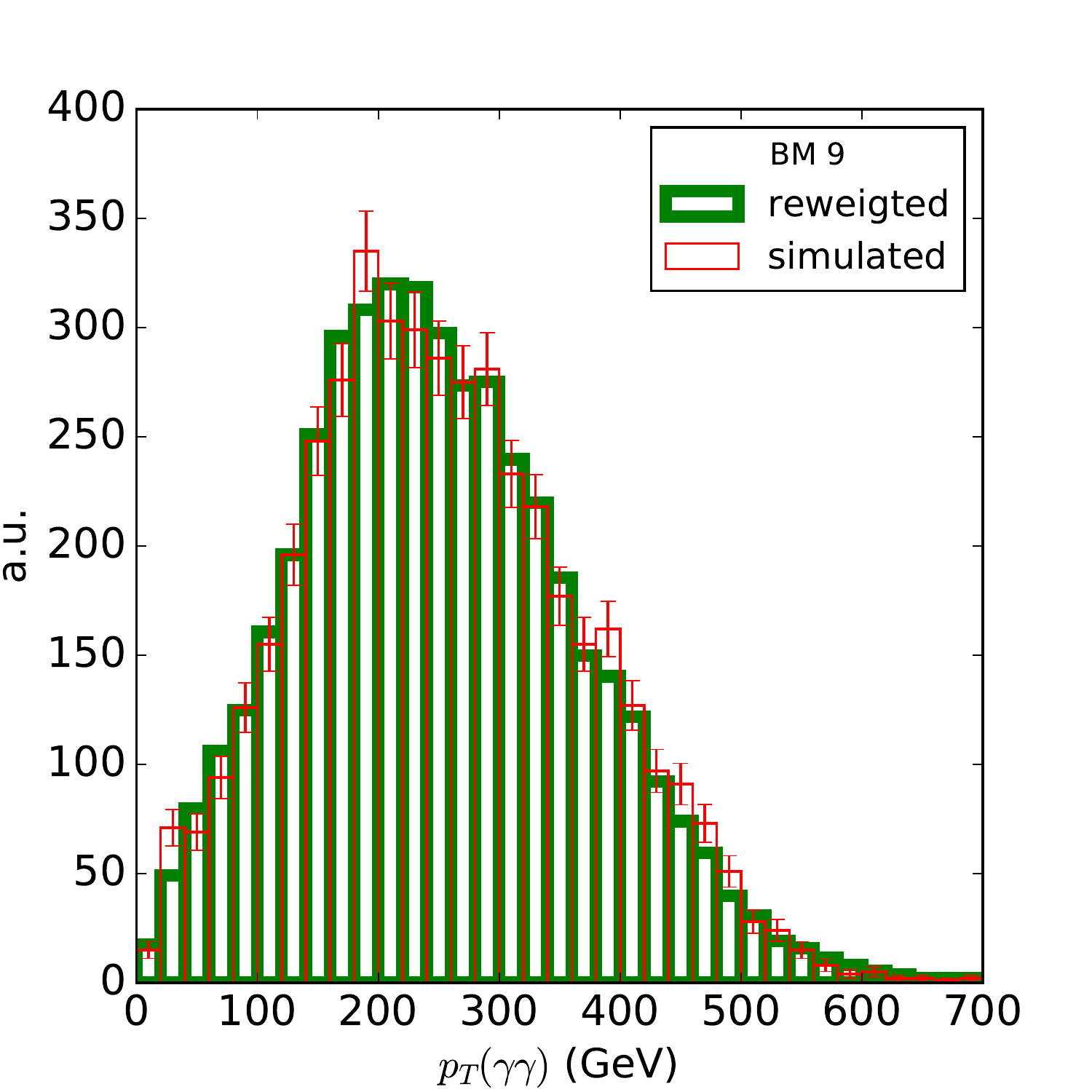}
  \includegraphics[width=0.32\textwidth]{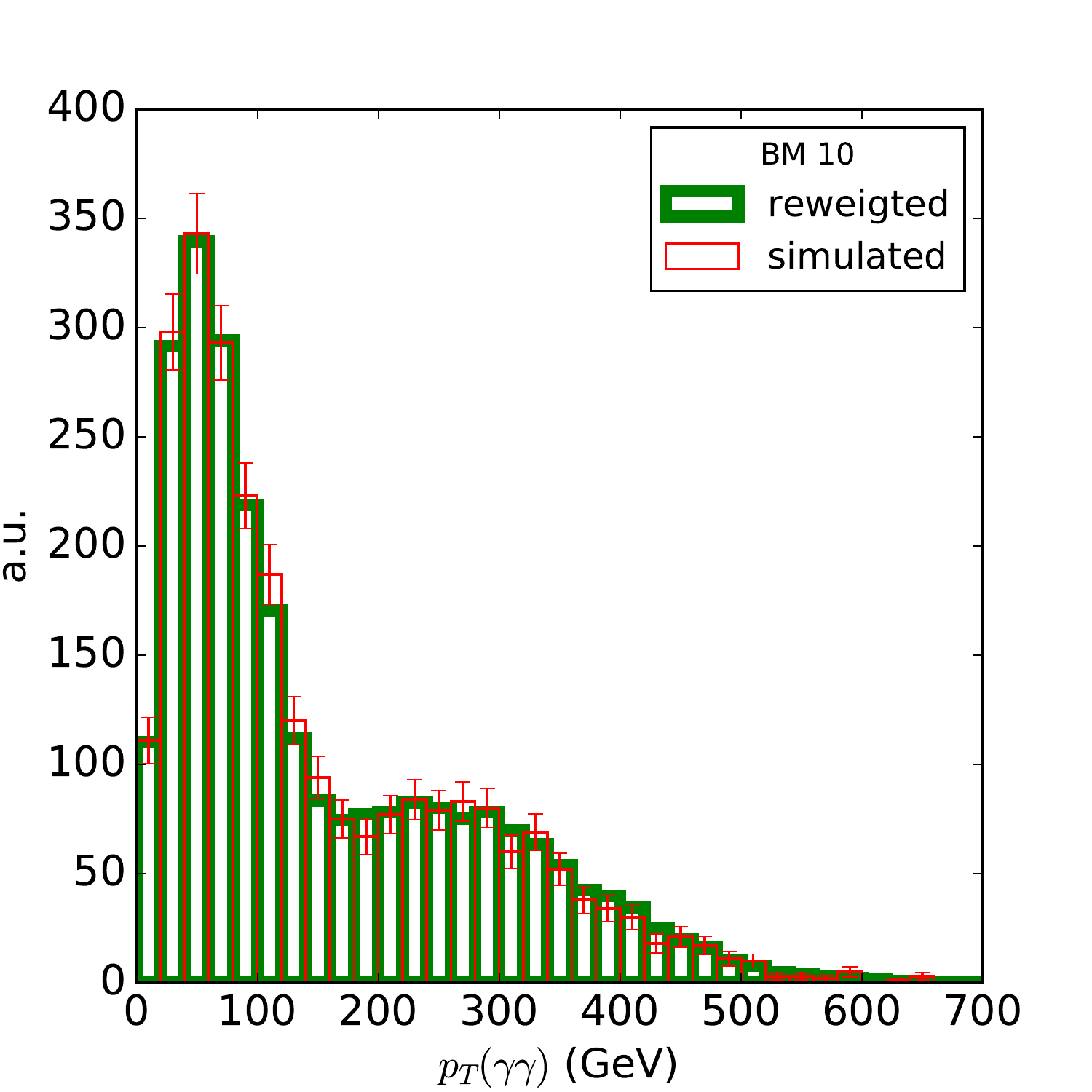}
  \includegraphics[width=0.32\textwidth]{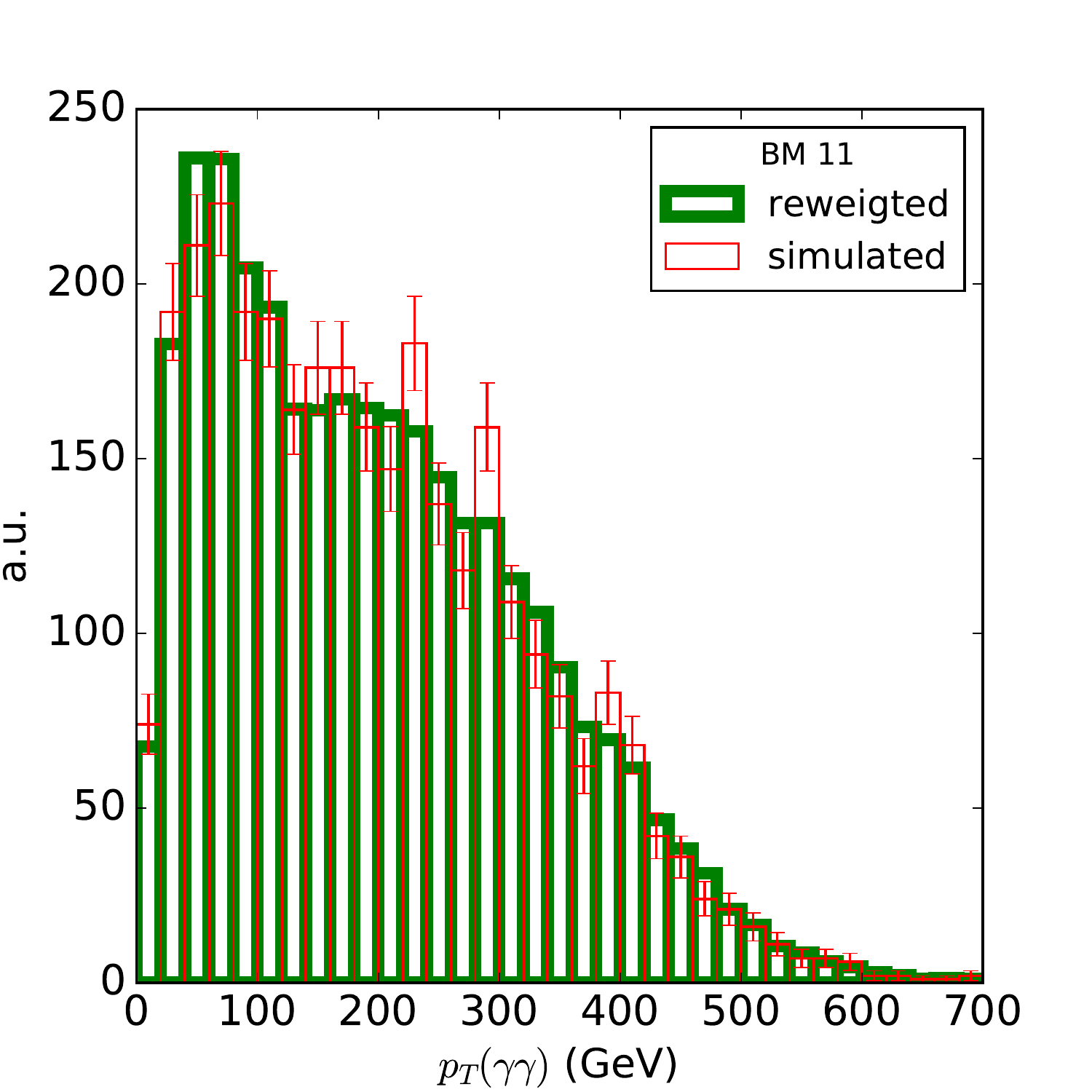}
  \includegraphics[width=0.32\textwidth]{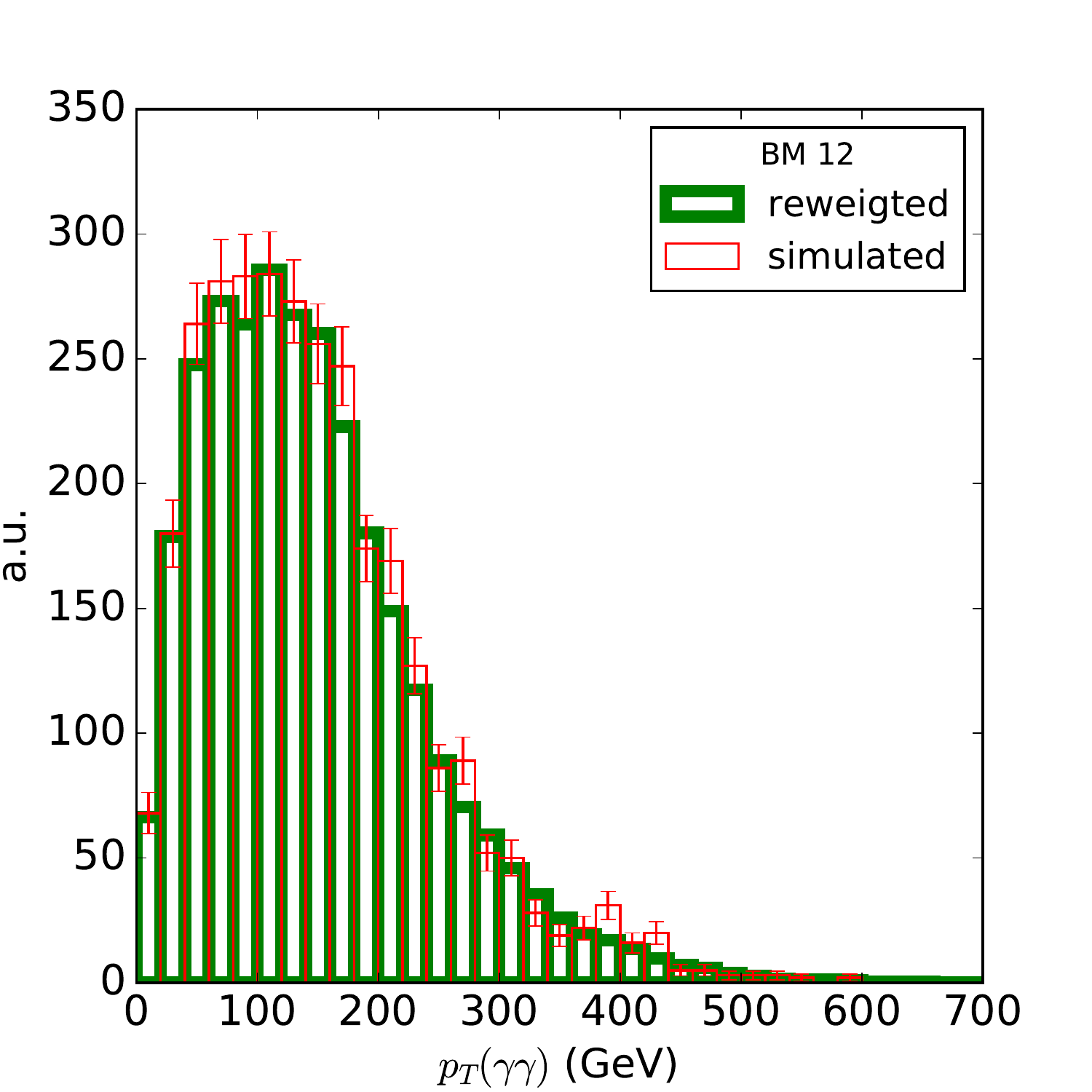}      %
  \caption{\footnotesize The reconstructed $p_T (\gamma\gamma)$ after ATLAS-like selection.  The histograms are normalized be signal efficiency times 100,000 events.  
    \label{fig:RecastBSM3}
  }
\end{figure}

\clearpage
\end{appendices}

\printbibliography
\end{document}